%% file: Dissertation.tex
\documentclass[12pt,oneside,openany,titlepage]{wmthesis}
\usepackage[intlimits]{amsmath}
\usepackage{amssymb}
\usepackage{amscd}
\usepackage{bm}
\DeclareMathAlphabet{\matheurm}{U}{eur}{m}{n}
\usepackage{color}
\usepackage[square,numbers,sort&compress]{natbib}
\usepackage{graphicx}
\graphicspath{{./figures}{.}}

\usepackage{epstopdf}

\usepackage{lscape}

\usepackage{enumerate}
\usepackage{dcolumn}
\usepackage{makeidx}
\usepackage{ifthen}
\input{Macros.tex}                      % Locate your macros here

\graphicspath{{./}{figures/}}

\newcommand{\comment}[1]{}

\def\Tr#1{\operatorname{Tr}\left[#1\right]}

\begin{document}                       % Begin the document
\normalsize                                % Return to Normal font size
\begin{spacing}{2.0}                    % Double Space the Text

% [asrich]  You should really use the templates provided at 
%
% http://www.wm.edu/as/graduate/Physical%20Standards.php
%
% for the Title page, the Abstract page, and the Approval page.
% For the rest, they only really care about the margins.

\frontmatter
\input{Title.tex}                          % Include the title page

\begin{abstract}
\thispagestyle{empty}                 % remove page number
\input{Abstract.tex}                    % Include the abstract
\end{abstract}

\input{Dedication.tex}                 % Include the dedication

\tableofcontents                         % Make a table of contents
\input{Acknowledgments.tex}     % Include the acknowledgements

% [asrich] I don't have any tables.  If you do, then uncomment the next line.
%\listoftables                             % Make a list of tables
\listoffigures                              % Make a list of figures

\mainmatter                              % Reset the Numbering to Arabic 1

% [asrich] I think this half title page is now optional.  
\makehalftitle\thispagestyle{empty}     % Make the half page title

% Input the main texts here: add more chapters if you want

\part{The Phase Space Theory of Resonant Mode Conversion}
\input{Chapter-Introduction.tex}

\input{Chapter-Theory.tex}

\input{Chapter-HigherOrder.tex}

\part{The Group Theoretical Foundations of Path Integrals}
\input{Chapter-PathInt-Intro.tex}

\input{Chapter-GroupTheory.tex}

\input{Chapter-SymbolTheory.tex}

\input{Chapter-PathIntegral.tex}

\input{Chapter-ModeConversion.tex}
\input{Chapter-Conclusion.tex}

\end{spacing}

% End of Main Thesis Text
%%%%%%
\newpage
%%%%%%
% Include the appendices here, if appropriate
\appendix

%\pdfbookmark[-1]{}{test}
%\input{Appendix.tex}           % Appendix A
\input{Appendix-Path-Integral.tex}         % Appendix B

%
%%%%%%
%%%%%%
% Put Bibliographic information last
%
\newpage
% TEMPORARY HACK for natbib package user ONLY: Remove this line if
\addcontentsline{toc}{chapter}{Bibliography}
\bibliography{refs}                         % Use a separate copy of my citation database
\bibliographystyle{apsrev}              % Indicate the sort by citation order style
% (find this file in REVTeX 4.0 package, download via publish.aps.org)
%
%%%%%%
%%%%%%
\input{Vita.tex}                        % Include the vita
%
%%%%%%
%%%%%%
% If you are indexing include the index format
%\begin{spacing}{1.0}
%\printindex                            % Print the Index
%\end{spacing}
%%%%%%
\end{document}

%% file: Macros.tex
\newcommand\eq[1]                              % single unlabeled equation
{
\begin{align*}
#1
\end{align*}
}

\newcommand\eql[2] % single labeled equation
{
\begin{equation}\label{#1}
\begin{split}
#2
\end{split}
\end{equation}
}

\newcommand\eqnl[2]        % single labeled equation (maunya sih, centered
{%                     equation array dg satu label saja (apakah bisa???))
\begin{equation}\label{#1}
#2
\end{equation}
}

\newcommand\eqsl[1]                            % multiply labeled equation
{
\begin{align}
#1
\end{align}
}

\newcommand\eqssl[2]                      % multiply labeled SUB-equations
{
\begin{subequations}\label{#1}
\begin{align}
#2
\end{align}
\end{subequations}
}

\newcommand\xcaption[4][] {
    \ifthenelse{\equal{#1}{}}
               {\caption[#3]{\label{#2}#3 #4}}
               {\caption[#1]{\label{#2}#3 #4}}
}
\providecommand\eqref[1] {(\ref{#1})}  %if your stylesheet doesn't have it

\definecolor{Red}{rgb}{0.8,0.2,0.2}%red
\definecolor{Green}{rgb}{0.08,0.60,0.08}%green
\definecolor{Blue}{rgb}{0.3,0.3,1.0}%blue
\definecolor{dkBlue}{rgb}{0.1,0.1,0.5}%dark blue
\definecolor{Black}{rgb}{0,0,0}
\definecolor{Gold}{rgb}{0.60,0.60,0.08}%gold-like color
\definecolor{Remarks}{rgb}{1,0.3,0.3}%red
\definecolor{Extra}{rgb}{0.2,0.2,1}%blue

\newcommand\COMMENTED[1] {}

%% file: Title.tex
%%%%%%%%%%%%%%%%%%%%%%%%%%%%%%%%%%%%%%%%%%%%%%%%%%%%%%%%%%%%%%%%%%%%%%%%%%
%
% Ph.D. dissertation manuscript
% Title Page
%
% Andrew Stephen Richardson (Fall 2007)
% College of William and Mary
% Department of Physics
% Prof. Eugene Tracy, advisor
%
% Based on Paul King and Andrew Norman's template (modified by Wirawan Purwanto)
%
%%%%%%%%%%%%%%%%%%%%%%%%%%%%%%%%%%%%%%%%%%%%%%%%%%%%%%%%%%%%%%%%%%%%%%%%%%

%%%%%%%%%%%%%%%%%%%%%%%%%%%%%%%%%%%%%%%%%%%%%%%%%%%%%%%%%%%%%%%%%%%%%%%%%%
\title{Topics in mode conversion theory and the group theoretical foundations of path integrals}
% Set the subtitle if you need to. Comment the following line if you don't
% want any subtitle (most of the students DON'T WANT subtitle):
%\subtitle{The New Way to Understand Physics}

\author{Andrew Stephen Richardson}
\authorhome{Fairfax, Virginia}
\authordegA{Master of Science, College of William and Mary, 2004}
\authordegB{Bachelor of Science, George Mason University, 2002}

% Committee members
% The first one must be your own advisor (the word "Chair" will be added
% automatically).  cmtyAUMI should be your advisor, name only, for the UMI page
\cmtyA{Professor Eugene Tracy, Physics}
\cmtyAUMI{Eugene Tracy}
% Sort the name of the rest of the committee members by their last name
% The external member must be put last. If he/she is from W&M, then
% the institution name "College of William and Mary" is not needed, the
% department name is sufficient.
\cmtyB{\parbox{3in}{\renewcommand{\baselinestretch}{1.0}\small
                    \center Professor William Cooke, Physics\\
The College of William and Mary}}
\cmtyC{\parbox{3in}{\renewcommand{\baselinestretch}{1.0}\small
                    \center Professor John Delos, Physics\\
The College of William and Mary}}
\cmtyD{\parbox{3in}{\renewcommand{\baselinestretch}{1.0}\small
                    \center Professor Dennis Manos, Physics\\
The College of William and Mary}}
\cmtyE{\parbox{3in}{\renewcommand{\baselinestretch}{1.0}\small
                    \center Professor Nahum Zobin, Mathematics\\
The College of William and Mary}}

\department{Department of Physics}

%%%%%%%%%%%%%%%%%%%%%%%%%%%%%%%%%%%%%%%%%%%%%%%%%%%%%%%%%%%%%%%%%%%%%%%%%%
\maketitle[January 2008]
\makeapproval[Approved by the Committee, November 2007]
%\makeapproval[Draft Version, \monthyear]  % Just remove this argument for
%%                                          % the final version.
%%%%%%%%%%%%%%%%%%%%%%%%%%%%%%%%%%%%%%%%%%%%%%%%%%%%%%%%%%%%%%%%%%%%%%%%%%

%% file: Abstract.tex
%%%%%%%%%%%%%%%%%%%%%%%%%%%%%%%%%%%%%%%%%%%%%%%%%%%%%%%%%%%%%%%%
%
% Ph.D. dissertation manuscript
% Dissertation abstract
%
% Andrew Stephen Richardson (Fall 2007)
% College of William and Mary
% Department of Physics
% Prof. Eugene Tracy, advisor
%
% $Id: Diss-vita.tex,v 1.4 2005/01/14 22:32:00 wirawan Exp $
%
% Based on Paul King and Andrew Norman's template (modified)
%
% This is to be used for BOTH the normal abstract and the UMI abstract.
%
% UMI abstract is printed as the last page of this document, but DO NOT
% include that page as the final dissertation.
% Include that separately from the three copies of the dissertation to
% the Office of the Dean of Graduate Studies.
%
%%%%%%%%%%%%%%%%%%%%%%%%%%%%%%%%%%%%%%%%%%%%%%%%%%%%%%%%%%%%%%%%

This dissertation reports research about the phase space perspective for solving wave problems, with particular emphasis on the phenomenon of mode conversion in multicomponent wave systems, and the mathematics which underlie the phase space perspective.  
Part I of this dissertation gives a review of the phase space theory of resonant mode conversion.  We describe how the WKB approximation is related to geometrical structures in phase space, and how in particular ray-tracing algorithms can be used to construct the WKB solution.  We further review how to analyze the phenomena of mode conversion from the phase space perspective.  By making an expansion of the dispersion matrix about the mode conversion point in phase space, a local coupled wave equation is obtained.  The solution of this local problem then provides a way to asymptotically match the WKB solutions on either side of the mode conversion region.  We describe this theory in the context of a pedagogical example; that of a pair of coupled harmonic oscillators undergoing resonant conversion.  Lastly, we present new higher order corrections to the local solution for the mode conversion problem which allow better asymptotic matching to the WKB solutions.  
The phase space tools used in Part I rely on the Weyl symbol calculus, which gives a way to relate operators to functions on phase space.  In Part II of this dissertation, we explore the mathematical foundations of the theory of symbols.  We first review the theory of representations of groups, and the concept of a group Fourier transform.  The Fourier transform for commutative groups gives the ordinary transform, while the Fourier transform for non-commutative groups relates operators to functions on the group.  We go on to present the group theoretical formulation of symbols, as developed recently by Zobin.  This defines the symbol of an operator in terms of a double Fourier transform on a non-commutative group.  We give examples of this new type of symbol, using the discrete Heisenberg-Weyl group to construct the symbol of a matrix.  We then go on to show how the path integral arises when calculating the symbol of a function of an operator.  We also show how the phase space and configuration space path integrals arise when considering reductions of the regular representation of the Heisenberg-Weyl group to the primary representations and irreducible representations, respectively.  We also show how the path integral can be interpreted as a Fourier transform on the space of measures, opening up the possibility of using tools from statistical mechanics (such as maximum entropy techniques) to analyze the path integral.  We conclude with a survey of ideas for future research and describe several potential applications of this group theoretical perspective to problems in mode conversion.

%% file: Dedication.tex
%%%%%%%%%%%%%%%%%%%%%%%%%%%%%%%%%%%%%%%%%%%%%%%%%%%%%%%%%%%%%%%%%%%%%%%%%%
%
% Ph.D. dissertation manuscript
% Dedication Page
%
% Andrew Stephen Richardson (Fall 2007)
% College of William and Mary
% Department of Physics
% Prof. Eugene Tracy, advisor
%
% Based on Paul King and Andrew Norman's template (modified by Wirawan Purwanto)
%
%%%%%%%%%%%%%%%%%%%%%%%%%%%%%%%%%%%%%%%%%%%%%%%%%%%%%%%%%%%%%%%%%%%%%%%%%%

% See the Thesis writer guide. There are two possible formats of the
% dedication page. The uncommented one below is the long one. If you want
% the shorter format, uncomment the part below

%\chapter*{DEDICATION}%%% DEDICATION
% FORMAT 1:
%\begin{spacing}{1.0}
%
% November 2003 guide requires this be single-spaced
%
%I dedicate this work to my parents, Andy and Diane Richardson.  They gave me a love for learning, and encouraged me to pursue that love.
%\end{spacing}

% FORMAT 2:
%\comment{% Begin commented text (remove if you want this one)
\newpage
\chapter*{}%%% DEDICATION (alternate, shorter fromat)
\begin{spacing}{1.25}
\vskip 7cm
\begin{center}
To my parents, Andy and Diane Richardson.  They gave me a love for learning, and encouraged me to pursue that love.
\end{center}
\end{spacing}
%}% End commented text (remove if you want this one)

%% file: Acknowledgments.tex
%%%%%%%%%%%%%%%%%%%%%%%%%%%%%%%%%%%%%%%%%%%%%%%%%%%%%%%%%%%%%%%%%%%%%%%%%%
%
% Ph.D. dissertation manuscript
% Acknowledgements
%
% Andrew Stephen Richardson (Fall 2007)
% College of William and Mary
% Department of Physics
% Prof. Eugene Tracy, advisor
%
%
% Based on Paul King and Andrew Norman's template (modified by Wirawan Purwanto)
%
%%%%%%%%%%%%%%%%%%%%%%%%%%%%%%%%%%%%%%%%%%%%%%%%%%%%%%%%%%%%%%%%%%%%%%%%%%

\chapter*{ACKNOWLEDGMENTS}
\addcontentsline{toc}{chapter}{Acknolwedgements}
\begin{spacing}{1.0}
% November 2003 guide requires that this be single-spaced

I would like to express my gratitude to Dr.~Eugene Tracy for his guidance, encouragement, and friendship to me over the course of my time as a graduate student at William and Mary.  I will always regard the time spent discussing physics at his whiteboard as the most fruitful and intellectually stimulating experience of my education.   

I would also like to thank Dr.~Nahum Zobin for the mathematical training, the helpful discussions, and his enthusiasm for mathematical physics.  His mathematical leadership, along with the continual support of Dr.~Dennis Manos, made the math-physics seminar series an invaluable component of my education, opening a door for me into the vast and fascinating world of mathematical physics.

A special thank you goes to the members of my defense committee: Drs.~Eugene Tracy, William Cooke, John Delos, Dennis Manos, and Nahum Zobin.  Their perseverance in reading through this dissertation, as well as their helpful comments and corrections, are much appreciated.

Finally I would like to thank my friends and family for all of their support during my time here.  Thank you to Dan, for being a great office-mate, and to Hendra, for the friendship, fellowship, and food.  To my family, Dad, Mom, Glenn and Lidia, Kelly, Travis, and Adriana, thank you for the many ways you have supported me during my graduate studies, and especially for all the faithful prayers.

% Add an empty line (I'm lazy to find out the best way)
{~}

This work was supported by the National Science Foundation.

\end{spacing}

%% file: Chapter-Introduction.tex
%%%%%%%%%%%%%%%%%%%%%%%%%%%%%%%%%%%%%%%%%%%%%%%%%%%%%%%%%%%%%%%%%%%%%%%%%%
%
% Ph.D. dissertation manuscript
% Chapter 1: Introduction
%
% Andrew Stephen Richardson (Fall 2007)
% College of William and Mary
% Department of Physics
% Prof. Eugene Tracy, advisor
%
% Based on Paul King and Andrew Norman's template (modified by Wirawan Purwanto)
%
%%%%%%%%%%%%%%%%%%%%%%%%%%%%%%%%%%%%%%%%%%%%%%%%%%%%%%%%%%%%%%%%%%%%%%%%%%

\chapter{Introduction}

\section{Introduction to Part I}

In the first part of this dissertation, we describe the phase space theory of mode conversion.  Chapter \ref{chp:PhaseSpaceIntro} introduces the phase space point of view for solving generic wave equations.  We will then review the WKB method for construction of approximate solutions, and show how WKB methods are connected to phase-space ray-tracing algorithms.  Then, in Chapter \ref{chp:coupled_osc}, the example of two coupled oscillators will be given.  If the natural frequencies of the oscillators are time dependant, and cross at some time, then this problem can be recast into a form which is mathematically very similar to a mode conversion problem.  The complete description of the solution of this coupled oscillator problem provides a pedagogical introduction to the phase space techniques used in the theory of mode conversion.

In Chapter \ref{chp:higher_order}, we will apply these tools to a standard avoided crossing mode conversion.  Usually, the solution of such a problem involves the linearization of the dispersion function about the mode conversion point.  A local solution is then constructed so that incoming and outgoing WKB solutions can be asymptotically matched, allowing the problem to be treated as a sort of ``ray splitting''.  We will analyze the effects of the next higher order terms, and show how to construct a new local solution which takes these quadratic terms into account.  By including the effects of the quadratic order terms, the region in which the matching can be performed will be enlarged substantially.

\section{Introduction to Part II}

The phase space theory described in Part I of this dissertation makes extensive use of the theory of symbols of operators.  In Part II, we describe the mathematical foundations of the theory of symbols.  Because these mathematical foundations rely heavily on the theory of representations of groups, we will first give a review of group theory in Chapter \ref{chp:GroupTheory}.  The example of the Heisenberg-Weyl group will be reviewed in detail, and we will show how the relationship between phase space and configuration space arises from the reduction of the regular representation of this group.  

Then, in Chapter \ref{chp:Symbols}, we will describe how the symbol of an operator can be calculated by a double Fourier transform.  The operator is first embedded into a section of the dual bundle, which is like an operator-valued ``function'' on the set of irreducible representations of a non-commutative group.  The non-commutative Fourier transform is then applied to convert this section into a function on the group, where the group is considered as a set.  Then, using a commutative group structure on this same set, we perform another Fourier transform.  The result is an ordinary complex-valued function, which is the Zobin symbol of the operator we started with.

Using this definition of the symbol, as developed recently by Zobin, we will calculate the symbol of a function of an operator in Chapter \ref{chp:PathIntegral}.  In particular, we will consider the exponential of an operator, defined using a power series.  The symbol of the $N^{th}$ power of an operator will be computed by using the $N^{th}$ star product of the symbol of the operator.  In the limit of large $N$, this repeated application of the star product can be written as a path integral, where the paths live in the dual to the commutative group.  We then illustrate this general theory of path integrals by explicitly calculating the repeated star product for the discrete Heisenberg-Weyl group.  This group has irreducible representations which are $n\times n$ matrices, so the calculation presented can be used to calculate functions of a matrix.  In particular, the exponential of a matrix will lead to a discrete ``path integral'', which by grouping similar paths can be written in terms of a multiplicity function.  This will lead to the consideration of the connections between path integrals and statistical mechanics.  In addition, the multiplicity function gives rise to a probability distribution, or measure, on the space of all possible paths, which will allow the path integral to be interpreted as a Fourier transform on the space of measures.  For the continuous Heisenberg-Weyl group this becomes an infinite-dimensional Fourier transform.  Considering the path integral for the Heisenberg-Weyl group will also show which aspect of group theory underlies the connection between the phase-space path integral and the configuration-space path integral.  Specifically, the reduction of the regular representation to the primary representations leads to consideration of functions on phase space.  This leads to the phase-space path integral.  Further reduction of the primary representations to irreducible representations involves functions on configuration space.  This reduces the phase-space path integral to the configuration-space path integral.

The new group-theoretical approach to path integrals which will be described in Chapter \ref{chp:PathIntegral} has many potential applications.  In Chapter \ref{chp:ModeConversion} we will outline several ways which this new point of view could be applied to mode conversion theory.  This chapter will point out several avenues of current and future research.  

We will first see how consideration of the star product formulation of the path integral may lead to using the diagonals of the dispersion matrix as ray hamiltonians for constructing WKB solutions to vector wave problems.  This requires the definition of a new ``normal form'' for the dispersion matrix, where the symbols of the diagonal elements Poisson commute with the symbols of the off-diagonal elements.  In addition to simplifying ray-tracing algorithms, this could also help provide physical insight for vector wave problems with non-standard mode conversion geometries.  

A second avenue of research based on our new group theory perspective involves calculating a ``double'' symbol for vector wave problems.  The wave operator for vector wave problems can be written as a matrix of (pseudo)differential operators.  The ordinary symbol of each element of the matrix can be calculated, giving the dispersion matrix as a function of phase space.  However, as we will see in Chapter \ref{chp:Symbols}, it is possible to calculate the ``symbol'' of a matrix.  So we can calculate the discrete ``symbol'' of the dispersion matrix at each point in phase space.  This will give a new ``double symbol'' of the wave operator, which is a function of several discrete variables in addition to being a function of the phase space variables.

As a final potential topic of further research, we will discuss in Chapter \ref{chp:ModeConversion} the possibility of using an averaged Wigner function to model the effects of turbulent plasma fluctuations on mode conversion.  This approach is based on the connection between the Wigner function and the density matrix in quantum mechanics.  In quantum mechanics, mixed state density matrices are used to model the decoherence of a quantum state due to interaction with the environment.  This decoherence makes the Wigner function for a mixed state look like a more classical probability distribution on phase space.  We will suggest that a ``mixed state'' Wigner function could be used to describe mode-converting waves in a turbulent plasma.  Preliminary calculations suggest that this would make the Wigner function appear more ``classical'', with amplitude confined to regions near the dispersion curves for the various wave modes.

\chapter{A Brief Introduction to Phase Space Methods for Wave Equations}
\label{chp:PhaseSpaceIntro}

%\section{Wave Equations and Their Solutions}

From simple waves on a string, to the wave-functions of quantum particles, to the tensor waves of relativistic gravity wave theory, waves and their behavior influence almost all physical systems.  In this work, the phenomenon of mode conversion is studied from the phase space perspective.  Mode conversion occurs in multicomponent wave problems, when variations in the background medium allow local resonances between different types of waves.  The resonances lead to energy transfer between the modes, which must be considered in order to understand the behavior of the system.  The phase space for a wave problem is composed of physical space and ``wave-vector'' space, just as classical phase space is composed of the position and momentum of a particle.

A generic wave equation can be written as an operator acting on a field.  The operator can vary in space and time, and the field may be multicomponent:
\begin{equation}\label{eq:wave}
\mathbf{\hat D}(\mathbf x,-i\nabla_{\mathbf x}; t, i \partial_t) \cdot \mathbf \Psi(\mathbf x,t) = 0.
\end{equation}
This equation, together with fitting to appropriate initial and boundary conditions, defines a generic wave problem.  Analysis of such a general wave equation is difficult, and in many practical situations, a general solution may not be useful.  Many methods have been developed which give approximate solutions for cases of physical interest.  Fourier methods are powerful for time-independent wave problems in a uniform medium.  Spectral techniques are powerful for analyzing stable configurations of quantum systems.  The asymptotic theory of Wentzel, Kramers, and Brillouin connects solutions of wave problems (involving partial differential equations) to the solution of classical mechanics problems (which involve ordinary differential equations).  This so-called WKB theory, or WKB method, brings together ideas from Hamiltonian mechanics, asymptotic wave theory, the mathematical theory of operators, and the theory of path integrals.  Further extension of WKB theory in the 1960's by Maslov and Arnold showed that underlying the theory are geometric structures in wave phase space.  These objects, called Lagrange manifolds, determine the phase fronts and amplitude variation of asymptotic solutions of the wave equation.  While originally developed for solving the scalar Schr\"odinger equation of quantum mechanics, the WKB method can also be employed to solve multicomponent problems, such as those which arise when solving Maxwell's equations in a plasma.  There is a large literature which shows that a ray-tracing approach based on the WKB method can be used to obtain asymptotic solutions to mode conversion problems, both in one spatial dimension \cite{citeulike:703463,PhysRevLett.58.1392,citeulike:784668,Kaufman:1999qy}, and in multiple spatial dimensions\cite{littlejohn:149,citeulike:472573,metaplectic_formulation,0741-3335-49-1-004,tracy:082102}.

In the following chapters, I first describe and illustrate this phase space theory with an analogy of a resonance crossing problem from classical mechanics; the problem of two coupled harmonic oscillators with time-varying natural frequencies (Chapter \ref{chp:coupled_osc}).  This serves to introduce basic notation and concepts we will need in later chapters.  Because of the introductory nature of this material, we will use a heuristic, one-dimensional, approach to this problem, and ignore the effects of caustics.  In addition, we will assume that all operators are self-adjoint.  The intuition gained from this pedagogic introduction is then applied in Chapter \ref{chp:higher_order} to the phase space theory of waves, and is used to study higher order corrections to solution of the mode conversion problem.

\section{Phase space theory for scalar waves\label{sec:scalar_phase_space_thy}}

The idea of using a ``phase space'' to analyze a wave problem can be introduced using the example of a one dimensional, dispersive, scalar wave equation.  Take, for example, the Schr\"odinger equation for a free particle in one dimension,
\begin{equation}
i\hbar \frac{\partial \psi}{\partial t} + \frac{\hbar^2}{2 m} \frac{\partial^2 \psi}{\partial x^2} =0.
\end{equation}
This equation has the form of an operator acting on a function, and the solutions are eigenfunctions with eigenvalue zero.  The wave operator is
\begin{equation}\label{eq:sch_op}
\hat D = i\hbar \frac{\partial}{\partial t} + \frac{\hbar^2}{2 m} \frac{\partial^2}{\partial x^2}.
\end{equation}
Solutions of the Schr\"odinger equation are plane waves, characterized by a frequency $\omega$ and wavenumber $k$.  Acting on the plane wave with the wave operator gives a relationship between frequency and wavenumber which must be satisfied in order for the wave to be a solution:
\begin{equation}\label{eq:sch_disp_function}
\hat D e^{i(kx - \omega t)} = \left(\hbar \omega - \frac{\hbar^2 k^2}{2 m} \right) e^{i(kx - \omega t)}.
\end{equation}
So, solutions must have 
\begin{equation}\label{eq:disp_zero}
D(k,\omega) \equiv \hbar \omega - \frac{\hbar^2 k^2}{2 m} =0,
\end{equation}
where the function $D(k,\omega)$ is called the {\em dispersion function}.  Solving this equation for $\omega$ gives us the {\em dispersion relation}
\begin{equation}
\omega(k) = \frac{\hbar k^2}{2 m}.
\end{equation}
Comparing the operator (\ref{eq:sch_op}) with the dispersion function (\ref{eq:disp_zero}) shows that they are equivalent if we can make a correspondence between operators and functions of $(\omega, k)$:
\begin{equation}\label{eq:operator_corresp}
\omega \rightarrow i\partial_t, \quad k \rightarrow -i\partial_x . 
\end{equation}
This correspondence is familiar from the semiclassical theory of quantum mechanics, but also forms the basis for analyzing generic wave equations.  While studying the equations of quantum mechanics, Wigner, Weyl and others developed a way to relate a generic operator with a function in such a way that the function contains all information about the original operator.  In the context of this theory, the dispersion function (\ref{eq:disp_zero}) is called the {\em symbol} of the wave operator (\ref{eq:sch_op}).  A generic dispersive wave equation, even one with non-constant coefficients, can be analyzed using symbol theory.  If we write the generic equation as
\begin{equation}\label{eq:generic_wave_eqn}
\hat D(x, -i\partial_x; t, i\partial_t) \psi(x,t) = 0,
\end{equation}
then the symbol of the wave operator is the dispersion function
\begin{equation}
D(x,k; t,\omega).
\end{equation}
As in Equation (\ref{eq:disp_zero}) above, eikonal solutions of the wave equation must have the relationship between, $x$, $k$, $\omega$, and $t$ given by
\begin{equation}
D(x,k; t,\omega)=0.
\end{equation}
This defines a three dimensional {\em dispersion surface} in the four-dimensional wave phase space $(x,k,t,\omega)$.  

In many problems that we will consider, the wave equation may be time independent.  This means that the $t$ dependence drops out of the dispersion function, and we can use the Fourier transform to separate frequency components.  The frequency is then a parameter, and the dispersion function is a function on the two-dimensional phase space $(x,k)$.  The dispersion surface is then a one-dimensional curve in phase space given by finding roots of 
\begin{equation}
D(x,k;\omega) = 0.
\end{equation}
If we solve for $\omega$, this gives $\omega = \Omega(x,k)$, which is the local dispersion relation.

If the dispersion function is only smoothly dependent on the phase space variables, then it might make sense to try using a Taylor's series expansion of the dispersion function to get an approximate wave equation.  We will always assume that this expansion will be possible.  Such an expansion will in fact give an approximate equation for the envelope of the wave when the carrier is given by the wavenumber and frequency about which we are expanding:
\begin{equation}\label{eq:disp_taylor_series}
D(x,k;\omega) = D(x_0,k_0;\omega_0) + \frac{\partial D}{\partial x} (x-x_0) + \frac{\partial D}{\partial k} (k-k_0) + \frac{\partial D}{\partial \omega} (\omega-\omega_0) + \ldots ,
\end{equation}
where all the derivatives are evaluated at $x_0, k_0,$ and $\omega_0$.
If we consider problems in a uniform medium, the $x$ derivative of the dispersion function is zero.  The rule for associating operators and functions (Weyl prescription) automatically symmetrizes terms involving products of $x$ and $k$ (for more details see Chapter \ref{chp:Symbols}).  In this case, since we have a uniform, time independent medium, the terms $\partial_k D$ and $\partial_\omega D$ are independent of $x$ and $t$, respectively, so we do not need to symmetrize.  
Using the correspondence in Equation (\ref{eq:operator_corresp}) to turn this back into an operator gives the wave equation
\begin{equation}
\left(  D(x_0,k_0;\omega_0) +\frac{\partial D}{\partial k} (-i\partial_x-k_0) + \frac{\partial D}{\partial \omega} (i\partial_t-\omega_0) \right) \psi(x,t) \approx 0.
\end{equation}
Setting the first term to zero gives the dispersion function for the carrier of our solution.  Insert this back to obtain an equation for the envelope.
\begin{eqnarray}
0 &=& \left(   \frac{\partial D}{\partial k} (-i\partial_x-k_0) + \frac{\partial D}{\partial \omega} (i\partial_t-\omega_0) \right) e^{i(k_0 x - \omega_0 t)}\tilde\psi(x,t) \\
&=&i e^{i(k_0 x - \omega_0 t)} \left(   -\left(\frac{\partial D}{\partial k}\right) \partial_x + \left(\frac{\partial D}{\partial \omega}\right) \partial_t \right)\tilde\psi(x,t) \label{eq:advection_equation}
\end{eqnarray}
This now looks like an advection equation, with the characteristic velocity
\begin{equation}
v_g = -\frac{\partial D / \partial k}{\partial D /\partial \omega}.
\end{equation}
This is an alternative form of the familiar group velocity $\frac{\partial \Omega}{\partial k}$, where $\Omega(k)$ is the dispersion relation.  We prefer to work with the dispersion function $D(x,k;\omega)$ rather than the dispersion relation because it is easier to interpret geometrically in higher dimensions.  The characteristic curves for Equation (\ref{eq:advection_equation}) are given in parametric form by \cite{Whitham:1974lr}
\begin{equation}\label{eq:ray_equations}
\frac{\partial t}{\partial \sigma} = \frac{\partial D}{\partial \omega}, \quad 
\frac{\partial x}{\partial \sigma} = -\frac{\partial D}{\partial k}.
\end{equation}
If we now consider the case where the background is weakly nonuniform, then the solution will locally look like a plane wave, but with the wavenumber slowly changing.  This means that we can approximate the solution as
\begin{equation}\label{eq:scalar_eikonal}
\psi(x,t) = A(x,t) e^{i (\Theta(x) - \omega_0 t)},
\end{equation}
where the envelope function $A(x,t)$ varies slowly compared to the phase.
Acting on this with $-i\partial_x$ brings down a derivative of the phase,
\begin{equation}
-i\partial_x \psi(x,t) = -i\partial_x A(x) e^{i (\Theta(x) - \omega_0 t)} = e^{i (\Theta(x) - \omega_0 t)} (-i\partial_x A(x) +A(x) \partial_x \Theta(x)  ).
\end{equation}
We can define the function
\begin{equation}
k(x) \equiv \partial_x \Theta(x),
\end{equation}
and then expand the dispersion function about $(x,k(x))$ to obtain
\begin{eqnarray}
0 &\approx&  \bigg(  D(x,k(x);\omega_0) + 
\overbrace{\frac{\partial D}{\partial k} (-i\partial_x-k(x)) + \frac{\partial D}{\partial \omega} (i\partial_t-\omega_0) }^{\text{symmetrize}}
\bigg) \psi(x,t) \\
&=& e^{i (\Theta(x) - \omega_0 t)}\bigg(  D(x,k(x);\omega_0) + \overbrace{\frac{\partial D}{\partial k} (-i\partial_x) + \frac{\partial D}{\partial \omega} (i\partial_t) }^{\text{symmetrize}}\bigg)A(x,t).  \label{eq:disp_expansion}
\end{eqnarray}
If the spatial gradient of $D$ is small, then, as before, we can write the derivative terms as a total derivative along some curve.  Assuming a time-independent medium, we can write
\begin{align}
\overbrace{\frac{\partial D}{\partial k} (-i\partial_x) + \frac{\partial D}{\partial \omega} (i\partial_t) }^{\text{symmetrize}} & = i(\partial_\omega D) \partial_t 
- \frac{i}{2}( \partial_x \partial_k D + 2 (\partial_k D)\partial_x  )  \\
&\approx i(\partial_\omega D) \partial_t -i (\partial_k D)\partial_x  .
\end{align}
Here we have assumed that $\partial_x \partial_k D$ is small compared to the other terms.  In the next section we will consider the effects of this term, but for now we have again obtained an advective derivative.  Equation (\ref{eq:ray_equations}) still gives the parametrization of the characteristic curves for this advective derivative.  We now need to ensure that the first term in Equation (\ref{eq:disp_expansion}) remains zero along the these characteristic curves:
\begin{eqnarray}
0 = \frac{d D}{d \sigma} &=& \frac{\partial D}{\partial x} \frac{\partial x}{\partial \sigma} +\frac{\partial D}{\partial k} \frac{\partial k}{\partial \sigma} \\
&=& \frac{\partial D}{\partial x} \left(-\frac{\partial D}{\partial k} \right) +\frac{\partial D}{\partial k} \frac{\partial k}{\partial \sigma}.
\end{eqnarray}
This implies that the function $k(x(\sigma))$ can be found as a function of $\sigma$ by solving
\begin{equation}\label{eq:ray_equations_k}
\frac{\partial k}{\partial \sigma} = \frac{\partial D}{\partial x}.
\end{equation}
Notice that this equation, together with the derivative of $x$ from Equation (\ref{eq:ray_equations}), has the form of Hamilton's equations, where $-D(x,k;\omega_0)$ acts as the hamiltonian function.  Solving for $(x,k)$ from these equations, with a family of initial conditions, gives us a surface in phase space called the {\em Lagrange manifold}.  Integrating $k$ along this surface allows us to find the function $\Theta(x)$, which can be used to find the phase of the solution as given in Equation (\ref{eq:scalar_eikonal}).  For a $2n$-dimensional phase space, the dispersion surface defined by $D(x,k)=0$ is $(2n-1)$-dimensional.  Embedded in this dispersion surface is the Lagrange manifold, which is $n$-dimensional, since the dimension of the Lagrange manifold is equal to the dimension of configuration space, i.e., half the dimension of phase space.  Each ray is only one-dimensional, and so in general a family of rays is required in order to trace out the Lagrange manifold.  Notice that in problems with only one spatial dimension, these geometric structures collapse.  Phase space is now only two-dimensional, and the dispersion surface, Lagrange manifold, and rays are all one-dimensional, and lie on the same curve in phase space.

The minus sign appearing in the ray equations when using the dispersion function $D(x,k;\omega_0)$ as the ray hamiltonian may seem somewhat unusual.  This is because we are using the dispersion function $D$ to generate the rays, instead of using the dispersion relation $\omega(k)$ as the ray hamiltonian.   In order to see how the dispersion relation gives us a different sign, consider the far field solution of a time independent problem in a uniform medium.  The wave equation is
\begin{equation}
\hat D(-i\partial_x,i\partial_t) \psi(x,t) = 0.
\end{equation}
The Fourier transform of this gives the integral equation
\begin{equation}
\int dk\, e^{i(kx - \omega(k) t)} D(k,\omega(k)) \tilde \psi(k) = 0.
\end{equation}
This means that the dispersion relation comes from solving $D(k,\omega(k))=0$.  We can then analyze the Fourier integral which gives the general solution:
\begin{equation}
\psi(x,t) = \int  dk\, e^{i(kx - \omega(k) t)} \tilde \psi(k).
\end{equation}
We can find the group velocity as a function of $k$ by using the stationary phase approximation to estimate this integral.  Substitute $x=v_g t$ into the integral, where $v_g$ is a free parameter, and consider the limit $t\rightarrow \infty$.  For a choice of $v_g$, the stationary phase point of this integral is given by
\begin{equation}
0 = \frac{\partial}{\partial k}(kv_g -\omega(k)) = v_g -\frac{\partial \omega(k)}{\partial k}.
\end{equation}
This fixes $k$ given $v_g$, and implies that
\begin{equation}
\frac{\partial x}{\partial t} = v_g = \frac{\partial \omega}{\partial k}.
\end{equation}
Comparing this with Equation (\ref{eq:ray_equations}) shows the relative difference in sign.  When this is substituted into the local dispersion relation $\omega(x,k(x))$ to obtain the parametric equation for $k$, we also find the sign difference:
\begin{eqnarray}
0 = \frac{d \omega}{d t} &=& \frac{\partial \omega}{\partial x} \frac{d x}{d t} +\frac{\partial \omega}{\partial k} \frac{d k}{d t} \\
&=& \frac{\partial \omega}{\partial x} \left(\frac{\partial \omega}{\partial k} \right) +\frac{\partial \omega}{\partial k} \frac{d k}{d t} \\
&\implies& \frac{d k}{d t} =-\frac{\partial \omega}{\partial x}.
\end{eqnarray}
Here $\frac{d}{dt}$ is the total derivative following a ray.
Although using the dispersion relation gives a sign difference in the ray equations as compared to the equations obtain from the dispersion function, the resulting solutions are the same.  This is due to the fact that the important physical object is the Lagrange manifold in phase space, not the rays which were used to find it.  The ray parameter $\sigma$ has no real physical significance.  However, care must be taken when using the ray parameter, and when relating it to real time via Equation (\ref{eq:ray_equations}), so that sign errors are not introduced.

\section{Eikonal ansatz and the WKB approximation\label{sec:WKB_intro}}

The phase space picture introduced in the previous section provides the setting for the WKB method of constructing approximate solutions to wave equations such as (\ref{eq:generic_wave_eqn}).  The WKB method starts with a trial solution in Equation (\ref{eq:scalar_eikonal}), which is called an {\em eikonal} wave:
\begin{equation}
\psi(x,t) = A(x,t) e^{i (\Theta(x) - \omega_0 t)}.
\end{equation}
We are assuming that the real amplitude $A(x,t)$ varies slowly compared to the phase $\Theta(x)$.  
Since the phase is the only quantity that varies rapidly, the action of the derivative is approximately the same as multiplication by the derivative of the phase.  In this paper, when we refer to the WKB approximation, this is what we mean; that the solution is in eikonal form, and the derivative goes to multiplication by the gradient of the phase.  
\comment{%Other authors use ``WKB approximation'' to mean a slightly different thing, see for example \cite{PhysRevA.6.709,PhysRevA.6.720}.
}

We now examine the expansion (\ref{eq:disp_expansion}) of the dispersion function more closely.  Since we are expanding about $(x,k(x))$, the partial derivatives of the dispersion function may still depend on $x$.  This means that when we convert the symbol of the dispersion function into an operator, we need to symmetrize terms containing derivatives with respect to $x$.  In particular, we get
\begin{eqnarray}
(\partial_k D ) k & \rightarrow & \frac{1}{2}\left( (\partial_k D ) \hat k + \hat k (\partial_k D ) \right) \\
&=& -\frac{i}{2} \left( 2(\partial_k D ) \partial_x + (\partial_x \partial_k D) \right).
\end{eqnarray}
We can now insert this symmetrized term into the expansion in Equation (\ref{eq:disp_expansion}).  Using the fact that $D(x,k(x);\omega_0)=0$ and that $\partial_x\Theta = k(x)$, we can simplify the equation to
\begin{equation}
i (\partial_\omega D) \partial_t A - \frac{i}{2} (\partial_x\partial_k D) A -i (\partial_k D) \partial_x A = 0.
\end{equation}
We can now look for solutions whose only time dependence is in the phase $e^{-i\omega_0 t}$.  This means that we can take $\partial_t A = 0$, and the two remaining terms in the above equation can be written as a logarithmic derivative:
\begin{equation}
-\frac{1}{2} \partial_x \ln (\partial_k D) = \partial_x \ln A.
\end{equation}
Integrating this along a ray gives the {\em WKB amplitude}:
\begin{equation}\label{eq:WKB_amp}
A(x) = A_0 \left( \left.\partial_k D\right|_{k=k(x)} \right)^{-1/2},
\end{equation}
where $A_0$ is a constant of integration.  The {\em WKB phase} is obtained by integrating $k(x)$ along a ray:
\begin{equation}\label{eq:WKB_phase}
\Theta(x) = \int^x k(x') \,dx' = \int k(\sigma) \frac{dx}{d\sigma} \, d\sigma.
\end{equation}

\section{Raytracing}

The WKB solution presented in Section \ref{sec:WKB_intro} depends on being able to calculate the Lagrange manifold.  From this surface, we can find $k(x)$, which is then integrated to give the phase function $\Theta(x)$.  Raytracing is the primary method used to find the Lagrange manifold.  In this section, we give a review of the raytracing method for the case where the configuration space $x$ is one dimensional.

\begin{figure}
\begin{center}
\includegraphics[scale=0.9]{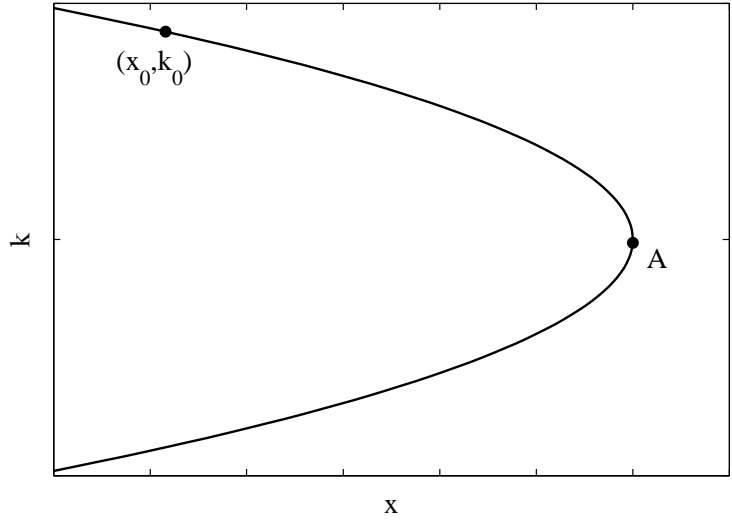}
\end{center}
\caption{\label{fig:raytracing}
Example of a dispersion surface in a two dimensional phase space.  In this case, one ray traces out the Lagrange manifold and the dispersion surface, starting from the point $(x_0,k_0)$.  At the turning point, A, the function $k(x)$ is multivalued, and the WKB amplitude diverges because $\dot x =0$.
}
\end{figure}

We start with an approximate solution near the point $x_0$ (see Figure \ref{fig:raytracing}).  The solution locally has the form
\begin{equation}
\psi(x,t) = A(x_0) e^{i(k_0 x - \omega_0 t)}.
\end{equation}
The wavenumber $k_0$ must satisfy the condition that the dispersion function is zero, $D(x_0,k_0;\omega_0)=0$.  In general, this equation may have several roots, which correspond to different wave modes, and so choosing $k_0$ determines which type of solution we will construct.

Now use the point in phase space $(x_0,k_0)$ as initial conditions for a ray.  The ray trajectory is determined by Equations (\ref{eq:ray_equations}) and (\ref{eq:ray_equations_k}).  These equations can be solved (numerically in most cases) to give the curve $(x(\sigma),k(\sigma))$.  Since phase space is only two dimensional in this calculation, this one ray traces out the Lagrange manifold, whose dimension is half that of phase space.  Also, the dispersion surface has dimension one less than the dimension of phase space, and so this one ray also traces out the dispersion surface.  In general, the dispersion surface is embedded in phase space, the Lagrange manifold is contained in the dispersion surface, and the ray is a curve in the Lagrange manifold.

Now that we have the solution $(x(\sigma),k(\sigma))$, we can use it to construct the WKB solution.  As in Equation (\ref{eq:WKB_phase}), the phase is determined by integrating $k(x)$.  Equation (\ref{eq:WKB_amp}) gives the amplitude, where the constant of integration is set by our initial conditions.  Putting these together gives the WKB solution
\begin{equation}
\psi(x,t) = A(x_0) \sqrt{\frac{|\partial_k D|_{(x_0,k_0)} }{|\partial_k D|_{(x,k(x))}} } \exp\left(i{\int_{x_0}^x k(x') \, dx'}\right).
\end{equation}
If the ray is parametrized such that $x(0) = x_0$, and $x(\sigma) = x$, then this equation can be written in terms of the ray parameter as
\begin{equation}\label{eq:WKB_solution_from_rays}
\psi(x,t) = A(x(0)) \sqrt{\frac{|\dot x(0)|}{|\dot x(\sigma)|} } \exp\left(i{\int_{0}^\sigma k(\sigma') \dot x(\sigma') \, d\sigma'}\right),
\end{equation}
with 
\begin{equation}
\dot x(\sigma) \equiv \frac{dx}{d\sigma}.
\end{equation}
From this form of the equation, it is evident that the WKB amplitude will diverge if $\dot x (\sigma) \rightarrow 0$.  Such points are called caustics, and occur, for example, at turning points.

\section{WKB Method and Raytracing for Vector Waves}

The WKB method for solving multicomponent wave problems is modeled on the asymptotic methods described in the previous sections for scalar waves.  Multicomponent wave problems arise in many different situations.  For example, waves in a fluid or gas can be written in terms of the (multicomponent) velocity field.  Depending on the types of waves being considered, we may also need to consider additional scalar fields such as the fluid density or temperature.  For plasmas, the electric and magnetic vector fields also need to be considered.  Gathering all these components together would in general give us a wave equation in the form of an $N\times N$ matrix operator acting on the $N$ components of the field.  For example, a two component wave equation can be written as
\begin{align}
\hat{\bf D} \cdot {\bm \psi}(x,t)=
\left(
\begin{array}{cc}
 \hat D_{11}  &  \hat D_{12}  \\
\hat D_{21}   &   \hat D_{22}
\end{array}
\right)
\left(
\begin{array}{c}
  \psi_1(x,t)   \\
  \psi_2(x,t) 
\end{array}
\right) =0,
\end{align}
where $\hat{\bf D}$ is the wave operator (a $2\times 2$ matrix of operators), and ${\bm \psi}(x,t)$ is the field (a 2-component vector).

We can now introduce a vector form of the eikonal wave, and include a slowly varying polarization vector ${\mathbf e}(x)$:
\begin{equation}\label{eq:vector_eikonal}
{\bm \psi}(x,t) = \mathbf e(x) A(x) e^{i(\Theta(x) - \omega_0 t)}.
\end{equation}
The wave operator for vector problems is a matrix of operators, and its symbol is called the dispersion matrix.  For the two-component case $N=2$, the dispersion matrix is a $2\times 2$ matrix:
\begin{align}
{\bf D}(x,k;\omega_0) = \left(
\begin{array}{cc}
 D_{11}(x,k;\omega_0)  &  D_{12}(x,k;\omega_0)  \\
 D_{21}(x,k;\omega_0)   &  D_{22}(x,k;\omega_0)
\end{array}
\right) .
\end{align}
This matrix-valued function on phase space has local eigenvectors and eigenvalues which are also functions of phase space:
\begin{align}
{\bf D}(x,k;\omega_0) \cdot {\bf e}_\alpha(x,k;\omega_0) = D_\alpha(x,k;\omega_0) {\bf e}_\alpha(x,k;\omega_0), \quad \alpha = 1,2,\ldots N.
\end{align}
The dispersion matrix can be expanded about a point just as the dispersion function was in Equation (\ref{eq:disp_taylor_series}):
\begin{equation}\label{eq:disp_matrix_taylor_series}
\mathbf D(x,k;\omega) = \mathbf D(x_0,k_0;\omega_0) + \frac{\partial \mathbf D}{\partial x} (x-x_0) + \frac{\partial \mathbf D}{\partial k} (k-k_0) + \frac{\partial \mathbf D}{\partial \omega} (\omega-\omega_0) + \ldots  .
\end{equation}
When converted back to an operator, the first term of this series gives the equation
\begin{equation}
\mathbf D(x_0,k_0;\omega_0) {\bm \psi}(x,t) = 0.
\end{equation}
This can be written in using the eigenvectors of $\mathbf D(x_0,k_0;\omega_0)$ as a basis, which gives (suppressing the $\omega_0$ dependence in $\mathbf D$ and $\mathbf e_\alpha$ for convenience)
\begin{equation}
\sum_{\alpha=1}^N \mathbf D(x_0,k_0) \mathbf e_\alpha (x_0,k_0) A_\alpha(x)e^{i(\Theta_\alpha(x) - \omega_0 t)} = \sum_{\alpha=1}^N D_\alpha (x_0,k_0) \mathbf e_\alpha (x,k_0) A_\alpha(x)e^{i(\Theta_\alpha(x) - \omega_0 t)}.
\end{equation}
This equation implies that in general, if $\bm \psi$ is going to be a nontrivial solution, then its polarization vector ${\bf e}(x)$ must be an eigenvector of the matrix $\mathbf D$ with zero eigenvalue.  

We now have a situation comparable to the scalar case discussed previously.  Each function $D_\alpha(x,k;\omega_0)$ defines a dispersion surface, and, as long as the dispersion surfaces for different polarizations are well separated in phase space, they each correspond to a different wave mode.  The eigenvalues $D_\alpha$ can then act as ray hamiltonians, and Equation (\ref{eq:WKB_solution_from_rays}) can be used to construct the WKB amplitude and phase corresponding to this polarization\footnote{The local eigenvectors of the dispersion matrix uniquely define the polarizations, modulo a non-holonomic phase \cite{friedland:3050,littlejohn:149}.}.

If the dispersion surfaces for different modes come near each other in phase space, this implies that the dispersion matrix is nearly degenerate.  Using the eigenvectors as polarizations for the different modes means that we are implicitly diagonalizing the dispersion matrix.  This diagonalization procedure would become undefined in the case of degenerate eigenvalues.  For the nearly degenerate case when the dispersion surfaces are close to each other, this implies that the eigenvectors could be rapidly varying.  If the eigenvectors vary rapidly, then the ordering assumptions used to construct the WKB solutions would no longer be valid.  A region where such a breakdown occurs is known as a mode conversion region, since energy initially in one of the wave modes can be transferred into the other wave mode.  This phenomenon is known by different names in different fields, e.g., a Landau-Zener level crossing or avoided crossing \cite{Zener:1932lr}, surface hopping \cite{Drukker:1999fk}, linear wave conversion \cite{tracy:2147}, and resonance crossing \cite{wave_emission}.  In the next chapter we will examine a simple model which exhibits this behavior, and describe the mathematical tools used to deal with the mode conversion region in the context of ray-tracing algorithms.

%% file: Chapter-Theory.tex
%%%%%%%%%%%%%%%%%%%%%%%%%%%%%%%%%%%%%%%%%%%%%%%%%%%%%%%%%%%%%%%%%%%%%%%%%%
%
% Ph.D. dissertation manuscript
% Chapter 2: Theory
%
% Andrew Stephen Richardson (Fall 2007)
% College of William and Mary
% Department of Physics
% Prof. Eugene Tracy, advisor
%
% Based on Paul King and Andrew Norman's template (modified by Wirawan Purwanto)
%
%%%%%%%%%%%%%%%%%%%%%%%%%%%%%%%%%%%%%%%%%%%%%%%%%%%%%%%%%%%%%%%%%%%%%%%%%%

\chapter{Coupled Oscillators: A Pedagogical Introduction\label{chp:coupled_osc}}

%
%\section{Introduction}
%\section{pedagogic}
%\section{metaplectic transformation}
%\section{symbols}

\section{The Wave -- Oscillator Analogy}

Problems exhibiting mode conversion arise in wave systems where the background medium is spatially varying.  Consider, for example, the wave equation in conservation form
\begin{equation}
\partial_t^2 \psi(x,t) - \partial_x \left(c^2(x) \partial_x \psi(x,t) \right) = 0.
\end{equation}
Here, the background variation is modeled via a wave speed which depends on position.  Since there is no explicit time dependence, we can Fourier transform from $t$ to $\omega$:
\begin{align}
-\omega^2 \tilde\psi_\omega(x) - \frac{d}{dx} \left( c^2(x) \frac{d}{dx} \tilde\psi_\omega(x) \right)=0
\end{align}
This is an equation for the spatial profile of each frequency component.  If we define the wavenumber $k(x) = \omega/c(x)$, we get
\begin{equation}
k(x)^2 \tilde\psi_\omega(x) + \frac{1}{c^2(x)} \left(\frac{d}{dx} c^2(x) \right) \left(\frac{d}{dx} \tilde\psi_\omega(x) \right) + \frac{d^2}{dx^2} \tilde\psi_\omega(x) =0
\end{equation}
If we can ignore the term with the derivative of $c(x)$, then this equation is similar in form to the equation for a harmonic oscillator with a varying natural frequency.
\begin{equation}\label{eq:classical_oscillator}
\ddot x(t) + \omega^2(t) x(t) = 0
\end{equation}
While the change of variables from $(x,k(x))$ to $(t,\omega(t))$ is somewhat awkward because of the previous use of the variables $t$ and $\omega$, it is worthwhile since it lets us think of the original wave problem in terms of a simple harmonic oscillator, which is a common and well understood system.

In this chapter we will analyze a pair of coupled oscillators with time-dependant frequencies.  This pair of oscillators provides a simple example of the phase space techniques used to solve the mode conversion problem.  There are two aspects of this setup which deserve special note, since they differ from standard harmonic oscillator problems.  

First, in this problem {\em phase space} is taken to mean the time-frequency $(t,\omega)$ plane.  Thus the natural frequencies of the oscillators describe curves in phase space.  Also, the Fourier transform of the solution 
\begin{equation}
\tilde x(\omega) = \frac{1}{\sqrt{2 \pi}} \int dt\,  e^{-i\omega t} x(t)
\end{equation}
is related to a rotation of phase space by $90^\circ$.  (This connection of the Fourier transform to a rotation in phase space will be discussed at length in following chapters.)  Unless otherwise noted, when the term {\em phase space} is used in the following, the $(t,\omega)$ plane is meant rather than the $(x, \dot x)$ phase space of classical mechanics problems.

The second aspect of this calculation which may be unusual is the use of complex values for $x(t)$.  The displacement of a physical oscillator is a real valued quantity.  However, if we use Fourier methods to solve the problem, the most general solutions will be complex valued.  The ``physical'' solution is then obtained by taking the real part of $x(t)$.  The Fourier method of expanding our field in complex valued plane waves arises naturally in this context.  In general, the wave equations we want to solve will take the form $\hat D(x,-i \partial_x; t, i\partial_t)\psi(x,t) = 0$.  Since $\hat D$ is a function of the operators $-i\frac{\partial}{\partial x}$ and $i\frac{\partial}{\partial t}$, we would like to expand it in eigenfunctions of these operators if possible.  In the best case scenario, this will diagonalize $\hat D$, and give us the solutions to our problem.  The eigenfunctions of these derivative operators are the complex functions $e^{ikx}$ and $e^{-i\omega t}$, so a solution written with these as basis functions will in general be complex valued.  

With these two caveats in mind, we now proceed to the actual calculation.

\section{Outline of Calculation}

The outline of this calculation is as follows.  First, we review the problem of two coupled oscillators, and discuss how the WKB approximation and raytracing methods can be used to solve the problem even when the natural frequencies of the oscillators are time dependent.

Then, we discuss how these methods break down when the frequencies of the oscillators become nearly equal for some range of values, allowing energy to transfer from one oscillator to the other during this resonance crossing.  We outline how we can use asymptotic methods to deal with this phenomena as if it were a scattering or ray splitting problem.

We then use two different methods to find the local solution in the resonance region.  We first show how a linearization near the resonance gives us the equation for the parabolic cylinder functions (this is a well known result).  Then, we show how a linear canonical transformation of $(t,\omega)$ phase space, along with the associated metaplectic transformation of our solutions (to be defined), give a more intuitive picture of the linearization procedure.  This phase space picture also allows the scattering parameters to be derived more easily from the local solutions. 

Finally, we compare our full raytracing solution (including the ray splitting at the mode conversion) with numerical solutions of the original equations.  

\section{Coupled Oscillator Review}

\subsection{Problem Setup\label{sec:setup}}

The problem we will examine is that of energy transfer between two coupled harmonic oscillators.  We start with the equation of motion for a harmonic oscillator, such as Equation (\ref{eq:classical_oscillator}), and introduce a second oscillator coupled to the first. 
\begin{eqnarray}
\ddot x_1 + \omega_1^2(t) \, x_1 +\eta \, x_2 &=&0 \label{eq:osc1} \\
\ddot x_2 + \omega_2^2(t) \, x_2 +\eta \, x_1 &=&0 \label{eq:osc2}
\end{eqnarray}
Here $x_i(t)$ is the state of the $i^{th}$ oscillator, $\omega_i(t)$ is the natural frequency of the $i^{th}$ oscillator given as a function of time, and the constant coupling between the oscillators is given by the coefficient $\eta$.  In many real coupled oscillator problems, the coupling would be proportional to a function of the difference between the oscillator's positions, e.g., $\eta\, (x_2-x_1)$.  However, since we are trying to model mode conversion in waves, we will use the form above, without the difference, since that is how the coupling appears in many mode conversion models \cite{metaplectic_formulation,wave_emission}.  This can be understood as a shift in the definitions of $\omega_1$ and $\omega_2$.

Equations \ref{eq:osc1} and \ref{eq:osc2} can be derived from an action principle \cite{Kaufman:PhysLettA1993}:  
\begin{equation}\label{eq:action}
\mathcal{A} = \int dt \; {\mathbf x}^\dagger(t) \cdot \hat{\mathbf D}(t,i\partial_t) \cdot {\mathbf x}(t) .
\end{equation}
Variation of this action with respect to ${\mathbf x}^\dagger$ gives the matrix form of the equations,
\begin{equation}
\label{eq:motion}
\hat{\mathbf D} \cdot  {\mathbf x} =
\left( 
\begin{array}{cc}
\omega_1^2 + \frac{d^2}{dt^2}  &  \eta \\
\eta & \omega_2^2 + \frac{d^2}{dt^2}
\end{array}
\right)
\left( 
\begin{array}{c}
x_1 \\
x_2
\end{array}
\right)
=0.
\end{equation}
The operator $\hat{\mathbf D}$ can be written in terms of functions of the phase space variables $(t,\omega)$ by calculating the ``symbol'' of the wave operator through the substitution\footnote{The sign in this substitution is set by considering the action of the derivative on a positive frequency plane wave.  Since a plane wave is usually written as $e^{-i\omega t}$, we get $i\partial_t e^{-i\omega t}=\omega e^{-i\omega t}$, which gives us the sign shown above.} $i\partial_t \rightarrow \omega$.  A full description of symbols of operators will be given in Chapter \ref{chp:Symbols}.
\begin{equation} \label{eq:osc_disp_matrix_symbol}
\mathbf{D}(t,\omega) =
\left( 
\begin{array}{cc}
\omega_1^2 - \omega^2  &  \eta \\
\eta & \omega_2^2 - \omega^2
\end{array}
\right)
\end{equation}
The equations of motion now have the form of an eigenvalue equation for a pair of coupled harmonic oscillators with slowly varying natural frequencies.  In terms of the mass matrix $\mathbf{M}$ and the potential matrix $\mathbf{V}$ from classical mechanics \cite{walecka}, Equation (\ref{eq:motion}) is 
\begin{equation}
\mathbf{D} \cdot \mathbf{x} =
(\mathbf{V} - \mathbf{M} \omega^2) \cdot \mathbf{x}  = 0 ,
\end{equation}
with
\begin{equation}
\mathbf{M} =
\left(
\begin{array}{cc}
 1  & 0  \\
  0 &   1
\end{array}
\right), \quad
\mathbf{V} =
\left(
\begin{array}{cc}
 \omega_1^2  & 0  \\
  0 &   \omega_2^2
\end{array}
\right)  + \eta
\left(\begin{array}{cc}
 0  & 1  \\
  1 &  0
\end{array}
\right).
\end{equation}

\subsection{Constant Natural Frequencies  \label{sec:const}}
Before proceeding to examine mode conversion in this problem, notice that the system in Equation (\ref{eq:motion}) can be solved exactly if the natural frequencies of the oscillators and the coupling are all constants.  In this case, there are two eigenmodes that can be found by setting $\det (\mathbf{D})=0$.  Solving for the eigenfrequencies gives
\begin{equation}\label{eq:eigenvalues}
\omega^2_\pm = \frac{1}{2}(\omega_1^2 + \omega_2^2) \pm \frac{1}{2} \sqrt{(\omega_1^2 - \omega_2^2)^2+4\eta^2} .
\end{equation}
There is an eigenvector associated with each of these modes.  These eigenvectors are
\begin{equation}\label{eq:eigenvectors}
\mathbf{e}_\pm=\mathcal{N}_\pm \left( 
\begin{array}{c}
\eta \; (\omega^2_\pm - \omega^2_1)^{-1}  \\
1
\end{array}
\right),
\end{equation}
where $\mathcal{N}_\pm$ is the normalization.
The most general solution for $\mathbf{x}(t)$ is then given by
\begin{equation}\label{eq:const_soln}
\mathbf{x}(t) = C_+ \mathbf{e}_+ e^{-i (\omega_+ t + \phi_+)} +
C_- \mathbf{e}_- e^{-i(\omega_- t + \phi_-)} ,
\end{equation}
where the real amplitudes $C_\pm$ and phases $\phi_\pm$ depend on the initial conditions.

Later in this calculation we will encounter the limiting case where the coupling is small and the natural frequencies of the oscillators are significantly different.  In this case, 
\begin{equation}
|\eta| \ll |\omega_1^2 - \omega_2^2| \equiv |\Delta|.
\end{equation}
The eigenfrequencies are then
\begin{eqnarray} \label{eq:approx_eigenvalue}
\omega_+^2 &\simeq& \omega_>^2 +\frac{|\eta|^2}{\Delta^2} \\
\omega_-^2 &\simeq& \omega_<^2 -\frac{|\eta|^2}{\Delta^2} ,
\end{eqnarray}
where $\omega_>$ ($\omega_<$) is the larger (smaller) of the natural frequencies $\omega_1$ and $\omega_2$.  If $\omega_1 = \omega_>$, then the eigenvectors are approximately
\begin{equation}
\mathbf{e}_+ = \left( 
\begin{array}{c}
1  \\
\xi^*
\end{array}
\right), \quad
\mathbf{e}_- = \left( 
\begin{array}{c}
-\xi  \\
1
\end{array}
\right);  \quad \omega_1 = \omega_> ,
\end{equation}
where 
\begin{equation}\label{eq:xi}
\xi = \eta / \Delta^2 .
\end{equation}
If $\omega_2 = \omega_>$, then we get
\begin{equation}
\mathbf{e}_+ = \left( 
\begin{array}{c}
\xi  \\
1
\end{array}
\right), \quad
\mathbf{e}_- = \left( 
\begin{array}{c}
1  \\
-\xi^*
\end{array}
\right); \quad \omega_2 = \omega_>.
\end{equation}

If we start with all the energy in the first oscillator, that is equivalent to specifying the initial condition
\begin{equation}
\mathbf{x}_0=\left( 
\begin{array}{c}
A \\
0
\end{array}
\right), \;\;
\dot {\mathbf{\;\,x_0}}=\left( 
\begin{array}{c}
0 \\
0
\end{array}
\right).
\end{equation}
Assuming that $\omega_1 = \omega_>$, we now decompose this onto the  eigenmodes, and find the initial amplitudes and phases
\begin{equation}
\mathbf{x}_0= A \left((1-|\xi|^2)\mathbf{e}_+ - \xi^* \mathbf{e}_- \right), \quad
\phi_+=\phi_-=0.
\end{equation}
This means that the second oscillator is driven by the first and its solution is
\begin{eqnarray}
x_2(t)&=& A \xi^* \left((1-|\xi|^2) e^{-i\omega_+ t} - e^{-i\omega_- t}\right).
\end{eqnarray}
The maximum amplitude that $x_2$ can take on is approximately
\begin{equation}
\max(x_2(t)) \simeq 2A|\xi| = \frac{4A\eta}{\Delta^2}.
\end{equation}
Since we are in the limit where $\xi \ll 1$, the magnitude of $x_2$ is always small, so there is not much energy transferred between the oscillators.  In this case, a good approximation to the solution is
\begin{equation}
\mathbf x (t) = \left(\begin{array}{c}
A \\
0
\end{array}\right)
e^{-i\omega_1 t}.
\end{equation}

\subsection{Varying the Frequencies $\omega_1$ and $\omega_2$}
\begin{figure}
\begin{center}
\includegraphics[scale=0.5]{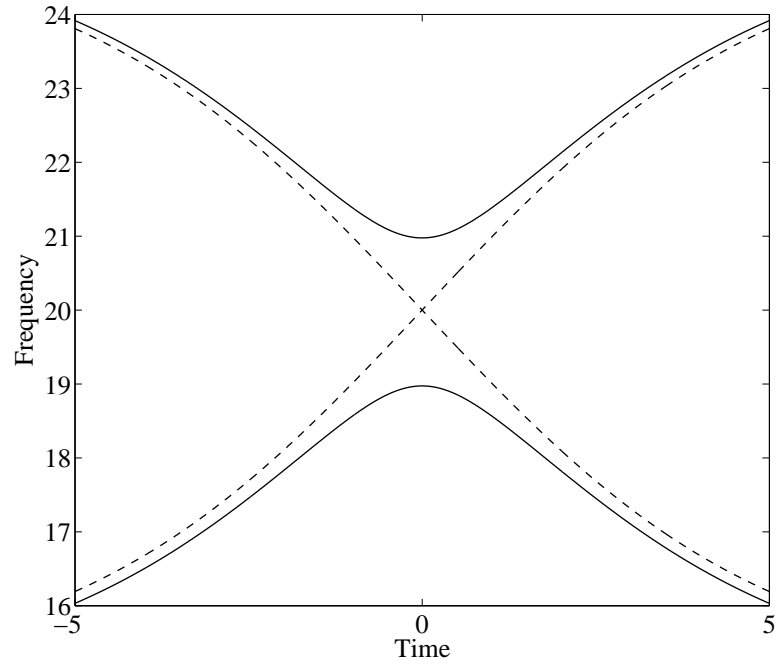}
\end{center}
\caption{\label{fig:adiabatic}
This plot show the natural frequencies of the oscillators (dashed lines) and the eigenfrequencies of the pair of coupled oscillators (solid lines).  The coupled frequencies are obtained by solving $\text{det}({\bf D}(t,\omega))=0$.  Far from the mode conversion, the natural frequencies are approximately equal to the eigenfrequencies.  In the mode conversion region, the effect of the coupling is more pronounced, giving rise to the avoided crossing structure.
}
\end{figure}

In order to model a mode conversion problem using coupled oscillators, they need to be set up so that their time-dependent natural frequencies become nearly equal at some time.  Figure \ref{fig:adiabatic} gives a plot of the frequencies for such a system.  In that figure, the dashed lines are the natural frequencies of the oscillators.  Near time $t=0$ the frequencies are approximately linear functions of time, and at time $t=0$ they are equal.  The dash-dotted lines are the eigenfrequencies, $\omega_\pm$, of the dispersion matrix.  Far from the crossing, the eigenfrequencies are approximately equal to the specified natural frequencies of the two oscillators.  Near $t=0$, the eigenfrequencies form a hyperbolic structure called an {\em avoided crossing}.

In any matrix problem, there is inherently a question as to which set of basis vectors should be used when formulating the problem and its solution.  Do you stick with the physical coordinates, or switch to a more appropriate set of generalized coordinates?  Or, asked a different way, which set of polarizations should be used to describe the wave?  In many problems, the eigenvectors of the matrix wave operator are the natural basis vectors to use.  For example, the ``up'' and ``down'' spin states correspond to energy eigenstates which are aligned with an applied magnetic field.  Any other spin state can be written as a combination of these.  Similarly, in a linear mode conversion problem, the two interacting modes provide a set of states which would seem natural to use as basis states.

In our example of two coupled harmonic oscillators, each oscillator $x_i$ has an amplitude that must be specified in order to give the complete state of the system.  Using these two amplitudes as components of the state vector gives expressions such as Equation (\ref{eq:motion}).  However, as illustrated by the calculation in Section \ref{sec:const}, it is often easier to solve the system if we choose a different set of basis vectors.  In that section, by using the eigenvectors of the dispersion matrix as a basis, the solution decoupled into normal modes with distinct frequencies.  It would seem that this normal mode basis would be the one to use when solving the full problem including mode conversion.  (An aside concerning nomenclature; in the atomic scattering literature, the eigenvector basis with coupling is called the {\em adiabatic} basis, and the original basis without coupling is called the {\em diabatic} basis.)

There are several problems with using the eigenvectors as a basis, however.  First, there is the question of where exactly does the mode conversion occur?  As seen in Figure (\ref{fig:adiabatic}), while the given natural frequencies ($\omega_1,\omega_2$) cross at a well defined point in phase space, the coupled frequencies ($\omega_+,\omega_-$) do not cross.  This {\em avoided crossing}  makes it more difficult to define the region in which mode conversion is occurring.  

Another problem with this basis arises when trying to define the polarization transport when using raytracing to solve the problem.  The WKB approximation assumes that the various quantities associated with the ray and the surrounding medium are slowly varying.  However, near a mode conversion point, the polarization starts to vary rapidly.  An involved calculation then becomes necessary to find the polarization of the transmitted and converted rays \cite{tracy:2147}.  Granted, the WKB approximation is not valid near the mode conversion anyway, but it would make the calculation simpler if the polarization varied smoothly and simply across the mode conversion region.

\section{The WKB approximation and Raytracing}

\subsection{Eikonal ansatz and the WKB approximation}

In the previous sections, the solution to our problem was obtained for coupled oscillators with fixed natural frequencies.  We now consider the possibility that these frequencies are not fixed, but could be slowly varying functions of time.  If the frequencies vary slowly enough, then we expect the solutions to look very similar to the plane wave solutions in Equation (\ref{eq:const_soln}), for at least a few oscillation periods.  A trial solution of this form is called an {\em eikonal ansatz} \cite{tracy:2147}.  Let's assume that the solution will have this eikonal form
\begin{equation}\label{eq:eikonal}
\mathbf x(t) = \mathbf e(t) A(t) e^{i\theta(t)}
\end{equation}
where the polarization $\mathbf e(t)$ and amplitude $A(t)$ vary slowly compared to the phase $\theta(t)$.  So the derivative of this with respect to time is approximately
\begin{equation}
i\frac{d}{d t} \mathbf x(t) \simeq -
\mathbf e(t) A(t) \left(\frac{d \theta(t)}{d t} \right) e^{i\theta(t)}.
\end{equation}
Note that this eikonal form has an ambiguity in the phase of the polarization.  Any change to the phase of the polarization can be absorbed by the phase $e^{i\theta(t)}$:
\begin{align}
e^{i\theta(t)} \mathbf e(t) = e^{i(\theta(t)+\gamma)} e^{-i\gamma}\mathbf e(t) 
=e^{i\theta'(t)} \mathbf e'(t).
\end{align}
We adopt the convention that there is an arbitrary initial phase in the polarization vector, but any change in phase appears through variations in $\theta(t)$.

We can now examine the action of our wave operator on our test solution.  First, assume that the coupling is turned off.  In this case, the polarization corresponding to either of the oscillators separately is an eigenvector of the matrix $\hat {\mathbf D}$.  In general, the time derivative brings down a derivative of the phase:
\begin{equation}\label{eq:wkb_approx}
0=\hat{\mathbf D}(t,i\partial_t) \cdot \mathbf e_\alpha(t) A(t) e^{i\theta(t)} \simeq
\mathbf e_\alpha(t) A(t) D_\alpha(t,-\dot\theta(t)) e^{i\theta(t)}
\end{equation}
Here, $D_\alpha(t,-\dot\theta)$ is the eigenvalue of the matrix $\hat {\mathbf D}$, where the time derivatives have been replaced by $\dot \theta$, turning the operator into a function of $t$.  In order to obtain a solution, we need 
\begin{equation}
D_\alpha(t, -\dot\theta(t))=0.
\end{equation}
 For example, if we are considering solutions for $x_1(t)$ in the limit $\eta \rightarrow 0$, then we take the first diagonal element of $\hat {\mathbf D}$, and replace the derivatives with $\dot \theta$ to obtain
\begin{equation}
{\bf e}_1^\dagger\cdot \hat{\bf D} \cdot {\bf e}_1 = D_{11}(t,-\dot\theta(t))=\omega_1^2(t)+\dot\theta^2(t) = 0.
\end{equation}
This implies that the phase is given by the integral of the natural frequency.
\begin{equation} \label{eq:wkb_phase}
\theta(t) = -\int^t \omega_1(t')dt'
\end{equation}

The amplitude as a function of time can now be derived from an approximate action principle for the wave equation.  Insert the approximation from Equation (\ref{eq:wkb_approx}) into the action in Equation (\ref{eq:action}) to obtain a new approximate action
\begin{equation} 
\tilde{\mathcal{A}} = \int_{t_1}^{t_2}  dt \; D_{11}(t,\omega=-\dot\theta) |A(t)|^2.
\end{equation}
This action depends on $\theta(t)$ only through its derivative, and so the action is not changed by adding an arbitrary constant to $\theta$.  This means that there is a Noether symmetry, and an associated conserved quantity.  Varying with respect to $\theta$ and assuming that $D_{11}$ is a smooth function of $\omega$ implies that 
\begin{equation}
\frac{d}{dt} \left( \left.\frac{\partial D_{11}}{\partial \omega}\right\vert_{t,\omega_1(t)}  |A(t)|^2 \right)=0.
\end{equation}
For the problem being considered here, Equation (\ref{eq:osc_disp_matrix_symbol}) gives us $D_{11} = (\omega_1^2-\omega^2)$, which means that
\begin{equation}
\omega_1(t) |A(t)|^2 = \text{const.}
\end{equation}
This means that the amplitude of $x_1$ must vary like
\begin{equation} \label{eq:wkb_amp}
A(t)= A_1^{\text{in} } \sqrt{\frac{\omega_1^{\text{in} }}{\omega_1(t)} },
\end{equation}
where $A_1^{\text{in}}$ and $\omega_1^{\text{in} }$ are constants.  Together, Equations (\ref{eq:wkb_amp}) and (\ref{eq:wkb_phase}) give the WKB solutions for the amplitude and phase of the oscillator.  The polarization vector ${\bf e}(t)$ is given by the instantaneous eigenvector ${\bf e}_\alpha(t)$ associated with the eigenvalue we are considering.  In order to fully define the polarization, the slowly-varying phase $\gamma$ of the polarization must also be defined:
\begin{align}
{\bf e}(t) = e^{i \gamma(t)} {\bf e}_\alpha(t).
\end{align}
The equation for the transport of this phase along a ray can be calculated from the next higher order terms which were dropped from Equation (\ref{eq:wkb_approx}).  For details of this calculation, see \cite{Kaufman:1987kx}.

\subsection{Raytracing}

The idea behind the raytracing method is to recast the wave problem into the form of a Hamiltonian mechanics problem.  There are many ways to do this.  We will use the tools of the ``symbol calculus'' (see Chapter \ref{chp:Symbols} and also S. MacDonald's excellent review paper \cite{McDonald:1988qy}).  In this section, we outline the basic ideas behind the ray-tracing method.

Start by considering the instantaneous eigenvectors ${\bf e}_\alpha(t)$, as in the previous section.  Use these vectors to decompose an eikonal wave:
\begin{align}
{\bm \psi}(t) = \sum_\alpha {\bf e}_\alpha(t) A_\alpha(t) e^{i\theta_\alpha(t)/\epsilon}.
\end{align}
Here we have introduced $\epsilon$ as an ordering parameter.  In the following we will consider the polarizations separately.  Instead of using an approximation as in Equation (\ref{eq:wkb_approx}), we can use the ordering parameter to calculate the action of a derivative:
\begin{equation}
\epsilon i\partial_t {\bf e}_\alpha(t) A_\alpha(t) e^{i\theta_\alpha(t)/\epsilon}= 
e^{i\theta_\alpha(t)/\epsilon} ( -\dot \theta_\alpha +\epsilon i\partial_t) {\bf e}_\alpha(t) A_\alpha(t).
\end{equation}
Substitute this expression into the wave equation:
\begin{align}
\hat{\bf D}(t,\epsilon i \partial_t) {\bf e}_\alpha(t) A_\alpha(t) e^{i\theta_\alpha(t)/\epsilon} = 
e^{i\theta_\alpha(t)/\epsilon} \hat{\bf D}(t,-\dot\theta+\epsilon i \partial_t) {\bf e}_\alpha(t) A_\alpha(t)=0.
\end{align}
Use the fact that ${\bf e}_\alpha$ is an eigenvector to construct a scalar wave operator:
\begin{align}
\hat D_\alpha \equiv {\bf e}_\alpha^\dagger \cdot \hat{\bf D} \cdot {\bf e}_\alpha.
\end{align}
This gives a scalar equation for the mode $\alpha$:
\begin{equation}\label{eq:amp_pol_eqn}
\hat{D}_\alpha (t,-\dot\theta + \epsilon i\partial_t) A_\alpha(t) =0.
\end{equation}
The WKB approximation says that the magnitude of $\dot \theta(t)$ is larger than the magnitude of any other derivatives that appear in this equation, and the ordering parameter $\epsilon$ makes this explicit.  If we could think of the derivative as a variable instead of as a function, then we could use the Taylor's series expansion to obtain an approximation to this equation.  This can in fact be done by using the symbol of the wave operator, $D_\alpha(t,\omega)$, as in Equation (\ref{eq:osc_disp_matrix_symbol}).  Expand the symbol about $\omega = -\dot\theta$ using the Taylor's series, and then convert it back into an operator.  Since $\partial_\omega D_\alpha$ is a function of time, the ordering of the multiplication and differentiation operators must be symmetrized.  We therefore have
\begin{align}
\hat{D}_\alpha(t,-\dot\theta + \epsilon i\partial_t) &= D_\alpha(t,-\dot\theta) + \frac{\epsilon}{2}\left( \left.\frac{\partial D_\alpha}{\partial \omega}\right|_{\omega=-\dot\theta} (i\partial_t) + i\partial_t \left.\frac{\partial D_\alpha}{\partial \omega}\right|_{\omega=-\dot\theta} \right)+ \mathcal{O}(\epsilon^2) \\
&= D_\alpha(t,-\dot\theta) + \frac{\epsilon i}{2}\left( \frac{d}{dt}\left( \left.\frac{\partial D_\alpha}{\partial \omega}\right|_{\omega=-\dot\theta}\right) + 2\left.\frac{\partial D_\alpha}{\partial \omega}\right|_{\omega=-\dot\theta} \partial_t  \right) + \mathcal{O}(\epsilon^2)
\end{align}
Truncating this series at $\mathcal{O}(\epsilon)$, and putting it into Equation (\ref{eq:amp_pol_eqn}) gives us
\begin{equation}
D_\alpha(t,-\dot\theta) A(t) + 2\epsilon \left(\left.\frac{\partial D_\alpha}{\partial \omega}\right|_{\omega=-\dot\theta}\right)^{1/2} \frac{d}{dt}\left( \left(\left.\frac{\partial D_\alpha}{\partial \omega}\right|_{\omega=-\dot\theta}\right)^{1/2}  A(t) \right)  \simeq 0.
\end{equation}
In order for this to equal zero, both terms must be zero.  This gives us equations for $\theta(t)$ and $A(t)$.
\begin{equation}
D_\alpha(t,-\dot\theta) = 0 
\end{equation}
\begin{equation}
\frac{d}{dt}\left( \left(\left.\frac{\partial D_\alpha}{\partial \omega}\right|_{\omega=-\dot\theta}\right)^{1/2}  A(t) \right) = 0
\end{equation}
The first of these equations is solved using ray-tracing techniques, while second equation gives us the standard WKB formula for the amplitude. 

Because $D_\alpha$ is a function on phase space, we are looking for the curve in the $(t,\omega)$ plane along which $D_\alpha(t,\omega)$ is zero.  If we write a parametric equation for this curve, then we will have the ``ray'' $(t(\sigma), \omega(\sigma))$, where $\sigma$ is the ray parameter, which plays the role of time from classical mechanics.

If we start with a point in phase space such that $D_\alpha(t_0,\omega_0)=0$, then the equations of motion for the ray can be derived by enforcing the condition that $D_\alpha$ remains constant along the ray.
\begin{equation}
0=\frac{dD_\alpha}{d\sigma}=\frac{\partial D_\alpha}{\partial t}\frac{dt}{d\sigma}+\frac{\partial D_\alpha}{\partial \omega}\frac{d\omega}{d\sigma}
\end{equation}
We can choose the parameterization as we like, so choose
\begin{equation}\label{eq:dt}
\frac{dt}{d\sigma}=\frac{\partial D_\alpha}{\partial \omega}.
\end{equation}
That leaves the equation
\begin{equation}
0=\frac{\partial D_\alpha}{\partial \omega}\left(\frac{\partial D_\alpha}{\partial t}+\frac{d\omega}{d\sigma}\right),
\end{equation}
or
\begin{equation}\label{eq:domega}
\frac{d\omega}{d\sigma}=-\frac{\partial D_\alpha}{\partial t}.
\end{equation}
Notice that Equations (\ref{eq:dt}) and (\ref{eq:domega}) are simply Hamilton's equations, with $D_\alpha(t,\omega)$ as the Hamiltonian function for the ray as with the scalar case discussed in Section \ref{sec:scalar_phase_space_thy}.  Once these equations have been solved for the ray, we can integrate along the ray to find the phase:
\begin{equation}
\theta(t') = -\int^{t'} \omega(t) \;dt = -\int^{\sigma_\text{final}} \omega(\sigma) \frac{dt}{d\sigma} \;d\sigma .
\end{equation}
Here, $\sigma_\text{final}$ is set so that $t(\sigma_\text{final}) = t'$.

We have now solved for the phase using the $\mathcal{O}(\epsilon^0)$ equation, and the amplitude using the $\mathcal{O}(\epsilon^1)$ equation.  There is also slow variation of the polarization, which comes in at $\mathcal{O}(\epsilon^2)$.  For this calculation, see Kaufman et al.~\cite{Kaufman:1987kx}.

If these equations are used to solve the {\em uncoupled} oscillator problem, we obtain the WKB results of Equations (\ref{eq:wkb_phase}) and (\ref{eq:wkb_amp}), since for this case the polarization vectors are constants.

\subsection{WKB with coupling}

If we return to the oscillator problem including the coupling, then an additional $\mathcal{O}(\eta^2)$ term will appear in the phase of the WKB solution.  In the coupled case, the eigenvalues are no longer equal to the diagonal elements of the dispersion matrix in Equation (\ref{eq:osc_disp_matrix_symbol}).  This means that we have to use the instantaneous eigenvalues and eigenvectors from Equations (\ref{eq:eigenvalues}) and (\ref{eq:eigenvalues}) when we construct the WKB solutions.

\begin{figure}
\begin{center}
\includegraphics[scale=0.5]{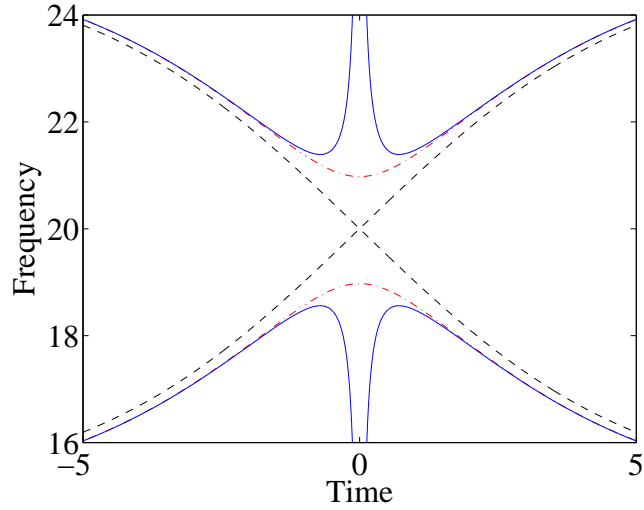}
\end{center}
\caption{\label{fig:adiabatic_approx}
The natural frequencies of the oscillators (dashed lines) are different from the eigenfrequencies of the pair of oscillators (dash-dotted lines), for finite coupling. 
 The solid lines show the linearized approximation to the coupled frequencies given in Equation (\ref{eq:lin_freq_approx}). 
}
\end{figure}

Assuming $\omega_1=\omega_>$, we can use the expansion for the eigenvalues given in Equation (\ref{eq:approx_eigenvalue}) to write $\omega_+$ as
\begin{equation}
\omega_+(t) = \omega_1(t) + \frac{\eta^2}{2 \omega_1(\omega_1^2 - \omega_2^2) } + \mathcal{O}(\eta^4).
\end{equation}
This expression for $\omega$ can be expressed directly as a function of time by linearizing $\omega_1$ and $\omega_2$ about $t=0$:
\begin{equation}
\omega_{1,2} \approx \omega_0 + t \; \dot  \omega_{1,2} (t=0).
\end{equation}
Putting this into the equation for $\omega_+$ gives an expression where the denominator is correct to $\mathcal{O}(t^2)$:
\begin{equation}\label{eq:lin_freq_approx}
\omega_+(t) \approx \omega_1(t) + \frac{\eta^2}{4 \omega_0^2(\dot \omega_1 -\dot \omega_2) t }
= \omega_1(t) + \frac{\tilde \eta^2}{t }.
\end{equation}
This approximation is good away from the mode conversion, but breaks down close to the mode conversion, where $t\rightarrow 0$.  See Figure \ref{fig:adiabatic_approx}.
Note that the denominator of the coupling term is the Poisson bracket of the diagonals of $\mathbf{D}$.  We will see in Section \ref{sec:normalized_coupling} that the bracket arises because of a change of variables in phase space which is performed in order to make the diagonal elements of $\mathbf{D}$ look like a canonical pair.

\section{Mode Conversion Setup}

\begin{figure}
\begin{center}
\includegraphics[scale=0.45]{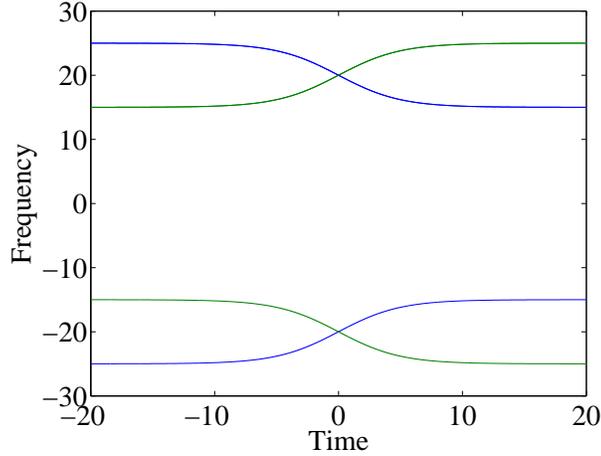}
\end{center}
\caption{\label{cap:nat}A plot of the natural frequency of the oscillators versus time, assuming an offset hyperbolic tangent form.}
\end{figure}

We make several assumptions in order to make this problem have the form of a linear mode conversion problem.  
\begin{enumerate}
\item $|\eta| \ll |\Delta| $ in regions far from the conversion region, so that incoming and outgoing waves have an eikonal form.
\item We assume we can ignore the negative frequency solutions during the calculations, and superimpose them with the positive solutions at the end.
\item The positive frequency solutions of interest are positive for all time.
\item In the time-frequency plane, the natural frequencies asymptote to constant values, and cross at only one point, $(\omega_0,t_0)$. See Figure (\ref{cap:nat}).
\item The timescale for changes in $\omega_i$ is given by $T$, as defined through $\dot \omega_i(t) \simeq \frac{\omega_i}{T}$. This $T$ is the same for both oscillators.
\item In order for WKB to be valid, $T \gg \tau_i = \frac{2\pi}{\omega_i}$, which implies that $T \omega_i \gg 2\pi$.
\end{enumerate}

\linespread{1}
\linespread{1.6}

We want to find solutions $x_i(t)$ for large time $t\gg t_0$, given the initial conditions in the form of an eikonal wave
\begin{eqnarray}
x_1(t\rightarrow-\infty)&=&A_1^{\text{in}} \exp(-i (t-t_0)\,\omega_1^{\text{in}})\\
x_2(t\rightarrow-\infty)&=&0.
\end{eqnarray}
The solutions at large time (long after the mode conversion) should be of the form:
\begin{eqnarray}
x_1 \rightarrow A_1^{\text{out}}\exp(-i (t-\phi_1)\,\omega_1^{\text{out}})\\
x_2 \rightarrow A_2^{\text{out}}\exp(-i (t-\phi_2)\,\omega_2^{\text{out}}).
\end{eqnarray}
The  question of finding the solutions for large time now becomes one of somehow solving for the amplitudes $A_i$ and the phase shifts $\phi_i$.  This can either be done by solving for $x_i(t)$ for all time (as in the numerical solution), or by treating the problem like a scattering problem, and solve it using matched asymptotics.  This approach uses WKB methods away from the resonance region, and a different asymptotic approximation near the resonance.  The regions where these approximations are valid will overlap; this is the matching region where the amplitude and phases of the far-field WKB solutions can be matched onto the local field.  This matching can then be used to construct an S-matrix which connects the incoming WKB wave on one side of the resonance to the outgoing WKB waves on the other side.  We can then use the S-matrix to directly match WKB solutions, effectively skipping over the resonance region.  Or, since we also have constructed the local solution, it can be evaluated to find the solution in the vicinity of the mode conversion, giving a solution $x_i(t)$ which is a valid approximation both far from the resonance and near it.

\subsection{Resonance Region}
The asymptotic solution to this problem divides into three components.  First, standard WKB methods are used to propagate the original solution from some early time into the vicinity of the mode conversion region.  This region is where the two oscillators have very similar natural frequencies, which causes the WKB approximation to break down.  In this region, a metaplectic transformation is used to convert the problem into a simpler form, which can be solved in terms of parabolic cylinder functions.  These solutions provide a connection between the incoming WKB solution, and the WKB solution on the other side of the mode conversion region.  The last component of the solution to this problem would be to again use WKB methods to find the solution long after the mode conversion.  The mathematical theories of group representations and the metaplectic transformation are covered in the appendix.

\begin{figure}
\begin{center}
\includegraphics[scale=0.5]{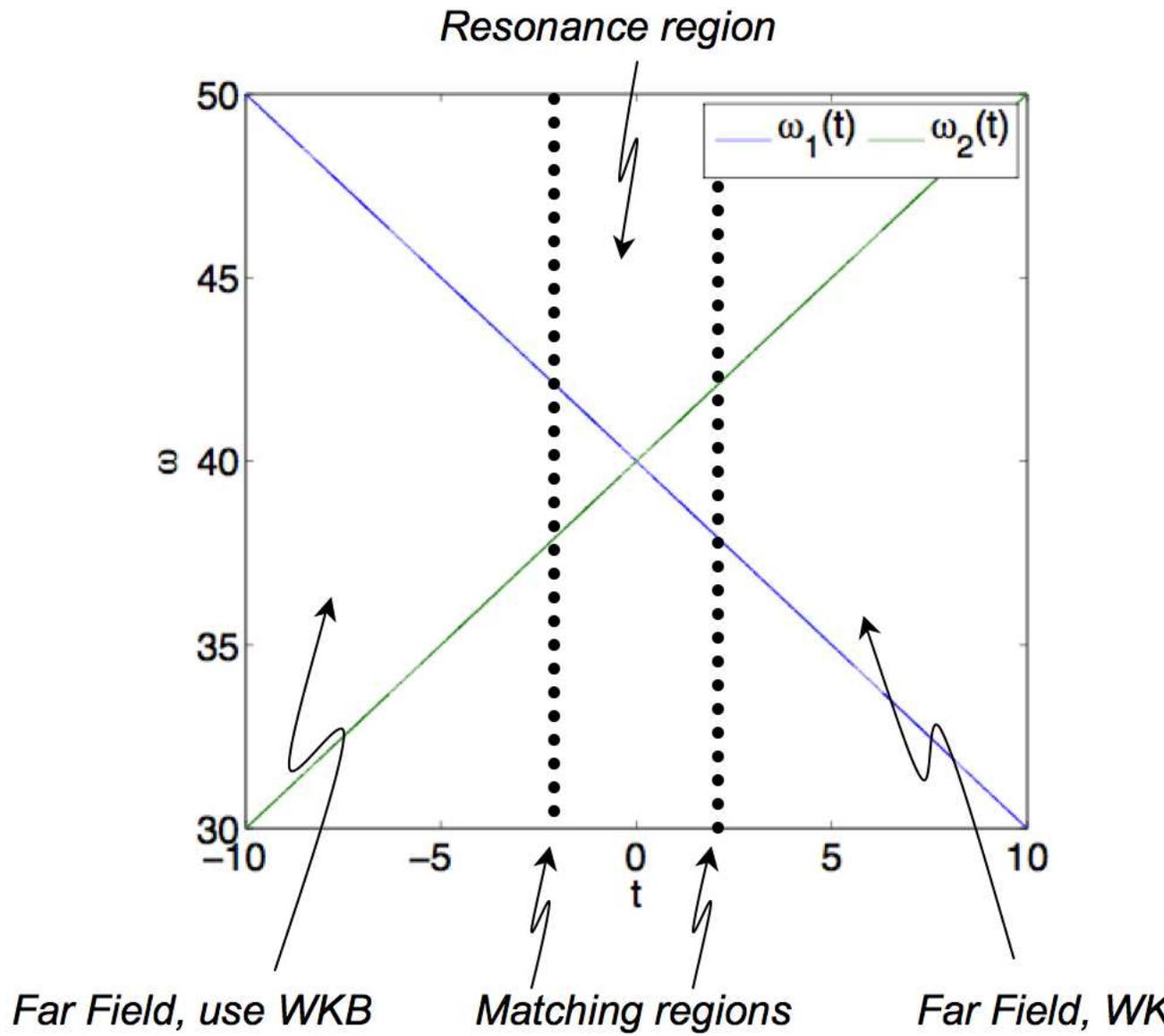}
\end{center}
\caption{\label{fig:matching_region}
The WKB solution is valid in the far field regions, while a local solution is necessary in the mode conversion region.  These solutions can be matched to each other in the matching region, where both types of solutions are approximately correct.}
\end{figure}

\section{Local Solution}

% end comment: local linearized solutions

We now need to find a way to solve the coupled oscillator Equations (\ref{eq:motion}) in the mode conversion region.  Since the natural frequencies of the oscillators become equal in this region (the dispersion surfaces cross), the WKB approximation is not valid here.  In order to find the solution in this region, we will perform a metaplectic transformation on the incoming data, and solve the equations in the new variables.  Once through this region, we can invert the transformation, and see that the outgoing data looks like another WKB solution.

The local solution is obtained in several steps:
\begin{itemize}
\item Simplify the dispersion matrix by expanding it in a Taylor's series about the mode conversion point, and truncating at linear order.  This is simpler if we first move the origin in phase space to the mode conversion point.  
\item Simplify the matrix further by performing a change of variables in phase space.  This change of variables is a combination of scaling transformation and a canonical transformation.  The transformations are chosen such that the diagonal elements of the matrix form a canonical pair of variables, $(q,p)$.  The new dispersion matrix that we obtain is referred to as the {\em normal form} of the dispersion matrix for the mode conversion problem.
\item Convert the dispersion matrix back into an operator in the $q$-representation.  This gives a set of equations which will contain at most first order derivatives, since the matrix is a linear function of the phase space variables.
\item Solve the simplified equations in the $q$-representation.  Then convert this local solution back to the original $t$-representation so that the matching to the WKB solutions can be performed.
\end{itemize}

\begin{equation}
\left(\begin{array}{c}
  x_1^{\text{out}}(q) \\ 
  x_2^{\text{out}}(q) 
\end{array} \right)
=
\left(\begin{array}{cc}
  \tau & \beta \\ 
  -\beta^* & \tau 
\end{array} \right)
\left(\begin{array}{c}
  x_1^{\text{in}}(q) \\ 
  x_2^{\text{in}}(q) 
\end{array} \right)
\end{equation}

\subsection{Linearized Dispersion Matrix}

In this section the dispersion matrix will be expanded about the mode conversion point $(t_0,\omega_0)$.  In order to simplify this calculation, first shift the origin in phase space to the mode conversion point.  This shift of the origin is performed by a change of the dependent variables $x_1(t)$ and $x_2(t)$:
\begin{equation}
\tilde{\mathbf x} (t') = e^{i\omega_0 t'} \mathbf x (t_0+t').
\end{equation}
This corresponds to the change of variables:
\begin{align}
t' = t - t_0, \quad  \nu = \omega - \omega_0.
\end{align}
From this we can see that the shift of the origin on the frequency axis corresponds to factoring out the carrier $e^{-i\omega_0 t}$ from the solution.  The shift of origin on the time axis is an ordinary shift.  In the following we drop the prime from $t'$ to simplify notation.

The Taylor's series for the dispersion matrix about the mode conversion point $(t_0,\omega_0)=(0,0)$ is
\begin{equation}\label{eq:taylor_matrix}
\mathbf D(t,\nu) = \mathbf D(0,0) + t \left.\frac{\partial \mathbf D}{\partial t}\right|_{0,0} + \nu \left.\frac{\partial \mathbf D}{\partial \nu}\right|_{0,0} +\ldots
\end{equation}
Truncating this expansion at linear order gives
\begin{equation}
\mathbf D(t,\nu) = 
\left( 
\begin{array}{cc}
2 \omega_0( \dot \omega_1 t - \nu)   &  \eta \\
\eta & 2 \omega_0(\dot \omega_2 t -\nu)
\end{array}
\right),
\end{equation}
where $\dot \omega_i$ is the time derivative of the $i^{\text{th}}$ natural frequency evaluated at $t=t_0=0$.

\subsection{The Normal Form\label{sec:normalized_coupling}}

We want the diagonals to look like a canonical pair, so first divide out a factor of $2 \omega_0$.  Then calculate the Poisson bracket of the diagonal terms:
\begin{equation}
\{ \dot \omega_1  t -\nu,  \dot \omega_2  t  - \nu \} = 
(\dot \omega_1 - \dot \omega_2) \{  t,  -\nu \} = \dot \omega_2 - \dot \omega_1 .
\end{equation}
The Poisson bracket is defined as
\begin{equation}
\{ f(t,\nu),g(t,\nu) \} = \frac{\partial f}{\partial t} \frac{\partial g}{\partial \nu} -\frac{\partial f}{\partial \nu} \frac{\partial g}{\partial t}.
\end{equation}
We want the bracket of the diagonals to be equal to one, when $\dot\omega_1<0$.  So, make the definition $\mathcal{B} = (\dot \omega_2 - \dot \omega_1) > 0$, and divide out a factor of $\mathcal{B}^{1/2}$.
\begin{equation}
\tilde{\mathbf D} (t,\nu) =
\left( 
\begin{array}{cc}
 (\dot \omega_1  t  -\nu) / \mathcal{B}^{1/2}  & \tilde  \eta \\
\tilde \eta & (  \dot \omega_2  t  -\nu) / \mathcal{B}^{1/2} 
\end{array}
\right), 
\end{equation}
where $\tilde \eta = \eta / (2 \omega_0 \mathcal{B}^{1/2} )$.  We can now perform a canonical transformation to the variables $(q,p)$ so that the matrix will be in normal form:
\begin{equation}\label{eq:disp_qp}
\mathbf D (q,p) = 
\left( 
\begin{array}{cc}
-p & \tilde  \eta \\
\tilde \eta & q 
\end{array}
\right).
\end{equation}
The operator associated with this normal form is
\begin{align}
\widehat{\bf D}^{\text{NF}} = \left( 
\begin{array}{cc}
-\hat p & \tilde  \eta \\
\tilde \eta & \hat q 
\end{array}
\right).
\end{align}
The linear symplectic (canonical) transformation from $(t,\nu)$ to $(q,p)$ is
\begin{equation}\label{eq:canonical_transform}
\left( 
\begin{array}{c}
q \\
p
\end{array}
\right)
= \frac{1}{\mathcal{B}^{1/2}}
\left( 
\begin{array}{cc}
\dot \omega_2 &  -1\\
-\dot \omega_1 & 1
\end{array}
\right)
\left( 
\begin{array}{c}
t \\
\nu
\end{array}
\right)
\end{equation}
This transformation on the operator induces a transformation of our oscillators.  The vector $\mathbf x (t)$ is transformed into a function of $q$ using the integral transform
\begin{equation}
\mathbf x (q) =\int  e^{iF_1(t,q)} \mathbf x (t) \; dt.
\end{equation}
Here, $F_1(t,q)$ is the generating function for the canonical transformation in Equation (\ref{eq:canonical_transform}).  This integral transform is called a metaplectic transformation.  In Section \ref{sec:SVN_theorem} we will show how this transform arises from the representation theory for the Heisenberg-Weyl group.  The generating function $F_1(t,q)$ has the property that the conjugate variables can be obtained from it by taking derivatives.
\begin{equation}\label{eq:generating_derivatives}
\nu = \frac{\partial F_1}{\partial t}, \quad p = -\frac{\partial F_1}{\partial q}
\end{equation}
For the transformation given in Equation (\ref{eq:canonical_transform}), the generating function is
\begin{equation}\label{eq:generating_function}
F_1(t,q) = \frac{1}{2} q^2 - \mathcal B^{1/2} t q + \frac{\dot \omega_2}{2} t^2.
\end{equation}

\subsection{$\mathbf{D}$ as an operator}

Now that we have changed variables in phase space, we need to convert the dispersion matrix (\ref{eq:disp_qp}) into an operator acting on functions of $q$.  This is achieved by making the substitution
\begin{equation}\label{eq:p_operator}
p \rightarrow \hat p = i\partial_q,
\end{equation}
since $p$ is the variable conjugate to $q$, just as $\omega$ was conjugate to $t$.  Note that this differs by a sign from the ordinary correspondence.  This is because $q$ is playing the role of time, not of a spatial coordinate.  The sign convention for plane waves in time is opposite that of plane waves in space, so this means that the correspondence between $p$ and the derivative with respect to $q$ will also differ by a sign from what we expect.
Performing this substitution gives us the equation
\begin{equation}\label{eq:wave_eqn_q_rep}
\left(
\begin{array}{cc}
- i\partial_q  &  \tilde\eta \\
 \tilde\eta  &   q
\end{array}
\right)\cdot
\mathbf{x}(q) = 0.
\end{equation}

Another way to check that the substitution in Equation (\ref{eq:p_operator}) is the one we want is to apply the metaplectic transformation on the $D_{11}$ element of the dispersion matrix in the $t$ representation.  We can us this to define the operator $\hat p$.
\begin{eqnarray}
-\hat p f(q) &\equiv & \int e^{iF_1(t,q)} (\hat D_{11} f(t)) \; dt \\
&=& \int e^{iF_1(t,q)} \left( \frac{1}{\mathcal B^{1/2}} (\dot\omega_1 t - i\partial_t) \right) f(t) \; dt
\end{eqnarray}
Use integration by parts to act with the derivative on the phase instead of on $f(t)$.  This changes the sign of the derivative term.
\begin{align}
-\hat p f(q) &= \int  \left( \frac{1}{\mathcal B^{1/2}} (\dot\omega_1 t + i\partial_t) e^{iF_1(t,q)} \right) f(t) \; dt \\
&= \int  \left( \frac{1}{\mathcal B^{1/2}} \left(\dot\omega_1 t - \frac{\partial F_1}{\partial t} \right) e^{iF_1(t,q)} \right) f(t) \; dt \\
&= \int  \left( \frac{1}{\mathcal B^{1/2}} \left(\dot\omega_1 t +\mathcal B^{1/2}q -\dot\omega_2 t \right) e^{iF_1(t,q)} \right) f(t) \; dt \\
&= \int \left(-\mathcal B^{1/2} t +q \right) e^{iF_1(t,q)}  f(t) \; dt \\
&= \int \left( \frac{\partial F_1}{\partial q} \right) e^{iF_1(t,q)}  f(t) \; dt \\
&= \int \left( -i\partial_q  e^{iF_1(t,q)} \right)  f(t) \; dt \\
&= - i\partial_q  \int e^{iF_1(t,q)} f(t) \; dt \\
&= - i\partial_q  f(q) 
\end{align}
This establishes the relationship $\hat p = i\partial_q$.

\subsection{Local solution in the $q$ representation}

We now want to solve Equation (\ref{eq:wave_eqn_q_rep}).  The second row is an algebraic equation which gives us
\begin{equation}
x_2(q) = -\frac{\tilde\eta}{q} x_1(q).
\end{equation}
Substituting this into the first row gives us a first order differential equation for $x_1(q)$.
\begin{equation}\label{eq:x1_q_rep_eqn}
\partial_q x_1(q) = \frac{i\tilde\eta^2}{q} x_1(q)
\end{equation}
This has the form of a logarithmic derivative, and can be integrated to give
\begin{equation}
x_1(q) = e^{i\tilde\eta^2 \text{Ln}(q)} = e^{i\tilde\eta^2 (\ln|q| + i\arg(q))}.
\end{equation}
The choice of branch cut for $\arg(q)$ will determine the solution we obtain.  For $q<0$ we have two choices, $i \arg(q) = \pm i \pi$.  This means that there are two options for the amplitude of the solution when $q < 0$, $|x_1| = e^{\pm \pi \tilde\eta^2}$.  In either case, when $q>0$, the magnitude is $|x_1|=1$.
With theses two options for the magnitude, we can define two solutions for $x_1(q)$, both of which are valid solution to Equation (\ref{eq:x1_q_rep_eqn}).  These are
\begin{equation}
x_1^a (q) = 
\left\{
\begin{array}{cc}
|q|^{i\tilde\eta^2}  & \quad \text{ if } q<0   \\
\tau |q|^{i\tilde\eta^2}  & \quad \text{ if } q>0       
\end{array}
\right.
\end{equation}
and
\begin{equation}
x_1^b (q) = 
\left\{
\begin{array}{cc}
\tau |q|^{i\tilde\eta^2}  & \quad \text{ if } q<0   \\
|q|^{i\tilde\eta^2}  & \quad \text{ if } q>0       
\end{array}
\right. ,
\end{equation}
where $\tau$ is called the transmission coefficient,
\begin{equation}
\tau \equiv e^{-\pi\tilde\eta^2} < 1.
\end{equation}
If the initial conditions are such that the energy starts out in the first oscillator, then $x_1^a(q)$ is the appropriate solution to use.  The factor of $\tau$ multiplying the solution for $q>0$ represents the energy lost to the second oscillator in the mode conversion.  The solution for the second oscillator is then
\begin{equation}
x_2(q) = 
\left\{
\begin{array}{cc}
\tilde\eta |q|^{i\tilde\eta^2-1}  & \quad \text{ if } q<0   \\
-\tau \tilde\eta |q|^{i\tilde\eta^2-1}  & \quad \text{ if } q>0       
\end{array}
\right. .
\end{equation}
Because of the negative power of $q$, this solution is localized near $q=0$, and its amplitude diverges as $q\rightarrow 0$.  We can combine these solutions into a vector,
\begin{equation}
\left(
\begin{array}{c}
 x_1(q)  \\
 x_2(q)
\end{array}
\right) = A(q) |q|^{i\tilde\eta^2}
\left(
\begin{array}{c}
1   \\
-\frac{\tilde\eta}{q}
\end{array}
\right),
\end{equation}
where $A(q) = 1$ for $q<0$ and $A(q)=\tau$ for $q<0$.

\subsection{Transform back to $t$ rep}

The local solution that has been obtained in the $q$ representation can now be converted back to the $t$ representation to compare with the WKB solutions.  This is achieved by an inverse metaplectic transformation.
\begin{equation}\label{eq:q_to_t_metaplectic_transform}
\left( 
\begin{array}{c}
x_1(t) \\
x_2(t)
\end{array}
\right)
=
\int dq\; e^{-iF_1(t,q)}
A(q)|q|^{i\tilde\eta^2}
\left( 
\begin{array}{c}
 1 \\
-\frac{\tilde\eta}{q}
\end{array}
\right) .
\end{equation}
This is an integral form of the parabolic cylinder equations, which are known to be the correct local form of the solution in the mode conversion region in the $t$ representation.  However, in order to match this local solution to the WKB modes far from $t=0$, we only need the asymptotic form of this expression for large $t$.  
Because of the singular nature of $x_2(q)$, this metaplectic integral is difficult to evaluate.  In \cite{tracy:082102}, Tracy et al.\ use the stationary phase approximation to obtain the asymptotics of this integral.  Brizard et al.\ describe in \cite{Brizard:1998rt} how to compute the Fourier transform of $x_2(q)$, which gives the $p$ representation of the lower channel.  In the $p$ representation, the singularity at the origin no longer appears, and the integral becomes easier to compute.

An alternative approach is to solve for $x_2(t)$ directly in the $t$ representation.  The metaplectic transformation of $x_1(q)$ gives $x_1(t)$ in terms of the parabolic cylinder function.  Insert this into the $t$ representation of the equation for $x_2(t)$, and use the  recurrence relations for the parabolic cylinder function allow us to evaluate the derivative:
\begin{align}
x_2(t)  &= \frac{1}{\eta \sqrt 2} (t -i\partial_t) x_1(t)   .
\end{align}
This representation of the solution can be easily compared to numerical simulations of the original system of equations, since methods for calculating the parabolic cylinder functions are readily available.  The particular form of parabolic cylinder functions which are obtained for $x_1(t)$ and $x_2(t)$ will depend on the parameters $\mathcal B$, $\dot\omega_j$, and $\eta$.  For a the plot of a specific example see Figure (\ref{cap:compare}).

\comment{%
These asymptotics can either be found in a handbook of functions, or the integral above can be evaluated using the stationary phase approximation.

, all we need to do is evaluate this integral asymptotically using the stationary phase approximation.  In order to use the stationary phase approximation, make the substitution $s=q/t$ in the integral.  The integral then has the form
\begin{equation}
\int e^{-i F_1(t,q)} f(q) \;dq = e^{-i\dot\omega_2 t^2/2} \int e^{-it^2 (-\mathcal B^{1/2} s + s^2/2)} f(st)\, t\;ds .
\end{equation}
}%

\subsection{The Transmission and Conversion Coefficients}
At this point, the transmission and conversion coefficients can be obtained from the solutions in $q$ and $p$ representation.  Comparing the incoming (negative $q$) and outgoing (positive $q$) values of $x_1$ gives the transmission coefficient $\tau$:
\begin{equation}
\tau = \frac{x_1(q)}{x_1(-q)} =  \left( \frac{q}{-q} \right)^{-i\tilde\eta^2} = (-1)^{-i\tilde\eta^2} =e^{-\pi\tilde\eta^2}.
\end{equation}
Because of the choice of branch cut that is needed to deal with the singularity at $q=0$, we take $-1=e^{-i\pi}$ in the above formula.  Calculation of the conversion coefficient $\beta$ is more difficult because of the singular nature of $x_2(q)$ at the origin, $q=0$.  However, the amplitude of $\beta$ can be calculated by action conservation:
\begin{align}
|\beta|^2 = 1 - |\tau|^2 = 1 - e^{-2\pi\tilde\eta^2}.
\end{align}
Getting the phase of $\beta$ correct is a more difficult calculation, which is performed in 
\cite{metaplectic_formulation}.  In that paper, $\beta$ is found to be given by 
\begin{align}
\beta = \frac{\sqrt{2\pi \tau}}{\tilde\eta \Gamma(-i|\tilde\eta|^2) } .
\end{align}
With these coefficients, we can write the scattering matrix as
\begin{align}
\left(
\begin{array}{c}
 x_1^{\text{out}}   \\
 x_2^{\text{out}}    
\end{array}
\right)
=
\left(
\begin{array}{cc}
 \tau  &  -\beta  \\
 \beta^*  & \tau  
\end{array}
\right)
\left(
\begin{array}{c}
 x_1^{\text{in}}   \\
 x_2^{\text{in}}    
\end{array}
\right), 
\end{align}
where $x_j^{\text{in}}$ and $x_j^{\text{out}}$ are the complex amplitudes of the WKB solutions to the left and right of the mode conversion.  This scattering matrix allows us to find the amplitude and phase for rays leaving the mode conversion, given the amplitude and phase on the incoming rays.

\section{Numerical Solution and Comparison with Theory\label{sec:numerical}}

\begin{figure}
\begin{center}
\includegraphics[scale=0.5]{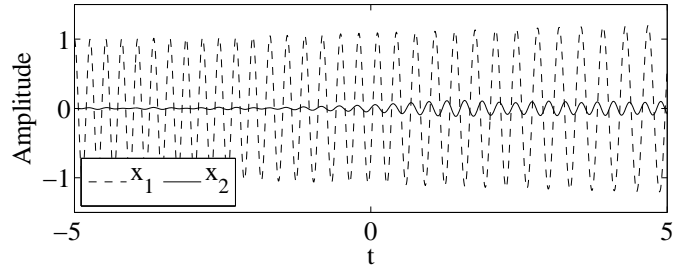}
\end{center}
\caption{\label{cap:nmc}
Numerical solution showing mode conversion between the two modes.
Simulation Parameters: $\omega_1(t) = 20-5\,\text{tanh}(t/5)$, $\omega_2(t) = 20+5\,\text{tanh}(t/5)$, $\mathbf {x}_0=(1,0)$, $\dot {\mathbf{\;\,x_0}}=(0,0)$
}
\end{figure}

The equations of motion in Equations (\ref{eq:osc1}) and (\ref{eq:osc2}), together with the appropriate initial conditions, can be solved numerically to find the output amplitudes $A_i^{\text{out}}$ and phase shifts $\phi_i$.  The ODE solver in {\sf MatLab} was used to perform this calculation.  The variation in the natural frequencies was taken to have a hyperbolic tangent form, as shown in Figure \ref{cap:nat}.  The numerical solution is shown in Figure \ref{cap:nmc}.

\comment{% I don't like this figure since it doesn't use the right matching
\begin{figure}
\begin{center}
\includegraphics[scale=0.5]{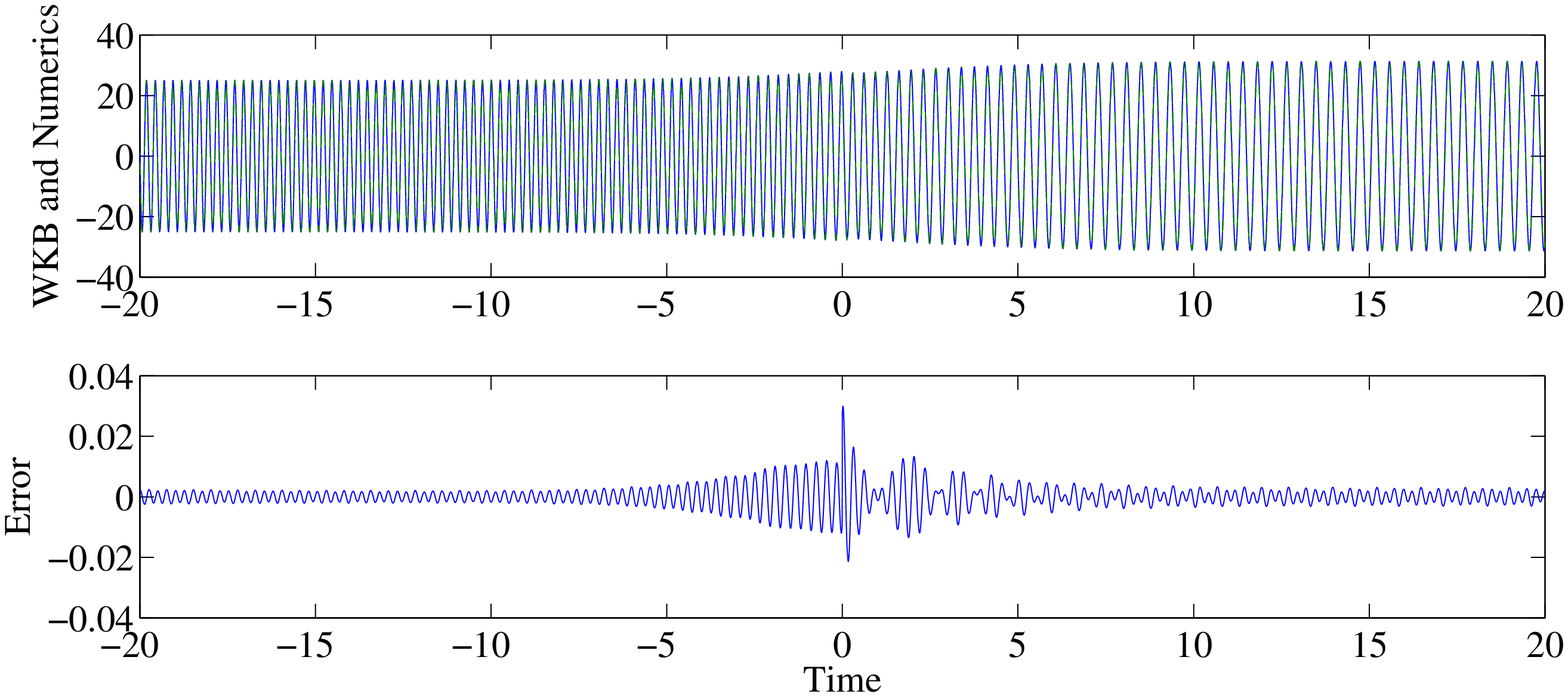}
\end{center}
\caption{\label{cap:err}A comparison of the WKB solution and the numerical output.  Note that the transmission coefficients (including phase shift) for the WKB solution shown here were estimated from the numerics.
}
\end{figure}
}%

\begin{figure}
\begin{center}
\includegraphics[scale=0.65]{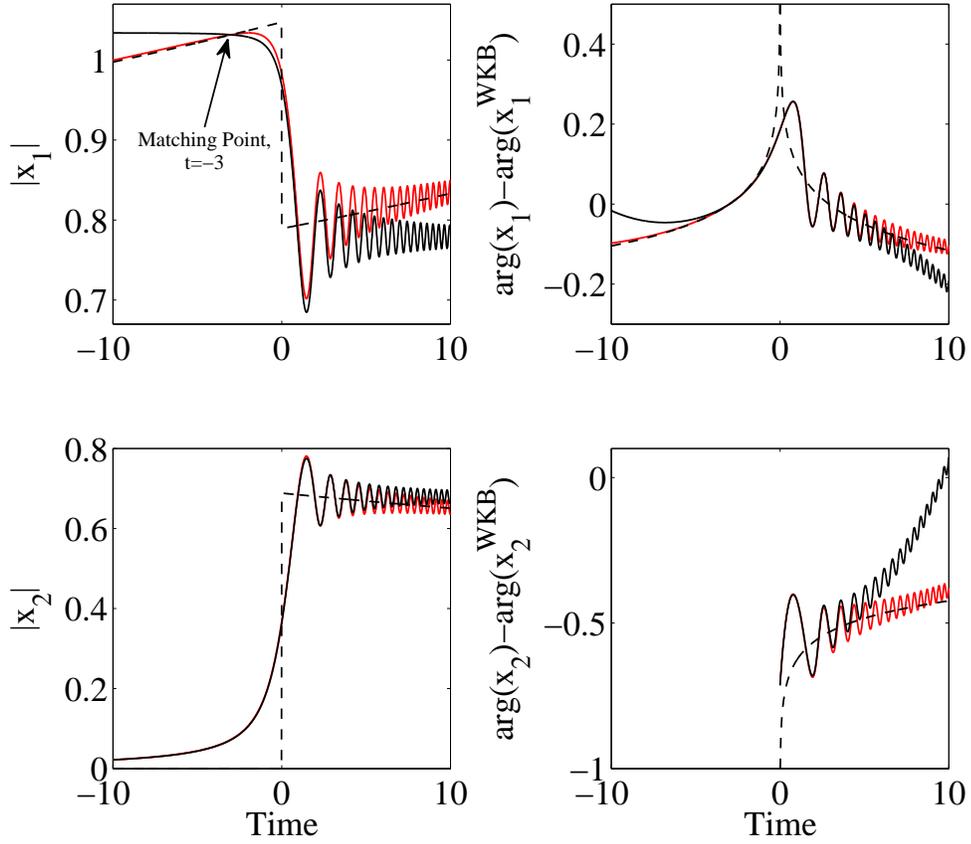}
\end{center}
\caption{\label{cap:compare}
Absolute value and relative phase of $x_1$ and $x_2$, from numerical solution (red), superimposed on the WKB values (dashed curves).  The exact parabolic cylinder functions (black) are used to set the amplitude and phase of the outgoing WKB solution.  The amplitude and phase of the local solution is matched to the incoming WKB solution at $t=-3$.  A phase mismatch of $\pi/4$ as been arbitrarily removed from the $x_2$, in order to better show the comparison between analytical and numerical solutions (see text for more details).  The simulation parameters for this figure are the same as in Figure (\ref{fig:compare_amp}).
}
\end{figure}

A more detailed comparison of the local solutions to the WKB solutions was also performed, using the linearized dispersion matrix in Equation (\ref{eq:disp_qp}) as a starting point.  The local solution, which is given by parabolic cylinder functions in the $t$ representation, is an exact solution to wave problem in this normal form.  We then added quadratic terms (such as $q^2$) to the normal form dispersion matrix, with small coefficients.  This will cause the numerical solution to deviate from the parabolic cylinder form.  Figure (\ref{cap:compare}) shows the results of one such calculation.  Far from the mode conversion, the WKB solution can correctly capture the effects of the quadratic terms, while near the mode conversion, the parabolic cylinder functions model the jumps in amplitude and phase.  The effects of these jumps are taken into account in the WKB solution shown in Figure (\ref{cap:compare}), by using the scattering coefficients $\tau$ and $\beta$ to modify the amplitude and phase at the origin.  For some reason, the phase of the second oscillator in our simulations was off from the expected phase by $\pi/4$.  We suspect that this could be due to the phase convention chosen for the metaplectic transformation in Equation (\ref{eq:q_to_t_metaplectic_transform}).  Other authors \cite{Brizard:1998rt} include a factor of $i^{-1/2}$ in the normalization of the metaplectic integral, which could be what we are seeing in these simulations.  Additionally, there is the possibility that this phase error is due to the fact that this calculation with oscillators uses functions of time, rather than functions of space as is the case for wave problems.  This leads to the opposite convention for the correspondence between derivatives and phase space variables, as shown in Equation (\ref{eq:p_operator}) and the subsequent discussion.

\section{Summary}

In this chapter we described how phase space techniques can be used to solve multicomponent wave problems, including problems which exhibit mode conversion.  We gave an example of a pair of coupled oscillators which undergo resonant conversion, and show how the phase space techniques can be applied to this problem.  The key idea is that the symbol of the wave operator (the dispersion matrix) is a smooth function on phase space.  We can then apply canonical transformations on phase space to simplify the form of the dispersion function.  Also, since the dispersion function is smooth, we can expand it about a point in phase space.  For the mode conversion problem, this local expansion gives an approximate wave equation, which can be solved to find the structure of the solution in the mode conversion region.  This local solution can then be matched onto WKB solutions, which are good approximate solutions far from the mode conversion.  This allows the mode conversion to be treated as a scattering problem, where the S-matrix connects the WKB solutions on opposite sides of the conversion.

In this chapter we also report numerical simulations, which show the matching between the local solution, the WKB solutions, and direct numerical simulation of the equations.  The matching shown in Figure (\ref{cap:compare}) is fairly good, giving the amplitude and phase jumps across the conversion.  However, higher order terms in the dispersion matrix have been neglected in the local analysis, which only used a linear approximation for the dispersion matrix.  The neglected terms cause the local solution to diverge from the WKB solution fairly rapidly.  In the next chapter, we will reintroduce the quadratic terms into the dispersion matrix, and derive a new local solution.  We will show that keeping these terms leads to much better matching between the corrected local solution and the far-field WKB solutions.

% end of comment: old chapter 3

%% file: Chapter-HigherOrder.tex
%%%%%%%%%%%%%%%%%%%%%%%%%%%%%%%%%%%%%%%%%%%%%%%%%%%%%%%%%%%%%%%%%%%%%%%%%%
%
% Ph.D. dissertation manuscript
% Chapter 3: Example of phase space calculations: Higher Order Corrections
%
% Andrew Stephen Richardson (Fall 2007)
% College of William and Mary
% Department of Physics
% Prof. Eugene Tracy, advisor
%
% Based on Paul King and Andrew Norman's template (modified by Wirawan Purwanto)
%
%%%%%%%%%%%%%%%%%%%%%%%%%%%%%%%%%%%%%%%%%%%%%%%%%%%%%%%%%%%%%%%%%%%%%%%%%%

\chapter{Higher Order Corrections\label{chp:higher_order}}

\section{Introduction and Motivation}

\begin{figure}[h]
\begin{center}
\includegraphics[scale=.46]{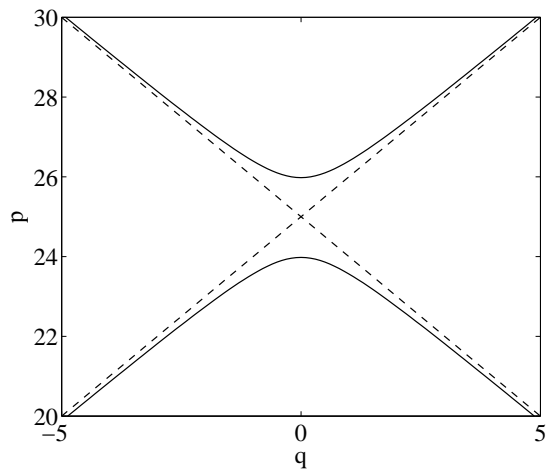}
\end{center}
\caption{\label{fig:crossing}
The phase space structure of a typical ``avoided crossing'' mode conversion.  The hyperbolic dispersion curves are the solid lines, and the dashed lines are the dispersion curves for the uncoupled modes.
}
\end{figure}

In previous work \cite{0741-3335-49-1-004, tracy:082102, metaplectic_formulation} it was shown that phase space ray-tracing techniques can be used to solve wave problems exhibiting mode conversion.  Such a problem can be written in matrix form.  We consider solutions at a given frequency, and write the two equations for the coupled wave modes together.
\begin{equation}\label{eq:wave}
\mathbf{\hat D}(q,-i\partial_q;\omega) \cdot \Psi(q) = 0
\end{equation}
While many problems of interest concern waves in multiple spatial dimensions, in this chapter we limit our analysis to the case where $q$ is one-dimensional.  Additionally, we will suppress the frequency dependence for brevity of notation.  Using the symbol calculus \cite{metaplectic_formulation,citeulike:703463}, we can define the symbol of the wave operator as a matrix valued function on wave phase space, $\mathbf{D}(q,p)$.  Here, the variable $p$ corresponds to the operator $-i\partial_q$, and products of operators are symmetrized (e.g., $qp \rightarrow -i(q \partial_q+\partial_q q)/2$).  In the vicinity of a mode conversion, there are two roots of the dispersion relation $\det(\mathbf{D}(q,p)) = 0$.  These two curves in phase space locally have a hyperbolic structure (an ``avoided crossing'', see figure \ref{fig:crossing}).  Linearizing the $q,p$-dependance of the dispersion matrix about the center of the hyperbola, and converting this linearized symbol back to an operator, gives a set of coupled equations which can be solved for the local wave fields.  Matching these local solutions onto uncoupled WKB solutions (which are a good approximation to the solutions far from the mode conversion region) gives transmission and conversion coefficients for the incoming and outgoing waves.  These coefficients can be used to treat the mode conversion as a ray-splitting process, where amplitude on the incoming ray is split onto the two types of outgoing rays.  

This ray-splitting approach captures the jump in amplitude caused by the coupling between the two modes at linear order in phase space variables.  However, higher order terms in the wave equation can lead to additional effects.  For example, the amplitude variation familiar from WKB theory is not captured by the linear solution.  This could cause difficulties when attempting to match the local wave fields onto the incoming and outgoing WKB solutions.  Figure (\ref{cap:compare}) illustrates this effect in the example of two coupled oscillators.  In this example, the local solution captures the jump in amplitude at the mode conversion, but misses the WKB amplitude variation.  This limits the matching region to a small range right near the mode conversion, which could make numerical ray-tracing algorithms somewhat unstable.  However, this example also suggests that there may be a simple correction to the linearized solution which will also capture the WKB amplitude variation.

In this chapter we consider the effect of adding generic quadratic terms to the wave equation.  These terms will modify the uncoupled dispersion relations, which will in turn modify the far-field WKB solutions for the incoming and outgoing waves.  When the new quadratic terms are added to the coupled equations, they will give us a new local solution for the wave fields.  This local solution will contain both non-propagating ``near-field'' contributions and phase modifications to the original, linear order, local solution.  The near-field terms do not propagate, and therefore do not modify the transmission and conversion coefficients.  The phase modifications, however, lead to phase corrections in the scattering coefficients at order $\epsilon |\eta|^2$.  Lastly, we give a comparison with numerical solutions for a simple example.

% end comment: local linearized solutions

\section{Extension to Higher Order}

In general, the Taylor's series expansion of the dispersion matrix in Equation (\ref{eq:taylor_matrix}) will contain terms of higher order than the linear terms kept in the approximation of the previous section.  These terms will have an effect on both the far-field WKB solutions, and on the local solution.  If we can calculate these effects, then we can obtain a better match of the local solution to the incoming and outgoing WKB solutions.  This will allow us calculate any corrections that there may be to the scattering coefficients.

\subsection{The Normal Form in One Spatial Dimension\label{sec:nf}}

As described in Section \ref{sec:normalized_coupling}, and shown in \cite{metaplectic_formulation,citeulike:784668}, the $2\times 2$ symbol of the dispersion matrix can be put into the following ``normal form'' at linear order:
\begin{equation}
\mathbf{D}_{\text{NF}}(q,p) = 
\left(
\begin{array}{cc}
-p & \eta \\
\eta^* & q
\end{array}
\right).
\end{equation}
Here, $\eta$ is a constant since we are working in a two-dimensional phase space.  The higher order corrections to this matrix appear at quadratic order in the phase space variables:
\begin{equation}
\mathbf{D}(q,p) = \mathbf{D}_{\text{NF}}(q,p) + 
\mathbf{D}_{2}(q,p)
 + \mathcal{O}(z^3).
\end{equation}
Each element of $\mathbf{D}_{2}$ can contain terms which are quadratic in the phase space variables $z=(q,p)$.

In this section, we will argue that it should be possible to perform a phase space dependent change of basis which puts the dispersion matrix into the form
\begin{equation}\label{eq:nf_diagonals}
\mathbf{D}'(z) = \mathbf{Q}^\dagger(z) \cdot \mathbf{D}(z) \cdot \mathbf{Q}(z) =
\mathbf{D}_{\text{NF}}(z) + 
\left(
\begin{array}{cc}
d^{(2)}_a(z) & 0 \\
0 & d^{(2)}_b(z)
\end{array}
\right) + \mathcal{O}(z^3) .
\end{equation}
This would imply that, in general, we only need to consider the effect of additional quadratic terms on the diagonals of $\mathbf{D}$, but not on the off-diagonals.

We are interested in removing the second order terms from the off-diagonals of $\mathbf{D}$.  We can assume that $\mathbf{D}$ is a hermitian matrix, so we will write the off-diagonal terms of $\mathbf{D}_2$ as
\begin{equation}
\mathbf{D}_2(q,p)=
\left(
\begin{array}{cc}
 \,  & d_1  \\
 d_1^*  &  \, 
\end{array}
\right) p^2 + 
\left(
\begin{array}{cc}
 \,  & d_2  \\
 d_2^*  &  \, 
\end{array}
\right) qp + 
\left(
\begin{array}{cc}
 \,  & d_3  \\
 d_3^*  &  \, 
\end{array}
\right) q^2.
\end{equation}
We will use a near-identity change of polarization basis, so we can write $\mathbf{Q}$ as
\begin{equation}
\mathbf{Q}(q,p) = \mathbf{1} + \mathbf a p + \mathbf b q.
\end{equation}
Some algebra shows that the matrices $\mathbf a$ and $\mathbf b$ should have the following form in order to remove the second order terms from the off-diagonals:
\begin{equation}
\mathbf a = 
\left(
\begin{array}{cc}
e^{i\phi}\sqrt{\alpha d_1^*}   &  d_1 \\
 \alpha  &  - e^{-i\phi}\sqrt{\alpha^* d_1}
\end{array}
\right),
\end{equation}
\begin{equation}
\mathbf b = 
\left(
\begin{array}{cc}
e^{i\phi}\sqrt{d_3^*(\alpha+d_2^*)}   &  d_2+\alpha^* \\
 -d_3^*  &  - e^{-i\phi}\sqrt{d_3(\alpha^*+d_2)}
\end{array}
\right).
\end{equation}
Here, $\phi$ is the phase of the coupling, $\eta = \vert \eta \vert e^{i\phi}$, and $\alpha$ is given by
\begin{equation}
\alpha=-\frac{d_2}{2} \pm \frac{1}{2} \sqrt{d_2^2 -4 d_1 d_3} .
\end{equation}
This transformation will make the off-diagonals in $\mathbf{D}'$ constants plus terms starting at $\mathcal{O}(z^3)$.  

In order to obtain the form of $\mathbf{D}$ given in equation (\ref{eq:nf_diagonals}), we need to correct the first order terms in the diagonals, since the change of basis will mix $q$ and $p$.  A canonical transformation of phase space will recover the normal form at linear order:
\begin{equation}
\left(
\begin{array}{c}
q'  \\
p'     
\end{array}
\right)
=\frac{1}{\vert\mathcal{B}\vert^{1/2}}
\left(
\begin{array}{cc}
   1-2\Re (\alpha \eta^*) &  2\Re (d_1 \eta^*)   \\
  2\Re (d_3 \eta^*)  & 1+2 \Re (\alpha \eta^*)  
  \end{array}
\right)
\left(
\begin{array}{c}
q  \\
p     
\end{array}
\right) .
\end{equation}
Here the normalization is introduced to make $(q',p')$ a canonical pair.  It is given by the Poisson bracket of the transformed diagonal elements,
\begin{equation}
\mathcal{B} = \{D_{11}',D_{22}'\} = \left(1 - 4 (\Re (\alpha \eta^*))^2 - 4\Re (d_1 \eta^*)\Re (d_3 \eta^*) \right) \{q,p\} = \{q,p\} (1+\mathcal{O}(\epsilon^2)).
\end{equation}

\subsection{Moyal Corrections for Phase Space Dependent Changes of Polarization\label{sec:moyal}}

In the previous section, a phase space dependent change of polarization is used to put the dispersion matrix into the form where its off diagonal elements are constants with a small perturbation that starts at order $z^3$.  This transformation is achieved through the conjugation
\begin{equation}
\mathbf{D}'(z) = \mathbf{Q}^\dagger(z) \cdot \mathbf{D}(z) \cdot \mathbf{Q}(z).
\end{equation}
However, since the matrices $\mathbf{D}$ and $\mathbf{Q}$ are actually matrix valued symbols of operators, we really need to use the Moyal star product to multiply the elements of the matrix.  The noncommutative star product is used in the symbol calculus so that the symbols (functions on phase space) maintain the commutation relations of the original operators.  So, we should actually use the expression
\begin{equation}
\mathbf{D}'(z) = \mathbf{Q}^\dagger(z) * \mathbf{D}(z) * \mathbf{Q}(z),
\end{equation}
where the matrices are multiplied in the usual way, but the elements of the matrices are multiplied using the star product.  In this section we will argue that the effect of the star product is to introduce additional terms into the expression for $\mathbf{D}'(z)$ which are of higher order than those that we are considering in this dissertation, and therefore can be neglected.

The Moyal star product of two symbols $A(z)$ and $B(z)$ is often written as a formal power series (see \cite{citeulike:703463} with $\hbar$ set to 1 since we are studying classical fields).  In general, this series can contain infinitely many terms:
\begin{align}\label{eq:moyal}
A(z) \!* \!B(z) &= A(z) \exp\left( \frac{i}{2}  \frac{\overleftarrow \partial}{\partial z_\alpha} J_{\alpha\beta} \frac{\overrightarrow \partial}{\partial z_\beta} \right) B(z) \\
& = A(z) \sum_{n=0}^\infty \frac{1}{n!}  \left( \frac{i}{2}  \frac{\overleftarrow \partial}{\partial z_\alpha} J_{\alpha\beta} \frac{\overrightarrow \partial}{\partial z_\beta} \right)^n  B(z) \\
& = A(z) B(z) + \frac{i}{2} \{A,B\} -\frac{1}{8} \left( \partial^2_q A \, \partial^2_p B -2\, \partial_q\partial_p A \, \partial_q\partial_p B + \partial^2_p A \, \partial^2_q B \right)  + \ldots
\end{align}
\comment{%
& = A(z) \cdot B(z) + \frac{i}{2} \{A,B\} -\frac{1}{8} \left( \{\partial_q A, \partial_p B \} -  \{\partial_p A, \partial_q B \}  \right)  + \ldots
\end{align}
}%
Here, $J_{\alpha \beta}$ is the symplectic matrix.  In our problem, we have an implicit small parameter because of the asymptotic nature of the calculation. This allows us to truncate this series at some order to obtain an approximation for the star product.  

We are assuming that the symbol of the dispersion matrix has a well defined Taylor's series in the mode conversion region, and that the lowest order terms in the series dominate.  This can be expressed mathematically by introducing a small parameter $\epsilon$.  Using the multi-index notation, we can expand the symbol in terms of all possible monomials:
\begin{equation}
A(z) = \sum_{|m| \ge 0} a_m (\epsilon z)^m = \sum_{M=0}^\infty \epsilon^M \sum_{|m| = M} a_m z^m .
\end{equation}
The ordinary product of two symbols is given by
\begin{eqnarray}
A(z) B(z) &=& \left(\sum_{M=0}^\infty \epsilon^M \sum_{|m| = N} a_m z^m\right) \cdot \left(\sum_{N=0}^\infty \epsilon^N \sum_{|n| = N} b_n z^n \right) \\
&=& \sum_{N,M=0}^\infty \epsilon^{M+N}      \sum_{|m| = M} \sum_{|n| = N} a_m b_n \, z^{m+n}\\
&=& \sum_{|l| \ge 0} \kappa_l \, (\epsilon z)^l .
\end{eqnarray}
The star product is given by a similar formula, where $\cdot$ is replaced by $*$:
\begin{equation}
A(z) * B(z) = \sum_{N,M=0}^\infty \epsilon^{M+N}  
    \sum_{|m| = M} \sum_{|n| = N} a_m b_n \, z^m * z^n .
\end{equation}
So we need to consider the star product of two generic monomials, $z^m * z^n$.  Because the star product can be written as a series in powers of derivatives, we can express the product as a polynomial of degree $L = |m|+|n|$.  Although all of the coefficients of this polynomial can be calculated from equation (\ref{eq:moyal}), we compute only the highest power:
\begin{equation}
z^m * z^n = z^{m+n} + \sum_{|l|=0}^{L-1} c_l z^l .
\end{equation}
This equation is significant for us because it means that the star product only introduces additional monomials of degree less than the degree of the ordinary product.
We can use this equation to write
\begin{align}
A(z) * B(z) &= A(z) B(z) + \sum_{N,M=0}^\infty \epsilon^{N+M}  
    \sum_{|m| = M} \sum_{|n| = N} a_m b_n  \sum_{|l|=0}^{L-1} c_l z^l     \\
&= A(z) B(z) + \sum_{L=0}^\infty \epsilon^{L}  
   \,  \sum_{|l|=0}^{L-1} c'_l z^l  \\
&= A(z) B(z) + \epsilon \left(\sum_{L=0}^\infty
 \sum_{|l|=0}^{L-1} \epsilon^{L-1-|l|}  \, c'_l \,  (\epsilon z)^l  \right)  .
\end{align}
This means that the corrections due to the star product are at least one order higher in epsilon than the regular commutative product, since the coefficients of $(\epsilon z)^l$ in the series for the corrections are at least $\mathcal{O}(\epsilon)$ while the coefficients in the series for $A\cdot B$ are $\mathcal O (1)$.

\subsection{Quadratic Order Terms}

As shown in the Section \ref{sec:nf}, the quadratic approximation to the dispersion matrix can be put into the form of equation (\ref{eq:nf_diagonals}), which, with the introduction of the small parameter $\epsilon$ in the quadratic terms, is
\begin{equation}\label{eq:disp_matrix}
\mathbf{D}(q,p)=
\left(
\begin{array}{cc}
-p + \epsilon (a_1 p^2 + b_1 pq + c_1 q^2) &  \eta  \\
 \eta^* & q   + \epsilon  (c_2 p^2 + b_2 pq + a_2 q^2)
\end{array}
\right).
\end{equation}
This second order matrix valued function on phase space can now be converted into a pair of coupled second order differential equations for the two modes:  
\begin{equation}\label{eq:2nd_order_wave_eqn}
\left(
\begin{array}{cc}
\widehat D_{11} &  \eta  \\
 \eta^* & \widehat D_{22}
\end{array}
\right)
\left(
\begin{array}{c}
\psi_1(q) \\
\psi_2(q)
\end{array}
\right) = 0,
\end{equation}
where the operators on the diagonals are 
\begin{align}
\widehat D_{11} = i\partial_q + \epsilon (a_1 (-i\partial_q)^2 -\frac{i b_1}{2} (\partial_q q + q\partial_q) + c_1 q^2),
\end{align}
and
\begin{align}
\widehat D_{22} = q   + \epsilon  (c_2 (-i\partial_q)^2 -\frac{i b_2}{2} (\partial_q q + q\partial_q)  + a_2 q^2).
\end{align}
Because of the form of these equations, it will be convenient to analyze the lower channel in Fourier space.  The Fourier transform of these equations gives the $p$-representation of the equations:
\begin{equation}\label{eq:2nd_order_wave_eqn_p_rep}
\left(
\begin{array}{cc}
\widehat D'_{11} &  \eta  \\
 \eta^* & \widehat D'_{22} 
\end{array}
\right)
\left(
\begin{array}{c}
\phi_1(p) \\
\phi_2(p)
\end{array}
\right) = 0,
\end{equation}
which has diagonal elements
\begin{align}
\widehat D'_{11} = -p + \epsilon (a_1 p^2 +\frac{i b_1}{2} (p\partial_p + \partial_p p) + c_1 (i\partial_p)^2),
\end{align}
and
\begin{align}
\widehat D'_{22} =  i\partial_p + \epsilon(a_2(i\partial_p)^2 +\frac{i b_2}{2}(p\partial_p + \partial_p p) +c_2 p^2 ).
\end{align}

We proceed by first finding the new, uncoupled, far-field WKB solutions, and then finding the coupled local solution.  Finally, we examine the local solution and match it onto the WKB solutions, which gives us the order $\epsilon$ corrections to the scattering coefficients.

\subsection{Uncoupled WKB Modes}

Because of the quadratic terms in the dispersion matrix [Equation (\ref{eq:disp_matrix})], the uncoupled WKB solutions will contain modifications of order $\epsilon$.  These corrections can be found by solving Equation (\ref{eq:2nd_order_wave_eqn}) [or, if we use the $p$-representation, Equation (\ref{eq:2nd_order_wave_eqn_p_rep})] with $\eta=0$.  

First, consider the second order derivatives, $-\epsilon a_1\partial_q^2$ and $-\epsilon c_2\partial_q^2$.  These terms introduce new solutions which have the form $e^{iq/a_1 \epsilon}$ and $e^{iq/c_2 \epsilon}$.  The dispersion surfaces for these solutions are far ($p \approx 1/a_1 \epsilon$) from the solution that we are interested in ($p \approx 0$).  The limit $\epsilon \rightarrow 0$ is a singular limit, since these new solutions cease to exist for $\epsilon=0$.  However, since these solutions are related to dispersion surfaces which are well separated in phase space from the mode conversion we are studying, we can treat the new solutions separately.  Effectively, we can ignore these solutions in the limit $\epsilon \rightarrow 0$, and only consider the solutions related to the mode conversion.

Neglecting the second order derivatives, the upper channel of equation (\ref{eq:2nd_order_wave_eqn}) becomes
\begin{equation}\label{eq:uncoupled_equation}
(1-\epsilon b_1 q )\left(i\partial_q\psi_1(q)\right) + \epsilon \left(\frac{-ib_1}{2} + c_1 q^2\right) \psi_1(q) = 0 .
\end{equation}
This is a logarithmic derivative of $\psi_1(q)$, which can be integrated to give us, up to an overall amplitude from the constant of integration, the solution to order $\epsilon$:
\begin{equation}\label{eq:uncoupled_WKB}
\psi_1(q) = \exp \left(\frac{\epsilon b_1 q}{2} + \frac{i \epsilon c_1 q^3}{3}  \right) =
 e^{i \epsilon c_1 q^3 / 3} \left(1+\frac{\epsilon b_1 q}{2} + \mathcal{O}(\epsilon^2) \right).
\end{equation}
These two terms have a very nice correspondence to standard WKB theory, which can be seen when we relate them back to the uncoupled dispersion function $D_{11}(q,p)$.  The WKB phase is given by the integral of the ``momentum'' $\mathcal{P}(q)$, which is the function which solves $D_{11}(q,\mathcal{P}(q))=0$.  Solving this to order $\epsilon$ gives $\mathcal{P}(q) = D_{11}(q,0) = \epsilon c_1 q^2$, which is the first term from the Taylor's series expansion of $D_{11}(q,p)$ about $p=0$.  This integrates up to give exactly the phase which appears in equation (\ref{eq:uncoupled_WKB}).

As in Equation (\ref{eq:WKB_amp}), the WKB amplitude comes from the next term in the expansion of $D_{11}(q,p)$ about $p=0$,
\begin{equation}
A(q) = \left| \frac{\partial D_{11} (q,p)}{\partial p} \right|_{p=0}^{-1/2} = |-1+\epsilon b_1 q|^{-1/2}.
\end{equation}
If we expand this amplitude as a series in $\epsilon$ it gives the same order $\epsilon$ correction to the amplitude which appears in equation (\ref{eq:uncoupled_WKB}).

An analogous calculation can be performed for the second channel in the $p$-representation.  Starting with equation (\ref{eq:2nd_order_wave_eqn_p_rep}) and setting $\eta$ to zero gives us a similar solution, except for the sign of the $b$ coefficient.  Our uncoupled WKB modes are then
\begin{equation}\label{eq:corrected_solution}
\psi_1(q) = \frac{e^{i\epsilon c_1 q^3/3}  }{\sqrt{1-\epsilon b_1 q}}  , \qquad 
\phi_2(p) = \frac{e^{i\epsilon c_2 p^3/3}  }{\sqrt{1+\epsilon b_2 p}} .
\end{equation}
These solutions will be used to match the incoming and outgoing WKB waves onto our local solution.

It remains to check that we do not introduce errors at order $\epsilon$ by neglecting the second order derivatives.  This check is straightforward since equation (\ref{eq:uncoupled_equation}) can be used to write the derivative of $\psi_1(q)$ as
\begin{equation}
\partial_q \psi_1(q) = \epsilon f(q) \psi_1(q).
\end{equation}
Since this derivative will be multiplied by $\epsilon a_1$ in the second order term, we have that the effect of this term is at least order $\epsilon^2$ when acting on our solution in (\ref{eq:corrected_solution}).  The same analysis holds for the second order derivative acting on the lower channel.

\begin{figure}
\begin{center}
\includegraphics[scale=0.78]{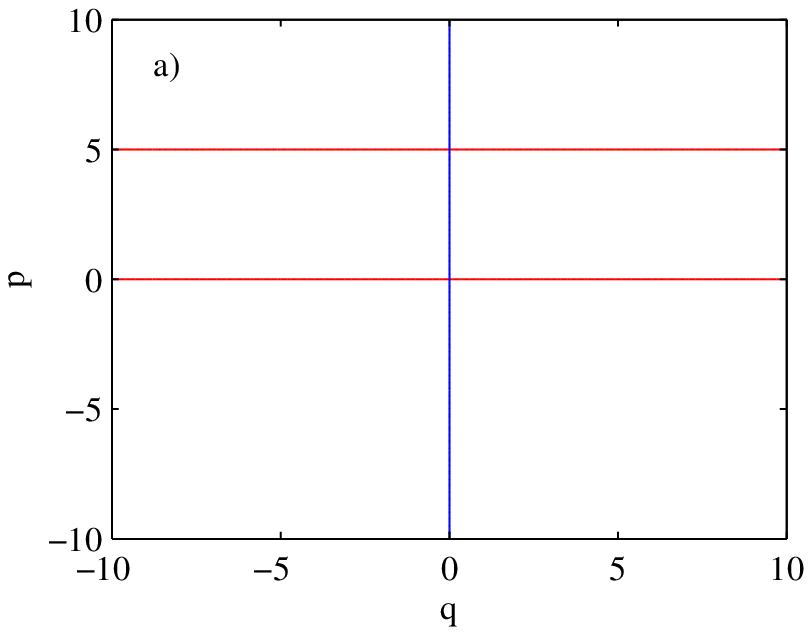}
\includegraphics[scale=0.78]{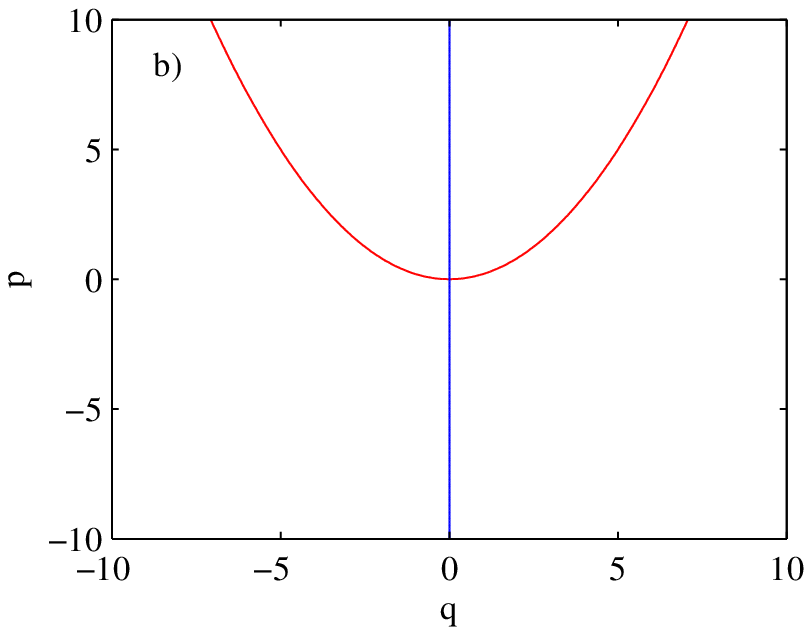}
\includegraphics[scale=0.78]{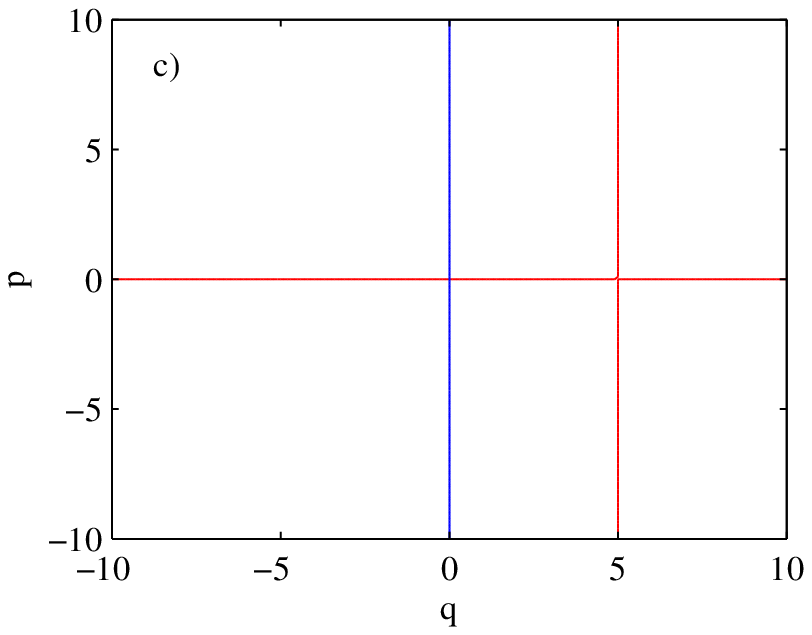}
\includegraphics[scale=0.78]{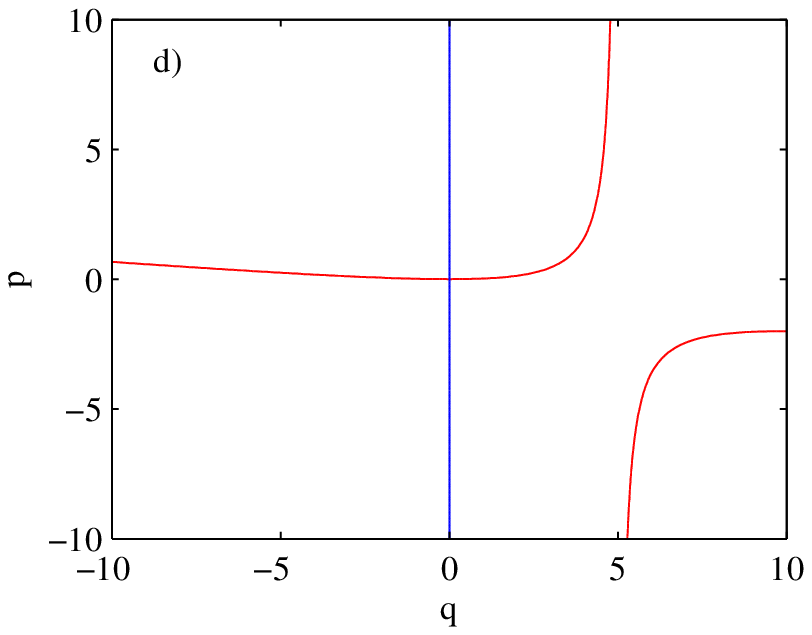}
\end{center}
\caption{\label{fig:dispersion_higher}
Dispersion curves for the uncoupled modes, showing the effect of the higher order terms ($\epsilon = 1/5$).  The red curves are solutions to $D_{11}=0$, and the blue curves are solutions to $D_{22}=0$.  The original crossing is at $q=p=0$.  The parameter values used for these plots are: (a) $a_1 = 1$:  This term introduces a new branch to the dispersion surface $D_{11}=0$ at $p=5$.  As $\epsilon\rightarrow 0$, this branch moves off to infinity.   (b) $c_1 = 1$:  This term introduces a curvature into the $D_{11}$ dispersion surface.  In the $p$ representation, an uncoupled solution in the upper channel would now look like an Airy function.  (c) $b_1 = 1$:  Although not obvious from the dispersion curves, this term introduces an amplitude variation which has a square-root form, and which matches the WKB amplitude variation due to action conservation.  (d) $b_1 = 1$, $c_1 = 0.1$:  In general, there will be several higher order terms, all of which affect the dispersion surfaces.
}
\end{figure}

\subsection{Coupled WKB Modes \label{sec:coupled_WKB}}

The uncoupled modes are generated by the eigenvalues of the dispersion matrix when we set the coupling parameter $\eta$ to zero.  If the dispersion matrix is in normal form, then the diagonals of the matrix can be interpreted as the generators of the uncoupled modes.  If, on the other hand, we are interested in the coupled modes, then we can use the eigenvalues of the dispersion matrix to generate the modes.  This is equivalent to using the determinant as the generator, since the diagonalized matrix has the eigenvalues on the diagonals.  So, in order to see the effect of the coupling on the WKB modes, we need to calculate the determinant of the dispersion matrix with $\eta \neq 0$.  First, we recall how this works in the case of the linear approximation to the dispersion matrix, and then extend the calculation to include the quadratic order terms.  We will need these expressions for the coupled WKB modes when we want to match incoming and outgoing rays across the mode conversion region.

The determinant of the linear part of the dispersion matrix is
\begin{equation}
\det \mathbf{D}(q,p) = -pq -|\eta|^2.
\end{equation}
With $\eta = 0$, solving for $\det \mathbf{D}(q,p) = 0$ gives the two uncoupled modes $p=0$ and $q=0$.  When we bring the coupling back in, then solving for $p$ as a function of $q$ gives us
\begin{equation}
p(q) = \frac{-|\eta|^2}{q}.
\end{equation}
This function goes into the phase of the WKB solution
\begin{equation}
\psi_1(q) = A_1(q) \exp \left( i \int^q p(q') \,dq'  \right) = A_1(q) \exp (-i|\eta|^2 \ln (q)).
\end{equation}
This logarithmic phase is what allows us to match the WKB solutions onto the local solution of the linearized equations.

When we include the second order terms in the dispersion matrix, we get 
\begin{equation}\label{eq:quad_det}
\det \mathbf{D}(q,p) = (-p + \epsilon \mathcal{D}_1(q,p))(q+\epsilon \mathcal{D}_2(q,p)) -|\eta|^2 =0.
\end{equation}
Here, $\mathcal{D}_1(q,p)$ and $\mathcal{D}_2(q,p)$ are the second order terms from the dispersion matrix
\begin{equation}
\mathcal{D}_1(q,p) = a_1 p^2 +b_1qp + c_1 q^2, \qquad
\mathcal{D}_2(q,p) = a_2 q^2 +b_2qp + c_2 p^2.
\end{equation}
We now want to solve this equation for $p(q)$.  Expand $p$ in powers of $\epsilon$
\begin{equation}
p(q) = p_0(q) + \epsilon p_1(q) + \ldots = \frac{-|\eta|^2}{q} + \epsilon p_1(q) + \ldots, 
\end{equation}
and insert this into equation (\ref{eq:quad_det}).  Keeping terms of order $\epsilon$, and solving for $p_1(q)$, we get
\begin{eqnarray}
p_1(q) &=& \frac{1}{q} \big( -p_0 \mathcal{D}_2(q,p_0) + q\mathcal{D}_1(q,p_0) \big) \\
&=& \frac{|\eta|^2}{q^2} \mathcal{D}_2(q,p_0) + \mathcal{D}_1(q,p_0) \\
&=& c_1 q^2 + |\eta|^2 (a_2 -b_1) + \mathcal{O}(|\eta|^4).
\end{eqnarray}
The term $c_1 q^2$ comes from the curvature of the uncoupled dispersion surface for the upper channel.  The additional two terms are effects of the coupling.  They enter the WKB solution at order $\epsilon |\eta|^2$:
\begin{align} \label{eq:wkb_coupled_1}
\psi_1(q) &= A_1(q) \exp \left( i \int^q p_0(q') + \epsilon p_1(q') \,dq'  \right) \\
&= A_1(q) \exp \left\{-i|\eta|^2 \ln (q) +i\epsilon\left(\frac{c_1 q^3}{3} + |\eta|^2 (a_2 -b_1) q \right)\right\}.
\end{align}
Here, the amplitude is computed from the derivative of the eigenvalue $D_\alpha$ which asymptotically approaches $D_{11}$, and which is the generator for ray evolution in the upper channel:
\begin{equation}
A_1(q) = \left| \frac{\partial D_{\alpha} (q,p)}{\partial p} \right|_{p=p(q)}^{-1/2} = \left|-1+\epsilon b_1 q + \frac{|\eta|^2}{q^2} + \ldots \right|^{-1/2}.
\end{equation}

The calculation for the coupled WKB mode in the lower channel proceeds along similar lines, and uses the $p$ representation of the equations.  The result is
\begin{equation}
\phi_2(p) = A_2(p) \exp \left\{-i|\eta|^2 \ln (p) +i\epsilon\left(\frac{c_2 p^3}{3} + |\eta|^2 (a_1 -b_2) p \right)\right\},
\end{equation}
with
\begin{equation}
A_2(p) = \left|1+\epsilon b_2 p + \frac{|\eta|^2}{p^2} + \ldots \right|^{-1/2}.
\end{equation}

We will use these expressions for the coupled WKB solutions when matching the local solution to the propagating modes.

\subsection{Local Coupled Solutions \label{sec:local_sol}}

Now that we know the form of the incoming and outgoing WKB waves, we need to solve the system of equations locally in order to find the scattering coefficients which connect the incoming to outgoing waves.  
We will find the higher order corrections by expanding the local fields in $\epsilon$, and then using the $\mathcal{O}(\epsilon)$ equations to get the local solution.
\begin{equation} \label{eq:psi1_q}
\psi_1(q) = q^{-i|\eta|^2} (1 + \epsilon \Theta_1(q) + \mathcal{O}(\epsilon^2) )
\end{equation}
\begin{equation} \label{eq:psi2_q}
\psi_2(q) = -\eta^* q^{-i|\eta|^2-1} (1 + \epsilon \Theta_2(q) + \mathcal{O}(\epsilon^2) )
\end{equation}
Because of the form of the equations, it is convenient to expand $\Theta_1$ and $\Theta_2$ as power series in $q$.
\begin{equation}
\Theta_1(q) = \sum_{n=-\infty}^\infty s_n q^n, \quad \Theta_2(q) = \sum_{n=-\infty}^\infty \tilde s_n q^n
\end{equation}
 
In order to compact the notation, and simplify the calculations, define $\beta_n$ as the coefficient obtained when Fourier transforming an arbitrary term in the series:
\begin{equation}\label{eq:fourier_mellin}
\int dq\, e^{-ipq} q^{-i|\eta|^2} q^n \equiv \beta_n p^{i|\eta|^2 - 1} p^{-n}.
\end{equation}
This definition is possible since it can be shown that 
\begin{align}
\int dq\, e^{-ipq} q^\alpha  \propto  p^{-\alpha -1}.
\end{align}
The inverse transformation will also involve $\beta_n$:
\begin{equation}
\int dp\, e^{ipq} p^{i|\eta|^2} p^{-n} = \frac{1}{\beta_{n-1}} q^{-i|\eta|^2 -1} q^n.
\end{equation}
Notice how positive and negative powers of $q$ and $p$ exchange roles.  This is most likely due to properties of the dilation group, since these functions are related to representations of that group.

Evaluating the Fourier integral by using the Hankel formula for the gamma function, we can find $\beta_0$, as in \cite{metaplectic_formulation}.  By analytic continuation of the complex gamma function, we find, for any integer $n$,
\begin{equation}
\beta_n \equiv \frac{-2\pi (-i)^{i|\eta|^2-n} }{\Gamma(i|\eta|^2-n)}
\end{equation}
We will also need ratios of $\beta$'s, which can be found using the properties of the gamma function.
\begin{equation}\label{eq:beta_ratio}
\frac{\beta_n}{\beta_{n-1}} = i (i|\eta|^2 - n)
\end{equation}
Our equations involve the operators $\hat p$, $\hat p^2$, $\hat q^2$, and $\widehat{pq}$.\footnote{By $\widehat{pq}$ we mean the operator whose Weyl symbol is $pq$.  Since $p$ and $q$ commute as variables in phase space, we could also have written this operator as $\widehat{qp}$}  Using the definitions above, we can evaluate the action of these operators for an arbitrary term in our series.  
\begin{align}
\hat p \, q^{-i|\eta|^2 + n} &= \hat p \, \beta_n \int dp\, e^{ipq} p^{i|\eta|^2-n-1} \\
&= \beta_n \int dp\, e^{ipq} p^{i|\eta|^2-n} \\
&= \frac{\beta_n}{\beta_{n-1}} q^{-i|\eta|^2 +n -1}
\end{align}
Applying this formula twice gives us
\begin{equation}
\hat p^2 \, q^{-i|\eta|^2 + n} = \frac{\beta_n}{\beta_{n-2}} q^{-i|\eta|^2 +n -2}.
\end{equation}
Similar calculations yield
\begin{equation}
\widehat{pq} \, q^{-i|\eta|^2 + n} = \left( -\frac{i}{2} + \frac{\beta_n}{\beta_{n-1}} \right) q^{-i|\eta|^2 + n}.
\end{equation}
and
\begin{equation}
\hat q^2 \, q^{-i|\eta|^2 + n} = q^{-i|\eta|^2 + n+2}
\end{equation}

We can now write out our system of equations in (\ref{eq:2nd_order_wave_eqn}) using the coefficients $\beta_n$ and our series expansions for the fields.  Keeping only terms of order $\epsilon$, we get
\begin{equation}
-\hat{p} \sum_n s_n q^{-i|\eta|^2 + n} + \mathcal{D}_1(\hat q, \hat p) q^{-i|\eta|^2} - |\eta|^2\sum_n \tilde s_n q^{-i|\eta|^2 -1 + n} =0
\end{equation}
and
\begin{equation}
\sum_n s_n q^{-i|\eta|^2 + n} - \mathcal{D}_2(\hat q, \hat p) q^{-i|\eta|^2-1} - q\sum_n \tilde s_n q^{-i|\eta|^2 -1 + n} =0
\end{equation}
Now evaluate the operators using the expressions above.
\begin{multline}
\sum_n \left( \frac{-\beta_n}{\beta_{n-1}} s_n - |\eta|^2 \tilde s_n \right) q^{-i|\eta|^2 -1 + n} \\
+ a_1 \left(\frac{\beta_0}{\beta_{-2}}\right) q^{-i|\eta|^2 -2} 
+ b_1\left( -\frac{i}{2} + \frac{\beta_0}{\beta_{-1}} \right) q^{-i|\eta|^2}
+ c_1 q^{-i|\eta|^2 +2} =0
\end{multline}
\begin{multline}
\sum_n \left( s_n - \tilde s_n \right) q^{-i|\eta|^2 + n} \\
- c_2 \left(\frac{\beta_{-1}}{\beta_{-3}}\right) q^{-i|\eta|^2 -3} 
 - b_2\left( -\frac{i}{2} + \frac{\beta_{-1}}{\beta_{-2}} \right) q^{-i|\eta|^2-1}
- a_2 q^{-i|\eta|^2 +1} =0
\end{multline}
These equations can now be solved for the coefficients $s_n$ and $\tilde s_n$.  For $n \not\in \{-3,-1,1,3 \}$ these can be solved to find $s_n = \tilde s_n =0$.  Otherwise, we have a set of linear equations which can be solved to give, after generous application of equation (\ref{eq:beta_ratio}) to simplify the coefficients,
\begin{eqnarray}
s_3 &=&  \frac{ic_1}{3} \\
\tilde s_3 &=& \frac{ic_1}{3} \\
s_1&=&  i(b_1-a_2)\frac{\beta_0}{\beta_{-1}} +\frac{b_1}{2} \\
\tilde s_1 &=&  i(b_1-a_2)\frac{\beta_1}{\beta_{0}} -\frac{b_1}{2} \\
s_{-1} &=&  i(b_2-a_1) \frac{\beta_0}{\beta_{-2}} +\frac{b_2}{2}  \frac{\beta_0}{\beta_{-1}} \\
\tilde s_{-1} &=& i(b_2-a_1) \frac{\beta_0}{\beta_{-2}} -\frac{b_2}{2} \frac{\beta_{-1}}{\beta_{-2}} \\
s_{-3} &=& \frac{ic_2}{3} \frac{\beta_0}{\beta_{-3}} \\
\tilde s_{-3} &=& \frac{i c_2}{3} \frac{\beta_{-1}}{\beta_{-4}}
\end{eqnarray}

Applying the Fourier transform in equation (\ref{eq:fourier_mellin}), we can find the local solutions in the $p$ representation. 
\begin{eqnarray}
\phi_1(p) &=& \int dq\, e^{-ipq} \psi_1(q) \\
&=& \beta_0 p^{i|\eta|^2-1} \left( 1 + \epsilon \sum_n \sigma_n p^n + \mathcal{O}(\epsilon^2) \right)
\end{eqnarray}
\begin{equation}
\phi_2(p) = -\eta^* \beta_{-1} p^{i|\eta|^2} \left( 1+ \epsilon \sum_n \tilde\sigma_n p^n+ \mathcal{O}(\epsilon^2) \right)
\end{equation}
The terms in the series are
\begin{equation}
\sigma_n \equiv s_{-n} \frac{\beta_{-n}}{\beta_0}, \quad \tilde\sigma_n \equiv \tilde s_{-n} \frac{\beta_{-n-1}}{\beta_{-1}},
\end{equation}
which means that the coefficients can be calculated explicitly to give
\begin{eqnarray}
\sigma_{-3} &=&  \frac{ic_1}{3} \frac{\beta_{3}}{\beta_0} \\
\tilde \sigma_{-3} &=& \frac{ic_1}{3} \frac{\beta_{2}}{\beta_{-1}}\\
\sigma_{-1}&=&  i(b_1-a_2)\frac{\beta_1}{\beta_{-1}} +\frac{b_1}{2} \frac{\beta_{1}}{\beta_0}\\
\tilde \sigma_{-1} &=&  i(b_1-a_2)\frac{\beta_1}{\beta_{-1}} -\frac{b_1}{2} \frac{\beta_0}{\beta_{-1}} \\
\sigma_{1} &=&  i(b_2-a_1) \frac{\beta_{-1}}{\beta_{-2}} +\frac{b_2}{2}   \\
\tilde \sigma_{1} &=& i(b_2-a_1) \frac{\beta_0}{\beta_{-1}} -\frac{b_2}{2}  \\
\sigma_{3} &=& \frac{ic_2}{3} \\
\tilde \sigma_{3} &=& \frac{i c_2}{3} .
\end{eqnarray}

\begin{figure}
\begin{center}
\includegraphics[scale=0.9]{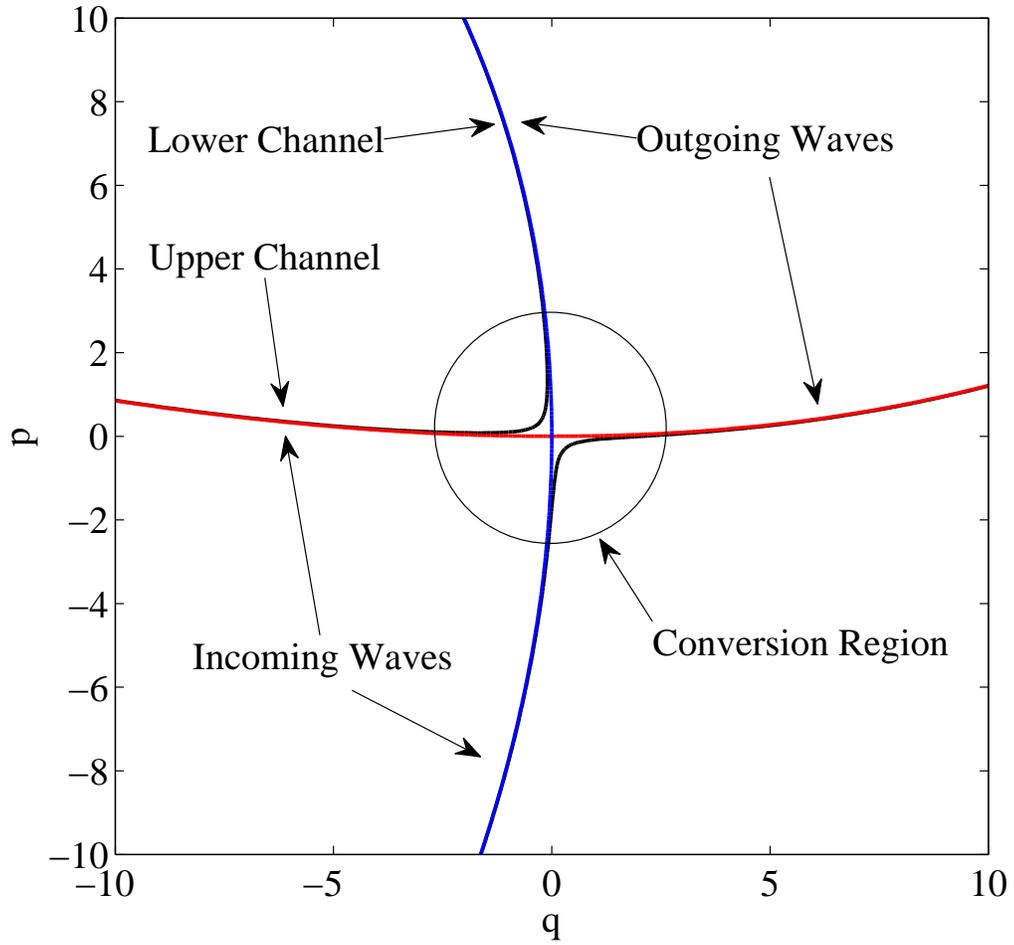}
\end{center}
\caption{\label{fig:uncoupled_modes}
Dispersion surfaces for the uncoupled WKB modes, defined by $D_{11}(q,p)=0$ and $D_{22}(q,p)=0$, cross at the mode conversion point.  The coupled dispersion surface, defined by solving $\det (\mathbf{D}(q,p))=0$, has a hyperbolic structure in the mode conversion region.  Its branches asymptote to the uncoupled dispersion surfaces.
}
\end{figure}

\subsection{Matching the Upper Channel}

In order to find the connection between the incoming and outgoing wave fields, we need to match the WKB solutions in Section \ref{sec:coupled_WKB} with the local solutions in Section \ref{sec:local_sol}.  This matching will allow us to find the scattering matrix for this mode conversion, which determines the amplitudes and phases of the outgoing waves given the amplitudes and phases of the incoming waves.  The various WKB waves can be identified with the dispersion surfaces from which their phases are calculated.  This lets us label the branches of the dispersion surface by the type of WKB mode which it generates, as in Figure \ref{fig:uncoupled_modes}.  

Matching a WKB wave incoming in the upper channel requires comparison of the uncoupled WKB solution and the local solution in the upper channel.  These are given by 
\begin{equation}
\psi_1^{(\text{WKB})}(q) = \frac{e^{i\epsilon c_1 q^3/3}  }{\sqrt{1-\epsilon b_1 q}} 
= \exp \left(  \frac{i\epsilon c_1 q^3}{3} + \frac{\epsilon b_1 q}{2} + \mathcal{O}(\epsilon^2) \right)
\end{equation}
and
\begin{equation}
\psi_1^{(\text{Local})} (q) = q^{-i|\eta|^2} \exp \left(   
 i\epsilon |\eta|^2 (a_2-b_1)q + \frac{\epsilon b_1 q}{2} + \frac{i \epsilon c_1 q^3}{3}
 \right).
\end{equation}
We have left out the terms with negative powers of $q$ from the phase of the local solution, since the effect of these terms is localized to the mode conversion region.  

A matching point $q_* < 0$ is now chosen where both of these expressions are valid.  The amplitude and phase of the incoming WKB mode at this point are used to set the incoming amplitude and phase of the local solution.
\begin{eqnarray}
\psi_1'(q) &=& A \frac{ \psi_1^{(\text{WKB})}(q_*)  }{\psi_1^{(\text{Local})} (q_*) } \psi_1^{(\text{Local})} (q) \\
&=& A q_*^{i|\eta|^2}  \exp \left(   i|\eta|^2 \epsilon b_1 q_*  - i|\eta|^2 \epsilon  a_2 q_*   \right)  \psi_1^{(\text{Local})} (q) 
\end{eqnarray}
where $A$ is the complex amplitude of the incoming wave.  We can put the term $q_*^{-i|\eta|^2}$ into the phase as a logarithm, if we first take the absolute value.  
\begin{equation}
q_*^{i|\eta|^2} = (-1)^{i|\eta|^2}\exp \left( i|\eta|^2\log(|q_*|) \right) = e^{-\pi |\eta|^2}\exp \left( i|\eta|^2\log(|q_*|) \right)
\end{equation}
Here we have used $-1 = e^{i\pi}$.  The $+$ sign is taken in this exponent since we know that the amplitude of the upper channel decreases across the mode conversion when it loses energy to the lower channel.  A more careful derivation of this sign can be found in Reference \cite{metaplectic_formulation}.  (We also are assuming that both waves are positive energy waves.  If one of the modes was a negative energy wave, then there is the possibility that the amplitude of both increase across the mode conversion.)  This gives us an expression for the local field, including the amplitude and phase of the incoming wave.
\begin{eqnarray} \label{eq:phi1_prime}
\psi_1'(q)&=& A e^{-\pi |\eta|^2} \exp \left( i|\eta|^2\log(|q_*|)  +i|\eta|^2 \epsilon (b_1-a_2) q_*  \right)  \psi_1^{(\text{Local})} (q)
\end{eqnarray}

Now choose a matching point $q_{**} > 0$ where this local solution can be matched to the outgoing WKB wave.  We then can write the outgoing field as
\begin{eqnarray}
\psi_1''(q) &=& \frac{\psi_1'(q_{**})}{\psi_1^{(\text{WKB})}(q_{**})}\psi_1^{(\text{WKB})}(q) \\
&=& A e^{-\pi |\eta|^2} \exp \left( i|\eta|^2\log(|q_*|)  +i|\eta|^2 \epsilon (b_1-a_2) q_*  \right)  \\
& & \times \exp \left(-i|\eta|^2\log(|q_{**}|)  -i|\eta|^2 \epsilon (b_1-a_2) q_{**}  \right) \frac{e^{i\epsilon c_1 q^3/3}  }{\sqrt{1-\epsilon b_1 q}}
\end{eqnarray}
This expression shows that, in addition to the amplitude jump found in the linear analysis, there is a phase shift of order $\epsilon |\eta|^2$ which is due to the presence of the quadratic terms.  This expression is valid for any matching points $q_*$ and $q_{**}$ in the appropriate matching regions.  However, it simplifies if we choose matching points symmetrically about the mode conversion point.  In this case, we get
\begin{equation}
\psi_1''(q) = A e^{-\pi |\eta|^2} \exp \left( 2 i|\eta|^2 \epsilon (a_2-b_1) q_\text{M}  \right) \frac{e^{i\epsilon c_1 q^3/3}  }{\sqrt{1-\epsilon b_1 q}},
\end{equation}
where we have defined $q_\text{M} \equiv -q_* = q_{**}$.  This expression includes both the linear order effects calculated previously (e.g., in \cite{metaplectic_formulation}), as well as the new effects of the quadratic order terms.  Theses effects are seen in the expression for the coupled WKB mode given in equation (\ref{eq:wkb_coupled_1}).

\subsection{Matching the Lower Channel}

Use the $p$ representation to match the lower channel.  The equations to match are
\begin{equation}
\phi_2^{(\text{WKB})}(p) = \frac{e^{i\epsilon c_2 p^3/3}  }{\sqrt{1+\epsilon b_2 p}} 
= \exp \left(  \frac{i\epsilon c_2 p^3}{3} - \frac{\epsilon b_2 p}{2} + \mathcal{O}(\epsilon^2) \right)
\end{equation}
and
\begin{equation}
\phi_2^{(\text{Local})} (p) = -\eta^* \beta_{-1} p^{i|\eta|^2} \exp \left(   
 i\epsilon |\eta|^2 (a_1-b_2)p - \frac{\epsilon b_2 p}{2} + \frac{i \epsilon c_2 p^3}{3}
 \right).
\end{equation}
The matching proceeds as in the first channel, except that the initial amplitude and phase are given by the incoming data in the upper channel.  Since the pair given in equations (\ref{eq:psi1_q}) and (\ref{eq:psi2_q}) have the correct relative amplitude and phase, we can apply the amplitude and phase corrections from (\ref{eq:phi1_prime}) to the $q$ representation for the lower channel.  After Fourier transforming into the $p$ representation, this gives us the local field for the lower channel.
\begin{equation}
\phi_2'(p) = A \alpha_1(q_*) \phi_2^{(\text{Local})}(p)
\end{equation}
The corrections from the first channel are defined as
\begin{equation}
\alpha_1(q_*) = e^{-\pi |\eta|^2} \exp (i|\eta|^2\ln |q_*| + i|\eta|^2 \epsilon (b_1 -a_2)q_*).
\end{equation}
We now pick a point $p_*>0$ where we match the local solution onto the outgoing WKB wave.  The outgoing wave is then
\begin{align}
\phi_2''(p) &= A \alpha_1(q_*)  \frac{\phi_2^{(\text{Local})}(p_*)}{\phi_2^{(\text{WKB})}(p_*)} \phi_2^{(\text{WKB})}(p) \\
&= -A \eta^* \beta_{-1} \alpha_1(q_*) \alpha_2(p_*) \frac{e^{i\epsilon c_2 p^3/3}  }{\sqrt{1+\epsilon b_2 p}} ,
\end{align}
where
\begin{equation}
\alpha_2(p_*) = \exp (i|\eta|^2\ln |p_*| - i|\eta|^2 \epsilon (b_2 -a_1)p_*).
\end{equation}
Putting these together means that the outgoing converted wave has the form
\begin{align}
\phi_2''(p) = -A \frac{2\pi e^{-\pi |\eta|^2/2}}{\eta \Gamma(i|\eta|^2)}  |q_*|^{i|\eta|^2} |p_*|^{i|\eta|^2} \exp \left( i\epsilon |\eta|^2 (  (b_1-a_2)q_* -(b_2-a_1)p_*) \right) \frac{e^{i\epsilon c_2 p^3/3}  }{\sqrt{1+\epsilon b_2 p}}.
\end{align}

\section{Comparison with Numerical Solution}

In order to verify the corrected solutions derived above, we compared both the corrected local fields and the matched WKB waves to the results of numerical simulations.  In order to avoid the singularities in the solutions, the numerical calculation was carried out in the $t$ representation, where the phase space coordinates $(t,\omega)$ are those defined by the linear canonical transformation in Equation (\ref{eq:canonical_transform}).
The associated metaplectic transformations of the WKB waves in equation (\ref{eq:corrected_solution}) are 
\begin{equation}\label{eq:met_1}
\psi_1^{(\text{WKB})}(t) = \int_{-\infty}^\infty e^{iF_1(t,q)} \frac{e^{i\epsilon c_1 q^3/3}  }{\sqrt{1-\epsilon b_1 q}} \, dq
\end{equation}
and
\begin{equation}\label{eq:met_2}
\psi_2^{(\text{WKB})}(t) = \int_{-\infty}^\infty e^{iF_2(t,p)} \frac{e^{i\epsilon c_2 p^3/3}  }{\sqrt{1+\epsilon b_2 p}} \, dp .
\end{equation}
where $F_1(t,q)$ and $F_2(t,p)$ are the generating functions for the linear canonical transformations:
\begin{eqnarray}
F_1(t,q) &=& \frac{1}{2} (t^2 - 2\sqrt{2} tq + q^2) \\
F_2(t,p) &=& -\frac{1}{2} (t^2 - 2\sqrt{2} tp + p^2) .
\end{eqnarray}
The conjugate variables are given by derivatives of the generating functions:
\begin{eqnarray}
\omega = \frac{\partial F_1}{\partial t}, &\qquad&  p = - \frac{\partial F_1}{\partial q}  \label{eq:gen_derivs_1} \\ 
\omega = \frac{\partial F_2}{\partial t}, &\qquad&  q = \frac{\partial F_2}{\partial p} .
\end{eqnarray}

In order to evaluate the metaplectic integrals in Equations (\ref{eq:met_1}) and (\ref{eq:met_2}), we write $q$ in terms of $t$ and possibly also derivatives with respect to $t$, i.e.\ we find the $t$ representation of the operator $\hat q$.  The properties of the generating functions allow us to do this.  First, use Equations (\ref{eq:canonical_transform}) and (\ref{eq:gen_derivs_1}) to write
\begin{equation}
q = \frac{1}{\sqrt 2} (t-k)  = \frac{1}{\sqrt 2} (t-\partial_t F_1(t,q)) .
\end{equation}
Now, notice that we can combine the derivative $\partial_t F_1(t,q)$ with the phase in the integral to write
\begin{equation}
\partial_t F_1(t,q) e^{i F_1(t,q)} = -i\partial_t e^{i F_1(t,q)}.
\end{equation}
We can now consider our corrections as a pseudodifferential operator acting on our function, by  making the substitution 
\begin{equation}
q \rightarrow \frac{1}{\sqrt 2} (t+i \partial_t ),
\end{equation}
wherever $q$ appears in our corrections.  In order to do this, write the corrections as a Taylor's series:
\begin{eqnarray}
S_1(q) & \equiv & \frac{e^{i\epsilon c_1 q^3/3}  }{\sqrt{1-\epsilon b_1 q}} \\
&=& \sum_n s_n q^n 
\end{eqnarray}
We associate this with a pseudodifferential operator by replacing $q$ with $(t+i \partial_t )/\sqrt 2$:
\begin{eqnarray}
S_1(q) &\rightarrow& \sum_n \frac{s_n}{\sqrt 2} (t+i\partial_t)^n \\
&=& S_1\left(\frac{1}{\sqrt 2}  (t+i\partial_t) \right) 
\end{eqnarray}

We can now use this to find the $t$ representation of the field.
\begin{eqnarray}
\psi_1^{(\text{WKB})}(t) &=& \int_{-\infty}^\infty e^{iF_1(t,q)} S_1(q) \, dq \\
&=&  S_1\left(\frac{1}{\sqrt 2}  (t+i\partial_t) \right) \int_{-\infty}^\infty e^{iF_1(t,q)} \, dq
\end{eqnarray}
The integral can be recast as a Gaussian integral, since the phase is quadratic in $q$.  
\begin{align}
\psi_1^{(\text{WKB})}(t) &=  S_1\left(\frac{1}{\sqrt 2}  (t+i\partial_t) \right) \sqrt{ 2\pi} 
e^{i\pi/4} e^{-i t^2/2} \\
&=\sqrt{ 2\pi} e^{i\pi/4} \left( 1 + \frac{\epsilon b_1}{2\sqrt{2}} (t+i\partial_t) + \frac{i\epsilon c_1}{3\sqrt{2}} (t+i\partial_t)^3 +\mathcal{O}(\epsilon^2) \right) e^{-i t^2/2} \\
&= \sqrt{ 2\pi} e^{i\pi/4} \left( 1 + \frac{\epsilon b_1}{2} (\sqrt{2}t) + \frac{i\epsilon c_1}{3} \left((\sqrt{2}t)^3+3i\sqrt{2} t \right) +\mathcal{O}(\epsilon^2) \right) e^{-i t^2/2} \\
&= \sqrt{ 2\pi} e^{i\pi/4} e^{-i t^2/2} \frac{e^{i\epsilon c_1 (\sqrt 2 t)^3/3}  }{\sqrt{1+\epsilon (-b_1 + 2 c_1 )\sqrt{2} t}} \label{eq:psi1_x_corr}
\end{align}
The appearance of the $c_1$ term in the amplitude is not too surprising.  Under the Fourier transform, the pure phase $\exp(ic_1q^3/3)$ turns into an Airy function, which has amplitude variations.  Converting from the $q$ representation to the $t$ representation is done with a metaplectic transformation, which is a sort of ``partial'' Fourier transform.  Therefore, we could expect that the cubic phase in $q$ would give rise to an amplitude variation in $t$.

A similar analysis can be computed for the lower channel.  Since our corrections are written in the $p$ representation, we need to make the substitution
\begin{equation}
p = \frac{1}{\sqrt 2} (t+k) \rightarrow \frac{1}{\sqrt 2} (t-i \partial_t ).
\end{equation}
Therefore, the WKB mode in the lower channel is
\begin{align}
\psi_2^{(\text{WKB})}(t) &=  S_2\left(\frac{1}{\sqrt 2}  (t-i\partial_t) \right) \sqrt{ 2\pi} 
e^{-i\pi/4} e^{i t^2/2} \\
&=\sqrt{ 2\pi} e^{-i\pi/4} \left( 1 - \frac{\epsilon b_2}{2\sqrt{2}} (t-i\partial_t) + \frac{i\epsilon c_2}{3\sqrt{2}} (t-i\partial_t)^3 +\mathcal{O}(\epsilon^2) \right) e^{i t^2/2} \\
&= \sqrt{ 2\pi} e^{-i\pi/4} \left( 1 - \frac{\epsilon b_2}{2} (\sqrt{2}t) + \frac{i\epsilon c_1}{3} \left((\sqrt{2}t)^3-3i\sqrt{2} t \right) +\mathcal{O}(\epsilon^2) \right) e^{i t^2/2} \\
&= \sqrt{ 2\pi} e^{-i\pi/4} e^{i t^2/2} \frac{e^{i\epsilon c_2 (\sqrt 2 t)^3/3}  }{\sqrt{1+\epsilon (b_2 - 2 c_2 )\sqrt{2} t}}.\label{eq:psi2_x_corr}
\end{align}

\comment{%
\begin{figure}
\begin{center}
\includegraphics[scale=0.7]{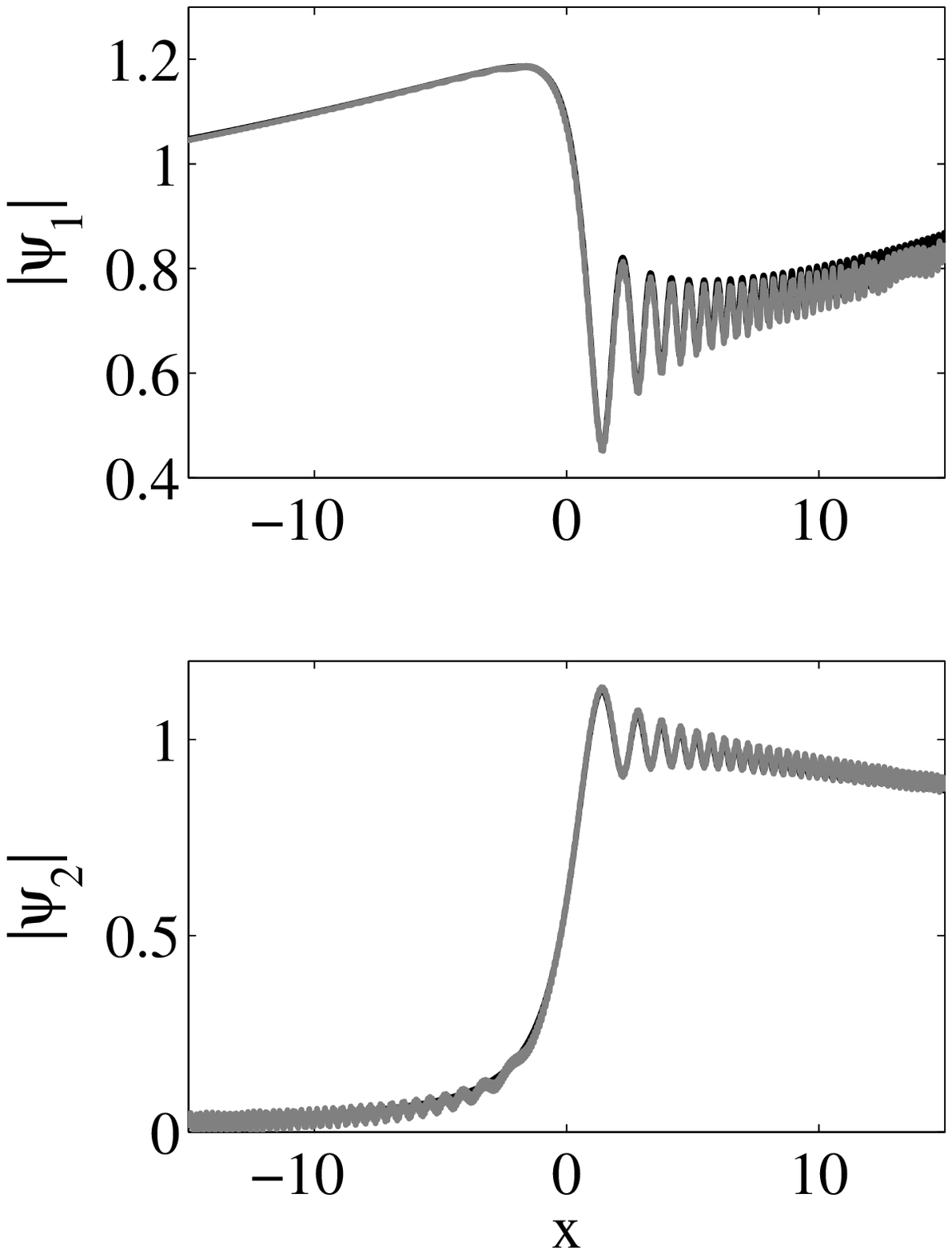}
\end{center}
\caption{\label{fig:compare_amp}
A comparison of the amplitudes of the corrected local solutions (Equations (\ref{eq:psi1_x_corr}) and (\ref{eq:psi2_x_corr}), black) with the amplitudes of a numerical simulation of the original equations (grey).  Notice how our new analytical solutions capture the slow variation due to action conservation.}
\end{figure}
\begin{figure}
\begin{center}
\includegraphics[scale=0.7]{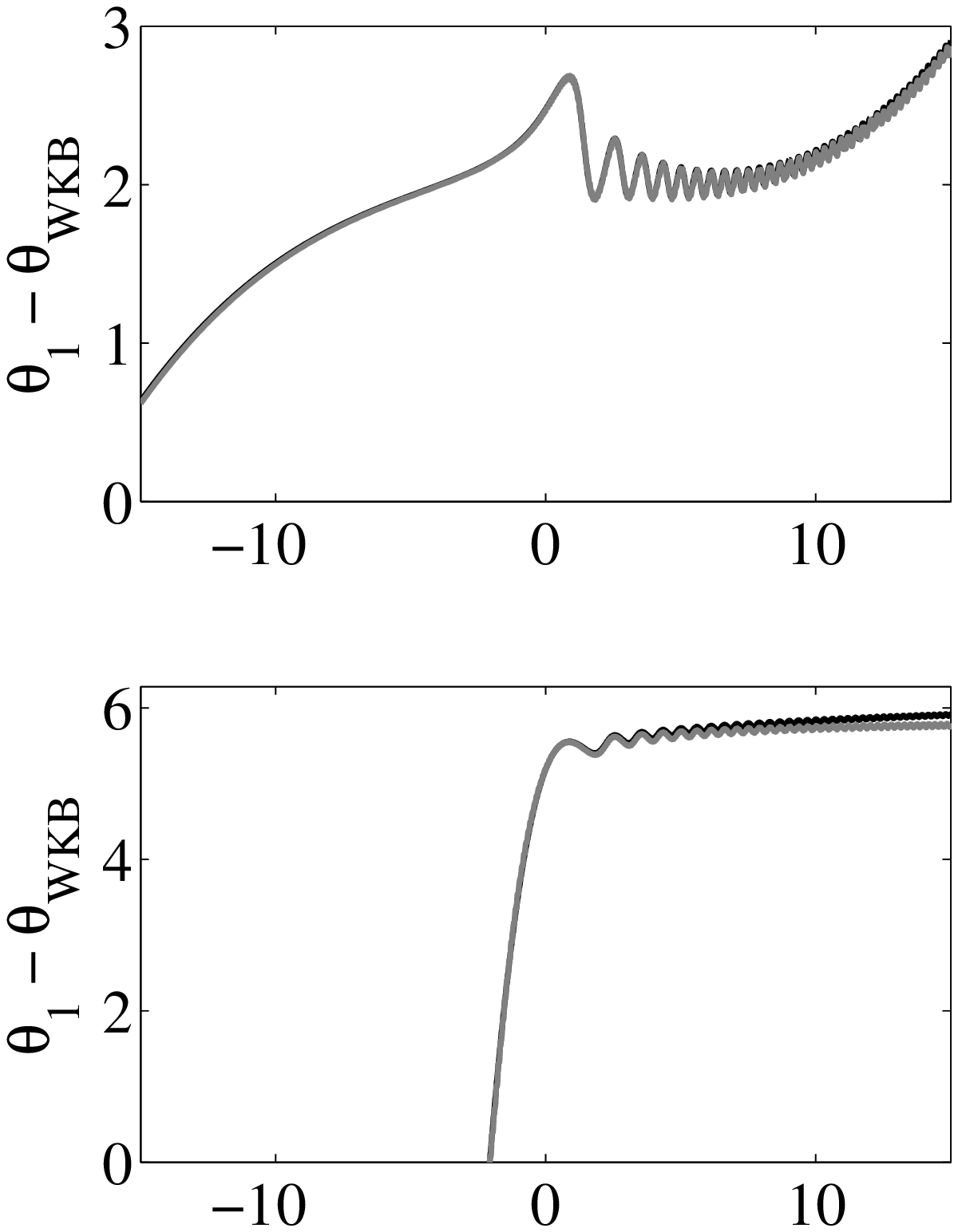}
\end{center}
\caption{\label{fig:compare_phase}
A comparison of the phases of the local analytical solution (black) and of the numerical solution (grey).  Here, $c_1 = 10^-3$, which gives the cubic variation in $\theta_1$.}
\end{figure}
}%

\begin{figure}
\begin{center}
\includegraphics[scale=0.6]{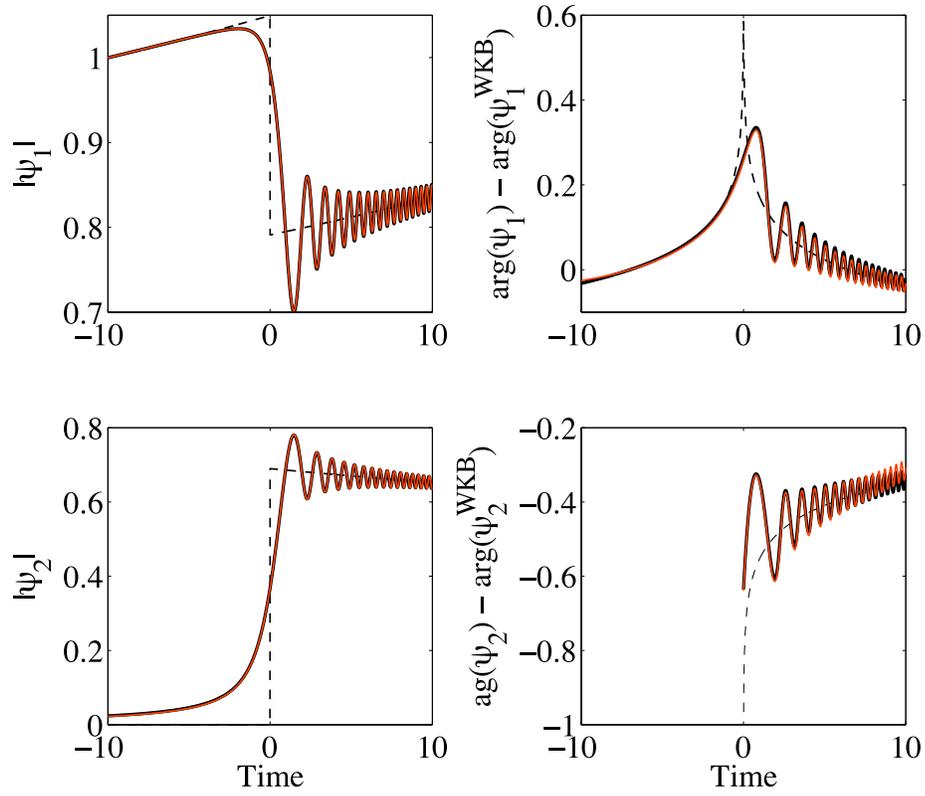}
\end{center}
\caption{\label{fig:compare_amp}
A comparison of the amplitudes and phases of the corrected local solutions (Equations (\ref{eq:psi1_x_corr}) and (\ref{eq:psi2_x_corr}), black) with the amplitudes of a numerical simulation of the original equations (red).  The coupled WKB solutions are shown by the dashed line.  Notice how our new analytical solutions capture the slow amplitude variation due to action conservation.  Simulation parameters: $a_1 = 3 \times 10^{-3}, b_1 = 7.4 \times 10^{-3}, c_1 = 1 \times 10^{-4}, a_2 = 3 \times 10^{-3}, b_2 = 7.4 \times 10^{-3}, c_2 = -5 \times 10^{-4}$.  As discussed in the text, the phase of the second channel contains an unaccounted-for factor of $\pi/4$, which has been corrected {\em ad hoc} in this plot.
}
\end{figure}

We can now compare our analytical expressions (Equations (\ref{eq:psi1_x_corr}) and (\ref{eq:psi2_x_corr})) and numerical simulations.  As seen in Figure (\ref{fig:compare_amp}), these corrected solutions correspond closely to the numerical simulations.  The amplitude of the analytical solutions now contain the square root variation which is due to action conservation, so they match the WKB solutions over a much wider range than before, cf. Figure (\ref{cap:compare}).  The phases of the solutions also show good agreement with the numerical simulations.  Notice that Figure (\ref{fig:compare_amp}) plots the phases relative to the WKB phase, since it changes rapidly compared with the variations due to the coupling and the higher order terms.  The calculation for the phase of the second channel is actually off by a factor of $\pi/4$.  This error is possibly related to the choice of phase in the normalization of the metaplectic integral, or perhaps due to the choice of sign convention for the plane waves.  Whatever the cause, the difference has been removed {\em ad hoc} in Figure (\ref{fig:compare_amp}), so that the smaller scale variations could be compared.

% end general corrections

\chapter{Summary}

In Part I of this dissertation, we gave a brief introduction to phase space methods for scalar wave equations.  In Chapter \ref{chp:PhaseSpaceIntro} we reviewed the connection between ray-tracing techniques and the WKB approximation.  This connection is most natural when considered from the point of view of $(x,k)$ phase space.  The Weyl symbol of the wave operator is the dispersion function on phase space, and the zero surfaces of the dispersion function define the dispersion relations for the problem being considered.  The dispersion function can then be used as a ray hamiltonian, and a family of rays generated by this hamiltonian can be used to reconstruct the phase and amplitude of an approximate solution to the wave equation.  

In Chapter \ref{chp:coupled_osc}, we consider the problem of multicomponent wave equations.  Phase space techniques can also be used for multicomponent problems, as long as the vector nature of the waves can be accounted for.  Generically, the eigenvectors of the dispersion matrix can be used as a polarization basis for the wave, and the multicomponent problem reduces to several scalar wave problems.  The eigenvalues of the dispersion matrix then generate the rays for each of the different modes, and the zeros of the eigenvalues give the dispersion relations for each mode.  However, the eigenvectors become nearly degenerate in mode conversion regions, and different modes can interact, exchanging energy.  We described this process by using an example of two coupled oscillators with slowly varying natural frequencies.  This pair of oscillators was analyzed using the phase-space tools which were described in the context of wave problems.  When the frequencies of the oscillators is nearly equal, the oscillators can resonantly exchange energy in a process analogous to mode conversion.  We described the matched asymptotic solution for the mode conversion problem, and showed numerical simulations for the coupled oscillator example.  Up to this point the material presented has been for the purpose of review.

We then moved on in Chapter \ref{chp:higher_order} to a new calculation.  The matched asymptotic techniques used to solve the mode conversion problem required the linearization of the dispersion matrix about the mode conversion point.  We showed how to extend the local analysis of the mode conversion to include quadratic order terms in the dispersion matrix.  This analysis gave us a new way to correct the local solution, which resulted in much better matching between the local solution and the far-field WKB solutions.  In the process of calculating these corrections, we showed that, for a mode conversion in one spatial dimension, the dispersion matrix can be put into normal form through second order.  This was done by a near-identity change of polarization basis, which made the off-diagonal terms constant (through second order).  We also showed that the Moyal corrections introduced by this change of polarization basis enter at a higher order, and therefore, at the order we are considering, can be neglected.

The phase space analysis which was described and used here in Part I are very useful.  The phase space perspective can provide geometrical insights, which can guide analytical calculations.  Also, phase space ray-tracing algorithms can be used to solve multicomponent wave problems which exhibit mode conversion, by treating the mode conversion as a ray splitting.  However, there are still problems which cannot be solved using the approach described in the previous chapters.  For example, mode conversions where the dispersion surfaces meet tangentially, instead of transversely, still need to be dealt with from a ray-tracing perspective.  Also, wave problems in multiple spatial dimensions can have rays with finite helicity, leading to new types of nonstandard mode conversions \cite{citeulike:472573}.  Problems such as these have led us to explore the mathematics which underlie the phase space analysis described in these chapters.  In particular, we will examine in Part II of this dissertation the connection between the phase space path integral, the Weyl symbol, and the theory of the irreducible representations of the Heisenberg-Weyl group.  These connections are best examined in the context of a new theory of symbols, which has recently been developed by Zobin, and which describes the symbol of an operator in terms of a double Fourier transform.  In Part II, we describe this new theory of symbols, and show how it suggests several new avenues of future research.

%% file: Chapter-PathInt-Intro.tex
%%%%%%%%%%%%%%%%%%%%%%%%%%%%%%%%%%%%%%%%%%%%%%%%%%%%%%%%%%%%%%%%%%%%%%%%%%
%
% Ph.D. dissertation manuscript
% Part II Introduction
%
% Andrew Stephen Richardson (Fall 2007)
% College of William and Mary
% Department of Physics
% Prof. Eugene Tracy, advisor
%
% Based on Paul King and Andrew Norman's template (modified by Wirawan Purwanto)
%
%%%%%%%%%%%%%%%%%%%%%%%%%%%%%%%%%%%%%%%%%%%%%%%%%%%%%%%%%%%%%%%%%%%%%%%%%%

\chapter{Introduction}
\label{chp:Intro}

In Part II of this dissertation, we will delve into the mathematical foundations upon which the phase space techniques used in Part I are built.  By doing so, the richness of these foundations are shown, and new areas of research are opened up.

In Chapter \ref{chp:GroupTheory} we review the theory of groups and their representations.  This groundwork, while well known to some mathematicians and physicists, is necessary for the subsequent chapters, and worthwhile reviewing.  In particular, we will describe the regular representation, and how it reduces uniquely to primary representations.  Further reduction of the primary representations to the irreducible representations will require a choice of basis.  We will describe this reduction in the case of the Heisenberg-Weyl group, and show how it is connected to choosing a representation (e.g., $q$-representation or $p$-representation) for the wavefunction in quantum mechanics.

Chapter \ref{chp:Symbols} continues the review of group theory by describing the Fourier transform in the context of groups.  We describe the Fourier transform for non-commutative groups, and show how it can be used to map operators (which are embedded into sections of the dual bundle) to functions on the group.  We then discuss Zobin's theory of symbols, which defines the symbol of an operator as a particular sequence of Fourier transforms.  First an inverse non-commutative Fourier transform takes an operator to a function on the group, where the group is considered as a set.  Since the group (considered as a set) also has a commutative group structure, there is also a commutative Fourier transform associated with the group.  This commutative transform is performed as the second step in calculating the symbol of an operator.  We then describe how the Weyl symbol is a special case of the Zobin symbol for the continuous Heisenberg-Weyl group, and how it is associated with the reduction of the regular representation to the primaries.  We end the chapter with the application of the Zobin symbol to the discrete Heisenberg-Weyl group.  This allows us to define the ``symbol'' of a matrix, which is a complex-valued function on a (discrete) space.  This novel type of symbol not only provides a new way to analyze matrices, but also shows how the group theoretical approach to the theory of symbols is widely applicable.

We next examine functions of operators in Chapter \ref{chp:PathIntegral}.  We show that calculation of the symbol of a function of an operator can be recast into the form of a path integral.  While the connection between Weyl symbols and path integrals has been examined previously \cite{Berezin:1991fk}, we will analyze this connection in the more general context of the Zobin symbol.  This new context for the path integral leads to several new insights.  First, when calculating the discrete ``path integral'' for the discrete Heisenberg-Weyl group, we see that there is a natural connection between the path integral and sums over probability densities (or measures).  This opens the door to the use of techniques from statistical mechanics and information theory for the analysis of path integrals; techniques such as using maximum entropy calculations for finding the most likely ``path''.  The path integral for a continuous group can be recast as an infinite-dimensional Fourier transform.  In the discrete case, the sum over all possible measures can be written in a way where it is evidently a finite dimensional Fourier transform.  Taking the limit of this to get to the continuous group turns the sum over measures into an integral over the space of measures, which then looks like an infinite-dimensional Fourier transform on the space of measures.  The second insight into path integrals that is obtained from this group theory perspective is the realization that the connection between the phase space path integral and the configuration space path integral stems from the reduction of the regular representation of the Heisenberg-Weyl group.  Reduction of the regular representation to the primary representations involves considering functions on phase space, rather than functions on the whole group.  The path integral then becomes an integral over paths in phase space.  The further reduction of the primary representations to irreducible representations requires choosing a ``configuration'' subspace in phase space, and only considering functions on configuration space.  In this case, the path integral is further restricted, and becomes an integral over paths in configuration space.

We conclude Part II with a survey in Chapter \ref{chp:ModeConversion} of several areas of research opened up by this new group theoretical perspective on path integrals.  First we describe how to define a new normal form for the dispersion matrix for vector wave problems.  The diagonal elements are now associated with ``uncoupled'' modes, and can be used as hamiltonian functions for generating rays.  This approach could simplify the geometry of the dispersion surfaces for mode conversion problems, turning ``avoided crossings'' into transverse intersections of the dispersion surfaces for the interacting modes.  A second idea suggested in Chapter \ref{chp:ModeConversion} is a proposal to create a ``double symbol'' for vector wave problems.  The standard approach computes the Weyl symbol of each element the wave operator individually.  This gives a matrix-valued function on phase space.  But, as shown in Chapter \ref{chp:Symbols}, the symbol of a matrix can also be computed.  This suggests combining the Weyl symbol with the symbol of a matrix, to get a new sort of ``double'' symbol for the wave operator.  The last idea presented in Chapter \ref{chp:ModeConversion} is related to dealing with uncertainties in mode conversion problems.  For example, turbulent fluctuations are often present in plasmas, and we would like to be able to analyze mode conversion problems even in the presence of such fluctuations.  We examine the possibility of introducing decoherence into mode conversion problems through an appropriate averaging of the Wigner function.  This procedure (which is done in the spirit of quantum decoherence of the density matrix) smooths out the complicated interference patterns in the Wigner function, and results in a much more ``classical'' distribution on phase space.

%% file: Chapter-GroupTheory.tex
%%%%%%%%%%%%%%%%%%%%%%%%%%%%%%%%%%%%%%%%%%%%%%%%%%%%%%%%%%%%%%%%%%%%%%%%%%
%
% Ph.D. dissertation manuscript
% Chapter 6: Theory
%
% Andrew Stephen Richardson (Fall 2007)
% College of William and Mary
% Department of Physics
% Prof. Eugene Tracy, advisor
%
% Based on Paul King and Andrew Norman's template (modified by Wirawan Purwanto)
%
%%%%%%%%%%%%%%%%%%%%%%%%%%%%%%%%%%%%%%%%%%%%%%%%%%%%%%%%%%%%%%%%%%%%%%%%%%

\chapter{Review of the Theory of Linear Representations of Groups\label{chp:GroupTheory}}

This chapter will provide a review of the theory of groups, and linear representations of groups.   Since it is intended to be a review, results will be presented without proofs.  Also, many general results will be stated for finite groups.  Such general results for infinite groups are difficult, if not impossible to come by.  However, the infinite groups that we are interested in (such as the Heisenberg-Weyl group) are nicely behaved, and the results obtained in the finite case can be shown to also hold in these specific infinite cases.  For more details and proofs, the textbooks by Serre \cite{Serre:1996lr} and Kirillov \cite{Kirillov:2004lr} are good references.

The main point of this review is to describe the regular representation, and how it naturally and uniquely reduces to the primary representations.  After a further (non-unique) choice of basis, this reduces to the irreducible representations.  For the Heisenberg-Weyl group, this series of reductions will take us from functions on the group, to functions on phase space, and then to functions on configuration space (more precisely to any Lagrange plane in phase space).  The choice of basis in this case corresponds to the choice of position and momentum coordinates in phase space.

\section{Groups and Group Actions}

A group is a set $G$ endowed with an associative product rule, or group multiplication, denoted here as $\diamond$.  There must be a neutral element in the group (the ``identity'' element), and each element in the group must have an inverse.

A simple example of a group is the set of all integers, with addition as the group product rule.  Zero is the identity element of this group, since adding zero to an integer gives the same integer back.  The negative integers are the inverse of the positive integers.

\subsubsection{Commutation}

An important characteristic of any given group is whether the group product law is commutative.  If the law is such that 
\begin{equation}\label{eq:comm}
g_1 \diamond g_2 = g_2 \diamond g_1, \quad \forall\, g_1, g_2 \in G
\end{equation}
then the group multiplication commutes, and the group is called commutative, or {\em Abelian}.  Normal addition is commutative, so, for example, the set of real numbers forms a commutative group, using addition as the group ``product''.  A {\em non-Abelian}, or noncommutative group has a group product law which is not commutative.  This makes the group structure much richer, and gives many groups their interesting and useful properties.

\subsubsection{Subgroups}

A {\em subgroup} $H$ of some group $G$ is a subset of the elements of $G$ which is a group under the multiplication inherited from $G$.  It is a smaller group which is embedded in the structure of the larger group $G$.  All products and inverse elements must be in the subset.
\begin{equation}
h, k \in H \, \implies \, h \diamond k ,  h^{-1}, k^{-1} \in H
\end{equation}

A particular subgroup which is of some importance is the so-called {\em center} of the group.  The center is the set of elements of the group which commute with any other element of the group, and is denoted $Z(G)$.  
\begin{equation}
Z(G) = \{ k \in G | \forall \, g \in G, k \diamond g = g \diamond k \}
\end{equation}
The identity element is definitely contained in the center, since multiplication from the left or from the right by the identity leaves all elements unchanged.  Also, direct computation shows that if an element $h$ is in the center, then its inverse $h^{-1}$ is also in the center.  Therefore, the center is a subgroup of $G$.

\subsubsection{Group Actions}

While groups are interesting as abstract entities, they become a powerful tool for the study of physics through their actions.  In order to properly define group actions, it is necessary to first define a specific group; the group of transformations of a set $X$.  An element of the group of invertible transformations is a mapping 
\begin{equation}
\phi : X \rightarrow X,
\end{equation}
and the product law in the group of transformations is simply the composition of mappings.
\begin{equation}
\phi_1 \diamond \phi_2 \equiv \phi_1 \circ \phi_2
\end{equation}
With this definition of the group of transformations, a generic group action can now be defined.  A {\em group action} $A$ of the group $G$ is a mapping of $G$ into the group of invertible transformations of some set $X$.
\begin{equation}
g \mapsto A_g, \text{ where } A_g : X \rightarrow X.
\end{equation}
This can also be written as
\begin{equation}
A_g(x) = x', \quad g \in G; \, x,x' \in X.
\end{equation}
In order for the action to satisfy the group product law, we must have the following property for the composition of actions.
\begin{equation}\label{eq:group_action}
A_{g_1 \diamond g_2} (x) = A_{g_1} (A_{g_2}(x))
\end{equation}

An example of an action is the translations generated by the group of integers, $G = \{\mathbb Z, + \}$.  The translation of a function on the real line can be written
\begin{equation}
(T_n f)(x) = f(n+x).
\end{equation}
This action satisfies the group product law, as can be seen by direct computation.
\begin{equation}
( T_{n+m} f) = f(n+m+x) = (T_n (T_m f))(x)
\end{equation}
So translation by integers is an action of the group $\{\mathbb Z, + \}$ on the set of functions $L_2(\mathbb R)$ on the real line.

\section{Group Actions and Linear Representations}

Since group actions respect the group product law, they give us a way to study properties of the group.  Because of the encoding of the product law in the action, the action provides us with a substitute for the group.  The most general group actions are not necessarily the best objects to study, however, since in general the group action could be given by a nonlinear transformation.  A simpler case would be an action which is a mapping into {\em linear} transformations.  Such a mapping is called a {\em linear representation} of the group, or often just a {\em representation} of the group.  A linear group representation $\rho$ is a mapping from the group to the space of operators acting linearly on some vector space $V$.
\begin{equation}
\rho: G \rightarrow \text{GL}(V)
\end{equation}
The operators $\rho(g)$ must satisfy the group product law in order to be a representation of the group.
\begin{equation}
\rho(g_1 \diamond g_2) v = \rho(g_1) \rho(g_2) v
\end{equation}
Although this notation is different, this is essentially the same as Equation (\ref{eq:group_action}) which gives the essential property of a group action.

Group actions can take many forms.  For example the nonlinear transformation given by the set of all possible linear fractional transformations which preserve the upper half plane can be thought of as a nonlinear representation of the group of $2\times 2$ symplectic matrices.  The linear fractional transformations map the complex plane onto itself.
\begin{equation}\label{eq:lft}
\sigma_1 (z) = \frac{a_1 z + b_1}{c_1 z + d_1}
\end{equation}
In order to preserve the upper half plane, we require the real parameters $a_1, b_1, c_1$, and $d_1$ satisfy the property
\begin{equation}\label{eq:lft_1}
a_1 d_1 - b_1 c_1 = 1.
\end{equation}
The composition of two linear fractional transformations is also a linear fractional transformation, given by
\begin{eqnarray}
\sigma_1(\sigma_2(z)) = \frac{a_3 z + b_3}{c_3 z + d_3},
\end{eqnarray}
where
\begin{equation}\label{eq:lft_2}
\left(
\begin{array}{cc}
 a_3  &  b_3 \\
c_3   & d_3  
\end{array}
\right)
= 
\left(
\begin{array}{cc}
 a_1  &  b_1 \\
c_1   & d_1  
\end{array}
\right)
\left(
\begin{array}{cc}
 a_2  &  b_2 \\
c_2   & d_2  
\end{array}
\right).
\end{equation}
Note: if we flip the overall sign of these matrices, we get the same $\sigma(z)$; this mapping from matrices to linear fractional transformations is 2 to 1.

Together, the properties in Equations (\ref{eq:lft_1}) and (\ref{eq:lft_2}) imply that the linear fractional transformations (which preserve the UHP) are a representation of the set of real $2\times 2$ matrices with determinant one, modulo an overall sign, called $\text{PSL}(2,\mathbb R) = \text{SL}(2,\mathbb  R) /  \{\pm 1\}$.

While nonlinear actions such as that in Equation (\ref{eq:lft}) are perfectly valid as group actions, it is often more convenient to use linear representations when studying a group.  A  representation $\rho$ is said to have dimension $n$, where $n$ is the dimension of the linear vector space $V$ on which $\rho$ acts.  The dimension could be infinite, as is the case if $V$ is an infinite-dimensional Hilbert space.

For linear representations which act in $n$-dimensional ($n<\infty$) vector spaces, the representation can be thought of as mapping elements of the group to $n\times n$ matrices.  Here are some properties of such a representation:
\begin{itemize}
\item $V$ = vector space of dimension $n$
\subitem Elements of $V$ are vectors, or can be thought of as a functions $f(x)$ on an $n$-point lattice, where $x$ labels the points of the lattice.
\item $GL(V)$ = group of automorphisms of $V$
\subitem $M \in GL(V)$ is an $n\times n$ matrix (an operator) acting on the space of functions on a lattice 
 \item $\rho(g)$ = a homomorphism of $G$ into $GL(V)$
\subitem $\rho(g)$ can be thought of as a matrix
\subitem $\rho(g_1 \diamond g_2)  = \rho(g_1) \rho(g_2)$ is given by matrix multiplication 
\end{itemize}  
 
Returning to the case of linear representations of arbitrary dimension, there are a few last notions worth discussing at this point.  The first is the notion of unitary representations.  If the vector space $V$ on which the representation acts is a Hilbert space, that implies that there is a well defined notion of lengths of vectors in $V$.  Operators which preserve the lengths of all vectors are called unitary operators, and a representation 
\begin{equation}
\rho : G \rightarrow U(V)
\end{equation}
which maps the group to unitary operators is called a {\em unitary representation}.  In this work we are mostly interested in unitary representations.  For the groups of interest here, unitary representations always exist.

Finally, there is a notion of equivalent representations.  If two representations $\rho_1$ and $\rho_2$ are related by a constant similarity transformation, they are said to be {\em unitarily equivalent}.
\begin{equation}
\rho_1(g) = U^{-1} \rho_2(g) U, \quad \forall \, g \in G \implies \rho_1 \sim \rho_2
\end{equation}
Two unitarily equivalent representations have the same dimension.  When discussing a representation of a group, we often really mean the equivalence class of the representation, since the particular choice of a representation in an equivalence class may not matter.  In cases where the distinction between a particular representation and its equivalence class is important, we note it in the text.

\section{Reductions of Linear Representations}

Let $\rho: G \rightarrow \text{GL}(V)$ be a linear representation of the group $G$ acting on the vector space $V$.  For the moment, let $G$ be a finite group, and $\rho$ be a finite unitary representation, in order to simplify this discussion.  Although the proof is not simple, these results also apply to the infinite groups we want to consider.

Let $W$ be a subspace of $V$.  The subspace $W$ is invariant under the action of the group if
\begin{equation}
\forall\, g \in G, \, w \in W, \; \rho(g) w \in W.
\end{equation}
If such an invariant subspace exists, then the restriction of $\rho$ to this subspace is itself a representation of the group.
\begin{equation}
\rho|_W : G \rightarrow \text{GL}(W)
\end{equation}
This is called a {\em subrepresentation} of $\rho$.  

For the groups and representations that we are interested in, the invariant subspace $W$ will have a complementary subspace $W_\perp$ which is also invariant.  This means that the representation $\rho$ can be put into block diagonal form, with the subrepresentations $\rho|_W$ and $\rho|_{W_\perp}$ on the diagonals, and zeros in the off-diagonal blocks.
\begin{equation}
\rho(g) = 
\left(
\begin{array}{c|c}
 \rho|_{W}(g)  &    \\ \hline
   &\rho|_{W_\perp}(g)  \\ 
\end{array}
\right)
\end{equation}
This breaking of $\rho$ into blocks can be continued if we can find invariant subspaces in $W$ or $W_\perp$.  If the space $V$ can be written as $V = W_1 + W_2 + \ldots$, where all of the $W_i$ spaces are invariant subspaces, then we get a block diagonalization which reduces $\rho$ into a direct product of representations acting in the smaller vector spaces.
\begin{equation}
\rho(g) = 
\left(
\begin{array}{c|c|c|c}
 \rho|_{W_1}(g)  & & &   \\ \hline
   &\rho|_{W_2}(g) & &   \\ \hline
   & &\rho|_{W_3}(g) &   \\ \hline
   & & & \ddots
\end{array}
\right)
\end{equation}
If a representation cannot be reduced any further, then it is called an {\em irreducible representation}.  The set of all (classes of unitarily equivalent) irreducible representations is called the dual to the group $G$, and is denoted $\widehat G$.  For the most interesting groups, any unitary representation of the group is made up of a direct product of irreducible representations.

\section{The Regular Representation and Observations About its Reduction\label{sec:obs}}

There are two important representations that can be formed for any group.  These are the right and left regular representations.  These representations map the elements of the group into shifts of functions on the group itself.  Consider $L_2(G,d\mu)$, the space of complex valued functions on the group, $f\in L_2(G,d\mu) : G \rightarrow \mathbb C$.  Here the measure $d\mu$ is the Haar measure, which is a shift-invariant measure which will be discussed later.
The {\em right regular representation} maps group elements to right shifts of functions in $L_2(G,d\mu)$:
\begin{equation}
(\rho_R(h) f)(g) = f(g\diamond h) .
\end{equation}
Some algebra shows that this is a representation of the group:
\begin{align}
(\rho_R(h_1 \diamond h_2) f)(g) &= f(g\diamond h_1 \diamond h_2).
\end{align}
To show this is a representation, we need to compare this to the composition of shifts.
\begin{align}
(\rho_R(h_1) \circ \rho_R(h_2) f)(g) &= (\rho_R(h_2) f) (g \diamond h_1) \\
& = f (g \diamond h_1 \diamond h_2)
\end{align}
So this is a representation.  The regular representation works by shifting the elements of the group according to the group multiplication law.   For a finite group, this can be pictured as a shuffling of the group elements in a way which is prescribed by the multiplication table for the group.  Each matrix of the regular representation $\rho_R(g)$ is then just a big permutation matrix, and knowing the set of all of these matrices for each $g \in G$ would allow you to reconstruct the multiplication table.

We can also form a representation using left shifts, which is called the {\em left regular representation}.
\begin{equation}
(\rho_L(h) f)(g) = f(h^{-1} \diamond g)
\end{equation}
Again we can check that this is a representation.
\begin{align}
(\rho_L(h_1 \diamond h_2) f)(g) &= f((h_1 \diamond h_2)^{-1} \diamond g)\\
&= f(h_2^{-1} \diamond h_1^{-1} \diamond g)
\end{align}
The composition of shifts is given by
\begin{align}
(\rho_L(h_1) \circ \rho_L(h_2) f)(g) &= (\rho_L(h_2) f) (h_1^{-1} \diamond g) \\
&= f(h_2^{-1} \diamond h_1^{-1} \diamond g)
\end{align}
These two regular representations are important because they can be constructed for any group.  

For the types of groups we want to consider, the space of functions has a measure which is invariant under left or right shifts. 
\begin{equation}
d(g \diamond h) = dg, \quad \text{ or } \quad d(h^{-1} \diamond g) = dg
\end{equation}
While the left invariant measure is not in general the same as the right invariant measure, it is the same for the groups we want to consider.  The unique (up to a constant) invariant measure on the group is called the {\em Haar measure} for the group.  Given the Haar measure, the notion of an $L_2$ norm can be defined.  With this norm, the regular representations are unitary representations.  
\begin{align}
\int_G \left\vert \Big(\rho_R(h)f\Big)(g) \right\vert^2 \, d\mu(g) 
&= \int_G \vert f(g \diamond h) \vert^2 \, d\mu(g) \\
&= \int_G \vert f(g') \vert^2 \, d\mu(g'\diamond h^{-1})  \\
&= \int_G \vert f(g') \vert^2 \, d\mu(g')
\end{align}
This substitution is possible because the measure is invariant under the shift $g' = g \diamond h$.  A similar calculation shows that the left regular representation is also unitary.

The regular representations are nice for another reason.  Even though the regular representations are not irreducible, they can be reduced, and in fact are made up of a direct product of copies of {\em all} of the irreducible representations of the group.  So, for the groups that we are interested in, one way to try to find all of the irreducible representations would be to start with the regular representation and reduce it as far as possible.

One way that the reduction of the regular representation could be achieved would be to consider the commutative subgroups $H$ of the group $G$.  Irreducible representations of these subgroups are all one dimensional, and so are straightforward to find.  Consider the following series of observations\footnote{As is the case with most of my knowledge of group theory, I am indebted to Prof.~Zobin for this series of observations, which he outlined for me.}
 about the regular representation and the subgroups of $G$.

{\bf Observation:}
Consider the following closed subspaces of $L_2(G)$.  Choose any {\em commutative} subgroup $H < G$, and any one dimensional irreducible representation $\tau$ of $H$.  Define the space of covariant functions as
\begin{equation}\label{eq:covariant_functions}
\mathcal S (H,\tau) = \{ f: \forall \, h \in H, \forall \, g \in G, f( g \diamond h) = \tau(h) f(g)\}
\end{equation}
If we use the subgroup $H$ to define equivalence classes in $G$, then we see that all functions $f \in \mathcal S (H,\tau)$ can be described by their values on a smaller set of elements in $G$.  If we pick one element $k$ from each equivalence class in $G / H$ (this is a set of {\em representatives} for the classes), then $f$ is determined by its values on the $k$'s.
\begin{equation}\label{eq:cov_factor_set}
\forall \, g = k \diamond h, f(g) = f(k\diamond h) = \tau(h) f(k)
\end{equation}
This means that the space $\mathcal S (H,\tau)$ is smaller than the original function space $L_2(G)$.  The dimension of $\mathcal S (H,\tau)$ is equal to the number of equivalence classes in $G / H$.

{\bf Observation:}
The subspace $\mathcal S (H,\tau)$ is invariant under the action of the left regular representation.
\begin{align}
(\rho_L(g_2) f)(g \diamond h) &= f(g_2^{-1} \diamond g \diamond h) \\
&= \tau(h) f(g_2^{-1} \diamond g) \\
&= \tau(h) (\rho_L(g_2) f)(g)
\end{align}
Since this space is a $\rho_L$ invariant subspace, it can be used to reduce the left regular representation.  

{\bf Observation:}
Larger commutative subspaces will reduce the regular representation more.  If $H_2> H_1$, then $\mathcal S (H_2,\tau) \subset \mathcal S (H_1,\tau)$.  This follows from the fact that the covariance condition in the definition of $\mathcal S (H_2,\tau)$ will impose more constraints, and so fewer functions will satisfy the condition.

{\bf Observation:}
Choose $H$ to be a maximal commutative subgroup.  Then $\mathcal S (H,\tau)$ will be minimal.  Very often (i.e., for all of the interesting groups) the restriction $\rho_L|_{\mathcal S (H,\tau)}$ of the regular representation to this subspace will be an irreducible representation.  Also very often, the restrictions for all maximal commutative $H$ and all irreducible representations $\tau$ will give you all of the irreducible representations of $G$.

{\bf Observation:}
Some of these restrictions will give unitarily equivalent representations.
\begin{equation}
\rho_L|_{\mathcal S (H,\tau)} \approx \rho_L|_{\mathcal S (H',\tau')} 
\quad \text{ for some } (H,\tau), (H',\tau')
\end{equation}
For Lie groups, the Kirillov orbit method \cite{Kirillov:2004lr} can be used to distinguish and label the various inequivalent irreducible representations.  For the Heisenberg-Weyl group, we can use the Stone--von Neumann theorem to check whether two irreducible representations are unitarily equivalent, as will be discussed.

\section{Primary and Irreducible Representations\label{sec:reduction}}

While the irreducible representations are in some sense the most fundamental representations of a group, when decomposing any given representation it is often more natural to work with {\em primary representations}.  In this section, we will describe how to reduce the regular representation of a group, and introduce primary representations in this context.  The reduction of the regular representation can proceed unambiguously to the level of the primaries.  This reduction is called the {\em canonical decomposition} \cite{Serre:1996lr}.  Each primary representation in turn is a direct sum of several copies of some particular irreducible representation.  Since each primary carries with it information about only one irreducible representation, the primary representations provide a nice decomposition.  However, since they are made up of several copies of the irreducible representation, there is still some redundancy.  At this point, it is necessary to introduce an arbitrary block basis into the vector space on which the primaries act in order to split the primary representations into irreducibles.  In some sense, the primaries are therefore more natural to work with than the irreducibles, since the irreducibles require this arbitrary choice of basis.

As will be discussed in later sections, the distinction between the primary representations and the irreducible representations for the Heisenberg-Weyl group roughly becomes a distinction between considering functions on classical phase space and functions on configuration space.  The choice of coordinates for configuration space is arbitrary, since a canonical transformation will take us from one set of coordinates to another.  Similarly, the reduction to irreducible representations from the primaries requires a choice of basis, or ``coordinates''.  This choice of coordinates splits the phase space into ``positions'' and ``momenta''.

\subsubsection{The Regular Representation}

The {\em regular representation} of a group arises when one considers the space of complex valued functions defined on the group itself; $f:G\rightarrow \mathbb C$.  Since the group is a set, or even, as is the case for Lie groups, a manifold, it is natural to think of functions which map this set to the complex numbers.  The space of all such functions forms a linear vector space, with addition of the vectors being defined as the point-wise addition of the complex values of the corresponding functions.
\begin{equation}
(f_1 + f_2)(g) = f_1(g) + f_2(g)
\end{equation}
This vector space, $\mathcal F_G$, is used to form (left) the regular representation of the group through the definition
\begin{equation}
(\rho_L(h) f)(g) = f(h^{-1} \diamond g)
\end{equation}
The computation in the previous section shows that this is a representation.

While the regular representation is perhaps the most natural representation of a group, it is not an irreducible representation.  In fact, for the groups we are considering, it contains copies of {\em all} of the irreducible representations of the group.  In order to reduce the regular representation, we need a way to find all of the invariant subspaces of $\mathcal F_G$.  This could be achieved by defining an arbitrary basis set for $\mathcal F_G$, and then writing the decomposition in terms of that basis.  However, a less arbitrary way to perform the decomposition could be achieved if we can define projection operators, each of which projects onto an invariant subspace.  It turns out that these projection operators can be found, but they do not project onto the subspaces associated with the irreducible representations, but rather those associated with the primaries.  Unlike the decomposition into irreducible representations, the decomposition into primaries is unique.

\subsubsection{Decomposition of the Regular Representation}

Let the irreducible representations of $G$ be $\rho_i: G\rightarrow \text{GL}(V_i)$.  For the calculations in this section, we let $G$ be a finite group.  The results also hold for the infinite groups that we want to study, but the derivations are much more complicated.

Assuming that the regular representation is a direct sum of irreducible representations, we can decompose the space $\mathcal F_G$ into a direct sum of the spaces $V_i$.
\begin{equation}
\mathcal F_G = \bigoplus_k V_k
\end{equation}
Since each irreducible representation may appear in $\rho_L$ more than once, there may be several copies of each $V_i$ in $\mathcal F_G$.  Let $m_i$ be the multiplicity of $\rho_i$ in $\rho_L$.  Then define the space $W_i$ as
\begin{equation}
W_i = m_i V_i = \underbrace{V_1 \oplus \ldots \oplus V_i}_{m_i \text{ times}} .
\end{equation}
This is the space on which the primary representation acts.  Define the primary representation as
\begin{equation}
\tilde\rho_i = \bigoplus_{k=1}^{m_i} \rho_i .
\end{equation}
This is a representation of the group $\tilde\rho_i : G \rightarrow GL(W_i)$.  Each primary representation contains $m_i$ copies of the irreducible representation $\rho_i$.  While the primary is a reducible representation, its reduction to irreducibles is not unique.  Its reduction requires a choice of basis for the decomposition of $W_i$, as we'll show.

We can then write the decomposition of the space $\mathcal F_G$ uniquely as
\begin{equation}
\mathcal F_G = \bigoplus_i W_i .
\end{equation}
The regular representation then decomposes into primaries.
\begin{equation}
\rho_L = \bigoplus_i \tilde\rho_i
\end{equation}
This reduction is easier to understand if we write the matrix associated with the representations.  The decomposition of the regular representation into primaries puts the primaries into blocks on the diagonal, with the proper arrangement of basis functions.
\begin{equation}
\rho_L  = 
\left(
\begin{array}{c|c|c|c}
 \tilde\rho_1  & & &   \\ \hline
   &\tilde\rho_2 & &   \\ \hline
   & &\tilde\rho_3 &   \\ \hline
   & & & \ddots
\end{array}
\right)
\end{equation}
Each of the primaries acts separately on the subspaces $W_i$.  The primaries could be reduced into block diagonal form, with $m_i$ copies of the irreducible representation on the diagonal, by choosing some block basis.
\begin{equation}
\tilde\rho_i  = 
\left(
\begin{array}{c|c|c|c}
 \rho_i  & & &   \\ \hline
   &\rho_i & &   \\ \hline
   & &\rho_i &   \\ \hline
   & & & \ddots
\end{array}
\right)
\end{equation}
The block basis does not require the choice of the entire basis, just one basis vector in each of the subspaces in $W_i$ on which the irreducibles act.  The rest of the basis vectors are then generated by the action of the representation on each of chosen vectors.

These two reductions ($\rho_L \rightarrow \tilde\rho_i$ and $\tilde\rho_i \rightarrow \rho_i$) can both be performed for a general group.  Perhaps the clearest way to do this is using projection operators, as in \cite{Serre:1996lr}.

\subsubsection{From Regular to Primaries}

To construct the projection operator, we need to first define the {\em character} of a representation.  The character of a representation is a function on $G$ defined for some representation $\rho$ as 
\begin{equation}
\chi_\rho(g) = \Tr { \rho(g)}
\end{equation}
This function has many nice properties\cite{Serre:1996lr}.  For example, the characters of irreducible representations satisfy an orthogonality relation as functions in $\mathcal F_G$, and therefore ``characterize'' a representation, in some sense.  The scalar product in $\mathcal F_G$ is defined as
\begin{equation}\label{eq:char_prod}
(f_1 | f_2) = \frac{1}{N_G} \sum_{g\in G} f_1(g) f_2^*(g) ,
\end{equation}
where $N_G$ is the order of the group $G$.  In the case of finite groups, the order is simply the number of elements in the group.
With this product, the orthogonality of the characters of the irreducible representations can be written
\begin{equation}
(\chi_i | \chi_j ) = \delta_{ij},
\end{equation}
where $\chi_i$ and $\chi_j$ are the characters of the irreducible representations $\rho_i$ and $\rho_j$, and $\delta_{ij}$ is the Kronecker delta.

As described in Section 2.6 of \cite{Serre:1996lr}, we can use the characters of the irreducible representation to construct the projection operator
\begin{equation}
P_i = \frac{n_i}{N_G} \sum_{g \in G} \chi_i^*(g) \rho_L(g) ,
\end{equation}
where $\chi_i(g)$ is the character of the irreducible representation $\rho_i$, and $n_i$ is the dimension of $\rho_i$.  Consider the action of this operator on an element $f$ of one of the subspaces $W_j$, i.e., $f$ is nonzero only on $W_j$ and zero everywhere else.  Then, since $W_j$ is invariant under the action of the regular representation, we get
\begin{eqnarray}
P_i f &=& \frac{n_i}{N_G} \sum_{g \in G} \chi_i^*(g) \rho_L(g) f \\
&=& \frac{n_i}{N_G} \sum_{g \in G} \chi_i^*(g) \rho_j(g) f 
\end{eqnarray}
Since $\rho_i$ is an irreducible representation, we can use Schur's Lemma to rewrite this sum in terms of the inner product in Equation (\ref{eq:char_prod}) to obtain
\begin{equation}
P_i f = \frac{n_i}{N_G} \frac{N_G}{n_i}  (\chi_i| \chi_j ) f  = \delta_{ij} f.
\end{equation}
So, the operator $P_i$ is a projection operator.  If the element $f$ lies in the subspace $W_i$, then $P_i$ leaves it unchanged, but if $f$ has no component in $W_i$, then it is sent to zero by $P_i$.  This projection operator lets us unambiguously define the subspaces on which the primary representations act.  So we have decomposed the regular representation into primary representations, and this decomposition is unique.

\subsubsection{From Primaries to Irreducibles}

The further reduction of the primary representations $\rho_i$ into irreducible representations requires an arbitrary choice of a block basis in the spaces $W_i$.  Instead of a complete discussion of the most general case, it is perhaps better to illustrate the procedure with a simple example.  (For another example, see Section \ref{sec:DHW}, where the reduction of the regular representation of the finite Heisenberg-Weyl group is carried out in detail.)

Consider the group with only one element.  This element must be the identity element, and the group product always maps back to this element.  $G = \{e\}, e\diamond e = e$.  This group has one unitary irreducible representation, which is multiplication by 1.
\begin{equation}
\rho : G \rightarrow GL(\mathbb R), \quad \rho(e) z = \text{id} z = z.
\end{equation}
Consider the reducible representation formed by the direct sum of two copies of $\rho$.  This is the identity operator on a 2-dimensional vector space.
\begin{equation}
\tilde\rho : G \rightarrow GL(\mathbb R^2)
\end{equation}
We can decompose the action of $\tilde\rho$ into its action on one dimensional subspaces simply by choosing two linearly independent vectors $(u,v)$ in the plane.  Scaling these vectors will form independent subspaces $U = \{ \lambda u | \lambda \in \mathbb R\}$ and $V = \{ \lambda v | \lambda \in \mathbb R\}$. 
Then, by restricting the action of $\tilde\rho$ to these spaces, we can write the decomposition 
\begin{equation}
\tilde\rho = \tilde\rho|_U \oplus \tilde\rho|_V.
\end{equation}
Each of the restrictions $\tilde\rho|_U$ and $\tilde\rho|_V$ are equivalent to the irreducible representation $\rho$, so we have reduced the primary representation to irreducible representations.  However, the reduction required the choice of the vectors $u$ and $v$, so this reduction is not unique.

\section{The Stone--von Neumann Theorem\label{sec:SVN_theorem}}

The essence of the Stone--von Neumann theorem is this; the multidimensional irreducible representations of the Heisenberg-Weyl group $\mathfrak H$ can be labeled by their action on the center $Z(\mathfrak H)$.  

Since elements of the center commute with all others, any representation of the center must also commute with all other elements of the representation.  This means that unitary representations of the center are given by a phase times the identity operator.  The Stone--von Neumann theorem says that this phase identifies the irreducible representation up to a unitary transformation.   Furthermore, the unitary transformation which relates the equivalent representations is induced by a symplectic group automorphism.  

\comment{%
As we'll see, the particular form of the irreducible representation (e.g., the particular matrices in the finite case) are related to ``coordinates''  that are used to label the group elements, once a splitting into positions and momenta has been chosen.} %

Any group automorphism will take an irreducible representation into another irreducible representation.  In general, the new irreducible representation will not be unitarily equivalent to the original representation.  We can ask what conditions are required such that the new and old representations {\em will} be unitarily equivalent.  The Stone--von Neumann theorem answers this question for the Heisenberg-Weyl group.

There are two types of automorphisms of the Heisenberg-Weyl group.  A {\em symplectic transformation} is an automorphism of $\mathfrak H$ which is the identity mapping for elements of the center $Z(\mathfrak H)$.  The other type of automorphism of $\mathfrak H$ is a group dilation.  The Stone--von Neumann theorem tells us how these automorphisms are related to the irreducible representations.  Symplectic transformations take an irreducible representation into an equivalent representation.
\begin{equation}\begin{CD}
\rho @>Sym >> \rho' \sim \rho
\end{CD}\end{equation}
However, group dilations take the representation into an inequivalent representation.
\begin{equation}\begin{CD}
\rho @>Dilation >> \rho' \not\sim \rho
\end{CD}\end{equation}
According to the Stone--von Neumann theorem, all inequivalent irreducible representations can be generated from one irreducible representation by application of the group dilations.

We can write these relationships in a diagram.  Here $\mathfrak H$ is the Heisenberg-Weyl group.  $U(\mathcal{H}_{\rho})$ is the set of unitary operators on the Hilbert space $\mathcal{H}_{\rho}$ associated with the irreducible representation $\rho$.
\begin{equation} \begin{CD}
 \mathfrak H @>\rho >> U(\mathcal{H}_{\rho})\\
@V{Sym}VV  		@VV{Met}V\\
 \mathfrak H @>\rho >> U(\mathcal{H}_{\rho})\\
\end{CD} \end{equation}
The unitary transformation which relates the old representation $\rho$ with the new representation $\rho'$ is called a {\em metaplectic transformation}.

A symplectic automorphism can be thought of as a relabeling of the group elements in the phase space component of the group, where the relabeling is done in a way consistent with the group product law.  A linear canonical transformation of the phase space components is such a transformation, since it preserves the symplectic product $\omega(z_1,z_2)$. 

\comment{%
$Met$ is the group of metaplectic transformations, a particular type of unitary transformation which relates representations to each other.  The Stone--von Neumann theorem says that if we apply a symplectic transformation to $G$, then the representation $\rho$ changes only by a metaplectic transformation.  Since the symplectic transformation leaves the center unchanged, the two representations will be unitarily equivalent irreducible representations.
}%

These ideas can be illustrated by specific examples.  In the following sections, we will show how the Fourier transform arises as a consequence of the Stone--von Neumann theorem, and how the metaplectic transformations can be thought of as a generalization of the Fourier transform.

\comment{ % analogy between Stone--von Neumann and rotations
Two reps of O(3)
\begin{itemize}
\item Rotation group acts on $\mathbf{R}^3$
\item Rotation operators act on functions in $\mathbf{R}^3$ (the Hilbert space)
\end{itemize}

Two reps of Met
\begin{itemize}
\item Acts on phase space as a linear canonical transformation: M is a 6x6 matrix $\in$ SP(6).  Preserves the  Poisson brackets
\item Also has a rep acting on Hilbert space: This is the metaplectic transformation which is a kind of generalized Fourier transform.
\end{itemize}
}% end comment

\section{Examples}

\comment{%%
\subsection{The Group \sf{SO(2)}}

\subsubsection{Definition of the group}
\subsubsection{Subgroups}
\subsubsection{Representations}
}% end comment

\subsection{The Discrete Heisenberg-Weyl Group\label{sec:DHW}}

\subsubsection{Definition of the group}
\begin{figure}
\begin{center}
\includegraphics[scale=0.5]{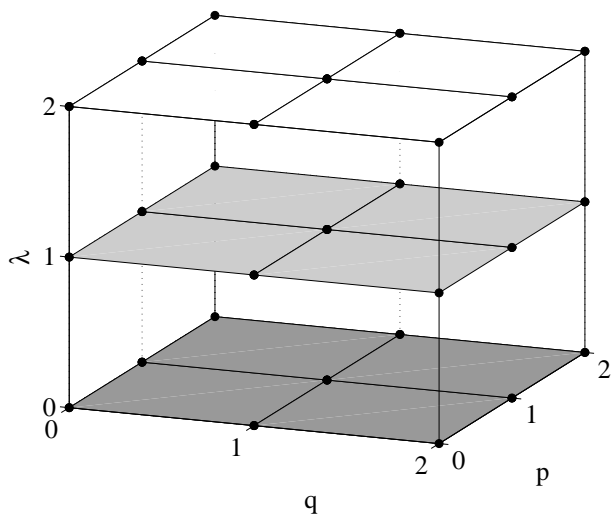}
\end{center}
\caption{\label{fig:dhw}
The elements of the discrete Heisenberg-Weyl group embedded as points in space.  The shaded plane at $\lambda = 0$ is the phase space, with elements $g=(q,p,0)$.
}
\end{figure}

The discrete Heisenberg group is made up of elements that each have three components.  Two are ``phase space'' components $z=(q,p)$ and the other is a phase $\lambda$.
\begin{equation}
\mathfrak H_n : \{ g = (z,\lambda)|z=(q,p) \in \mathbb Z_n \times \mathbb Z_n; \lambda \in \mathbb Z_n \}
\end{equation}
For concreteness, we will use $\mathfrak H_3$ to give explicit examples and to draw pictures.  Results are quoted for general $\mathfrak H_n$ when the generalization is straightforward, as it often is.  Figure \ref{fig:dhw} shows the points in the group embedded in space, for the case where $n=3$.  The group multiplication law for this group is defined as
\begin{equation}
g_1 \diamond g_2 = (z_1 + z_2,\lambda_1 + \lambda_2 + \omega(z_1,z_2) )\text{ mod } n
\end{equation}
where $\omega$ is the symplectic product
\begin{equation}
\omega(z_1,z_2) = q_1 p_2 - q_2 p_1 = z_1\cdot J^{-1} \cdot z_2
\end{equation}
The phase component $\lambda$ of the group product records the symplectic product of the phase space components, and makes this a noncommutative group.  Note that if $n=2$, this reduces to a commutative product law, which is why we use $n=3$ and not $n=2$ in our examples.

Now, for $n=3$, there are nine possibilities for $z=(q,p)$, and three for $\lambda$.  We have the following list of 27 group elements.
{\begin{spacing}{1.0}
\begin{equation}(q,p,\lambda)=\left(
\begin{array}{c}
(0,0,0)\\
(1,0,0)\\
(2,0,0)\\
(0,1,0)\\
(1,1,0)\\
(2,1,0)\\
(0,2,0)\\
(1,2,0)\\
(2,2,0)\\
(0,0,1)\\
(1,0,1)\\
(2,0,1)\\
(0,1,1)\\
(1,1,1)\\
(2,1,1)\\
(0,2,1)\\
(1,2,1)\\
(2,2,1)\\
(0,0,2)\\
(1,0,2)\\
(2,0,2)\\
(0,1,2)\\
(1,1,2)\\
(2,1,2)\\
(0,2,2)\\
(1,2,2)\\
(2,2,2)
\end{array} \right)
\end{equation}
\end{spacing}
} % end single space
{~}
This listing of the group elements can be used in a lexicographical way to write the multiplication table for the group.  The list gives labels to each element, $1=(0,0,0), 2=(1,0,0), 3=(2,0,0), \ldots$.  A multiplication table for the group lists the group elements in a row and column, and then fills in the table with the group element obtained by multiplying the row and column elements.
{\begin{spacing}{1.0}
\begin{equation}
\begin{array}{c|cccc}
\diamond & e & a & b& \ldots \\ \hline
e & e & a & b&  \\
a & a & \cdot & \cdot & \\
b & b &  \cdot & \cdot  & \\
\vdots & & & & \ddots 
\end{array}
\end{equation}
\end{spacing} } 
{~}

Since there are 27 elements, the order of the group is $N_{\mathfrak H_3} = n^3=27$.  If we apply the multiplication rule to these elements, then we get the following multiplication table.
{\small \begin{spacing}{1.0}
\begin{equation}\label{eq:disc_group_multi_table}
\setlength{\arraycolsep}{.45\arraycolsep}
\left(
\begin{array}{ccccccccccccccccccccccccccc}
1&2&3&4&5&6&7&8&9&10&11&12&13&14&15&16&17&18&19&20&21&22&23&24&25&26&27\\
2&3&1&14&15&13&26&27&25&11&12&10&23&24&22&8&9&7&20&21&19&5&6&4&17&18&16\\
3&1&2&24&22&23&18&16&17&12&10&11&6&4&5&27&25&26&21&19&20&15&13&14&9&7&8\\
4&23&15&7&26&18&1&20&12&13&5&24&16&8&27&10&2&21&22&14&6&25&17&9&19&11&3\\
5&24&13&17&9&25&20&12&1&14&6&22&26&18&7&2&21&10&23&15&4&8&27&16&11&3&19\\
6&22&14&27&16&8&12&1&20&15&4&23&9&25&17&21&10&2&24&13&5&18&7&26&3&19&11\\
7&17&27&1&11&21&4&14&24&16&26&9&10&20&3&13&23&6&25&8&18&19&2&12&22&5&15\\
8&18&25&11&21&1&23&6&13&17&27&7&20&3&10&5&15&22&26&9&16&2&12&19&14&24&4\\
9&16&26&21&1&11&15&22&5&18&25&8&3&10&20&24&4&14&27&7&17&12&19&2&6&13&23\\
10&11&12&13&14&15&16&17&18&19&20&21&22&23&24&25&26&27&1&2&3&4&5&6&7&8&9\\
11&12&10&23&24&22&8&9&7&20&21&19&5&6&4&17&18&16&2&3&1&14&15&13&26&27&25\\
12&10&11&6&4&5&27&25&26&21&19&20&15&13&14&9&7&8&3&1&2&24&22&23&18&16&17\\
13&5&24&16&8&27&10&2&21&22&14&6&25&17&9&19&11&3&4&23&15&7&26&18&1&20&12\\
14&6&22&26&18&7&2&21&10&23&15&4&8&27&16&11&3&19&5&24&13&17&9&25&20&12&1\\
15&4&23&9&25&17&21&10&2&24&13&5&18&7&26&3&19&11&6&22&14&27&16&8&12&1&20\\
16&26&9&10&20&3&13&23&6&25&8&18&19&2&12&22&5&15&7&17&27&1&11&21&4&14&24\\
17&27&7&20&3&10&5&15&22&26&9&16&2&12&19&14&24&4&8&18&25&11&21&1&23&6&13\\
18&25&8&3&10&20&24&4&14&27&7&17&12&19&2&6&13&23&9&16&26&21&1&11&15&22&5\\
19&20&21&22&23&24&25&26&27&1&2&3&4&5&6&7&8&9&10&11&12&13&14&15&16&17&18\\
20&21&19&5&6&4&17&18&16&2&3&1&14&15&13&26&27&25&11&12&10&23&24&22&8&9&7\\
21&19&20&15&13&14&9&7&8&3&1&2&24&22&23&18&16&17&12&10&11&6&4&5&27&25&26\\
22&14&6&25&17&9&19&11&3&4&23&15&7&26&18&1&20&12&13&5&24&16&8&27&10&2&21\\
23&15&4&8&27&16&11&3&19&5&24&13&17&9&25&20&12&1&14&6&22&26&18&7&2&21&10\\
24&13&5&18&7&26&3&19&11&6&22&14&27&16&8&12&1&20&15&4&23&9&25&17&21&10&2\\
25&8&18&19&2&12&22&5&15&7&17&27&1&11&21&4&14&24&16&26&9&10&20&3&13&23&6\\
26&9&16&2&12&19&14&24&4&8&18&25&11&21&1&23&6&13&17&27&7&20&3&10&5&15&22\\
27&7&17&12&19&2&6&13&23&9&16&26&21&1&11&15&22&5&18&25&8&3&10&20&24&4&14
\end{array} \right)
\end{equation}
\end{spacing}
} % end single space
An important feature of this table is that it is not symmetric, which reflects the fact that the group is not commutative (non-abelian).

\subsubsection{Commutative Subgroups}
This group has at least two types of commutative subgroups, the cyclic subgroups, and the maximal commutative subgroups.  The commutative cyclic subgroups are generated by powers of a given element of the group.
\begin{equation}
S_g = \{ g^j = \overbrace{g\diamond g\diamond \ldots \diamond g}^{j \text{ times}}  | j = 1,2,\ldots, n \}
\end{equation}
Because of the way in which it was constructed, this is a commutative subgroup.  
\begin{align}
g^{j_1} \diamond g^{j_2} & = \overbrace{g\diamond g\diamond \ldots \diamond g}^{j_1 \text{ times}} \diamond \overbrace{g\diamond g\diamond \ldots \diamond g}^{j_2 \text{ times}} \\
&= \overbrace{g\diamond g\diamond \ldots \diamond g}^{j_1+j_2 \text{ times}} \\
&= \overbrace{g\diamond g\diamond \ldots \diamond g}^{j_2+j_1 \text{ times}} \\
& = \overbrace{g\diamond g\diamond \ldots \diamond g}^{j_2 \text{ times}} \diamond \overbrace{g\diamond g\diamond \ldots \diamond g}^{j_1 \text{ times}} \\
&= g^{j_2} \diamond g^{j_1}
\end{align}
Notice that if $n$ is not a prime number, then it is possible to construct cyclic subgroups with less than $n$ elements.  For example, if $n$ is even, then it is possible to create a cyclic subgroup with $n/2$ elements, by the proper choice of the generating element $g$.  In the following, we assume this is not the case, and that the cyclic subgroups that we use do in fact have $n$ elements.

Two examples of cyclic subgroups include the ``$q$-axis'' and ``$p$-axis'' in phase space.  If you think of the $z$ components as forming a lattice in phase space, then the $q$ and $p$ axes of phase space form two separate subgroups.  These subgroups  $S_{(1,0,0)}$ and $S_{(0,1,0)}$ are generated by the elements ${(1,0,0)}$ and ${(0,1,0)}$.  For example, the $q$-axis is
\begin{equation}
S_{(1,0,0)} = \{(0,0,0), (1,0,0), (2,0,0), (3,0,0), \ldots (n-1,0,0) \}
\end{equation}
Another example of a cyclic subgroup of this group is the {\em center} of the group.  The center is set of elements which commute with all elements of the group, and in this case is formed by the $\lambda$ axis, which is generated by the element $(0,0,1)$.
\begin{equation}
S_\lambda = Z(\mathfrak H_n) = \{ (0,0,\lambda) | \lambda \in \mathbb Z_n \}
\end{equation}
This subgroup is in fact a {\em normal} subgroup of $\mathfrak H_n$.  This means that 
\begin{equation}
\forall \, h \in \mathfrak H_n, \quad h S_\lambda h^{-1} = S_\lambda,
\end{equation}
or in terms of a specific element of the center,
\begin{equation}
h \diamond \lambda \diamond h^{-1} \in S_\lambda , \quad \forall \, \lambda \in S_\lambda.
\end{equation}

\begin{figure}
\begin{center}
\includegraphics[scale=0.5]{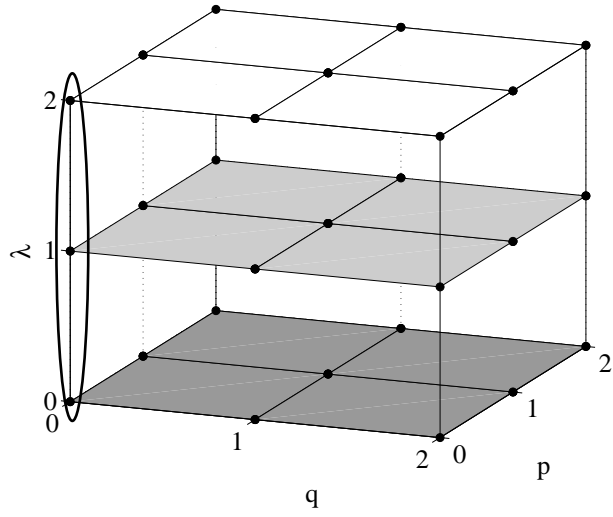}
\end{center}
\caption{\label{fig:center}
The elements of the subgroup $S_\lambda$ are circled.  These elements commute with all other elements of the group, and thus form the center of the group.  
}
\end{figure}

The second type of commutative subgroup is the maximal commutative subgroup.  A {\em maximal commutative subgroup} is one to which to which you cannot add any more elements of the group, and have the subgroup remain commutative.  One can prove that, for the group we are considering, a maximal commutative subgroup is a direct product of a cyclic subgroup and the center.  These sets have the form
\begin{equation}
\tilde S_g = S_g \oplus S_\lambda, \quad g \not\in S_\lambda.
\end{equation}
Any additional element that one might add to this subgroup will not commute with all of the elements already in the subgroup.  Interestingly, this subgroup is also a normal subgroup, $\forall \, h \in \mathfrak H_n, \, h \tilde S_g h^{-1} = \tilde S_g$.

\comment{ % cosets of the subgroups
Now that we have some specific examples of subgroups, we can construct cosets of elements in our group $\mathfrak H_n$.  As our first example, consider the subgroup $S_p=\{(0,p,0)|p=0,1,2\}$.  Then the right coset of an element not in $S_p$, e.g.~$g=(1,0,0)$, is  
\begin{align}
R&=\{(0,p,0)\diamond(1,0,0)\}\\
&=\{ (1,p, \omega((0,p),(1,0)) ) \text{ mod }3 \}\\
&=\{ (1,p, p) \}\\
&=\{ (1,0,0), (1,1,1), (1,2,2)\}
\end{align}
The left coset turns out to be different, because of the antisymmetry of $\omega$.
\begin{align}
L&=\{(1,0,0)\diamond(0,p,0)\}\\
&=\{ (1,p, \omega((1,0),(0,p)) )\text{ mod }3 \}\\
&=\{ (1,p, -p)\text{ mod }3 \}\\
&=\{ (1,0,0), (1,1,2), (1,2,1)\}
\end{align}
} % cosets

\comment{  % begin comment: factor group of S_l + S_z
As another example, consider the maximal commutative subgroup $\tilde S_z=\{(z^j,0)\diamond(0,0,\lambda)|j=1,\ldots, n; \lambda \in \mathbb Z_n\}$.  For this subgroup, the left and right cosets of any element $g\not\in \tilde S_z$ are the same, since $\tilde S_z$ is a normal subgroup.  There are $n$ distinct right and left cosets, and they are, for $g=(z',\lambda')$,
\begin{align}
R=L&=\{(z^j,0)\diamond(0,0,\lambda) \diamond (z',\lambda') \}\\
&=\{ (z^j + z',\lambda + \lambda' + j\, \omega(z,z') )\text{ mod } n \}\\
&=\{ (z^j + z', \lambda) \text{ mod } n \} .
\end{align}
These sets can also be thought of as equivalence classes of $\tilde S_z \backslash \mathfrak H_n$.  Since $\tilde S_z$ is a normal subgroup of $\mathfrak H_n$, the factor space 
\begin{equation}
F=\tilde S_z \backslash \mathfrak H_n
\end{equation}
is itself a group.  Elements of $F$ are equivalence classes, with respect to the equivalence
\begin{equation}
g_1 \sim g_2 \iff \exists s \in \tilde S_z \text{ s.t. } g_1 = s g_2.
\end{equation}
This factor group is order $n = (\text{order of }\mathfrak H_n)/(\text{order of }\tilde S_z) = n^3/n^2$.

The factor group $F$ is important because there is a natural action of the group $\mathfrak H_n$ on $F$.
\begin{equation}
\forall\, \bar k \in F, g \in \mathfrak H_n : \, \bar k g \equiv \overline{k\diamond g}
\end{equation}
This action can be used, along with a representation of $\tilde S_z$, to form an induced representation of $\mathfrak H_n$.  
} % end comment: factor group of S_l x S_z

\comment{ % cosets 
As another example, consider the center of the group $S_\lambda=\{(0,\lambda)| \lambda \in \mathbb Z_n\}$.  For this subgroup, the left and right cosets of $g\not\in S_\lambda$ are the same, since $S_\lambda$ is a normal subgroup.  There are $n^2$ distinct cosets, and they are, for $g=(z',\lambda')$,
\begin{align}
R=L&=\{(0,0,\lambda) \diamond (z',\lambda') \}\\
&=\{ (z', \lambda) \text{ mod } n \} .
\end{align}
Each coset is labeled by a point $z \in \mathbb Z_n \times \mathbb Z_n$.
These sets can also be thought of as equivalence classes of $S_\lambda \backslash \mathfrak H_n$.  Since $S_\lambda$ is a normal subgroup of $\mathfrak H_n$, the factor space 
\begin{equation}
F=S_\lambda \backslash \mathfrak H_n
\end{equation}
is itself a group.  Elements of $F$ are equivalence classes, with respect to the equivalence
\begin{equation}
g_1 \sim g_2 \iff \exists s \in S_\lambda \text{ s.t. } g_1 = s \diamond g_2.
\end{equation}
That is to say, two elements are the equivalent if their phase space components are the same.
This factor group is order $n^2 = (\text{order of }\mathfrak H_n)/(\text{order of }S_\lambda) = n^3/n$.
} % cosets

\comment{ %%%
The factor group $F$ is important because there is a natural action of the group $\mathfrak H_n$ on $F$.
\begin{equation}
\forall\, \bar k \in F, g \in \mathfrak H_n : \, \bar k g \equiv \overline{k\diamond g}
\end{equation}
This action can be used, along with a representation of $S_\lambda$, to form an induced representation of $\mathfrak H_n$.  
} %%%%

\subsubsection{Irreducible Representations}

The finite Heisenberg-Weyl group $\mathfrak H_n$ has $n^2$ inequivalent 1-dimensional irreducible representations, and $n-1$ inequivalent $n$-dimensional irreducible representations. 

We can check that these are all of the irreducible representations by using the equation $\sum n_i^2 = N_{\mathfrak H_n}$ (Section 2.5, Corollary 2(a) in \cite{Serre:1996lr}), i.e., the sum of the squared dimensions of the irreducible representations equals the order of the group.  So for the group $\mathfrak H_n$ we have
\begin{equation}\label{eq:dimension_count}
\sum n_i^2 = n^2 \cdot 1^2 + (n-1)\cdot n^2 = n^3,
\end{equation}
which is equal to the number of elements in this group, $N_{\mathfrak H_n}=n^3$.  These are therefore all of the possible unitary irreducible representations.

The 1-dimensional irreducible representations are labeled by a pair of integers $u,v \in \mathbb{Z}_n^2$
\begin{equation}\label{eq:1dim_rep}
\varrho_{u,v}(q,p,\lambda) = \exp \left( \frac{2 \pi i}{n}( u q + v p) \right) .
\end{equation}
The representation with $(u,v) = (0,0)$ gives the trivial representation, where every element of the group maps to 1.  There are $n^2$ of these representations, so for $n=3$, we have $3\times 3=9$ of these one dimensional irreducible representations.

The $n$-dimensional irreducible representations involve shifts and multiplication by a phase.  In matrix form, these can be written as
\begin{equation}\label{eq:ndim_rep}
\rho_\alpha(q,p,\lambda) = \exp \left( \frac{2 \pi i \alpha}{n}(\lambda + qp) \right)
\mathbf{T}^{2 \alpha p} \mathbf{S}^{q}
\end{equation}
where $\alpha \in \mathbb Z_n \backslash \{0\}$ and $\mathbf{S}$ is the shift matrix, and $\mathbf{T}$ is the diagonal matrix of $n^\text{th}$ roots of unity.  For example, if $n=3$, then we have
\begin{equation}\label{eq:matrix_rep}
\mathbf{S}=\left(\begin{matrix}
0 & 1 & 0\\
0 & 0 & 1\\
1 & 0 & 0\end{matrix}\right)
\qquad
\text{and}
\qquad
\mathbf{T}=\left(\begin{matrix}
1 & 0 & 0\\
0 & e^{2\pi i/3} & 0\\
0 & 0 & e^{4\pi i/3}\end{matrix}\right).
\end{equation}
These matrices act on the vector space $\mathcal{H}_{\rho_\alpha} = \mathbb C^3$.  There are, for $n=3$, two choices for $\alpha \neq 0$, so there are two 3-dimensional irreducible representations.  The case $\alpha = 0$ was excluded because with $\alpha = 0$, this representation reduces to a direct sum of one dimensional representations.

Direct computation shows that these are in fact representations of the group, and that they are irreducible.  Since these representations do not coincide on the center, they must be inequivalent representations.  (This is a consequence of the Stone--von Neumann theorem.)  Also, by the dimension count in Equation (\ref{eq:dimension_count}), these, together with the one dimensional irreducible representations, must be all of the irreducible representations of this group.  

\comment{ %%%
The $n$ dimensional representations given in Equation (\ref{eq:ndim_rep}) can be induced by the one dimensional irreducible representations of the subgroup $S_\lambda$.  These irreducible representations are
\begin{equation}
\rho_\alpha(0,0,\lambda) = \exp \left( \frac{2 \pi i}{n}( \alpha \lambda) \right), \quad \alpha \in \mathbb Z_n.
\end{equation}
}%%%

\subsubsection{The Regular Representation for $\mathfrak H_n$}

We can also write out the regular representation of this group.  The regular representation acts of the space of functions on the group.
\begin{equation}
(\rho_L(h)f)(g) = f(h^{-1} \diamond  g), \quad f:\mathfrak H_n \rightarrow \mathfrak H_n
\end{equation}
For the group we are considering, the function $f$ can be thought of as a vector in an $n^3$ dimensional space.  We can take as basis vectors delta functions on each of the elements of the group.  
\begin{equation}\label{eq:group_delta_fn}
\delta_{h}(g) = \left\{
\begin{array}{cc}
1   &  \text{ if } h=g  \\
0   &  \text{ otherwise} 
\end{array}
\right.
\end{equation}
The regular representation can then be written as a set of $n^3\times n^3$ matrices with this basis.  By looking at the multiplication table, we can figure out how some element $g$ maps the elements of the group back into the group, and construct a matrix $\rho_L(g)$ that permutes the basis vectors in the same way.  For example, if $g = (1,0,0)$, then from the multiplication table in Equation (\ref{eq:disc_group_multi_table}) we get that (leaving out the zeros)
{\begin{spacing}{0.8}
\begin{equation}
\setlength{\arraycolsep}{.6\arraycolsep}
\rho_L(1,0,0)=\left(
\begin{array}{ccc|ccc|ccc|ccc|ccc|ccc|ccc|ccc|ccc}
 & &1& & & & & & & & & & & & & & & & & & & & & & & & \\
1& & & & & & & & & & & & & & & & & & & & & & & & & & \\
 &1& & & & & & & & & & & & & & & & & & & & & & & & & \\ \hline
 & & & & & & & & & & & & & & & & & & & & & & &1& & & \\
 & & & & & & & & & & & & & & & & & & & & &1& & & & & \\
 & & & & & & & & & & & & & & & & & & & & & &1& & & & \\ \hline
 & & & & & & & & & & & & & & & & &1& & & & & & & & & \\
 & & & & & & & & & & & & & & &1& & & & & & & & & & & \\
 & & & & & & & & & & & & & & & &1& & & & & & & & & & \\ \hline
 & & & & & & & & & & &1& & & & & & & & & & & & & & & \\
 & & & & & & & & &1& & & & & & & & & & & & & & & & & \\
 & & & & & & & & & &1& & & & & & & & & & & & & & & & \\ \hline
 & & & & &1& & & & & & & & & & & & & & & & & & & & & \\
 & & &1& & & & & & & & & & & & & & & & & & & & & & & \\
 & & & &1& & & & & & & & & & & & & & & & & & & & & & \\ \hline
 & & & & & & & & & & & & & & & & & & & & & & & & & &1\\
 & & & & & & & & & & & & & & & & & & & & & & & &1& & \\
 & & & & & & & & & & & & & & & & & & & & & & & & &1& \\ \hline
 & & & & & & & & & & & & & & & & & & & &1& & & & & & \\
 & & & & & & & & & & & & & & & & & &1& & & & & & & & \\
 & & & & & & & & & & & & & & & & & & &1& & & & & & & \\ \hline
 & & & & & & & & & & & & & &1& & & & & & & & & & & & \\
 & & & & & & & & & & & &1& & & & & & & & & & & & & & \\
 & & & & & & & & & & & & &1& & & & & & & & & & & & & \\ \hline
 & & & & & & & &1& & & & & & & & & & & & & & & & & & \\
 & & & & & &1& & & & & & & & & & & & & & & & & & & & \\
 & & & & & & &1& & & & & & & & & & & & & & & & & & &  
 \end{array} \right)
\end{equation}
\end{spacing}
} % end single space
{~\\~}
Other 27-dimensional matrices can be constructed for each other element of the group.

\subsubsection{Decomposition of the left regular representation}

Recall that the left regular representation is a shift operation on functions on the group.  It is defined as
\begin{equation}
(\rho_L(h) f)(g)=f(h^{-1} \diamond g),
\end{equation}
where $f(g)$ is a function on the group.  The set of such functions can be thought of as an $n^3$ dimensional vector space $\mathcal F = \{ f: \mathfrak H_n \rightarrow \mathbb C \}$.
As described in Section \ref{sec:reduction}, we can reduce this representation into a direct sum of the irreducible representations, with multiplicities equal to the dimension of the particular irreducible representation.  This reduction is performed by constructing a projection operator which projects functions onto the $\rho$-invariant subspace associated with the particular irreducible representation.  This procedure divides the Hilbert space into the primary components.  These primary components can be further reduced to the irreducible components by an arbitrary choice of ``basis'' in the $\rho$-invariant subspace.  As we'll see the primary representations act on functions on phase space, while the irreducible representations act on functions on a ``coordinate'' subspace.

\subsubsection{Canonical Decomposition: the reduction to primaries}

This procedure can be carried out explicitly in this case, using the formulas for the irreducible representations in Equations (\ref{eq:1dim_rep}) and (\ref{eq:ndim_rep}), and the formula for the projector \cite{Serre:1996lr}.  Let's use the characters of the irreducible representation $\rho_\alpha$ defined above, with $\alpha=1$, and also specialize to the case $n=3$.  The projector is 
\begin{equation}
P_1=\frac{1}{3} \sum_{h\in G} \overline{\chi_{\rho_1}(h)}\rho_L(h)
\end{equation}
Using the explicit formula above for the irreducible representation $\rho_1$, we can calculate its character
\begin{equation}
\chi_{\rho_1}(g)=3\left(\delta_{(0,0,0)}+e^{\frac{2\pi i}{3}}\delta_{(0,0,1)}+e^{\frac{4\pi i}{3}}\delta_{(0,0,2)}\right)(g),
\end{equation}
where the delta functions are functions on the group defined in Equation (\ref{eq:group_delta_fn}).
If we think of this character as a function on the group, we see that it is only nonzero on the center $S_\lambda$, where its functional form is given by an irreducible representation of the subgroup $\rho: S_\lambda \rightarrow e^{\frac{2\pi i}{3} \lambda}$.

\begin{figure}
\begin{center}
\includegraphics[scale=0.6]{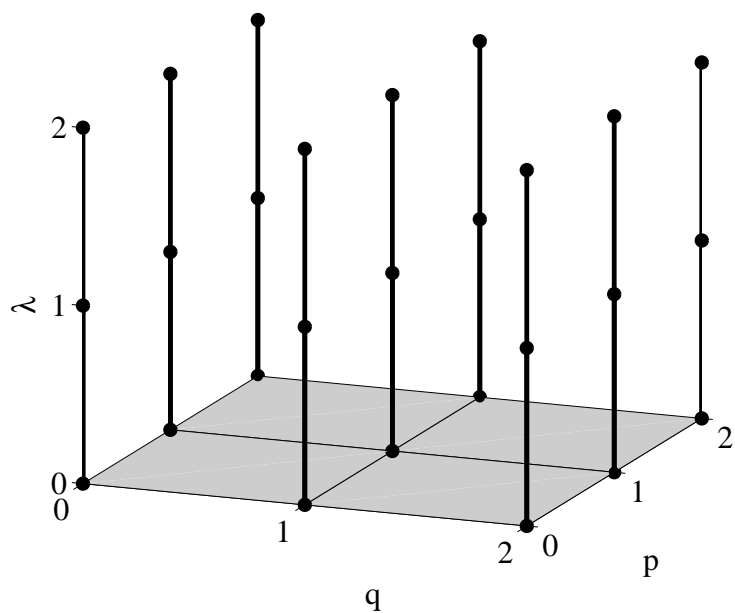}
\end{center}
\caption{\label{fig:primary_basis}
The basis functions of the primary subspace are delta functions along fibers over points in phase space.  Functions in this subspace can have arbitrary values for different points in phase space, but their values in the $\lambda$ direction are determined by the particular irreducible representation associated with this invariant subspace.
}
\end{figure}

Using the formula for the character of $\rho_1$, we see that the projector acting on a given function on the group is 
\begin{align}
(P_1 f)(g)&= \frac{1}{3} \sum_{h\in G} \overline{\chi_{\rho_1}(h)}(\rho_L(h) f)(g) \\
&= \sum_{h\in G} \left(\delta_{(0,0,0)}+e^{\frac{4\pi i}{3}}\delta_{(0,0,1)}+e^{\frac{2\pi i}{3}}\delta_{(0,0,2)}\right)\!(h) \, f(h^{-1} \diamond g) \\
&= f(g) + e^{\frac{4\pi i}{3}} f( (0,0,1)^{-1} \diamond g) + e^{\frac{2\pi i}{3}} f( (0,0,2)^{-1} \diamond g) \\
&= f(g) + e^{\frac{4\pi i}{3}} f( (0,0,2) \diamond g) + e^{\frac{2\pi i}{3}} f( (0,0,1) \diamond g) 
\end{align}
This projector should project any function onto a 9-dimensional subspace of $\mathcal F$ which is invariant under the action of the regular representation.  Since the delta functions form a basis in $\mathcal F$, we can use them to find a basis of the 9-dimensional subspace.
\begin{align}
(P_1 \delta_h)(g)&=\delta_h(g) + e^{\frac{4\pi i}{3}} \delta_h((0,0,2) \diamond g) + e^{\frac{2\pi i}{3}} \delta_h((0,0,2)\diamond g) \\
&=\delta_h(g) + e^{\frac{4\pi i}{3}} \delta_{(0,0,-2)\diamond h}(g) + e^{\frac{2\pi i}{3}} \delta_{(0,0,-1)\diamond h}(g) \\
&= \sum_{\lambda \in \mathbb{Z}_3} e^{\frac{2\pi i}{3} \lambda} \delta_{(h_q,h_p,h_\lambda-\lambda)}(g)\\
&= \sum_{\lambda \in \mathbb{Z}_3} e^{\frac{2\pi i}{3} (h_\lambda-\lambda)} \delta_{(h_q,h_p,\lambda)}(g) \\
&= e^{\frac{2\pi i}{3} h_\lambda} \sum_{\lambda \in \mathbb{Z}_3} e^{-\frac{2\pi i}{3}\lambda} \delta_{(h_q,h_p,\lambda)}(g)
\end{align}
From this expression it is clear that the delta functions for all possible values of $h$ project onto only nine linearly independent functions on the group, one associated with each point $(h_q, h_p)$ in the phase space.  Each of the delta functions with the same phase space component map to the same function, up to a complex coefficient given by the $\lambda$ component (see Figure \ref{fig:primary_basis}).  

We can use this projection of the delta functions to define a basis for the 9-dimensional invariant subspace.
\begin{equation}
f_{(q,p)}(g) = (P_1 \delta_{(q,p,0)})(g)=
\sum_{\lambda \in \mathbb{Z}_3} e^{-\frac{2\pi i}{3} \lambda} \delta_{(q,p,\lambda)}(g)
\end{equation}
These functions are an example of the covariant functions described in Section \ref{sec:obs}.  They satisfy the covariance relationship in Equation (\ref{eq:covariant_functions}) for the commutative subgroup $S_\lambda$.  This can be checked by calculating how the functions vary under a right shift by an element of the center $S_\lambda$.
\begin{align}
f_{(q,p)}(g\diamond (0,0,\lambda)) &= \sum_{\lambda' \in \mathbb{Z}_3} e^{-\frac{2\pi i}{3} \lambda'} \delta_{(q,p,\lambda')}(g\diamond (0,0,\lambda))
\end{align}
We can use the delta function to move the element $(0,0,\lambda)$ from the argument into the phase.
\begin{align}
g\diamond (0,0,\lambda) = (q,p,\lambda')  \, \implies  \,
g = (q,p,\lambda' - \lambda) \equiv (q,p,\lambda'')
\end{align}
Use the new variable $\lambda'' =\lambda' - \lambda$ to simplify the expression above.
\begin{align}
f_{(q,p)}(g\diamond (0,0,\lambda)) &= \sum_{\lambda'' \in \mathbb{Z}_3} e^{-\frac{2\pi i}{3} (\lambda''+\lambda)} \delta_{(q,p,\lambda'')}(g) \\
&= e^{-\frac{2\pi i}{3} \lambda} \sum_{\lambda'' \in \mathbb{Z}_3} e^{-\frac{2\pi i}{3} \lambda''} \delta_{(q,p,\lambda'')}(g) \\
&= e^{-\frac{2\pi i}{3} \lambda} f_{(q,p)}(g)
\end{align}
So the functions $f_{(q,p)}(g)$ are elements of the subspace $\mathcal S (S_\lambda,\tau)$,
\begin{equation}
\mathcal{F} \supset \mathcal S (S_\lambda,\tau) = \{ f: \forall \, h \in S_\lambda, \forall \, g \in G, f( g \diamond h) = \tau(h) f(g)\},
\end{equation}
where $\tau$ is the irreducible representation of the center given by 
\begin{equation}
\tau((0,0,\lambda)) = e^{-\frac{2\pi i}{3} \lambda}.
\end{equation}
As shown in Equation (\ref{eq:cov_factor_set}), any function in the space $\mathcal S (S_\lambda,\tau)$ can be described by the values that it takes on the elements of the factor space $\mathfrak{H}_n/S_\lambda$.  This can be written fairly simply by using the functions $f_{(q,p)}(g)$ as a basis.  For any $\phi \in \mathcal S (S_\lambda,\tau)$, we can expand it on the basis.
\begin{equation}
\phi(g) = \sum_{q,p} c_{(q,p)} f_{(q,p)}(g)
\end{equation}
The points $(q,p)$ label the equivalence classes in $\mathfrak{H}_n/S_\lambda$, and we can think of the expansion coefficients $c_{(q,p)}$ as a function on the space of equivalence classes.

\comment{
These functions are each confined to the ``fiber'' of $\lambda$ values over a particular point in phase space (see Figure \ref{fig:primary_basis}).  The behavior of the function in the $\lambda$ direction is determined by the phase of the irreducible representation $\rho_1$ which we have chosen.  (These functions can also be thought of as functions on the factor group $G / S_\lambda$.  Elements in the factor group are equivalence classes of points, where two group elements are said to be equivalent if their phase space components are equal.  The irreducible representation $\rho_1$ that we are considering can be induced from an irreducible representation of the center, and these functions on $G / S_\lambda$ are just the ones that arise when considering the induced representation.  This construction of the irreducible representations using induction from irreducible representations of the center is related to Kirillov's orbit method.  Interested readers should refer to Kirillov's book \cite{Kirillov:2004lr} for a deeper discussion of this point of view.)
}

The invariance of the subspace $\mathcal S (S_\lambda,\tau)$ under the action of the left regular representation can be shown by computing the action of $\rho_L(h)$ on an arbitrary basis function, for any $h$ in the whole group.
\begin{eqnarray}
(\rho_L(h) f_{(q,p)})(g)&=&   f_{(q,p)}(h^{-1}\diamond g)\\
&=& \sum_{\lambda \in \mathbb{Z}_3} e^{-\frac{2\pi i}{3} \lambda} \delta_{(q,p,\lambda)}(h^{-1} \diamond g) \\
&=& \sum_{\lambda \in \mathbb{Z}_3} e^{-\frac{2\pi i}{3} \lambda} \delta_{(h_q+q,h_p+p,h_\lambda+\lambda +(h_q p - h_p q) )}(g) \\
&=&\exp\left(\frac{2\pi i}{3}(h_\lambda + h_q p - h_p q)\right) f_{(q+h_q,p+h_p)}(g) \label{eq:dhw_primary}
\end{eqnarray}
So, the regular representation permutes the functions $f_{(q,p)}$ among themselves, and multiplies them by a phase.  Therefore,  $\mathcal S (S_\lambda,\tau)$ is invariant under the action of the regular representation.  Even though we used an explicit choice of basis functions (the $f_{(q,p)}(g)$ functions), this is actually not necessary in order to reduce the regular representation.  All the information about this invariant subspace is contained in the projection operator $P_1$ which we used to construct this basis.  The restriction of the left regular representation to the subspace $\mathcal S (S_\lambda,\tau)$ is a primary representation associated with the irreducible representation $\rho_1$.  This primary representation acts on functions on phase space, i.e., it acts on functions which are defined by their values $c_{(q,p)}$ at each point $(q,p)$.

The 9-dimensional invariant subspace $\mathcal S (S_\lambda,\tau)$ can be further reduced into its three invariant 3-dimensional components.  However, this reduction is arbitrary, since there is a choice of basis for the 3-dimensional subspaces which must be made.  This choice is related to the choice of a larger commutative subgroup (see Section \ref{sec:obs}) which is necessary for the following reduction.

% end of old construction of irreducible representations

\subsubsection{Reduction of the regular representation to irreducibles}

As observed in Section (see Section \ref{sec:obs}), if we want to reduce the regular representation the most, we can do this by finding the largest commutative subgroup.  For the group $\mathfrak{H}_n$ that we are considering, the maximal commutative subgroups are formed as a direct sum of the center and another cyclic subgroup.  There is some flexibility in choosing the cyclic subgroup, which is reflected in fact that reduction of the regular representation to irreducibles is not unique.  In this section, we will choose a cyclic subgroup which lies in the phase space component of the group.  Pick an element of the group which is in the phase space component, $g_z = ({\bf z},0)$.  The point ${\bf z} \neq (0,0)$ can now be used to generate a cyclic subgroup.  
\begin{equation}
S_z = \{ g_z, g_z \diamond g_z, \ldots , g_z^n \}
\end{equation}
This choice of an element in phase space splits the phase space into separate components.  If for example we think of the element ${\bf z}$ as a momentum, then the subgroup $S_z$ would be the momentum axis, or ``p''-axis.  All other elements of phase space $({\bf z'},0)$ will have some ``position'' component, so they will not commute with the ``momenta'' in $S_z$.

\begin{figure}
\begin{center}
\includegraphics[scale=0.5]{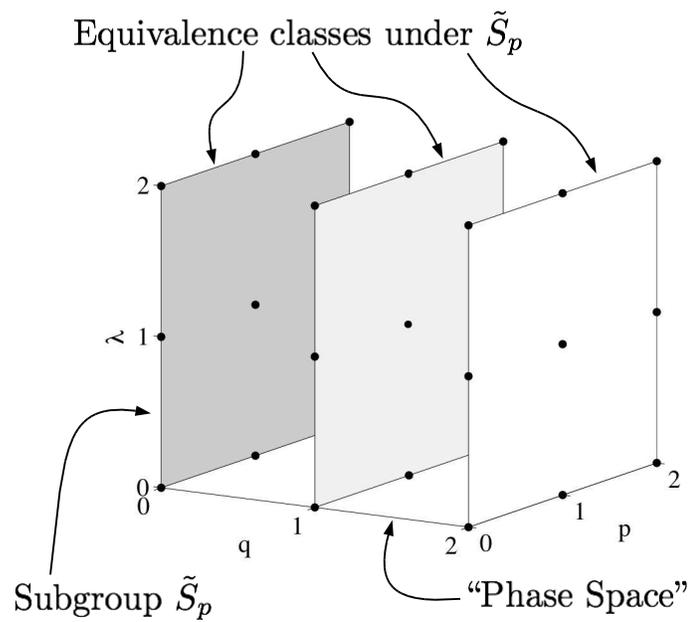}
\end{center}
\caption{\label{fig:max_subgroup}
The elements of the subgroup $\tilde S_p = S_p \oplus S_\lambda$ lie in the $\lambda p$-plane, which is shaded dark grey.  The two planes parallel to it are the equivalence classes in the factor group $\mathfrak H / \tilde S_z$.  These classes are labeled by the value of $q$, which is distinct in each class.  Note that in  contrast to the $qp$-plane, the $\lambda p$-plane is not a ``phase space'' since all elements in the $\lambda p$-plane mutually commute.}
\end{figure}

We can now use the subgroup $S_z$ to construct a maximally commutative subgroup.  Form the direct sum subgroup $\tilde S_z = S_z \oplus S_\lambda, {\bf z}\neq 0$.  This subgroup is a maximal commutative subgroup, with $n^2$ elements.  It can be thought of as the set of elements in the group which form the plane through the group which lies over the ``$\bf z$''-axis.  For example, if ${\bf z} = (0,1)$, this set is the $\lambda p$-plane, as shown in Figure (\ref{fig:max_subgroup}).

This commutative subgroup has 1-dimensional irreducible representations
\begin{equation}
\rho_{(\alpha,\beta)}( j z, \lambda) = \exp \left( \frac{2\pi i}{n} (\alpha j + \beta \lambda) \right),
\end{equation}
where $(\alpha,\beta) \in \mathbb Z_n \oplus \mathbb Z_n$.  We would like to use one of these representations as a starting point for constructing an irreducible representation of the group.

In the reduction of the regular representation to primaries, we got functions that only depended on the phase space $(q,p)$ component of $h$.  That means that they only depended on the class of $h$ in $\mathfrak H_n / S_\lambda$.  Similarly, we will now consider functions which only depend on the class of $h$ in $\mathfrak H_n / \tilde S_z$.  Such functions can be found by considering the space of function which are covariant with respect to right shifts by elements of the subgroup $\tilde S_z$.
\begin{equation}\label{eq:cov_for_irreps}
\mathcal S (\tilde S_z,\rho_{(\alpha,\beta)}) = \{ f: \forall \, h \in \tilde S_z, \forall \, g \in G, f( g \diamond h) =  \rho_{(\alpha,\beta)}(h) f(g)\}
\end{equation}
This space is invariant under the action of $\rho_L$, and it is a subspace of the space $\mathcal S (S_\lambda,\tau)$, if we choose the representation $\rho_{(\alpha,\beta)}$ of $\tilde S_z$ which corresponds to the representation $\tau$ of $S_\lambda$.

We would now like to define a set of basis functions for $\mathcal S (\tilde S_z,\rho_{(\alpha,\beta)})$.  As in the reduction to the primary representations, these basis functions can be labeled by the right equivalence classes in $\mathfrak H_n / \tilde S_z$.  First pick a set of representatives $k$ for the classes.  This set is a choice of $n$ group elements, one in each of the different equivalence classes.  In particular, we will choose $k$'s that form a cyclic subgroup.  This can easily be done by taking $k \in S_{{\bf z}_\perp}$, where $\omega({\bf z},{\bf z}_\perp) \neq 0$.  This means that we are choosing the the elements $k$ to be in the phase space, i.e., they have $\lambda =0$.  While this choice is not necessary, it makes the following construction simpler.

Now define the basis functions analogous to the functions $f_{(q,p)}$ in the previous section.
\begin{align}\label{eq:basis_for_cov_for_irreps}
\phi_{k} (g) &= \sum_{h \in \tilde S_z}  \rho_{(\alpha,\beta)}(h)  \delta_{k\diamond h}(g ) 
\end{align}
Since there are $n$ different equivalence classes, there are only $n$ of these functions, so the regular representation restricted to this subspace is actually an $n$ dimensional irreducible representation, as long as we can show that this is an invariant subspace.  Actually, all we need to do is show that these functions satisfy the covariance relationship in Equation (\ref{eq:cov_for_irreps}).  If they do, then by construction, they belong to a subspace which is invariant under the action of the left regular representation.

Let's check the covariance of these functions by directly computing how behave under a right shift by an element of the subgroup $\tilde S_z$.
\begin{align}
\phi_k(g\diamond h') &=  \sum_{h \in \tilde S_z}  \rho_{(\alpha,\beta)}(h)  \delta_{k\diamond h}(g \diamond h')  
\end{align}
Now use the delta function to rearrange the group elements.
\begin{equation}
k\diamond h = g\diamond h' \quad \implies \quad k \diamond h'' = g
\end{equation}
Here we made the change of variables $h'' = h\diamond (h')^{-1} = h - h'$.  Since these are elements of a commutative subgroup, we are allowed to turn the multiplication by the inverse into ordinary subtraction.  Insert these results back into the summation.
\begin{align}
\phi_k(g\diamond h') &=  \sum_{h'' \in \tilde S_z}  \rho_{(\alpha,\beta)}(h''+h')  \delta_{k\diamond h''}(g ) \\
&= \rho_{(\alpha,\beta)}(h'') \sum_{h'' \in \tilde S_z}  \rho_{(\alpha,\beta)}(h')  \delta_{k\diamond h''}(g ) \\
&= \rho_{(\alpha,\beta)}(h'') \phi_k(g)
\end{align}
This shows that the functions $\phi_k(g)$ do actually belong to the set $\mathcal S (\tilde S_z,\rho_{(\alpha,\beta)})$.  Also, since there are $n$ of these functions, and they are linearly independent (a property inherited from the delta functions) they form a basis of the $n$-dimensional space $\mathcal S (\tilde S_z,\rho_{(\alpha,\beta)})$.

Now that we have identified the invariant subspace $\mathcal S (\tilde S_z, \rho_{(\alpha,\beta)})$, we can construct an irreducible representation $\tilde \rho = \rho_L \vert_{\mathcal S (\tilde S_z, \rho_{(\alpha,\beta)})}$ by restricting the left regular representation to this subspace.  The exception to this is the case where $(\alpha, \beta) = (\alpha,0)$.  In this case, the reduced representation $\tilde \rho$  is a reducible permutation of the elements of $\mathcal S (\tilde S_z, \rho_{(\alpha,\beta)})$.  This could be shown by direct calculation, but is easier to see using the Stone--von Neumann theorem.  When $\beta=0$, the representation $\tilde \rho$ maps the elements of the center to the identity operator.  Then, by the Stone--von Neumann this representation is unitarily equivalent to the trivial representation, where all elements map to the $n$-dimensional identity operator.  Therefore, since the $n$-dimensional identity operator is reducible, $\tilde \rho$ is also reducible.

The reduction of the regular representation to $n$-dimensional irreducible representations performed above contains an arbitrary choice of basis.  This can be seen in the choice of the subgroup we used to construct this basis.  There is no reason to choose one element $z$ over any other, and a different choice of $z$ component can give us a different reduction.  However, according to the Stone--von Neumann theorem, this different reduction will give the same irreducible representations, up to a unitary transformation, since it maps the center of the group to the same set of operators.

\subsubsection{Example of the reduction to irreducible representations}

As a concrete example of the reduction to irreducible representations, take the discrete Heisenberg-Weyl group with $n=3$.  Consider the subgroup $\tilde S_p$, shown in Figure (\ref{fig:max_subgroup}).  This subgroup contains the 9 points in the $p \lambda$-plane, $(0,p,\lambda)$.    Take the representation with $(\alpha, \beta) = (0,2)$, which, because $n=3$, is equivalent to $(\alpha, \beta) = (0,-1)$.
\begin{equation}
\rho_{(0,-1)}  (0,p,\lambda) = \exp \left( -\frac{2\pi i}{3}  \lambda \right)
\end{equation}
The basis functions are written as a sum over all values of $p$ and $\lambda$, and can be labeled by the $q$ value of points in the equivalence class.
\begin{align}\label{eq:dhw_q_basis}
\phi_q (g) &= \sum_{h\in \tilde S_p} \rho_{(0,-1)}(h) \delta_{k\diamond h}(g) \\
&= \sum_p \sum_\lambda \exp \left( -\frac{2\pi i}{3} \lambda \right) \delta_{(q,0,0)\diamond (0,p,\lambda)}(g) 
\end{align}
Here we have used the three elements on the $q$ axis to represent the classes, 
\begin{equation}
S_q = \{(0,0,0), (1,0,0), (2,0,0) \}.
\end{equation}
Using these as representatives of the classes $\mathfrak H / \tilde S_p$ means that an arbitrary element of the group can be written as
\begin{equation}
h = (h_q, h_p, h_\lambda) = (h_q, 0,0) \diamond (0, h_p, h_\lambda -h_q h_p).
\end{equation}
We now want to see how the regular representation decomposes on the basis we have chosen.  The decomposition of group elements given above is used several times in the following in order to move the element $h$ from the argument of the delta function into the parameter of the delta function.
\begin{align}
(\rho_L(h)\phi_q)(g) &= \phi_q(h^{-1}\diamond g) \\
&= \sum_p \sum_\lambda \exp \left( -\frac{2\pi i}{3} \lambda \right) \delta_{(q,0,0)\diamond (0,p,\lambda)}(h^{-1}\diamond g) \\
&= \sum_p \sum_\lambda \exp \left( -\frac{2\pi i}{3} \lambda \right) \delta_{(q+h_q,0,0)\diamond (0,p+h_p,\lambda+h_\lambda -h_p(h_q+2q))}(\diamond g) 
\end{align}
Now change the variables in the summation to simplify this expression.
\begin{align}
(\rho_L(h)\phi_q)(g) &= \sum_{p'} \sum_{\lambda'} \exp \left( -\frac{2\pi i}{3} (\lambda' - h_\lambda +h_p(h_q+2q) ) \right) \delta_{(q+h_q,0,0)\diamond (0,p',\lambda')}(g) \\
&= \exp\left( \frac{2\pi i}{3} (h_\lambda - h_p (h_q + 2 q) ) \right) \phi_{q+h_q}(g) \label{eq:disc_sch_rep} 
\end{align}
This reduction of the regular representation cannot be reduced any further.  
This representation is a discrete version of the standard Schr\"odinger representation of the continuous Heisenberg-Weyl group. 

We can write this representation as a matrix representation, where we use the functions $\phi_q$ as the basis vectors.
\begin{equation}
\phi_0(g) = \left(\begin{array}{c} 1\\0\\0\end{array}\right), \,
\phi_1(g) = \left(\begin{array}{c} 0\\1\\0\end{array}\right), \,
\phi_2(g) = \left(\begin{array}{c} 0\\0\\1\end{array}\right)
\end{equation}
Using this basis, the matrices $\bf S$ and $\bf T$ from Equation (\ref{eq:matrix_rep}) can be used to write the reduced representation as
\begin{equation}\label{eq:qr_matrix}
({\bf R}_Q (h) )\phi_q = e^{ \frac{2\pi i}{3} (h_\lambda +h_p h_q )} {\bf T}^{-2 h_p} {\bf S}^{-h_q} \phi_q .
\end{equation}
So here it is, an concrete example of how the regular representation can be reduced to an irreducible representation.

\subsubsection{Metaplectic Transformations and the Discrete HW Group}

The discrete Heisenberg-Weyl group that we are considering is an ideal context in which to illustrate the Stone--von Neumann theorem.  From the perspective of a physicist, perhaps the most interesting thing about the Stone--von Neumann theorem is the importance it gives to phase space instead of configuration space.  Mathematically, phase space is related to the primary representations of the Heisenberg-Weyl group, and this is good since the reduction of the regular representation to primaries is well defined.  On the other hand configuration space is related to the reduction of primary representations to irreducible representations.  This reduction involves an arbitrary choice of a particular commutative subgroup in phase space to use as the configuration space.  Since this choice of configuration space is not unique, it is not the most natural space to consider.  A unitary transformation of the representation will take you from one choice of configuration space to a different choice, so any choice you make is not really any better than any other.

In the case of the discrete Heisenberg-Weyl group these ideas are readily illustrated.  The primary representation acts on functions on phase space.  Writing the phase space component as $z=(q,p)$, the primary representation in Equation (\ref{eq:dhw_primary}) becomes 
\begin{equation}
(\rho(z',\lambda) f_{(z)} ) (g) = \exp\left( \frac{2\pi i}{3} (\lambda + \omega(z,z'))\right) f_{(z-z')} (g).
\end{equation}
Further reduction of the regular representation required choosing a maximal commutative subgroup $\tilde S_z$ of $\mathfrak H$ and a set of labels for the equivalence classes in $\mathfrak H / \tilde S_z$.  In Equation (\ref{eq:dhw_q_basis}) we chose $\tilde S_p$ as the maximal commutative subgroup, and we chose the $q$-axis as the labels for the classes in $\mathfrak H / \tilde S_p$.  This gave us a set of functions labeled by $q$ on which the regular representation could be reduced.  We obtained the representation in Equation (\ref{eq:disc_sch_rep}).
\begin{equation}
(\rho_L(h)\phi_q)(g) = \exp\left( \frac{2\pi i}{3} (h_\lambda - h_p (h_q + 2 q) ) \right) \phi_{q+h_q}(g) 
\end{equation}
If instead we use the subgroup $\tilde S_q = S_\lambda \oplus S_{(q,0,0)}$ to decompose the group, and label our functions by elements along the $p$-axis, then we would get the representation
\begin{equation}
(\rho_L(h)\phi_{p})(g) = \exp\left( \frac{2\pi i}{3} (h_\lambda + h_q (h_p + 2 p) ) \right) \phi_{p+h_p}(g).
\end{equation}
Using the $\{\phi_{p}\}$ basis, the matrix form of this representation is
\begin{equation}\label{eq:pr_matrix}
({\mathbf R}_P (h) ) \phi_p = \exp\left( \frac{2\pi i}{3} (h_\lambda -h_q h_p) \right){\bf T}^{2 h_q} {\bf S}^{-h_p} \phi_p .
\end{equation}

\comment{
In matrix form, this representation is
\begin{equation}
\rho_R(h) = e^{ \frac{2\pi i}{3} (h_\lambda - h_q h_p)} \mathbf T^{2 h_q} \mathbf S^{h_p}.
\end{equation}
}

The representations ${\bf R}_Q$ and ${\bf R}_P$ both map elements $(0,0,\lambda)$ of the center to multiplication by the phase $\exp\!\left( \frac{2\pi i}{3} \lambda \right)$.  The Stone--von Neumann theorem then tells us that there is a unitary transformation which converts one of these representations into the other.  In this case the transformation which does this is related to the discrete Fourier transformation.  We can write the discrete Fourier matrix as
\begin{equation}\label{eq:ft_matrix}
[\mathbf W]_{j,k} = \exp\left( \frac{2\pi i}{3} (jk) \right).
\end{equation}
The Fourier transform takes the $q$-axis to the $p$-axis by rotating phase space by $90^\circ$.  Another application of the Fourier transform would give another rotation, which is the same as flipping the sign on both axes.  Since we don't want to change the sign of the axes, we might expect that we need to use the inverse Fourier transform to relate the two representations ${\bf R}_Q$ and ${\bf R}_P$.  Direct calculation confirms that this is correct.

Using the definition of $\bf W$, it can be shown that the $3\times 3$ matrices $\bf W^{-1}$, $\bf S$,  and $\bf T$ satisfy certain commutation relations.
\begin{align}
{\bf S}^{-1} \cdot{\bf W}^{-1} &=  {\bf W}^{-1} \cdot{\bf T}^{-2} \\
{\bf T}^{-1} \cdot{\bf W}^{-1} &=  {\bf W}^{-1} \cdot{\bf S} \\
{\bf S} \cdot {\bf T} &= e^{\frac{2\pi i}{3}}{\bf T}\cdot {\bf S}
\end{align}
Using these relations, we can show that the matrix representations ${\bf R}_Q$ and ${\bf R}_P$ are related as follows
\begin{equation}
\mathbf{R_P} \cdot \mathbf W^{-1} = \mathbf W^{-1} \cdot \mathbf{R_Q}.
\end{equation}
So the inverse Fourier transform given by the matrix $\mathbf{W}^{-1}$ relates the two representations.  It is the unitary transformation which we knew must exist as a consequence of the Stone--von Neumann theorem.  If the  representations ${\bf R}_Q$ and ${\bf R}_P$ were constructed using different maximal commutative subgroups, they would still be related by a unitary transformation.  Instead of the Fourier transform matrix, however, the unitary transformation would be a general metaplectic transformation.

% end generic metaplectic

\subsection{The Continuous Heisenberg-Weyl Group\label{sec:cont_HW}}

The continuous Heisenberg-Weyl group, $\mathfrak H$ is a generalization of the discrete group discussed in the previous section.  As a set, $\mathfrak H$ is composed of a phase space component, and a additional one-dimensional degree of freedom which encodes the commutation relations.  So its elements are points in $\mathbb R^{2n+1}$.
\begin{equation}\label{eq:cont_HW}
\mathfrak H : \{ g=({\bf z},\lambda) | {\bf z = (q,p)} \in \mathbb R^n \times \mathbb R^n ; \lambda \in \mathbb R \}
\end{equation}
Here $n$ is the dimension of the configuration space.  For an $N$ particle system in 3 dimensional space, $n= 3N$.  The group product law analogous to the discrete case, except that the operations are now vector operations.
\begin{equation}
g' \diamond g = ({\bf z' + z},\lambda' + \lambda + \frac{1}{2} \omega({\bf z',z}) )
\end{equation}
and $\omega$ is again the symplectic product
\begin{equation}
\omega({\bf z',z}) = {\bf q'^{\rm T} \cdot p - q^{\rm T} \cdot p' = z'^{\rm T}\cdot J^{\rm -1} \cdot z }.
\end{equation}

\subsubsection{Commutative Subgroups}

As in the discrete case, the continuous Heisenberg-Weyl group has a variety of commutative subgroups.  The center of the group is again formed by the $\lambda$ axis.  Each element $(0,\lambda)$ commutes with all elements of the group, and the set of such elements forms a commutative subgroup which is isomorphic to the group of real numbers with addition, $(\mathbb R, +)$.

There are also subgroups which can be generated by a given element of the group.  Take any element $({\bf z},\lambda)$.  Then consider the set generated scalar multiples of this element. 
\begin{equation}
S_{g} = \{ (c {\bf z}, c \lambda) | c \in \mathbb R \}
\end{equation}
This set is a line through the origin in $\mathbb R^{2n +1}$, and it is a commutative subgroup.

More interesting commutative subgroups can be formed by taking direct products of subgroups of the form $S_{g}$, where $g=({\bf z}, 0)$ has no $\lambda$ component.  This group then is a line in the $2n$-dimensional phase space.   Consider a maximally commuting, linearly independent, set of vectors ${\bf z}_i$ in phase space.  Since phase space is $2n$-dimensional, we can pick $n$ different vectors such that $\omega({\bf z}_i,{\bf z}_j) =0$.  These vectors can be used as basis vectors for a Lagrange plane in phase space.  A {\em Lagrange plane} is a maximal commutative subgroup contained in phase space, and which can be used as a ``configuration space''.  Then form the direct sum of the commutative subgroups they generate,
\begin{equation}
S_L = \bigoplus_{i=1}^n S_{ {\bf z}_i}.
\end{equation}
Because $(c_i {\bf z}_i, 0 ) \diamond  (c_j {\bf z}_j,0) = (c_i {\bf z}_i + c_j {\bf z}_j,0)$, this is also a commutative subgroup of $\mathfrak H$.  We can think of this subspace as a configuration space, and the vectors $\{ {\bf z}_i \}$ as the various directions in this configuration space.  For example, any point in the subgroup $S_L$ could be written 
\begin{equation}
{\bf q} = \sum_i q_i {\bf z}_i  .  
\end{equation}
Any additional linearly independent vector ${\bf z}'$ that we might try to add to this set will have a nonzero symplectic product with at least one of the $ {\bf z}_i$'s.  This means that any additional vector contains a contribution from the conjugate momentum of at least one of the ${\bf z}_i$'s.

\subsubsection{Irreducible Representations}

The irreducible representations of the Heisenberg-Weyl group $\mathfrak H$ come in two flavors.  First, as with the discrete group, there are one dimensional representations.
\begin{equation}\label{eq:con_1D_irrep}
\varrho_{({\bf u},{\bf v})}({\bf q, p},\lambda) = \exp(i ( {\bf u \cdot q + v \cdot p} ))
\end{equation}
The parameters $\bf u$ and $\bf v$ are $n$-dimensional vectors, so these representations form  a $2n$-dimensional (continuously infinite) family of one dimensional irreducible representations.

There is also a family of infinite dimensional irreducible representations, which are analogous to the $n$-dimensional irreducible representations of the finite Heisenberg-Weyl group.  These infinite dimensional irreducible representations give the standard Schr\"odinger representation of the Heisenberg-Weyl group, which maps the group to operators acting on functions on configuration space.
\begin{equation}\label{eq:con_inf_irrep}
\Big(\rho_\alpha({\bf q', p'},\lambda) \psi \Big) ({\bf q}) = e^{i \alpha (\lambda - \frac{1}{2}{\bf p' \cdot q'} -{\bf p' \cdot q} ) }  \psi({\bf q+q}')
\end{equation}
Here, the function $\psi$ is a complex function of $n$ variables.

\subsubsection{Reduction of the Regular Representation}

The reduction of the regular representation to primaries gives shift operators on phase space.
\begin{equation}
\Big(S_{\alpha}({\bf z}', \lambda) f\Big) ({\bf z}) = e^{i \alpha (\lambda+\frac{1}{2}\omega({\bf z',z})) } f({\bf z+z'})
\end{equation}
Reducing this gives the irreducible representations $\rho_\alpha$ in Equation (\ref{eq:con_inf_irrep}), which act on functions of $\bf q$.  An alternate basis for the decomposition would give operators which act on functions on momentum space, $\bf p$.
\begin{equation}\label{eq:con_irrep_p}
\Big(\tilde \rho_\alpha({\bf q', p'},\lambda) \tilde\psi \Big) ({\bf p}) \equiv e^{i \alpha (\lambda + \frac{1}{2}{\bf p' \cdot q'} +{\bf p \cdot q'} ) }  \tilde\psi({\bf p+p}')
\end{equation}
These representations are related to each other by a Fourier transform (and an inversion and scaling of the $p$ axis), which can be shown by explicit calculation.
\begin{align}
&\int d{\bf q}\, e^{i{\bf p\cdot q}} \Big(\rho_\alpha({\bf z'},\lambda) \psi \Big)({\bf q}) \\
 &=\int d{\bf q}\, e^{i{\bf p\cdot q}} e^{i \alpha (\lambda - \frac{1}{2}{\bf p' \cdot q'} -{\bf p' \cdot q} ) }  \psi({\bf q+q}') \\
&= e^{i \alpha (\lambda - \frac{1}{2}{\bf p' \cdot q'})} \int d{\bf q}\, e^{i{\bf p\cdot q}} e^{-i\alpha{\bf p' \cdot q}  }  \psi({\bf q+q}')  \\
&= e^{i \alpha (\lambda - \frac{1}{2}{\bf p' \cdot q'})} \int d{\bf q''}\, e^{i ({\bf p}- \alpha {\bf p}') \cdot {\bf (q''-q')}}  \psi({\bf q}'') \\
&= e^{i \alpha (\lambda - \frac{1}{2}{\bf p' \cdot q'})}  e^{-i ({\bf p}- \alpha {\bf p}') \cdot {\bf q'}} \int d{\bf q''}\, e^{i ({\bf p}- \alpha {\bf p}') \cdot {\bf q''}}  \psi({\bf q}'') \\
&= e^{i \alpha (\lambda + \frac{1}{2}{\bf p' \cdot q'})}  e^{-i {\bf p} \cdot {\bf q'}}      \tilde\psi({\bf p}- \alpha {\bf p}') \\
&= e^{i \alpha (\lambda + \frac{1}{2}{\bf p' \cdot q'} +  {\bf p} \cdot {\bf q'})} \tilde\psi({\bf p}+ {\bf p}')
\end{align}
In the last line, the substitutions $\bf p \rightarrow - {\rm \alpha} p$, and $\tilde \psi ({\bf p}) \rightarrow \tilde \psi(-{\bf p}/\alpha)$ put this in the form of Equation (\ref{eq:con_irrep_p}).

This transformation is an example of the Stone--von Neumann theorem applied to the continuous Heisenberg-Weyl group.  Since the representations in Equations (\ref{eq:con_inf_irrep}) and (\ref{eq:con_irrep_p}) have the same action on the center of the group, then, by the Stone--von Neumann theorem, these are equivalent irreducible representations.  This means that there must exist a unitary transformation which relates them, and the previous calculation shows the transformation.

In general, there are infinitely many ways to express the irreducible representation from Equation (\ref{eq:con_inf_irrep}).  Each realization requires a choice of the position and momentum subspaces (the Lagrange manifold) in phase space.  As is known from classical mechanics, different choices of coordinates in phase space are related to each other by linear canonical transformations.  Explicitly, the linear canonical transformation can be written in terms of a $2n \times 2n$ transformation matrix with unit determinant.
\begin{equation}
\left(
\begin{array}{c}
\bf  Q  \\
\bf  P  
\end{array}
\right) = \left(
\begin{array}{cc}
\bf a  & \bf b \\
\bf c & \bf d 
\end{array}
\right)
\left(
\begin{array}{c}
\bf q   \\
\bf p  
\end{array}
\right)
\end{equation}
Alternatively, this transformation can be written in implicit form, using the generating functions for canonical transformations.  The generating function $F_1({\bf Q, q})$ for this transformation is 
\begin{equation}
F_1({\bf Q, q}) = \frac{1}{2{\bf b}} \left( {\bf d}  |{\bf Q}|^2 - 2 {\bf q}^{\text{T}} \cdot {\bf Q} +{\bf a} |{\bf q}|^2 \right) .
\end{equation}
The unitary transformation which relates the irreducible representations is called the {\em metaplectic transformation}, and, using this generating function, can be written
\begin{equation}\label{eq:metaplectic_definition}
\tilde \psi ({\bf Q}) = \mathcal{N} \int d{\bf q} \, e^{iF_1({\bf Q, q}) } \psi({\bf q}).
\end{equation}
We can check that this transformation gives the correct transformation of the operators $\hat q$ and $\hat p$.  The new ``position'' operator can be written in terms of the old position and momentum operators.
\begin{equation}
\hat Q = {\bf a} \hat q + {\bf b} \hat p
\end{equation}
In the $q$ representation, this is a combination of multiplication by ${\bf q}$ and derivative with respect to ${\bf q}$.
\begin{equation}
\hat Q = {\bf a} \cdot {\bf q} + {\bf b}\cdot (-i \nabla_{\bf q})
\end{equation}
Put this into the integral in Equation (\ref{eq:metaplectic_definition}) in order to find $\hat Q$ in the $Q$ representation.
\begin{align}
\hat Q \tilde\psi({\bf  Q }) &= \mathcal{N} \int d{\bf q} \, e^{iF_1({\bf Q, q}) } ( {\bf a} \cdot {\bf q} -i {\bf b}\cdot  \nabla_{\bf q}) \psi({\bf q})
\end{align}
Use integration by parts to act with the derivatives on the phase.
\begin{align}
\hat Q \tilde\psi({\bf  Q }) &= \mathcal{N} \int d{\bf q} \, \psi({\bf q}) ( {\bf a} \cdot {\bf q} +i {\bf b}\cdot  \nabla_{\bf q}) e^{iF_1({\bf Q, q}) } \\
&= \mathcal{N} \int d{\bf q} \, \psi({\bf q}) \left( {\bf a} \cdot {\bf q} - {\bf b}\cdot  (\nabla_{\bf q} F_1)\right) e^{iF_1({\bf Q, q}) } 
\end{align}
Evaluating the derivative of the generating function gives the result we wanted.
\begin{align}
\hat Q \tilde\psi({\bf  Q }) 
&= \mathcal{N} \int d{\bf q} \, \psi({\bf q}) \left( {\bf a} \cdot {\bf q} + {\bf  Q } - {\bf a} \cdot {\bf q}\right) e^{iF_1({\bf Q, q}) }  \\
&= \mathcal{N} \int d{\bf q} \, \psi({\bf q}) \left(  {\bf  Q } \right) e^{iF_1({\bf Q, q}) }  \\
&= {\bf  Q } \, \tilde \psi({\bf  Q } )
\end{align}
Similar calculations show that the new ``momentum'' operator is the expected derivative.
\begin{equation}
\hat P = {\bf  c } \hat q + {\bf  d } \hat p \rightarrow -i \nabla_{\bf Q}
\end{equation}

% end comment (induced reps?)

% end Mumford construction

\comment{ % dilations and fft
\subsubsection{Dilations and the FFT (Incomplete!)}

The phase of the element $g_\lambda = (\lambda,0,0)$ is simply $e^{\frac{2 \pi i \lambda}{n}}$.  The rest of the phases can be constructed by considering the action of dilations $\mathcal{D}_{\gamma}(q,p,\lambda)=(\gamma q,\gamma p,\gamma^2\lambda)$ on the group.  Since the dilations take $\lambda \rightarrow \gamma^2 \lambda$, a dilation will take the rep $e^{\frac{2 \pi i \lambda}{n}}$ to the different rep $e^{ \frac{2 \pi i \lambda\gamma^2}{n}}$.  Since we take the phase $\lambda$ mod $n$, there are $n$ distinct phases that are produced by the dilations.  However, the $\gamma=0$ dilation simply gives the trivial representation, which we have already counted.  So there are $n-1$ of these $n$-dimensional reps.  
} % dialations

\comment{ %
This can be reduced further to give the irreducible representations, which can be written in terms of the Heisenberg operators,
\begin{equation}
T_z=e^{-i\omega(\hat z,z)}.
\end{equation}

Some properties of the Heisenberg operators are
\begin{equation}
\Tr{T_z^\dagger T_{z'}} = (2\pi)^N \delta(z-z')
\end{equation}
\begin{equation}
T_z^\dagger=T_z^{-1}=T_{-z}
\end{equation}
and
\begin{equation}
T_z T_{z'} = e^{\frac{i}{2} \omega(z,z')}T_{z+z'}.
\end{equation}
The third property derives from the CBH theorem, and the definitions of $T_z$ and $\omega(z,z')$.  The meaning of the operator $\hat z$ depends on the particular choice of decomposition of the primary representation is chosen.  

}%

%% file: Chapter-SymbolTheory.tex
%%%%%%%%%%%%%%%%%%%%%%%%%%%%%%%%%%%%%%%%%%%%%%%%%%%%%%%%%%%%%%%%%%%%%%%%%%
%
% Ph.D. dissertation manuscript
% Chapter 7: Theory
%
% Andrew Stephen Richardson (Fall 2007)
% College of William and Mary
% Department of Physics
% Prof. Eugene Tracy, advisor
%
% Based on Paul King and Andrew Norman's template (modified by Wirawan Purwanto)
%
%%%%%%%%%%%%%%%%%%%%%%%%%%%%%%%%%%%%%%%%%%%%%%%%%%%%%%%%%%%%%%%%%%%%%%%%%%

\chapter{Symbols of Operators and the Noncommutative Fourier Transform}
\label{chp:Symbols}

\section{From Operators to Functions}

In Part I, the phase space perspective of wave problems was described and shown to provide a powerful perspective for analytical solution of wave problems.  At its core, this perspective relied on being able to convert a generic wave equation from an operator into a function on phase space.  For wave problems, this gives the dispersion function, which encodes all information that is contained in the wave operator.  With the dispersion function it is possible to define the dispersion surfaces.  These surfaces give a graphical representation of the allowed frequencies and wavenumbers for the various wave modes that could be solutions to the wave equation.  Also, for multicomponent problems, the dispersion surfaces can help to identify mode conversion regions through ``avoided crossing''  type structures, and other types of near-degeneracies.

In this chapter, we review the mathematical tools needed to convert an arbitrary operator into a function on phase space.  This function is called the {\em symbol} of the operator, and the mathematics of symbols is sometimes called the {\em symbol calculus}.  Various types of symbols have been introduced in the literature, each using different orderings and complexification of the position and momentum operators. \cite{Lee:1995lr} 
\begin{equation}
\begin{array}{ccclc}
   &  & e^{z \hat a^\dagger} e^{-z^* \hat a} & \!\!\!\!\!\!\!\!\!\!\text{\scriptsize normal-ordered}  & \\
   &  & \begin{CD} @AAA\end{CD}  & & \\
 e^{i\eta \hat p} e^{i \xi \hat q}  &  \begin{CD}@<\text{antistandard-ordered}<<\end{CD} & 
 e^{i\eta \hat p+ i \xi \hat q}=e^{z \hat a^\dagger -z^* \hat a} & \begin{CD}@>\text{standard-ordered}>>\end{CD} & e^{i \xi \hat q} e^{i\eta \hat p} \\
   &  & \begin{CD} @VVV\end{CD}  &  & \\
   &  & e^{-z^* \hat a} e^{z \hat a^\dagger} & \!\!\!\!\!\!\!\!\!\!\text{\scriptsize antinormal-ordered} & \\
\end{array}
\end{equation}
The Weyl symbol uses a symmetrized ordering.  It thus treats $\hat q$ and $\hat p$ on an equal footing, and takes the central place in the above diagram describing how the different types of symbols relate to each other (see Berezin and Shubin \cite{Berezin:1991fk} for a discussion of a similar diagram).  While these various symbols have been extensively studied, and their relations to each other have been worked out, the resulting formula are often presented in a mysterious way as the result of brute-force calculations.  By recasting the theory of symbols in terms of Fourier transforms on groups, the proper context for a deeper understanding can be obtained.

We start this chapter with a review of the symbol theory developed by Weyl in the context of quantum mechanics.  We then proceed to a more general theory which defines symbols as a type of Fourier transform on noncommutative groups, a perspective introduced by Zobin.  Finally, the Zobin perspective is used to construct the ``symbol'' of a matrix.  Since a matrix is really an operator, it is possible to compute a function on a discrete phase space which is the symbol of this operator.  This example illustrates the ideas of this chapter in a very concrete way.

\section{The Weyl Symbol}

In quantum mechanics, the state of a system is given as an abstract vector in some Hilbert space.  Measurable quantities are then calculated as the expectation values of particular operators.  The Schr\"odinger approach to quantum mechanics is to write the state as a wavefunction, with differential operators representing, e.g., the momentum of the system. 
\begin{equation}
\langle \hat p \rangle = \int dq\, \psi^\dagger(q) (-i \partial_q) \psi(q)
\end{equation}
Wigner, Weyl, and others developed an alternative approach to quantum mechanics.  Wigner realized that the information about the state of the system could be written in what looks like a more classical way, but expressing the state as a function on phase space.  The {\em Wigner function} of a state $\psi(q)$ is a function on phase space, and is defined as
\begin{equation}
W(q,p)=\int ds \; e^{ips} \psi^\dagger\left(q+\frac{s}{2}\right) \psi\left(q-\frac{s}{2}\right) .
\end{equation}

It turns out that the Wigner function is simply the {\em Weyl symbol} of the density matrix of the state $\vert \psi \rangle$.  Given the integral kernel $K_A(q,q')$ of some operator $\hat A$, the Weyl symbol of $\hat A$ is defined as \cite{dealmeida-1992}
\begin{equation}\label{eq:weyl_symbol}
A(q,p)=\int ds \; e^{ips} K_A\left(q+\frac{s}{2},  q-\frac{s}{2}\right). 
\end{equation}
For the density matrix, the integral kernel is 
\begin{equation}
K_\psi(q,q') = \langle q \vert \psi \rangle \langle \psi \vert q' \rangle = \psi^\dagger(q) \psi(q').
\end{equation}
Inserting this into the equation for the symbol gives us the Wigner function.

The Weyl symbol has many nice properties which motivate its use in the analysis of operators, especially when compared to the integral kernel of the operator.  As an example, consider the derivative operator.  Its kernel is a distribution, not a smooth function.
\begin{eqnarray}
\frac{d}{dx} f(x) &=& \int dx'\, \delta(x-x') \frac{d}{dx'} f(x') \\
&=& -\int dx'\, \left( \frac{d}{dx'}\delta(x-x') \right) f(x') \\
&=& \int dx'\, \delta^{(1)}(x-x')  f(x')
\end{eqnarray}
Now, if we calculate the symbol of this kernel, we will get a smooth function on phase space.  This is not too surprising since the definition of the symbol looks like a Fourier transform, and we know that the Fourier transform of a delta function is a constant.
\begin{eqnarray}
D(q,p) &=& \int ds\, e^{ips} \delta^{(1)}(q+s/2 -q +s/2) \\
&=& \int ds\, e^{ips} \frac{d}{ds} \delta(s) \\
&=& - \int ds\, \delta(s) \frac{d}{ds} e^{ips}  \\
&=& -ip \int ds\, \delta(s) e^{ips} =-ip 
\end{eqnarray}
This result generalizes.  The integral kernel of local, differential operators will generically be some singular distribution.  However, since the symbol is a kind of Fourier transform, the singularities get turned into smooth functions.

This calculation of the symbol of the derivative also illustrates another feature of the Weyl symbol calculus.  It enables us to relate quantum operators like $\hat p = i \partial_q$ with classical functions on phase space, like the classical momentum $p$.  It is also possible to go the other way.  We can ``quantize'' a function on phase space by inverting the formula for the symbol.  This will give us an operator whose action is essentially the same as the original operator $\hat A$.
\begin{equation}
\hat A' = \int \exp(i\omega(z,\hat z)) A(z) dz
\end{equation}
Here the phase space shift operator is given by
\begin{equation}
T_z=e^{-i\omega(\hat z,z)},
\end{equation}
where $\hat z = (\hat q, \hat p)$ is composed of the position operator, $\hat q f(q) = q f(q)$, and the momentum operator $\hat p f(q) = -i \partial_q f(q)$.

\comment{%
To get an operator from a symbol, we use the symbol as a kind of expansion coefficient for a ``basis set'' of operators.  For the Weyl symbol, the right basis set of operators to use is the set of Heisenberg operators.
\begin{equation}
T_z=e^{-i\omega(\hat z,z)}.
\end{equation}
Here the operator $\hat z$ is composed of the position operator, $\hat q f(q) = q f(q)$, and the momentum operator $\hat p f(q) = -i \partial_q f(q)$.

Some properties of the Heisenberg operators are
\begin{equation}
\Tr{T_z^\dagger T_{z'}} = (2\pi)^N \delta(z-z')
\end{equation}
\begin{equation}
T_z^\dagger=T_z^{-1}=T_{-z}
\end{equation}
and
\begin{equation}
T_z T_{z'} = e^{\frac{i}{2} \omega(z,z')}T_{z+z'}.
\end{equation}
The third property derives from the CBH theorem, and the definitions of $T_z$ and $\omega(z,z')$.  The meaning of the operator $\hat z$ depends on the particular choice of decomposition of the primary representation is chosen.  
}%

\comment{%
\subsubsection{Properties of the Wigner Function}

``Magical'' properties of the Wigner function.
\begin{enumerate}
\item Transformation properties.  $\hat T_z \psi$ gives rigid shift of Wigner function of $\psi$.  

\item $\hat M \psi$ gives canonical transformation of phase space $W(M^{-1} z)$ (Left quasi-regular rep of the metaplectic group?)

\item Moyal product gives phase space path integral

\end{enumerate}
These properties are easily explained once the group theory background is understood.
}%

\comment{%  some stuff for this section?

\subsection{General Operators and the Weyl Symbol}
The Wigner function, the Weyl Symbol, the Moyal Star Product.  Operator ordering.  Integral and differential forms of the star product.  The Moyal bracket, the commutator, and the Poisson bracket.

Show Wigner function as a special case of the Weyl symbol.  Show how the Weyl symbol is usually formed from the integral kernel of an operator (with full details).  $\hat A$ goes to A(z) goes to $\hat A' = \int \exp(i\omega(z,\hat z)) A(z)$.  Claim: $\hat A$ and $\hat A'$ have the same action in Hilbert space up to some set of measure zero, so they are ``the same''.

Alternatively, the Weyl symbol is simply a decomposition of a given operator onto the Heisenberg operators (non-commutative shifts in phase space).  Related to the irreducible representations of the HW group.

Why do we care?  Because we can Taylor expand the  symbol A(z) around some point in phase space, and then form an approximation to the operator.  This gives us something ``local'' in phase space. 

For waves, local in phase space is ``approx plane wave near x with approx wavevector k''.
}%

\section{The Fourier Transform: A Group Theory Perspective}

\subsection{Introduction\label{sec:ft_intro}}

This section is based on a series of seminars given by N. Zobin at William and Mary in the spring of 2004.  The lectures described the group-theoretical basis of Fourier transforms, for both commutative and non-commutative groups.  In the case of commutative groups, the transform that arises is the ordinary Fourier transform from harmonic analysis.  However, when the construction described here is used for non-commutative groups, a different sort of Fourier transform is obtained.  This non-commutative Fourier transform provides the group theoretical basis for understanding the symbol of an operator.  This group theory point of view, which was developed by Zobin, provides the proper context for understanding the various types of symbols (Weyl, Wick, anti-Wick, normal, anti-normal, etc.) that have been defined in the literature.

\subsubsection{Motivational Example}

The group theory perspective of the Fourier transform is perhaps best explained by first starting with a simple example.  The ideas presented in this example will then extend to a general definition of the Fourier transform.

Consider the group of cyclic shifts on a set of four points.  These shifts form a group, $G$.  There are four different shifts which can be applied to the four points.  They are shifts by zero, one, two, or three elements.  As group elements, they can be written as
\begin{equation}
G= \{ e, g, g^2, g^3\}.
\end{equation}
Here $e$ is the identity element of the group, which corresponds to shifts by zero points.  The one-point shift corresponds to the group element $g$.  Repeated applications of shifts by one point will give all the other shifts, so they are written as powers of $g$.

This is a commutative group, so its irreducible representations are simply phases.  There are four different irreducible representations, which are labeled by the integers $k \in \{0,1,2,3\}$.
\begin{equation}
\rho_k (g^j) = e^{\frac{2\pi i}{4} k j}
\end{equation}
These phases can be written out in the character table, which relates the group elements and the characters of the representations.
\begin{equation}
\begin{array}{|c|cccc|}\hline
& e & g & g^2 & g^3  \\ \hline
\rho_{0}&1&1&1&1 \\
\rho_{1}&1&e^{\pi  i/2}&e^{\pi  i}&e^{3\pi  i/2} \\
\rho_{2}&1&e^{\pi  i}&1&e^{\pi  i} \\
\rho_{3}&1&e^{3\pi  i/2}&e^{\pi  i}&e^{\pi  i/2} \\ \hline
\end{array}
\end{equation}
This table has the form of a matrix, where each row corresponds to a representation, and each column corresponds to an element of the group.  As a matrix, this could multiply a vector with four elements.  By looking at the labels in the table above, it is clear that the elements of the vector naturally correspond to the element of the group.  So the vector could be thought of as a function on the group, when it is considered as a set.
Each element in the vector resulting from the multiplication corresponds to a row from the matrix, and so can be labeled by the representation.  This new vector is then a function on the set of irreducible representations of the group.
\begin{equation}\label{eq:dis_ft_ex}
\left(
\begin{array}{c}
\hat f(\rho_0) \\
\hat f(\rho_1) \\
\hat f(\rho_2) \\
\hat f(\rho_3)
\end{array}\right) = 
\left(\begin{array}{cccc}
1&1&1&1 \\
1&i&-1&-i \\
1&-1&1&-1 \\
1&-i&-1&i \\ 
\end{array}
\right)
\left(
\begin{array}{c}
f(e) \\
f(g) \\
f(g^2) \\
f(g^3)
\end{array}\right)
\end{equation}
This matrix equation is in fact the discrete Fourier transform written explicitly as a matrix.  These considerations apply to the continuous case as well.  If we start with the group of shifts on the real line, construct its the character table, and define a multiplication like that in Equation (\ref{eq:dis_ft_ex}) above, we obtain the continuous Fourier transform.

\begin{figure}
\begin{center}
\includegraphics[scale=0.8]{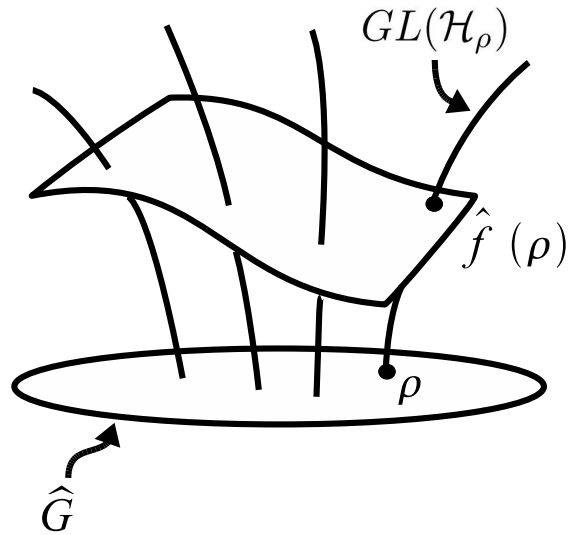}
\end{center}
\caption{\label{fig:dual_bundle}
A ``section'' of the dual bundle.
}
\end{figure}

It turns out that this type of construction can be extended to the case of non-commutative groups.  Consider a finite, but non-commutative group $G$.  For this group we can construct the {\em dual}, $\widehat G$, which is the set of all equivalence classes of irreducible representations.  Since the group is non-commutative, some of these representations will be matrix representations.   However, we can still list them in a table like the one above.  All groups have the trivial representation where everything maps to multiplication by 1.  Call that representation $\rho_0$ and put it on the first row of the table.  The rest of the representations $\rho_i$ are $n_i \times n_i$ matrices.  So, the rows of the table will be filled with matrices.  Each row will have matrices which are all the same size, but different rows will have different size matrices. 
\begin{equation}
\begin{array}{|c|ccccc|l}\cline{1-6}
& e & g_1 & g_2 & g_3 & \ldots  \\ \cline{1-6}
\rho_{0}&1&1&1&1 & \ldots & \} \, 1 \times 1\\
\rho_{1}& id_{n_1 \times n_1} & \rho_1(g_1) & \rho_1(g_2) & \rho_1(g_3) & \ldots &  \Bigr\} \, n_1 \times n_1\\ [1ex]
\rho_{1}& id_{n_2 \times n_2} & \rho_2(g_1) & \rho_2(g_2) & \rho_2(g_3) & \ldots & \biggr\} \, n_2 \times n_2\\ [2ex]
\rho_{1}& id_{n_3 \times n_3} & \rho_3(g_1) & \rho_3(g_2) & \rho_3(g_3) & \ldots & \Biggr\} \, n_3 \times n_3\\ 
\vdots  &  \vdots & \vdots & \vdots & \vdots & \\ \cline{1-6}
\end{array}
\end{equation}
In the commutative case, this table had an obvious interpretation as a matrix, which then multiplied a function over the group considered as a set.  Now this is not a matrix, but it can still be ``multiplied'' by a function on the group.  
%\begin{landscape}
\begin{equation}\label{eq:nc_ft}
\begin{array}{r}
1 \times 1 \, \{ \\
n_1 \times n_1 \, \Bigl\{ \\ [1ex]
n_2 \times n_2 \, \biggl\{ \\ [2ex]
n_3 \times n_3 \, \Biggl\{ \\
 \,
\end{array}
\left(
\begin{array}{c}
\hat f (\rho_0) \\
\hat f (\rho_1) \\ [1ex]
\hat f (\rho_2) \\ [2ex]
\hat f (\rho_3) \\
\vdots
\end{array}
\right)
=
\left(
\begin{array}{ccccc}
1&1&1&1 & \ldots \\
 id_{n_1 \times n_1} & \rho_1(g_1) & \rho_1(g_2) & \rho_1(g_3) & \ldots \\ [1ex]
 id_{n_2 \times n_2} & \rho_2(g_1) & \rho_2(g_2) & \rho_2(g_3) & \ldots \\ [2ex]
 id_{n_3 \times n_3} & \rho_3(g_1) & \rho_3(g_2) & \rho_3(g_3) & \ldots  \\ 
 \vdots & \vdots & \vdots & \vdots & 
\end{array}
\right)
\left(
\begin{array}{c}
f (e) \\
f (g_1) \\ [1ex]
f (g_2) \\ [2ex]
f (g_3) \\
\vdots
\end{array}
\right)
\end{equation}
%\end{landscape}
This equation is meant to be interpreted in a way analogous to matrix multiplication.  The ``vector'' on the left hand side is actually a list of matrices of different sizes.  An element is of this list is computed by summing over the matrices in the same row in the table of representations, with the weights of the matrices in the sum given by the appropriate element of the vector $f(g)$.  This computation is used as the definition of the Fourier transform for a non-commutative group.
\begin{equation}
\hat f(\rho) = \sum_{g \in G} f(g) \rho(g) 
\end{equation}
In the commutative case, this summation resulted in a complex-valued function on the set of irreducible representations.  In this non-commutative case, we have something analogous.  However, instead of a function, the Fourier transform is operator valued.  Not only that, but for each $\rho$, the dimension of the operator is most likely going to vary, as seen in Equation (\ref{eq:nc_ft}).  Such an object is called a {\em section} of the fiber bundle over the set $\widehat G$ of irreducible operators.  This object is illustrated in Figure (\ref{fig:dual_bundle}).

The dual fiber bundle is constructed as follows.  There is a space of operators associated to each point $\rho$ in the dual, $\widehat G$.  This space of operators is $GL(\mathcal H_\rho)$, where $\mathcal H_\rho$ is the Hilbert space on which the representation $\rho$ acts.  Put each of these spaces of operators over the set $\widehat G$ as fibers, with the base point of each fiber at the appropriate point $\rho$.  Even though the fiber is generically a multidimensional space, we can think of it in a one-dimensional way, as drawn in Figure (\ref{fig:dual_bundle}).  The collection of all of these fibers is the {\em fiber bundle}, $\Lambda(\widehat G)$.  If we now ``slice'' the fiber by choosing one operator from each fiber, this will create a ``section'' of the fiber bundle.  A {\em section} is a mapping of the dual $\widehat G$ into the bundle $\Lambda(\widehat G)$.
\begin{equation}
\hat f : \widehat G \rightarrow \Lambda(\widehat G)
\end{equation}
If we put the elements of $\widehat G$ into a list, then the section will look like a list of operators, each of different dimension.  This is exactly the object that we obtain from the equation for the Fourier transform of a non-commutative group.

In the following sections we will define the commutative Fourier transform from the group theory perspective.  We will then give examples of various types of Fourier transforms for commutative groups, showing how the traditional equations for Fourier transforms agree with the group theory perspective.  We then define the Fourier transform for non-commutative groups and give the example of the non-commutative Fourier transform for the Heisenberg-Weyl group.

\subsection{Pontryagin Theorem}

The Fourier transform on commutative groups was summarized in the 1930's by Pontryagin in the following theorem \cite{pontrjagin_thm}.

Let $G$ be a locally compact commutative group.  Let $\widehat G$ be its Pontryagin dual group defined by
\begin{equation}
\rho \in \widehat G \iff \rho : G \rightarrow \mathbb{T}
\end{equation}
where $\rho$ is a continuous homomorphism.  Since all of the irreducible representations of $G$ are one-dimensional, they are all just characters.  So the dual group is the set of characters of $G$.  Let $\mu$ be the Haar measure on $G$ and let $P$ be the Haar measure on $\widehat G$.  Consider the Fourier transform operator
\begin{equation}
\mathcal F : L_2(G,d\mu) \rightarrow L_2(\widehat G,dP)
\end{equation}
defined by
\begin{equation}
(\mathcal F f) (\rho) \equiv \hat f(\rho) = \int_G \, \rho(g)f(g) \,dg
\end{equation}
This operator has the following properties:
\begin{enumerate}
\item The transform $\mathcal F$ is an isometry
\begin{equation}
\int_G \, f_1^*(g) f_2(g) \, dg = \int_{\widehat G} \, \hat f_1^*(\rho) f_2(\rho)  \, dP
\end{equation}
\item The inverse of this operator, $\mathcal F^{-1}$, is also a Fourier transform
\item A convolution can be defined using the Fourier transform so that the transform of the convolution is the product of the transforms
\begin{equation}
\widehat{(f_1 * f_2)}(\rho) = \hat f_1 (\rho) \hat f_2(\rho)
\end{equation}
Using this and the definition of the Fourier transform, we obtain an equation for the convolution
\begin{equation}\label{eq:nc_conv}
(f_1 * f_2)(g) = \int_G \, f_1(g \diamond h^{-1}) f_2(h) \, dh
\end{equation}
\end{enumerate}

\subsection{Examples of Fourier Transforms for Commutative Groups}

\subsubsection{The Classical Fourier Transform}

The classical Fourier transform is an operator that acts on functions in $L_2(\mathbb{R},dx)$.  Here, $\mathbb{R}$ is the group of reals under addition.  The Haar measure, $dx$, for this group is simply the Lebesgue measure.  The Fourier transform operator is
\begin{equation}
\mathcal F : L_2(\mathbb{R},dx) \rightarrow L_2(\mathbb{R},dx)
\end{equation}
defined by 
\begin{equation}\label{eq:classical_FT}
\hat f (k) = \int\displaylimits_\mathbb{R}{e^{-ikx} f(x) dx}
\end{equation}
with the appropriate normalization.  There are several things to notice about this transform.

This transform is an isometry, which in this case means that it is a unitary transformation.   This isometry property is commonly know as the Plancharel theorem, or the Parseval theorem.  For functions $f_1, f_2 \in L_2(\mathbb R, dx)$ the Fourier transform preserves the inner product of the functions.
\begin{equation}
\int_\mathbb{R} \, f_1(x) f_2(x) \,dx = \int_\mathbb{R} \, \hat f_1(k) \hat f_2(k) \,dk 
\end{equation}

Another thing to notice is that the inverse of this operator $\mathcal F^{-1}$ is also a Fourier transform.  In this case, the space dual to the group $\mathbb R$ is also $\mathbb R$.  This means that it also has a Fourier transform, which, up to a sign in the phase, is the inverse of the transform in Equation (\ref{eq:classical_FT}).  This Fourier transform takes functions from $L_2(\mathbb{R},dx)$ back to the same set of functions.  It is quite a special situation, since the Fourier transform does not usually do this.

\comment{%
\item There is a convolution theorem related to this transformation.  The shift operator for this group is $(S_a f)(x)=f(x+a)$, so 
\begin{equation}
f(x+a)=e^{i \lambda a}\hat f (\lambda) 
\end{equation}
So the shift becomes multiplication by a character of the group $\mathbb{R}$.
}%

\subsubsection{Fourier Series}

The Fourier series is another example of a commutative Fourier transform.  It can be expressed as the Fourier transform on the torus, which is the group of real numbers with addition modulo $2\pi$.
\begin{equation}
\mathcal F : L_2(\mathbb{T},d\phi) \rightarrow L_2(\mathbb{Z},dN)
\end{equation}
Here functions on the circle with the Lebegue measure $d\phi$ are taken to functions on the integers with the counting measure $dN$.

\comment{% Need to expand on this if I want to include it.
\subsubsection{Almost Periodic Functions}

As another example, we can define the Fourier transform on the discrete real numbers $\mathbb{R}_d$, which is the set where every real number is distance one from every other number.  Here we get
\begin{equation}
\mathcal F : L_2(\mathbb{R}_d,dN) \rightarrow L_2(B,d\mu)
\end{equation}
where $B$ is the Bohr compactification and  $L_2(B,d\mu)$ is the space of almost periodic functions.
}%

\subsubsection{Discrete Fourier Transform}

The finite, discrete Fourier transform also falls into the category of Fourier transforms on commutative groups.  Here, we use the group of integers with addition modulo an integer $n$.  The Fourier transform is then
\begin{equation}
\mathcal F : L_2(\mathbb{Z}_n,dN) \rightarrow L_2(\widehat{\mathbb{Z}_n},dN)
\end{equation}
Here we use the notation that $\widehat{\mathbb{Z}_n}$ is the set of characters of the group $\mathbb{Z}_n$.  It can be shown that the characters of $\mathbb{Z}_n$ are given by
\begin{equation}
\lambda : \mathbb{Z}_n \rightarrow \mathbb{T}
\end{equation}
where 
\begin{equation}
\lambda(x)=e^{2\pi i \lambda x}, \quad \lambda \in \mathbb{Z}_n
\end{equation}
Therefore, the set of characters $\widehat{\mathbb{Z}_n}$ is also a group, and is in fact the group $\mathbb{Z}_n$.  Since this group is finite, the Fourier transform can be written in matrix form as
\begin{equation}
\mathcal F_{k,l} = [e^{2\pi i k l}]_{k,l\in\mathbb{Z}_n}
\end{equation}
This is the matrix that was obtained in the motivational example for the case where $n=4$, and shown in Equation (\ref{eq:dis_ft_ex}).

An interesting side note about the discrete Fourier transform for the group $\mathbb{Z}_n$ where $n=2k$.  Consider the homomorphism $\Theta : \mathbb{Z}_n \rightarrow \mathbb{Z}_n$ given by
\begin{equation}
\Theta x = 2 x \mod k
\end{equation}
This is a dilation of the group that effectively breaks the group into the even terms and the odd terms.  The Fourier transform of $\mathbb{Z}_{2k}$ can then be reduced to the Fourier transform on $\mathbb{Z}_k$ and the Fourier transform on the group of dilations.  This reduction is what gives rise to the fast Fourier transform algorithms.  An understanding of the group theory origins of the FFT algorithms could help lead to the development of new types of fast transforms for other groups.

\section{Fourier Transforms of Non-commutative Groups}

\begin{figure}
\begin{center}
\includegraphics[scale=0.6]{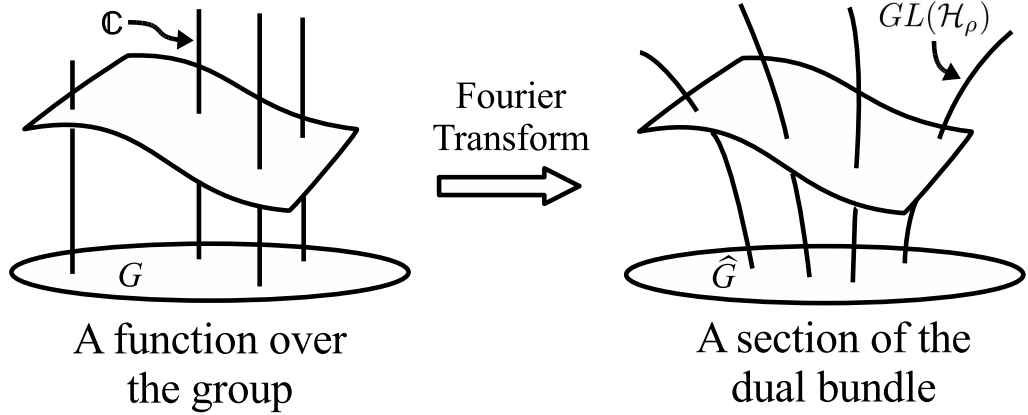}
\end{center}
\caption{\label{fig:FT_idea}
The group theoretical basis for the Fourier transform.  The Fourier transform maps a function over a group, considered as a set, to a section of the dual bundle.  If the group is commutative, all its irreducible representations are one-dimensional, and so the section of the dual bundle can be thought of as an ordinary complex-valued function.  Otherwise the section is a listing of operators, one for each fiber.}
\end{figure}

The Pontryagin theorem nicely defines the Fourier transform for commutative groups, but for non-commutative groups the definitions need to be extended, as discussed in Section \ref{sec:ft_intro}.  Instead of mapping functions over the group, considered as a set, to functions over the dual, the Fourier transform will now map functions to sections of the dual bundle.  [See Figure (\ref{fig:FT_idea})]  The rest of the theory proceeds in a way completely analogous to the commutative theory described in the last section.

\subsection{Definitions}

Let $G$ be a locally compact, but non-commutative group.  Let $\widehat G$ be the set of equivalence classes of continuous irreducible unitary representations of $G$.  So
\begin{equation}
\rho \in \widehat G \implies \rho : G \rightarrow GL(\mathcal H_\rho),
\end{equation}
where $GL(\mathcal H_\rho)$ is the set of operators on some Hilbert space $\mathcal H_\rho$.  For the case when $\mathcal H_\rho$ is not just a finite-dimensional vector space, we should actually consider Hilbert-Schmidt operators $HS(\mathcal H_\rho)$.

Now $\widehat G$ is not a group as it was for commutative $G$, since the tensor product of irreducible representations is usually reducible.  However, sections of the dual bundle are still well defined, as described in Section \ref{sec:ft_intro}.
\begin{equation}
\hat f(\rho) \in \Gamma(\widehat G,dP) : \widehat G \rightarrow HS(\mathcal H_\rho)
\end{equation}
where $HS(\mathcal H_\rho)$ is the set of Hilbert-Schmidt operators that act on the Hilbert space $\mathcal H_\rho$ associated with the element $\rho$ in $\widehat G$.  The norm for a section $\hat f$ is defined as by
\begin{equation}
\Vert \hat f\Vert _{L_2}^2 = \int_{\widehat G}{tr(\hat f(\rho)\hat f^\dagger(\rho)) \, dP(\rho)}
\end{equation}

The Fourier transform can now be defined as a mapping from functions on the group to sections of the dual bundle, with the appropriate measures in each case.  
\begin{equation}
L_2(G,dg)  \buildrel \mathcal F \over \longrightarrow \Gamma(\Lambda(\widehat G),dP)
\end{equation}
As we might expect from the example in Section \ref{sec:ft_intro} and from the group theory interpretation of the commutative Fourier transform, the non-commutative Fourier transform is defined as
\begin{equation}
\hat f (\rho) = \int_G \, \rho(g)f(g) \,dg
\end{equation}
where now we have that $\hat f$ is in $\Gamma(\Lambda(\widehat G),dP)$.  As in the commutative case, this mapping is an isometry, it is invertible, and it can be used to define a convolution of functions on the group.  
\begin{enumerate}
\item The transform $\mathcal F$ is an isometry
\begin{equation}
\int_G \, f_1^*(g) f_2(g) \, dg = \int_{\widehat G} \, tr(\hat f_1^\dagger(\rho) \hat f_2 (\rho)) \, dP(\rho)
\end{equation}
\item The inverse of this operator $\mathcal F^{-1}$ is also a FT, and it is defined as follows for $F \in \Gamma(\Lambda(\widehat G),dP)$,
\begin{equation}\label{eq:inverse_FT}
(\mathcal{F}^{-1} F)(g) = \int_{\widehat G} \, {tr(\rho^\dagger(g)F(\rho)) \, dP(\rho)}
\end{equation}
\item The convolution is defined just as in the commutative case
\begin{equation}
(f_1 * f_2)(g) = \int_G \, f_1(g \diamond h^{-1}) f_2(h) \, dh
\end{equation}
In the commutative case the convolution was commutative, since the group elements commute.  However, this is no longer true in the non-commutative case.
\begin{equation}
(f_1 * f_2)(g) \neq (f_2 * f_1)(g)
\end{equation}
This point will become important later on when we define the symbol of an operator, and will give rise to a non-commutative ``star'' product for functions on phase space.
\end{enumerate}

\subsection{Fourier Transform on the 3-dimensional Heisenberg-Weyl Group}

The 3-dimensional Heisenberg-Weyl (HW) group $\mathfrak H $ is defined as in Section \ref{sec:cont_HW}.
\begin{equation}
\mathfrak H = \{({\bf z},\lambda) | {\bf z} \in \mathbb{R}^2,  \lambda \in \mathbb{R} \}
\end{equation}
The measure $dg$ on this group is the Lebegue measure $dg=d^2z \, d\lambda$.
The set of irreducible representations, $\hat{\mathfrak H}$, includes one-dimensional representations, shown in Equation (\ref{eq:con_1D_irrep}), as well as a family of infinite-dimensional representations, shown in Equation (\ref{eq:con_inf_irrep}).  So a section of the dual bundle is a collection of operators which come in two different types.  The part of the section associated with the one-dimensional representations is a mapping of the dual $\widehat G$ to the complex numbers.  The other part of the section is associated with the infinite-dimensional representations.  This part maps $\widehat G$ to operators acting on $L_2(\mathbb R)$.

\begin{figure}
\begin{center}(a)
\includegraphics[scale=0.5]{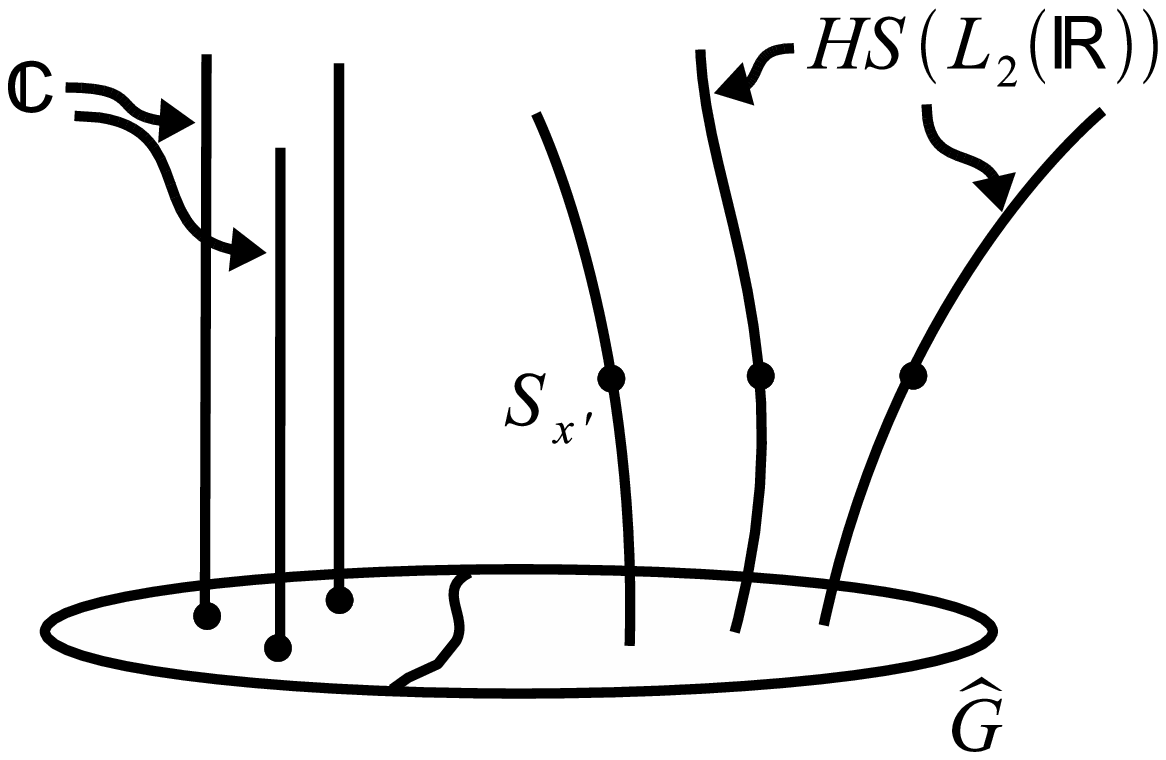} \quad (b)
\includegraphics[scale=0.5]{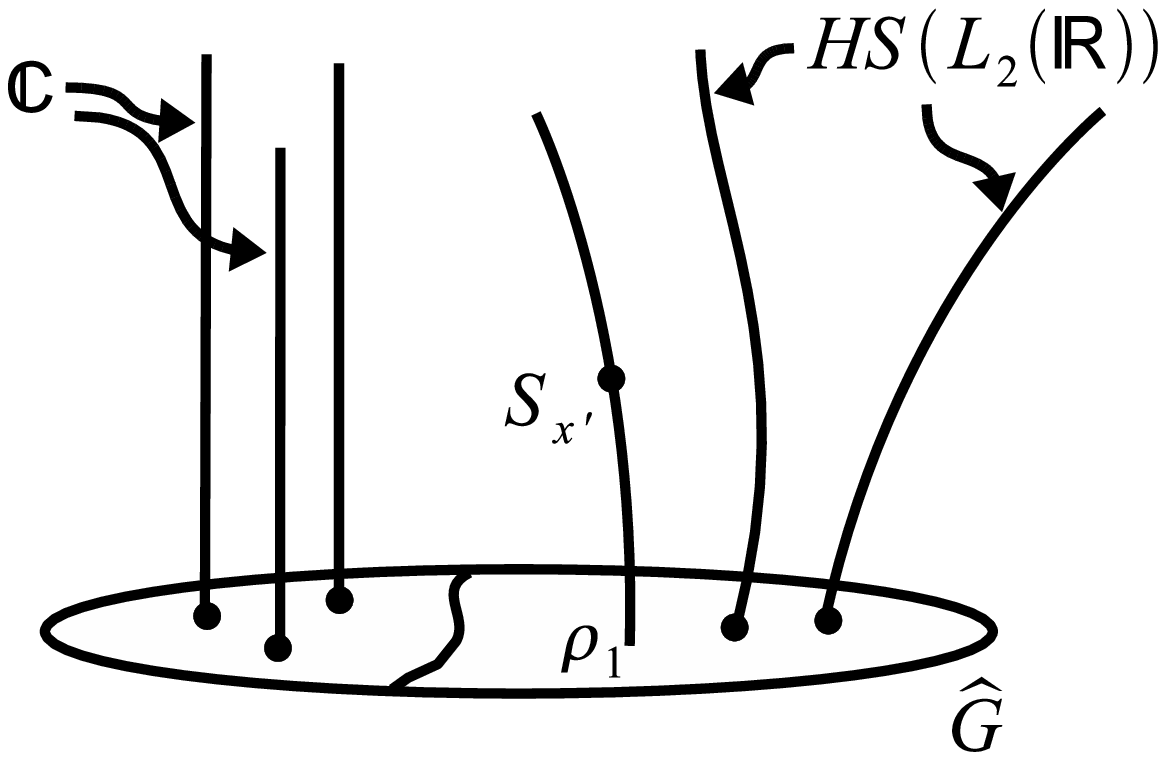}
\end{center}
\caption{\label{fig:embedding}
Possible embeddings of an operator into a section of the dual bundle of the Heisenberg-Weyl group.  In (a) the operator is assigned to all representations which have the right type of fiber, while those with the wrong type of fiber take the value zero.  In (b) the operator is assigned to only one representaion $\rho_1$.  All other representations get zeros.
}
\end{figure}

As an illustration of the sorts of calculations that can be done using the non-commutative Fourier transform, we can compute the inverse transform of a specific operator.  Consider the shift operator acting on complex functions of one real variable.  
\begin{equation}
(S_{x'} f)(x) = f(x-x')
\end{equation}
This operator lives in the same Hilbert space as the infinite-dimensional irreducible representations of the HW group.  So, we can use the Fourier transform of the HW group to take the inverse transform of this operator.  In order to do this, we need to embed the operator in a section of the dual bundle.  Since there is a family of representations which all live in the space of operators acting on $L_2(\mathbb R)$, there are many ways to embed the shift operator into a section.  The two simplest embeddings would be a constant embedding over all possible representations, and a delta function embedding over one specific representation.  Figure (\ref{fig:embedding}) illustrates these embeddings with a simple cartoon.

These two example embeddings can be explicitly written out.  The constant embedding is a section where all of the infinite-dimensional representations get mapped to the operator $S_{x'}$, while all of the finite-dimensional representations get mapped to zero.
\begin{equation}
A_1(\rho) = \begin{cases}
S_{x'} &  \rho = \rho_\alpha, \forall \, \alpha \\
0 & \rho = \varrho_{(u,v)}
\end{cases}
\end{equation}
The delta function section maps everything to zero, except for one of the infinite-dimensional representations, which it maps to the operator $S_{x'}$.  To make it definite, choose the representation $\rho_1$ to map to $S_{x'}$.
\begin{equation}\label{eq:HW_embedding}
A_2(\rho) = \begin{cases}
S_{x'} &  \rho = \rho_1 \\
0 & \text{otherwise}
\end{cases}
\end{equation}
We can now use the inverse Fourier transform in Equation \eqref{eq:inverse_FT} to calculate the transform of these sections.  Also, we will make use of the fact that the representations look like shifts to write
\begin{equation}
S_{x'} = \rho_\alpha(x',0,0) .
\end{equation}
This gives us, with $g=(q,p,\lambda)$,
\begin{align}
(\mathcal F^{-1} A_1)(g) &= \int_{\widehat G} \, tr(\rho(g) A_1(\rho)) \, dP(\rho) \\
& = \int_{\widehat G} \, tr(\rho_\alpha(g) \rho_\alpha(x',0,0) ) \, dP(\rho) \\
& = \int_{\widehat G} \, tr(\rho_\alpha(q + x',p,\lambda ) \, dP(\rho) \\
&  \quad  \vdots \\
& = \mathcal{N} \delta(q-x') \delta(p) \delta(\lambda)
\end{align}
So the ``constant'' section gives a delta function on the group.  

The transform of the alternative embedding can also be calculated.
\begin{align}
(\mathcal F^{-1} A_2)(g) &= \int_{\widehat G} \, tr(\rho(g) A_2(\rho)) \, dP(\rho) \\
& = \int_{\widehat G} \, tr(\rho_\alpha(g) \rho_1(x',0,0) ) \delta(\alpha - 1) \, dP(\rho) \\
& = tr(\rho_1(q + x',p,\lambda )  \\
& = \chi_{\rho_1}(g \diamond (x',0,0)) \\
& \quad \vdots \\
& = \mathcal{N} e^{i\lambda}  \delta(q-x') \delta(p)
\end{align}
So our section which started out localized transforms to a function spread out in $\lambda$.

\section{The Non-commutative Fourier Transform and Zobin's Symbol Theory}

The Fourier transform for non-commutative groups provides the mathematical basis for the theory of symbols of operators.  While it has been understood that ideas from group theory play an important role in the theory of symbols, it was Zobin who recognized that the mapping from operators to functions could be described as a type of double Fourier transform on groups.  
\begin{equation} \label{eq:zobin_transform} \begin{CD}
\Gamma(\Lambda(\widehat G),dP)
@> \mathcal{F}_{G}^{-1} >>
L_2(G,dg)
\stackrel{\text{as a set}}{=}
L_2(G_0,dg) 
@> \mathcal{F}_{G_0} >> 
L_2(\widehat{G}_0,d\hat g)
\end{CD} \end{equation}
The idea of this transformation is this:  Start with an operator of interest, $\hat A$.  This operator lives in some space of operators, $HS(\mathcal H_{\hat A})$.  There is often then a non-commutative group $G$ which has an irreducible representation which also lives in the space $HS(\mathcal H_{\hat A})$.  So, the operator $\hat A$ can be embedded in a section of the dual bundle, call the section $S_A(\rho)$.  We can now use the theory of Fourier transforms for non-commutative groups to transform this section of the dual bundle into a function on the group, where the group is considered as a set.  The inverse Fourier transform $\mathcal F_G^{-1}$ is the first transform on the way from calculating the symbol of the operator.  We now have a function on a group.  But, this is really just a function on a set, since the function itself is perfectly well defined without the group structure.  Now consider a new group $G_0$.  This is the group which has the same set of elements as the group $G$.  However, instead of the non-commutative group product rule from $G$, the group product for $G_0$ is a commutative rule.  So the function that was obtained from the inverse Fourier transform can actually be thought of as a function on the commutative group $G_0$.  The commutative version of the Fourier transform for the group $G_0$ can now be applied to obtain a function on the dual, $\widehat{G}_0$.  This function on the dual is the Zobin symbol for the section $S_A(\rho)$.

Since this procedure relates functions to operators, it can be thought of as a type of ``quantization'' mapping.  A function is a ``classical'' object, and this double Fourier transform lets us find a corresponding operator, or ``quantum'' object.  We will sometimes refer to this double transform as a quantization, and write it as
\begin{equation}
Q = \mathcal{F}_G \circ \mathcal{F}_{G_0}^{-1} : L_2(\widehat G_0, d\hat g) \rightarrow \Gamma(\Lambda(\widehat G), dP) .
\end{equation}
The Zobin symbol of an operator is then just the application of the inverse of this quantization mapping.  For an operator $\hat A$ embedded in a section $S_A(\rho)$, write
\begin{equation}
A(g) = (Q^{-1} S_A) (g)
\end{equation}
for the symbol of $\hat A$.

\subsection{The Zobin Symbol Theory for the Heisenberg-Weyl Group}

A straightforward and useful example of the Zobin transform is the case of the double Fourier transform for the continuous Heisenberg-Weyl group.  As a set, this group is composed of a $2n$-dimensional phase space, and one additional dimension which is used to record the canonical commutation relations for the elements of phase space. 
\begin{equation}
\mathfrak H \stackrel{\text{as a set}}{=} \mathbb R^{2n+1}
\end{equation}
As described in Section (\ref{sec:cont_HW}), the group product law is almost just the commutative addition of elements of $\mathbb R^{2n+1}$.
\begin{equation}
g_1 \diamond g_2 = \left( {z}_1 + {z}_2, \, \lambda_1 + \lambda_2 + \frac{1}{2} \omega({z}_1, {z}_2) \right)
\end{equation}
The set $\mathbb R^{2n+1}$ also has a natural commutative group structure, that of ordinary addition.  So, the double Fourier transform will be an inverse transform from the dual $\widehat{\mathfrak H}$ to the Heisenberg-Weyl group, followed by an ordinary $2n+1$-dimensional Fourier transformation.
\begin{equation} \begin{CD}
\Gamma(\Lambda(\widehat {\mathfrak H}),dP)
@> \mathcal{F}_{\mathfrak H}^{-1} >>
L_2({\mathfrak H},dg)
\stackrel{\text{as a set}}{=}
L_2(\mathbb R^{2n+1},d^{2n+1}x) 
@> \mathcal{F}_{\mathbb R^{2n+1}} >> 
L_2(\widehat{\mathbb R}^{2n+1},d^{2n+1}\hat x)
\end{CD} \end{equation}

The irreducible representations of the Heisenberg-Weyl group are given in Equations (\ref{eq:con_1D_irrep}) and (\ref{eq:con_inf_irrep}).  There, a splitting of phase space has been chosen, which specifies the ``position'' and ``momentum'' coordinates in $\mathbb R^{2n}$.  Since there are two types of representations, the dual space $\widehat{\mathfrak H}$ has two components.  The first, which is composed of the one-dimensional representations from Equation (\ref{eq:con_1D_irrep}), turns out to have zero Plancharel measure, and so we can neglect it when forming sections of the dual bundle for transforming.  The second component of the dual is given by the one parameter family of infinite-dimensional representations in Equation (\ref{eq:con_inf_irrep}), reproduced here for convenience.
\begin{equation}
\rho_\alpha({\bf q', p'},\lambda) \psi({\bf q}) = e^{i \alpha (\lambda - \frac{1}{2}{\bf p' \cdot q'} -{\bf p' \cdot q} ) }  \psi({\bf q+q}')
\end{equation}
Since these representations are parametrized by real, nonzero $\alpha$, the dual is isomorphic as a set to the real line $\mathbb R / \{0\}$.  These representations are operators which act on functions on configurations space, so for each $\alpha$ the fiber is the space of operators $HS(L_2(\mathbb R^n))$.  A section of the dual bundle is therefore, up to the set of measure zero associated with the one-dimensional representations, a choice of Hilbert-Schmidt operators, one for each $\alpha \neq 0$.

\subsection{The Weyl Symbol as a special case of the Zobin Transform\label{sec:weyl-zobin}}

The Weyl symbol of an operator can be thought of as a special case of the Zobin transform for the Heisenberg-Weyl group.  If the operator is embedded in a specific section of the Heisenberg-Weyl dual, then its transform will be restricted to a $2n$-dimensional space which can be thought of as phase space.  This relationship between the Weyl symbol and the double non-commutative Fourier transform can be demonstrated by an explicit calculation, performed as follows.

Take an operator of interest, $\hat A$, and embed it in a section of the dual bundle, as in Equation (\ref{eq:HW_embedding}).
\begin{equation}
A_2(\rho) = \begin{cases}
\hat A & \text{ if $\alpha = 1$} \\
0 & \text{ otherwise} 
\end{cases}
\end{equation}
With this embedding the section is only non-zero for the representation with $\alpha=1$, which in the $q$ representation, is the operator given in Equation (\ref{eq:con_inf_irrep}).
\begin{equation}
\rho_1({\bf q',p'},\lambda') \psi({\bf q}) = e^{i (\lambda' - \frac{1}{2}{\bf p' \cdot q'} -{\bf p' \cdot q} ) }  \psi({\bf q+q}')
\end{equation}
Then, the non-commutative inverse Fourier transform of $A_2(\rho)$ is
\begin{equation}
\tilde A(g') = \int_{\hat G}  \text{tr}(\rho_1^\dagger(g') S_A(\rho)) \, dP(\rho)= \text{tr}(\rho_1^\dagger(g') \hat A)
\end{equation}
Then the symbol of the section $A_2(\rho)$ is given now given by the commutative, $2n+1$-dimensional Fourier transform of $\tilde A(g')$.  We think of the elements $g'$ of the group as simply points in $2n+1$-dimensional space, and write the Fourier transform in such a way as to be invariant under changes of splitting.  Also, the trace in $\tilde A (g')$ can be explicitly written as an integral in the $q$ representation.
\begin{align}
A(z,\lambda) & =  \int_G d^{2n}z' d\lambda' e^{i\omega(z,z') +i\lambda \lambda'}
\tilde A(z', \lambda') \\
&= \int_G d^n q' d^n p' d\lambda' \, e^{i( {\bf q\cdot p' - p\cdot q'} +\lambda \lambda' )} e^{-i\lambda'} 
\int d^n q'' \langle {\bf q''+q'} \vert  e^{i(\frac{1}{2}{\bf p' \cdot q'} +{\bf p' \cdot q''}  )}  \hat A  \vert {\bf q''} \rangle \\
&= \int e^{i\lambda'(\lambda - 1)} d\lambda'  
\int e^{i( {\bf q\cdot p' - p\cdot q'}+\frac{1}{2}{\bf p' \cdot q'} +{\bf p' \cdot q''}  )}  
\langle {\bf q''+q' } \vert  \hat A  \vert {\bf q''} \rangle  d^n q'' d^n q' d^n p' \\
&=\delta(\lambda-1) 
\int \delta\left( {\bf  q}  + \frac{\bf q'}{2} +{\bf q''}  \right)  e^{-i{\bf  p\cdot q'} }  
\langle {\bf q''+q' } \vert  \hat A  \vert {\bf q''} \rangle  d^n q'' d^n q' \\
&=\delta(\lambda-1) 
\int e^{-i{\bf  p\cdot q'} }  
\langle {\bf -q +q' }/2   \vert  \hat A  \vert {\bf - q - q'}/2  \rangle  d^n q' 
\end{align}
From this we see that, up to an inversion of the $\bf q$ axis, the symbol as calculated by using the double Fourier transform agrees with the definition of the Weyl symbol [Equation (\ref{eq:weyl_symbol})] on the plane $\lambda = 1$.
\begin{equation}
A_\text{symbol}({\bf -q},{\bf p},1) = A_\text{Weyl}({\bf q},{\bf p})
\end{equation}
So the theory of Weyl symbols can be reinterpreted from the point of view of Zobin's double non-commutative Fourier transform for the Heisenberg-Weyl group.  Also, the Moyal star product for symbols of operators is simply the convolution for the non-commutative Fourier transform on the Heisenberg-Weyl group.

\subsection{The Star Product}

In the previous sections the Zobin symbol of an operator was defined, using the double Fourier transform for non-commutative groups [Equation (\ref{eq:zobin_transform})].  This works well for finding the symbol of one operator.  Often, a particular calculation will involve products of operators.  In this case, the group theory point of view becomes very useful.  The non-commutative Fourier transform of a product can be found using the convolution theorem [Equation (\ref{eq:nc_conv})].  Application of the commutative Fourier transform then gives an equation for the symbol of a product.  This symbol can be used to define a generalization of the Moyal star product for symbols.

The convolution theorem gives the transform of a product of sections.  
\begin{equation}
S_1(\rho) S_2(\rho) \stackrel{\mathcal{F}_G^{-1}}{\rightarrow} (f_1 * f_2)(g) 
= \int_G \, f_1(g\diamond h_2^{-1}) f_2(h_2) \, dh_2
\end{equation}
Here the functions $f_1(g)$ and $f_2(g)$ are the inverse transforms of the sections $S_1(\rho)$ and $S_2(\rho)$, respectively.  Application of the commutative Fourier transform gives us the symbol, which can then be used to define the star product, denoted $\star$, of two symbols.
\begin{align}
S_1(\rho) S_2(\rho)  \rightarrow (\tilde f_1\star \tilde f_2 )(\tau) &\equiv \int_G \int_G  \tau(h_1) f_1(h_1\diamond h_2^{-1}) f_2(h_2) \, dh_1 \, dh_2 \\
&= \int_G \int_G  \tau(h_1 \diamond h_2) f_1(h_1) f_2(h_2) \, dh_1 \, dh_2 \label{eq:star_int}
\end{align}
Here $\tilde f_1(\tau)$ and $\tilde f_2(\tau)$ are the symbols of $S_1(\rho)$ and $S_2(\rho)$.  
\begin{equation}
\tilde f_j(\tau) = \int_{{G}_0}  \tau(h) f_j(h) \, dh
\end{equation}
The argument $\tau$ is one of the irreducible representations of the commutative group $G_0$.  We can insert the inverse of this relationship into Equation (\ref{eq:star_int}) in order have a definition of the star product that involves $\tilde f_j(\tau)$ instead of $f_j(h)$.
\begin{align}
(\tilde f_1\star \tilde f_2 )(\tau) &= \int_G \int_G  \tau(h_1 \diamond h_2) 
\left(\int_{\widehat{G}_0} \tau_1^\dagger(h_1) \tilde f_1(\tau_1) d\tau_1 \right)
\left(\int_{\widehat{G}_0} \tau_2^\dagger(h_2) \tilde f_2(\tau_2) d\tau_2 \right)
  dh_1  dh_2 \\
&= \int_{\widehat{G}_0}\int_{\widehat{G}_0} \tilde f_1(\tau_1) \tilde f_2(\tau_2)
\left(\int_G \int_G \tau(h_1 \diamond h_2) \tau_1^\dagger(h_1) \tau_2^\dagger(h_2)dh_1  dh_2 \right) d\tau_1 d\tau_2 \\
&= \int_{\widehat{G}_0}\int_{\widehat{G}_0} \tilde f_1(\tau_1) \tilde f_2(\tau_2)
 \, K_\star(\tau,\tau_1,\tau_2) \, d\tau_1 d\tau_2
\end{align}
In the last line, the integrals over the group have been combined, and used to define the integral kernel $K_\star(\tau,\tau_1,\tau_2)$ of the star product.  
\begin{equation}\label{eq:star_kernel}
K_\star(\tau,\tau_1,\tau_2) \equiv \int_G \int_G \tau(h_1 \diamond h_2) \tau_1^\dagger(h_1) \tau_2^\dagger(h_2)dh_1  dh_2
\end{equation}
This kernel form of the star product is useful when evaluating the star product of more than two terms, e.g., for finding $(\tilde f_1 \star \tilde f_2 \star \tilde f_3)(\tau)$ in terms of the symbols $\tilde f_j(\tau)$.

If the group multiplication $\diamond$ were commutative, then from Equation (\ref{eq:star_int}) we can see that the star product would reduce to the ordinary product of the symbols.
\begin{align}
(\tilde f_1\star \tilde f_2 )(\tau) &= \int_G \int_G  \tau(h_1) \tau(h_2) f_1(h_1) f_2(h_2) \, dh_1 \, dh_2 \\
& = \left( \int_G \tau(h_1) f_1(h_1)  \right) \left( \int_G \tau(h_2) f_1(h_2)  \right)
\end{align}
But the group product is not commutative, and so the integral fails to simplify to an ordinary product.  In the case when $G$ is the Heisenberg-Weyl group, we can write $h_j$ in component form as $h_j = (z_j, \lambda_j)$, and the product as
\begin{equation}
h_1 \diamond h_2 = h_1 + h_2 + \left(0,\frac{1}{2}\omega(z_1,z_2)\right).
\end{equation}
Each of these terms is an element of the commutative group $G_0$, and so the representation $\tau$ can be separated. 
\begin{equation}
\tau(h_1 \diamond h_2) = \tau(h_1) \, \tau(h_2) \, \tau\!\left( \left(0,\frac{1}{2}\omega(z_1,z_2)\right) \right)
\end{equation}
The integral which defines the star product is then left with an extra phase, since we can write $\tau$ as an exponent, $\tau(h) = \exp( i \langle a_\tau , h \rangle)$.
\begin{equation}
(\tilde f_1\star \tilde f_2 )(\tau) = \int_G \int_G e^{i \langle a_\tau , (0,\omega(z_1,z_2)) \rangle /2}  \tau(h_1)  f_1(h_1) \tau(h_2) f_2(h_2) \, dh_1 \, dh_2
\end{equation}
The phase in the exponent is what gives the star product its non-commutivity, and as we will see later, will provide the $\int (q\, dp - p \, dq)$ term in the phase space path integral which gives rise to Hamilton's equations in the semiclassical limit.

The structure of this star product is why the double Fourier transform is used to define the symbol of an operator.  If the non-commutivity of $G$ depends on some small parameter, then, as the parameter goes to zero, the star multiplication of the symbols becomes ordinary multiplication of functions.  This will find direct application in the semiclassical limit of quantum mechanics, and in the small wavelength limit of more generic wave equations.

\section{The ``Symbol'' of a Matrix}

The non-commutative double Fourier transform lets one associate a complex scalar function with a given operator.  Here we show how this can be done for the example of an n$\times$n matrix.  The matrix is embedded in a section of the dual bundle of the discrete Heisenberg-Weyl group $\mathfrak H_n$, and two group Fourier transforms are applied to find the symbol of the matrix.  
\begin{equation} \label{eq:matrix_double_transform}
\begin{CD}
\Gamma(\Lambda(\widehat {\mathfrak H_n}),dP)
@> \mathcal{F}_{\mathfrak H_n}^{-1} >>
L_2({\mathfrak H_n},dg)
\stackrel{\text{as a set}}{=}
L_2(\mathbb Z^3/n,dN) 
@> \mathcal{F}_{\mathbb Z^3/n} >> 
L_2(\widehat{\mathbb Z}^3/n,dN)
\end{CD} \end{equation}
The first, inverse, Fourier transform takes the section of the dual bundle to a function on the group.  But the group is, as a set, just a lattice of points.  So the second Fourier transform is an ordinary discrete Fourier transform.  

In this section, the symbols of specific matrices are calculated, and compared with the intuition developed by considering the Wigner function.  Whenever possible, the dimension $n$ of the matrix will be left arbitrary.  Specific examples and calculations will be performed for the case $n=3$.

\subsection{Irreducible Representations of $\mathfrak H_n$}

The group theory for the discrete Heisenberg-Weyl group $\mathfrak H_n$ is described in Section \ref{sec:DHW}.   The group consists of $n^3$ points labeled by integers mod $n$
\begin{equation}
\mathfrak H_n=\{g=(q,p,\lambda) \in \mathbb{Z}/n \oplus \mathbb{Z}/n \oplus \mathbb{Z}/n \}, 
\end{equation}
and so, as a set, this group is equivalent to the integer lattice $\mathbb{Z}^3/n$.

This group has $n^2$ inequivalent one-dimensional irreducible representations, and $n-1$ inequivalent $n$-dimensional irreducible representations.  The one-dimensional irreducible representations are labeled by a pair of integers $u,v \in \mathbb{Z}^2/n$
\begin{equation}
\varrho_{u,v}(q,p,\lambda) = \exp \left( \frac{2 \pi i}{n}( u q + v p) \right) .
\end{equation}
The $n$-dimensional irreducible representations involve shifts and multiplication by a phase.  In matrix representation, these can be written as
\begin{equation}
\rho_\alpha(q,p,\lambda) = \exp \left( \frac{2 \pi i}{n}( \alpha\lambda + \alpha qp) \right)
\mathbf{T}^{2 \alpha p} \mathbf{S}^{q}
\end{equation}
where $\alpha \in \mathbb Z_n / 0$,  $\mathbf{S}$ is the shift matrix, and $\mathbf{T}$ is the diagonal matrix of $n^\text{th}$ roots of unity, as in Equation (\ref{eq:matrix_rep}).

Given these irreducible representations of $\mathfrak H_n$, we can construct the dual $\widehat{\mathfrak H}_n$.  The dual has $n^2 + n -1$ elements, one for each irreducible representation.  The $n^2$ one-dimensional representations are multiplications by a complex phase, so their associated fibers are just the space of complex numbers, $\mathbb C$.  The $n-1$ $n$-dimensional representations are given by matrices, so their fibers are the space $n\times n$ of matrices, $GL(n)$.

\subsection{Sections of the dual bundle}

A section of the dual fiber bundle is an assignment of an operator to each point in $\widehat{\mathfrak H}_n$.  For $n=3$, this means picking two 3$\times$3 matrices, and 27 complex numbers.
\begin{equation}
S = \left( \begin{array}{c}
\rho_1 \rightarrow \mathbf{A} \\
\rho_2 \rightarrow \mathbf{B} \\
\varrho_{0,0} \rightarrow c_{0,0}\\ 
\varrho_{0,1} \rightarrow c_{0,1}\\
                \vdots
\end{array}\right)
\end{equation}
Since we are interested in the symbol of a single matrix, we will consider sections of the form
\begin{equation}\label{eq:matrix_embed}
S_A = \left( \begin{array}{c}
\rho_1 \rightarrow \mathbf{0} \\
\rho_2 \rightarrow \mathbf{A} \\
\varrho_{0,0} \rightarrow 0\\ 
\varrho_{0,1} \rightarrow 0\\
                \vdots
\end{array}\right)
\end{equation}
where $\mathbf{A}$ is the matrix of interest.

\subsection{Representations of the commutative group $\mathbb{Z}_n^3$}

In order to construct the symbol of a matrix, we will need the representations of the commutative group that is associated with $\mathfrak H_n$.  This is the group formed by the finite lattice $\mathbb{Z}_n^3 = \mathbb{Z}/n \oplus \mathbb{Z}/n \oplus \mathbb{Z}/n $.  The group product law for this group is normal addition mod $n$, so the irreducible representations are all one-dimensional.  They are labeled by three integers $u,v,w \in \mathbb{Z}^3/n$, and are given by
\begin{equation}
\tau_{u,v,w}(q,p,\lambda)=\exp\left(\frac{2\pi i}{n}(uq+vp+w\lambda) \right) .
\end{equation}
Since these are labeled by triples in $\mathbb{Z}^3/n$, the dual, $\widehat{\mathbb{Z}}_n^3$, is isomorphic as a set to $\mathbb{Z}^3/n$.  Also, because the representations are all one-dimensional, the fibers are all just $GL(1) = \mathbb{C}$.

\subsection{The symbol of a section}

We can now calculate the symbol of the section $S_A$ given in Equation (\ref{eq:matrix_embed}).  The first step in the double transform given in Equation (\ref{eq:matrix_double_transform}) is to perform the inverse transform of the section to a function on the group.  The second step is to perform a discrete Fourier transform.
Together, these transforms define the symbol of the section $S_A(\rho)$.
\begin{equation}
A(\tau)=\sum_{g\in \mathbb{Z}_n^3} 
\sum_{\rho \in \widehat{\mathfrak{H}}_n^3} 
\text{tr}\left(S_A(\rho) \rho^{\dagger}(g)\right) \nu_{\rho} \tau(g)
\end{equation}
Here $\nu_{\rho}$ is a normalization (the Plancharel measure) associated with each representation, $\nu_{\rho}=1/\text{dim}(\rho)$.  The inverse of this mapping can be used to reconstruct a section from a given symbol.
\begin{equation}
S_A (\rho) = \sum_{g\in \mathbb{Z}_n^3} 
\sum_{\tau \in \widehat{\mathbb{Z}}_n^3} 
A (\tau)  \tau(g) \rho(g)
\end{equation}
These equations can now be used to calculate specific examples.

\subsection{Delta functions in the symbols}
From the equations for the irreducible representations of $\mathfrak{H}_n$ and $\mathbb{Z}_n^3$, we can derive some useful general properties of the symbol of a section.  In the case that $n=3$, each of the irreducible representations of $\mathfrak{H}_n$ are associated with a different subspace of $\widehat{\mathbb{Z}}_n^3$.  In particular, with some algebra the following can be derived.

If the section is only nonzero for the one-dimensional irreducible representations, then
\begin{align}
A(\tau) = A(u,v,w) &= \sum_g \sum_{a,b} S_A(\varrho_{a,b}) \varrho_{a,b}^\dagger(g) \tau_{u,v,w}(g) \\
&= \sum_g \sum_{a,b} c_{a,b} \exp\left(\frac{2\pi i}{n} (-aq-bp +uq + vp + w\lambda)\right) \\ 
&= \sum_{a,b} c_{a,b} \delta_{u,a}\delta_{v,b}\delta_{w,0}
=c_{u,v}\delta_{w,0}
\end{align}

If the section is only nonzero for one of the $n$-dimensional irreducible representations $\rho_\alpha$ with $\alpha \neq 0$, then
\begin{align}
A(u,v,w) &= \sum_g \nu_{\rho_\alpha}\text{tr}\left(S_A(\rho_\alpha) \rho_\alpha^{\dagger}(g) \right) \tau_{u,v,w}(g)  \\
&= \sum_{q,p,\lambda} \nu_{\rho_\alpha}\text{tr}\left(S_A(\rho_\alpha) {\bf T}^{2\alpha p} {\bf S}^q \right)  \exp\left(\frac{2\pi i}{n} (-\alpha \lambda -\alpha q p +uq + vp + w\lambda)\right) \\
&= \delta_{w,\alpha} \times \sum_{q,p} \nu_{\rho_\alpha}\text{tr}\left(S_A(\rho_\alpha) {\bf T}^{2\alpha p} {\bf S}^q \right)  \exp\left(\frac{2\pi i}{n} ( -\alpha q p +uq + vp )\right) 
\end{align}

This calculation shows that the different values of $w$ in the symbol correspond to different components of the section $S_A$.  To each value of $w$ there is associated a plane in the space $\widehat{\mathbb Z}^3_n$ which looks like a phase space.  Sections which look like delta functions over $\widehat{\mathfrak H}_n$ transform into symbols constrained to one of these ``phase space'' components of $\widehat{\mathbb Z}^3_n$.  This is a specific example of the calculation described in Section \ref{sec:weyl-zobin}, where it was shown how the Weyl symbol can be calculated by transforming a particular embedding of an operator.

\subsection{Example: Wigner function analogy}

It is instructive to calculate the symbol for matrices that are constructed from the outer product of a vector with itself.  Such a matrix is a projection matrix, and is the discrete version of the quantum mechanical projection operator.  The Weyl symbol of the quantum projection operator is the Wigner function, so we might expect the symbol of this projection matrix to behave like the Wigner function of a quantum state.  

\begin{figure}
\begin{center}
\includegraphics[scale=0.35]{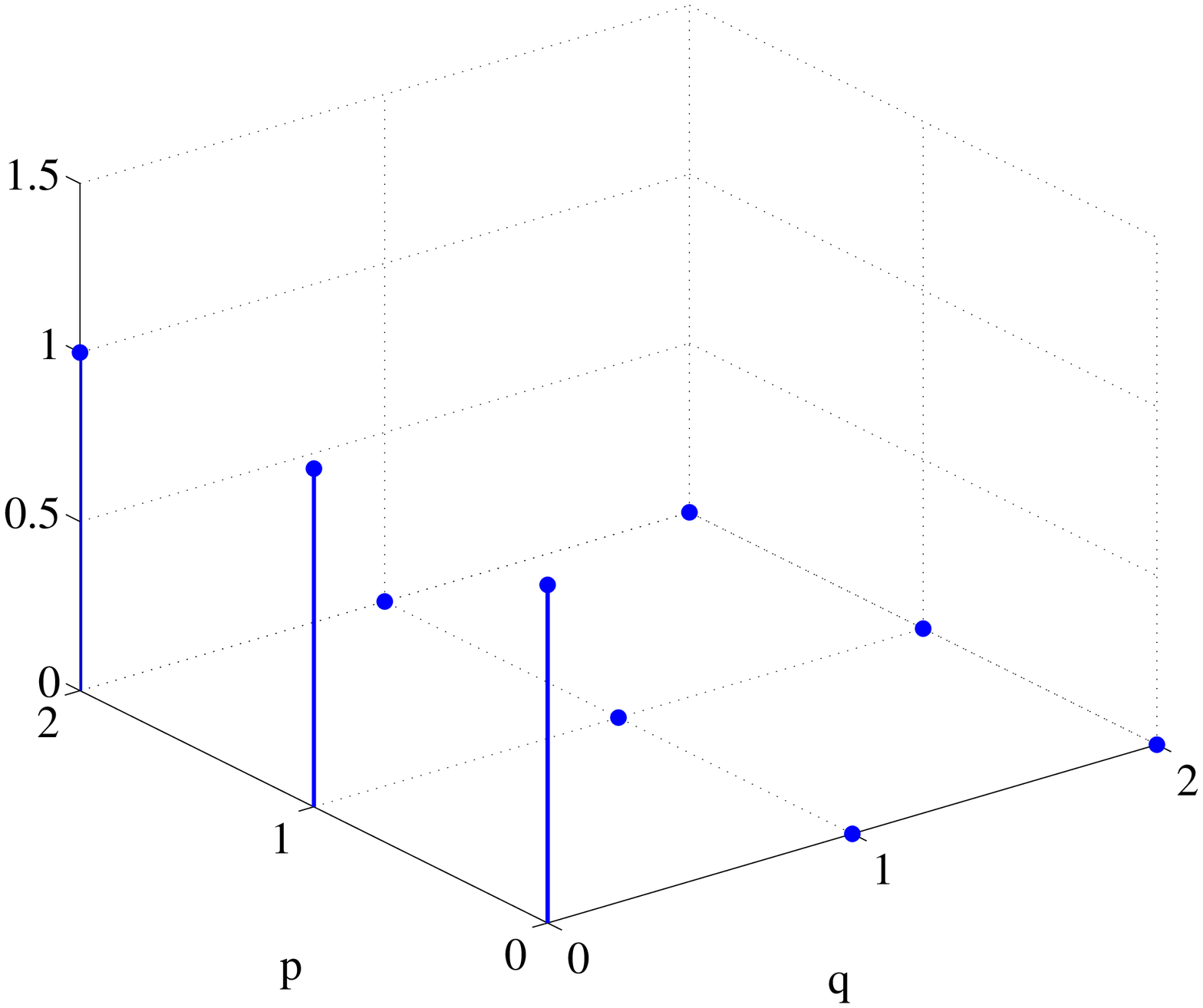}
\includegraphics[scale=0.35]{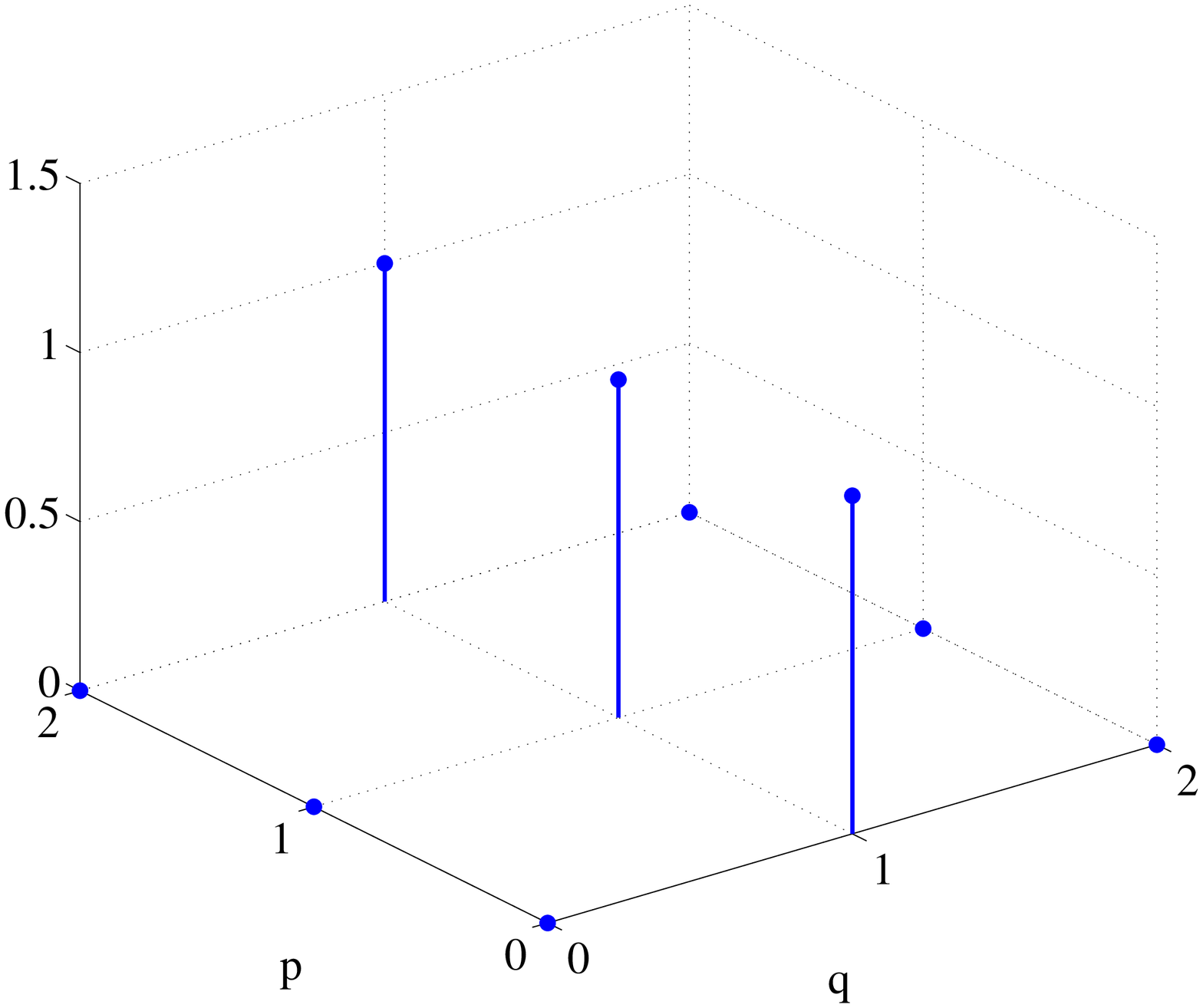}
\includegraphics[scale=0.35]{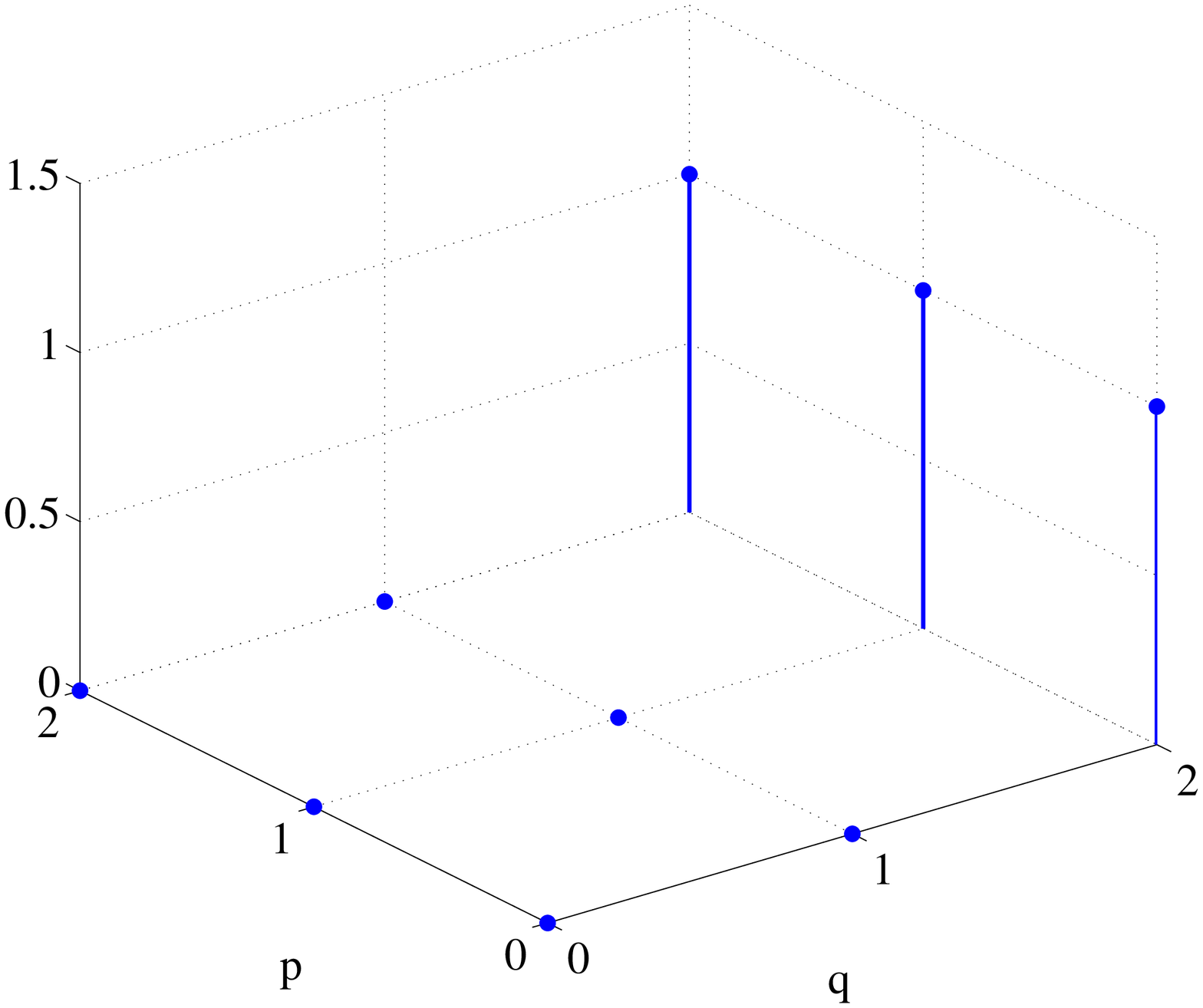}
\end{center}
\caption{\label{cap:xproj}
Symbol of projectors onto the discrete position eigenstates, $(1,0,0)$, $(0,1,0)$ , and  $(0,0,1)$.  Since the matrix is embedded in a delta-like way as a section of the dual bundle, the symbol is restricted to a $3\times 3$ ``phase space''.  The values of the symbol on this phase space are shown here, and they resemble discrete versions of the Wigner function for continuous position eigenstates.
}
\end{figure}

In particular, the Wigner function of a position eigenstate is a delta function in the position coordinate, and a constant in momentum.  The discrete version of a position eigenstate is a vector with only one nonzero element, e.g.~$(1,0,0)$. Figure (\ref{cap:xproj}) shows that the symbol of this discrete version is also a delta function in position and constant in momentum. 

\begin{figure}
\begin{center}
\includegraphics[scale=0.35]{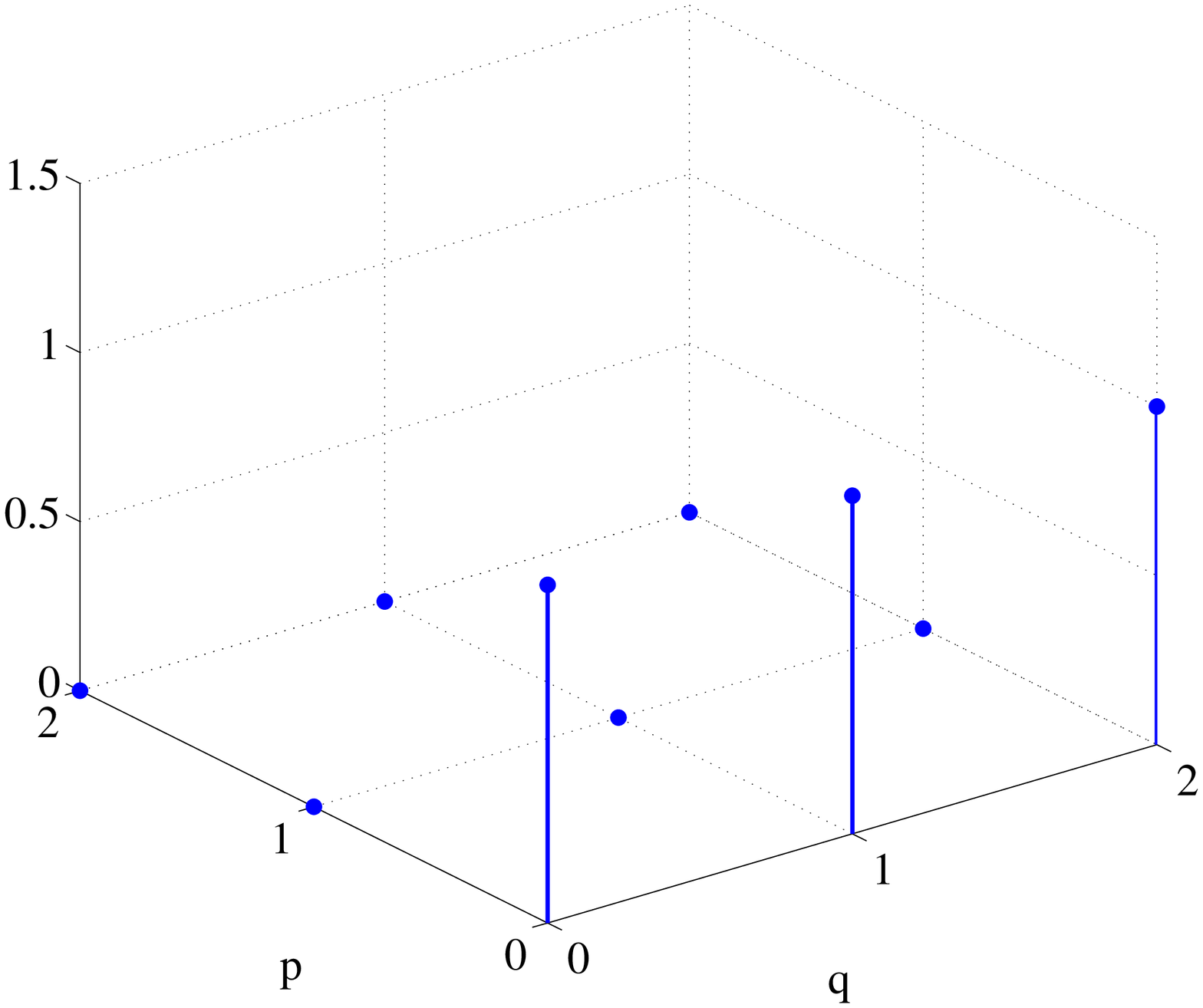}
\includegraphics[scale=0.35]{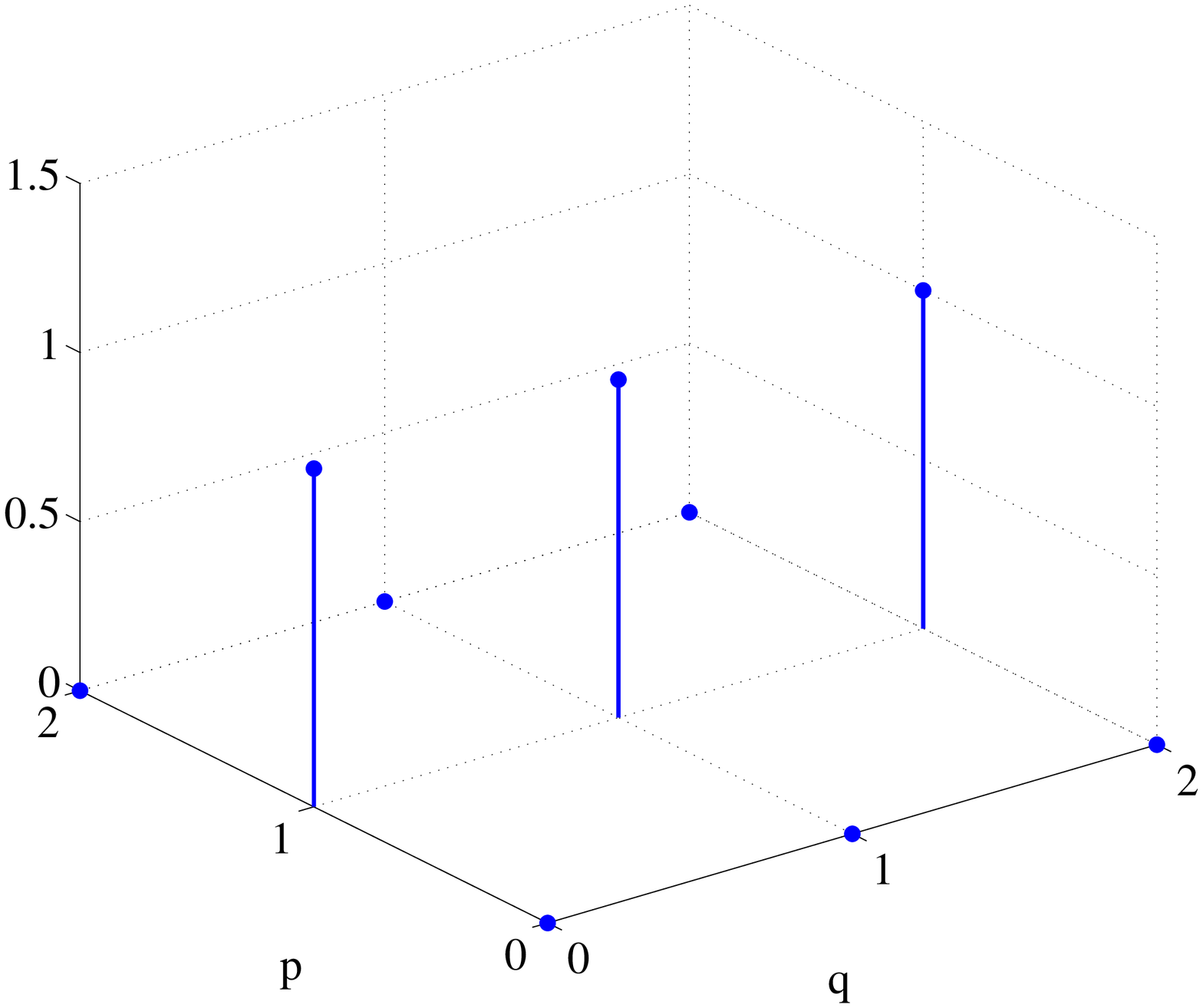}
\includegraphics[scale=0.35]{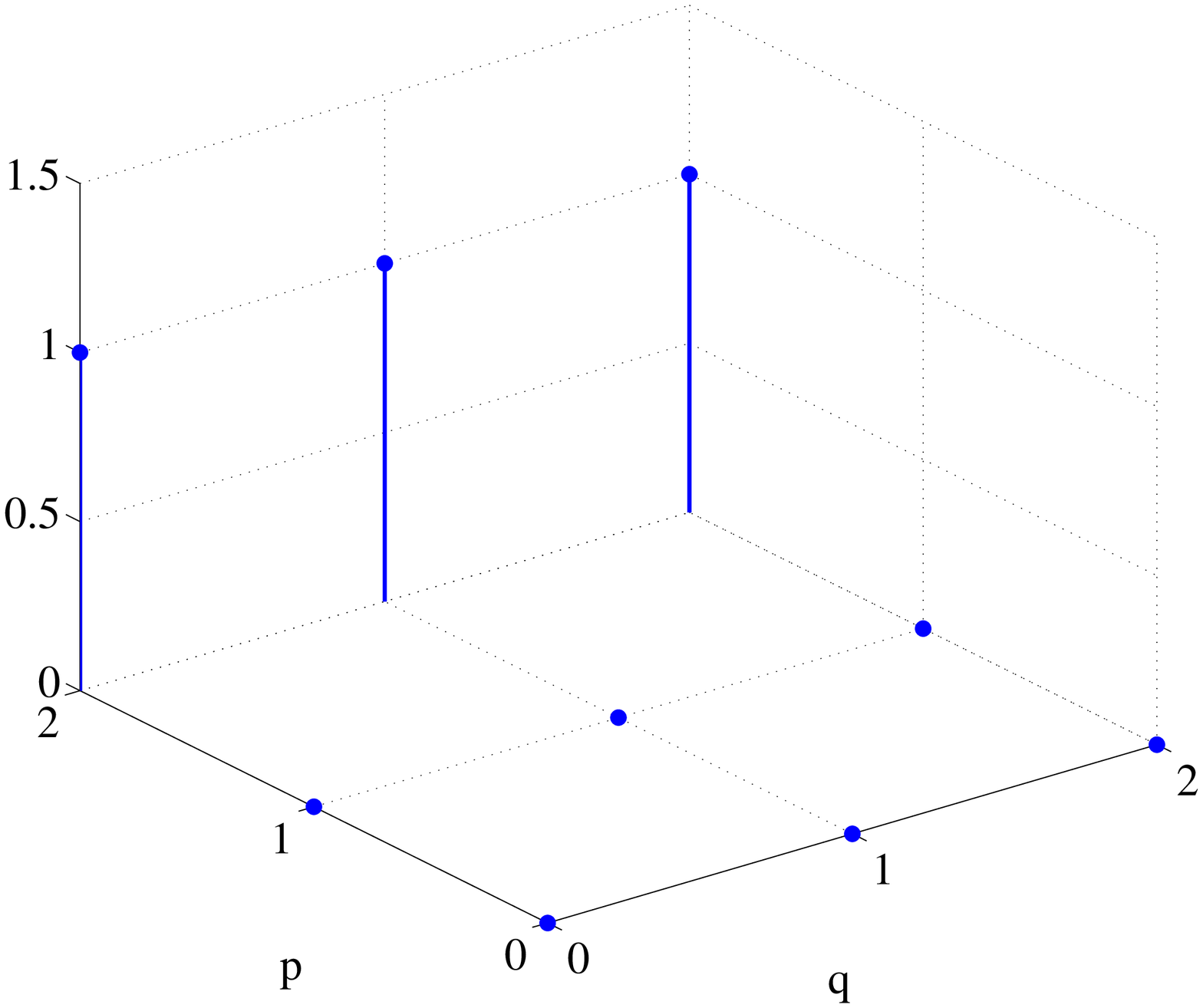}
\end{center}
\caption{\label{cap:yproj}
Symbol of projectors onto the discrete plane waves, $\exp\left(-\frac{2 \pi i}{3}  q p\right)$, for $p=0,1,2$.
}
\end{figure}

Similarly, the Wigner function of a plane wave (a momentum eigenstate) is a delta function in momentum and a constant in position.  This is exactly what we see in Figure (\ref{cap:yproj}), where the symbol of the projector onto a discrete version of the plane wave is shown.

\subsection{Star Product for Matrices\label{sec:matrix_star_product}}
When using the symbol calculus, it is often necessary to calculate the symbol of a product of operators.  In the case of symbols of matrices, this calculation can be done once for all by finding the kernel of the star product, which will be a discrete version of Equation (\ref{eq:star_kernel}).  This kernel can then be used to compute the star product rule for symbols of matrices. 

The discrete version of the star product looks like a contraction of a rank three tensor.
\begin{equation}
[A \star B](\tau) = \sum_{\tau_1} \sum_{\tau_2} K_\star(\tau, \tau_1, \tau_2) A(\tau_1) B(\tau_2)
\end{equation}
Using the equation for the representations $\tau$, and the equation for the kernel, we can compute the elements of $K_\star$.
\begin{equation}\label{eq:matrix_kernel}
K_\star(\tau, \tau_1, \tau_2) = \left(\frac{1}{n^3}\right)^2 \sum_{h_1} \sum_{h_2} \tau(h_1 \diamond h_2) \tau_1(-h_1) \tau_2(-h_2)
\end{equation}
This summation can be directly computed, and used to create a numerical algorithm for calculating the star product.  For example, this calculation was performed using {\sf Matlab}, and the results can be used to quickly calculate the star product of two sections.  As an example, here is the matrix $K_\star(\tau,\tau_1,\tau_2)$ for the particular choice of $\tau = (0,0,1)$.
{\begin{spacing}{1.0}
\begin{equation}\label{eq:group_multi_table}
\setlength{\arraycolsep}{.6\arraycolsep}
\left(
\begin{array}{ccccccccccccccccccccccccccc}
0&0&0&0&0&0&0&0&0&0&0&0&0&0&0&0&0&0&0&0&0&0&0&0&0&0&0\\
0&0&0&0&0&0&0&0&0&0&0&0&0&0&0&0&0&0&0&0&0&0&0&0&0&0&0\\
0&0&0&0&0&0&0&0&0&0&0&0&0&0&0&0&0&0&0&0&0&0&0&0&0&0&0\\
0&0&0&0&0&0&0&0&0&0&0&0&0&0&0&0&0&0&0&0&0&0&0&0&0&0&0\\
0&0&0&0&0&0&0&0&0&0&0&0&0&0&0&0&0&0&0&0&0&0&0&0&0&0&0\\
0&0&0&0&0&0&0&0&0&0&0&0&0&0&0&0&0&0&0&0&0&0&0&0&0&0&0\\
0&0&0&0&0&0&0&0&0&0&0&0&0&0&0&0&0&0&0&0&0&0&0&0&0&0&0\\
0&0&0&0&0&0&0&0&0&0&0&0&0&0&0&0&0&0&0&0&0&0&0&0&0&0&0\\
0&0&0&0&0&0&0&0&0&0&0&0&0&0&0&0&0&0&0&0&0&0&0&0&0&0&0\\
0&0&0&0&0&0&0&0&0&0&0&0&0&0&0&0&0&0&0&0&0&0&0&0&0&0&0\\
0&0&0&0&0&0&0&0&0&0&1&1&e^{\frac{4\pi i}{3}}&e^{\frac{4\pi i}{3}}&e^{\frac{4\pi i}{3}}&e^{\frac{2\pi i}{3}}&e^{\frac{2\pi i}{3}}&e^{\frac{2\pi i}{3}}&0&0&0&0&0&0&0&0&0\\
0&0&0&0&0&0&0&0&0&0&1&1&e^{\frac{2\pi i}{3}}&e^{\frac{2\pi i}{3}}&e^{\frac{2\pi i}{3}}&e^{\frac{4\pi i}{3}}&e^{\frac{4\pi i}{3}}&e^{\frac{4\pi i}{3}}&0&0&0&0&0&0&0&0&0\\
0&0&0&0&0&0&0&0&0&0&e^{\frac{2\pi i}{3}}&e^{\frac{4\pi i}{3}}&1&e^{\frac{2\pi i}{3}}&e^{\frac{4\pi i}{3}}&1&e^{\frac{2\pi i}{3}}&e^{\frac{4\pi i}{3}}&0&0&0&0&0&0&0&0&0\\
0&0&0&0&0&0&0&0&0&0&e^{\frac{2\pi i}{3}}&e^{\frac{4\pi i}{3}}&e^{\frac{4\pi i}{3}}&1&e^{\frac{2\pi i}{3}}&e^{\frac{2\pi i}{3}}&e^{\frac{4\pi i}{3}}&1&0&0&0&0&0&0&0&0&0\\
0&0&0&0&0&0&0&0&0&0&e^{\frac{2\pi i}{3}}&e^{\frac{4\pi i}{3}}&e^{\frac{2\pi i}{3}}&e^{\frac{4\pi i}{3}}&1&e^{\frac{4\pi i}{3}}&1&e^{\frac{2\pi i}{3}}&0&0&0&0&0&0&0&0&0\\
0&0&0&0&0&0&0&0&0&0&e^{\frac{4\pi i}{3}}&e^{\frac{2\pi i}{3}}&1&e^{\frac{4\pi i}{3}}&e^{\frac{2\pi i}{3}}&1&e^{\frac{4\pi i}{3}}&e^{\frac{2\pi i}{3}}&0&0&0&0&0&0&0&0&0\\
0&0&0&0&0&0&0&0&0&0&e^{\frac{4\pi i}{3}}&e^{\frac{2\pi i}{3}}&e^{\frac{4\pi i}{3}}&e^{\frac{2\pi i}{3}}&1&e^{\frac{2\pi i}{3}}&1&e^{\frac{4\pi i}{3}}&0&0&0&0&0&0&0&0&0\\
0&0&0&0&0&0&0&0&0&0&e^{\frac{4\pi i}{3}}&e^{\frac{2\pi i}{3}}&e^{\frac{2\pi i}{3}}&1&e^{\frac{4\pi i}{3}}&e^{\frac{4\pi i}{3}}&e^{\frac{2\pi i}{3}}&1&0&0&0&0&0&0&0&0&0\\
0&0&0&0&0&0&0&0&0&0&0&0&0&0&0&0&0&0&0&0&0&0&0&0&0&0&0\\
0&0&0&0&0&0&0&0&0&0&0&0&0&0&0&0&0&0&0&0&0&0&0&0&0&0&0\\
0&0&0&0&0&0&0&0&0&0&0&0&0&0&0&0&0&0&0&0&0&0&0&0&0&0&0\\
0&0&0&0&0&0&0&0&0&0&0&0&0&0&0&0&0&0&0&0&0&0&0&0&0&0&0\\
0&0&0&0&0&0&0&0&0&0&0&0&0&0&0&0&0&0&0&0&0&0&0&0&0&0&0\\
0&0&0&0&0&0&0&0&0&0&0&0&0&0&0&0&0&0&0&0&0&0&0&0&0&0&0\\
0&0&0&0&0&0&0&0&0&0&0&0&0&0&0&0&0&0&0&0&0&0&0&0&0&0&0\\
0&0&0&0&0&0&0&0&0&0&0&0&0&0&0&0&0&0&0&0&0&0&0&0&0&0&0\\
0&0&0&0&0&0&0&0&0&0&0&0&0&0&0&0&0&0&0&0&0&0&0&0&0&0&0
\end{array}
\right)
\end{equation}
\end{spacing}
} % end single space

%% file: Chapter-PathIntegral.tex
%%%%%%%%%%%%%%%%%%%%%%%%%%%%%%%%%%%%%%%%%%%%%%%%%%%%%%%%%%%%%%%%%%%%%%%%%%
%
% Ph.D. dissertation manuscript
% Chapter 7: Path Integrals 
%
% Andrew Stephen Richardson (Fall 2007)
% College of William and Mary
% Department of Physics
% Prof. Eugene Tracy, advisor
%
% Based on Paul King and Andrew Norman's template (modified by Wirawan Purwanto)
%
%%%%%%%%%%%%%%%%%%%%%%%%%%%%%%%%%%%%%%%%%%%%%%%%%%%%%%%%%%%%%%%%%%%%%%%%%%

\chapter{Functions of Operators and the Path Integral\label{chp:PathIntegral}}

\subsubsection{Introduction}

In this chapter, we describe how to use the Zobin symbol theory to calculate the symbol for a function of an operator.  While functions of operators are generally important, one particular example stands out, and that is the exponential of an operator.  In order to connect this section to a problem in physics, we start with an example of how the exponential of an operator arises in the context of a generic wave equation.  We then return to the context of the Zobin symbol theory, and describe how the path integral arises when calculating the symbol of a function of an operator.

Consider a generic wave equation in a time-independent medium, for example the equation for the electric field in a nonuniform plasma.  Such an equation can be written in integral form as
\begin{equation}
\int {\bf D}(x,x'; t-t') \cdot {\bf E}(x',t')  \, dx' \, dt'= 0.
\end{equation}
Because this has the form of a convolution in time, we can Fourier transform in time to obtain
\begin{equation}
\int {\bf D}(x,x'; \omega) \cdot {\bf E}_\omega(x')  \, dx'= 0,
\end{equation}
where the frequency now appears as a parameter.  The integral kernel ${\bf D}(x,x';\omega)$ encodes the physics of the problem, and solutions of this equation describe waves of frequency $\omega$ in the plasma.
Alternatively, this equation can be written in pseudodifferential form, for example by using the Weyl symbol calculus.  Abstractly, such an equation can be written as an operator acting on some state:
\begin{equation}
\hat D_\omega \vert \psi \rangle = 0 .
\end{equation}
The parameter  $\omega$ appears because the time dependence has been removed by Fourier transforming with respect to time.  If we are looking for narrow-banded solutions, then we can expand about $\omega_0$, the carrier frequency, to find:
\begin{equation}
\left( \hat D(\omega_0) + (\partial_\omega \hat D ) (\omega - \omega_0) + \ldots \right) \vert \psi \rangle=0 .
\end{equation}
If the variation of $\hat D(\omega)$ is slow enough, then we can ignore the higher order terms in this series.  Performing an inverse Fourier transform of the truncated series gives the equation
\begin{align}
\left( \hat D(\omega_0) + (\partial_\omega \hat D) (i\partial_t - \omega_0) \right) \vert \psi \rangle =0
\end{align}
If we write the field as a carrier $e^{-i\omega_0 t}$ times an envelope, the $\omega_0$ term can be removed: 
\begin{align}
0&=\left( \hat D(\omega_0) + (\partial_\omega \hat D) (i\partial_t - \omega_0) \right) e^{-i\omega_0 t} \vert \tilde\psi \rangle \\
&=e^{-i\omega_0 t} \left( \hat D(\omega_0) + ( \partial_\omega \hat D )( i\partial_t) \right) \vert \tilde\psi \rangle .
\end{align}
We can now drop the phase $e^{-i\omega_0 t}$.  Continuing the analysis of this equation with a generic operator involves inverting $\partial_\omega \hat D$.  However, since this calculation is only meant to illustrate the usefulness of the exponential of an operator, we make the simplifying assumption that $\partial_\omega \hat D$ is proportional to the identity:
\begin{align}
\partial_\omega \hat D = {\bf id}.
\end{align}
This operator can be inverted, and we can now isolate the time derivative by multiplying through with the inverse:
\begin{equation}
0 = \left(  \hat D(\omega_0) +   i\partial_t \right) \vert \tilde\psi \rangle .
\end{equation}
Now define a new operator 
\begin{equation}
\hat H = - \hat D(\omega_0).
\end{equation}
With this definition, the equation for the envelope of the wave looks like the Schr\"odinger equation.
\begin{equation}
\left(\hat H - i\partial_t \right) \vert \tilde \psi \rangle = 0
\end{equation}
So a wave equation with the form of the Schr\"odinger equation is actually very general.  We can expect evolution equations of this form to arise whenever we consider the dynamics of the envelope of a narrow-banded wavepacket.  

This equation, which has one derivative in time, can be solved by exponentiating the operator $it \hat H$.
\begin{equation}
\vert\psi(t) \rangle = e^{it\hat H} \vert \psi(0) \rangle
\end{equation}
The exponent of an operator can be computed in two different ways.  The first, which usually only works in special cases, would be to find the eigenvalues and eigenfunctions of $\hat H$.  If the eigenfunctions are used as a basis, the the operator $it\hat H$ is diagonal with the eigenvalues on the diagonal.  The exponential of the operator can then be calculated by exponentiating the eigenvalues.  This is how we usually solve quantum problems, where the eigenvalues are related to the energies of the stationary states.  If $\hat H$ is Hermitian, this diagonalization can always be done in principle, but in practice  can be difficult.  Also, physical insight might be lacking which could be gained using other asymptotic methods.

The second option for finding the exponent is to define $e^{it\hat H}$ in terms of a series involving powers of $\hat H$.  Finite powers are perfectly well defined, but may become cumbersome to compute for large orders.  If instead we can compute the symbol of the operator $\hat H$, and use it to find the symbol of $e^{it\hat H}$, we would have an alternative approach to calculating $e^{it\hat H}$.
\begin{equation}
\begin{CD}
\hat H @>\text{power series}>> e^{it \hat H} \\
@V{Q^{-1}}VV      @VV{Q^{-1}}V \\
\text{Symbol of } \hat H @>\text{path integral}>> \text{Symbol of }  e^{it \hat H}
\end{CD}
\end{equation}
As we will see in this chapter, finding the symbol of the exponent requires calculating repeated star products.  As the number of repeated star products goes to infinity, the formula that is obtained is a path integral expression for the symbol of the exponent of an operator.  While this path integral may be no easier to compute than the eigenvalues and eigenfunctions of $\hat H$, it can still be a useful tool.  For example, since the phase which appears in the path integral is related to the classical action, various methods of semiclassical analysis can be applied to the path integral, providing physical insight into the systems being studied.

In addition to describing how the path integral arises when calculating a function of an operator, we will discuss two interesting observations about the path integral.  The first is that when calculating the path integral solution for vector wave equations, the function which arises as the ray hamiltonian is not one of the eigenvalues of the dispersion matrix, but rather one of the diagonal elements of the matrix.  This could lead to a new approach for ray-tracing approximations, which might be able to handle novel types of mode conversion that are not of the familiar ``avoided crossing'' form.  For example, in multidimensional problems, mode conversion can occur in regions where the rays have non-zero helicity \cite{PhysRevLett.91.130402,citeulike:472573}.  This implies that the mode conversion cannot be recast as an avoided crossing, and the standard tools used for calculating the effect of the mode conversion cannot be used.  While we do not claim to have solved these novel mode conversion problems, the new theoretical tools described in this dissertation may be useful for obtaining such solutions.  This is a topic for future research.

The second interesting observation arises when considering the ``path integral'' that is obtained when calculating the exponent of a finite matrix.  In this case, the symbol is a function over a discrete set of points, and the path integral is naturally interpreted as a sum over the space of probability distributions on this set of points.  This suggests calculating the sums by using a maximum entropy argument instead of the integral along a path.  If this interpretation can be extended to the continuous case, it could provide a new perspective on the meaning and use of the path integral.  This measure-space interpretation will also provide a way to define and calculate the Fourier transform of functions in infinite-dimensional spaces.

\section{Previous Work on Path Integrals}

The use of path integrals in quantum mechanics was pioneered by Feynman \cite{Feynman:1948lr}, though key ideas were anticipated by Wentzel \cite{Antoci:1998lr} and Wiener \cite{wiener_harmonic_analysis}.  The treatment of classical wave equations, such as those in plasma physics, by path integral methods is mathematically identical to the quantum case, with the classical limit corresponding to the ray or WKB limit.  The Feynman approach uses an integral over paths in configuration space, and this configuration-space path integral has become a standard formulation of quantum mechanics.  Zee gives a nice introduction along the following lines \cite{Zee:2003qy}.  Suppose we want to find the probability of transitioning from the state $\vert q_I \rangle$ to the state $\vert q_F \rangle$ in the time $T$.  This is given by the matrix element
\begin{align}
\langle q_F \vert e^{-iT \hat H} \vert q_I \rangle.
\end{align}
By inserting $N-1$ complete sets of states into this expression, we can write this transition probability in terms of the transition probability for short time steps $\delta t = T/N$:
\begin{align}
\langle q_F \vert e^{-iT \hat H} \vert q_I \rangle &= 
\int \prod_{j=1}^{N-1} dq_j \, \langle q_F \vert e^{-i\delta t \hat H} \vert q_{N-1} \rangle 
\langle q_{N-1} \vert e^{-i\delta t \hat H} \vert q_{N-2} \rangle \ldots \langle q_1 \vert e^{-i\delta t \hat H} \vert q_I \rangle .
\end{align}
Since the Hamiltonian will in general involve both position $\hat q$ and momentum $\hat p$ operators, some care needs to be taken when calculating the $j^{th}$ term in this integral.  Evaluating the $\hat p$ terms in the $p$ representation involves calculating a Gaussian integral, and (for a standard kinetic plus potential Hamiltonian) results in
\begin{align}
\langle q_{j+1} \vert e^{-i\delta t \hat H} \vert q_j \rangle = \left( - \frac{2\pi i \,m}{\delta t} \right)^{1/2} \exp\!\left( \frac{i\delta t \,m}{2} \left(\frac{q_{j+1} -q_j }{\delta t}\right)^2 -i\delta t \,V(q_j) \right).
\end{align}
Inserting this back into the previous equation, and taking the limit $N\rightarrow \infty$, gives us the path integral in configuration space:
\begin{align}
\langle q_F \vert e^{-iT \hat H} \vert q_I \rangle &= 
\int e^{i\int_0^T \frac{1}{2} m\dot q^2(t) -V(q(t)) \, dt} \, Dq(t).
\end{align}

While this path integral was explicitly constructed from the perspective of configuration space, it is possible to form a path integral from the perspective of phase space.
Berezin \cite{Berezin:1991fk} showed that the phase-space path integral arises whenever calculating the Weyl symbol of a product of many operators.  Calculating the symbol of the operator $e^{iT\hat H}$ using the phase-space path integral then gives
\begin{align}\label{eq:traditional_phase_space_path_integral}
\int e^{i\int_0^T \left[\frac{1}{2} (q(t)\dot p(t) - p(t)\dot q(t))   - H(q(t),p(t)) \right]\, dt} \, Dq(t) \,Dp(t).
\end{align}
This path integral is much more in the spirit of classical Hamiltonian mechanics, since $p$ and $q$ are treated on an equal footing.

In this chapter we show how the configuration space and phase space path integrals are related, using group theoretic ideas from Chapter \ref{chp:GroupTheory}.  We start by writing the path integral for a generic function of an operator, using the Zobin symbol theory.  The example of the symbol for the finite Heisenberg-Weyl group is worked through in detail.  We then outline how this path integral on the dual group $\widehat G_0$ reduces to the phase-space path integral.  This reduction is related to the reduction to the primary representations.  Further reduction to the irreducible representations requires the selection of a ``configuration space'', and leads to the configuration-space path integral.  To our knowledge the connection between the reduction of group representations (primary to irreducible) and the reduction of the phase-space path integral to the configuration-space path integral is new.

\section{Using Symbols to Calculate Functions of Operators} 

Calculating a function of an operator can be difficult in many cases.  However, the symbol of the operator can be used to find the symbol of the function of the operator.  This symbol of the function can then be converted back into an operator, giving the desired function of the operator.  In some cases, the symbol of the function can be used to develop asymptotic approximations to the operator.  It is of great interest that many of these asymptotic approximations are ``semiclassical'' in flavor, even when the problems being considered are not quantum mechanical.

For the exponential function, this procedure can be illustrated in the following diagram.
\begin{equation}
\begin{CD}
\hat H      @.\!\!\!\!\!\!\!\!\!\!\!\!\!\!\!\!\!\!\!\!\!\!\!\!\!\!\!\!\!\!\!\!\!\!\!\!\!\!\!\!\!\!\!\!\!\!\!\!
\overset{\text{series definition}}{\bf{-\,-\,-\,-\,-\,\rightarrow}} \,\,\,\,\,\,\,\,\,\,\, e^{it \hat H} \\
@V{Q^{-1}}VV         @AA{Q}A \\
\text{Symbol of } \hat H @>\text{ path integral }>> \text{Symbol of }  e^{it \hat H}
\end{CD}
\end{equation}
Instead of trying to go directly from the operator to the exponential (following the dashed arrow $\bf{-\,-\,-\,\rightarrow}$), we first go down to its symbol using the inverse quantization mapping, $Q^{-1}$.  As we will show, the symbol of the exponent can be calculated from the symbol of the operator, by using a path integral.  The result of the path integral is another function, which we can quantize using $Q$ to obtain the operator $\exp(it\hat H)$.

In order to find the symbol of $\exp(it\hat H)$ from the symbol of $\hat H$, we need to define what we mean by the exponential function.  The ordinary exponential of a number can be defined using the limit
\begin{equation}
e^x \equiv \lim_{N\rightarrow \infty} \left( 1 + \frac{x}{N} \right)^N.
\end{equation}
Since we know how to calculate products of operators, we could use this expression to define the exponential of an operator:
\begin{equation}\label{eq:operator_exponent}
e^{it \hat H} \equiv \lim_{N\rightarrow \infty} \left( 1 + \frac{it \hat H}{N} \right)^N .
\end{equation}
The operator $\hat H$ commutes with itself, so it acts like a c-number in this formula.
For any fixed value of $N$, this is simply the $N^{th}$ power of the operator $1+it\hat H /N$. 
Repeated application of the star product can be used to calculate the symbol of this power from the symbol of the operator:
\begin{align}
Q^{-1} \left[ \left( 1 + \frac{it \hat H}{N} \right)^N \right](\tau) &= Q^{-1} \left[ \overbrace{\left( 1 + \frac{it \hat H}{N} \right)\cdot \left( 1 + \frac{it \hat H}{N} \right) \cdot \ldots \cdot  \left( 1 + \frac{it \hat H}{N} \right)}^{N \text{ products of operators}}  \right](\tau) \\
&=  \overbrace{Q^{-1} \left[ 1 + \frac{it \hat H}{N} \right](\tau) \star \ldots \star Q^{-1} \left[ 1 + \frac{it \hat H}{N} \right](\tau) }^{N \text{ star products}} \\
&\equiv  \left( Q^{-1} \left[ 1 + \frac{it \hat H}{N} \right]  \right)^{\star N}(\tau).
\end{align}
This {\em defines} the $\star$ exponent, $\star N$.
Taking the large $N$ limit of this expression gives us a way to calculate the symbol of the exponential of the operator.
\begin{equation}\label{eq:star_exponent}
Q^{-1}\left[ e^{it \hat H}\right](\tau) = \lim_{N\rightarrow \infty}\left( Q^{-1} \left[ 1 + \frac{it \hat H}{N} \right]  \right)^{\star N}(\tau)
\end{equation}
As we will show, evaluating this expression in the limit leads to a path integral on the space of $\tau$'s.  Since $\tau$ is an irreducible representation of the commutative group $G_0$, this will be an integral over paths in $\widehat G_0$.  While the path integral result is fairly generic for symbols defined for continuous groups, the limiting form of this expression looks quite different for discrete groups.  In the discrete case there is no way to define continuous paths in the space $\widehat G_0$, and we get a sum over densities instead of a path integral.

Instead of computing the limiting form of Equation (\ref{eq:star_exponent}) for a generic group, we will give two examples.  In the following sections we will describe how to calculate the exponent of a matrix using the repeated star product.  This will illustrate the sum over densities.  Then we will show how the path integral arises for continuous groups, by considering the example of the Heisenberg-Weyl group.

\section{The ``Path Integral'' for a Discrete Group}
% Primary vs irreducible
% No "path" but rather a density
% A "measure space" Fourier transform
% entropy type calculations in the limit

% 1+eA ...  spec approx =  spec id...  calculate spec for HW example -> trpl degenerate because it has 3 copies of the irrep 

\begin{figure}
\begin{center}
\includegraphics[scale=0.8]{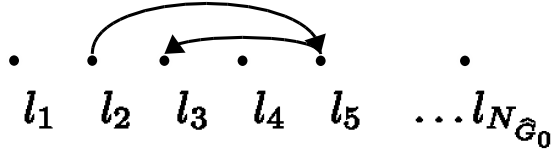}
\end{center}
\caption{\label{fig:path_example}
Example of a discrete path in a lattice of points.  The path of three elements is given by$(\tau_1,\tau_2,\tau_3) = (l_2, l_5, l_3)$.
}
\end{figure}

The ``path integral'' for the discrete Heisenberg-Weyl group $\mathfrak{H}_n$ is obtained by evaluating the expression in Equation (\ref{eq:star_exponent}) using the star product defined in Section \ref{sec:matrix_star_product}.  Using the kernel $K_\star(\tau,\tau_1,\tau_2)$ from Equation (\ref{eq:matrix_kernel}), we recall that the star product can always be written as a tensor contraction.
\begin{equation}
[A \star B](\tau) = \sum_{\tau_1,\tau_2} K_\star(\tau, \tau_1, \tau_2) A(\tau_1) B(\tau_2)
\end{equation}
The symbol of powers of a matrix can also be computed using this kernel.  If $A(\tau)$ is the symbol of the section into which our matrix of interest was embedded, then the appropriate symbol for the $k^{th}$ power of the matrix can be derived by induction from repeated application of the star product.
\begin{align}
[A \star A \star A](\tau) &= \sum_{\tau_1,\tau_2} K_\star(\tau, \tau_1, \tau_2) A(\tau_1) [A\star A](\tau_2) \\
&= \sum_{\tau_1,\tau_2} \sum_{\tau_3,\tau_4} K_\star(\tau, \tau_1, \tau_2) A(\tau_1)
K_\star(\tau_2, \tau_3, \tau_4) A(\tau_3) A(\tau_4)
\end{align}
From this expression it is fairly evident what the $N^{th}$ power would give.  In order to simplify the expressions, define a new function 
\begin{equation}\label{eq:def_M_A}
M_A(\tau_1,\tau_2) \equiv \sum_{\tau'} K_\star(\tau_1,\tau', \tau_2) A(\tau').
\end{equation}
If we consider $A(\tau)$ as a vector indexed by $\tau$, and $K_\star$ as a rank three tensor, then $M_A$ is a matrix with elements
\begin{align}
[M_A]_{\tau_1,\tau_2} = \sum_{\tau'} [K_\star]_{\tau_1, \tau', \tau_2} [A]_{\tau'}.
\end{align}
This lets us write the $\star$-powers $A^{\star N}(\tau)$ in terms of powers of the matrix $M_A$.  For example,
\begin{align}
[A \star A \star A](\tau) &= \sum_{\tau_1,\tau_2}  M_A(\tau,\tau_1) M_A(\tau_1,\tau_2) A(\tau_2) \\
&= \left[ M_A^2 \cdot A \right]_{\tau}
\end{align}
Extending this to the $N^{th}$ power is straightforward, and gives
\begin{align}
A^{\star N}(\tau) &\equiv \overbrace{(A \star A \star \ldots \star A)}^{N\text{ star products}} (\tau)\\
&=\left[ M_A^{N-1} \cdot A \right]_\tau .
\end{align}
As with any matrix multiplication, this can be written in terms of the elements of the matrices.  Each multiplication requires the sum over the inner indexes, and so $N-1$ multiplications will give sums over $N-1$ indexes,
\begin{align}\label{eq:matrix_product_as_path_sum}
\left[ M_A^{N-1} \cdot A \right]_\tau = \sum_{\tau_1}\sum_{\tau_2}\ldots \sum_{\tau_N} [M_A]_{\tau,\tau_1}[M_A]_{\tau_1,\tau_2} \ldots [M_A]_{\tau_{N-1},\tau_N} [A]_{\tau_N}.
\end{align}
Note that since this is a power of a matrix, it will become dominated by the largest eigenvalue of $M_A$ in the limit of large $N$.  This gives an alternative approach to the calculation of the powers of the matrix, but we have not pursued this approach.
This form of the matrix multiplication is completely general.  Because each of the variables $\tau_j$ will take on all possible values, this means that all possible ``paths'' of length $N$ will appear in this summation.  It is worthwhile to take a moment to describe the notation that will be used for these paths.  Because we are working with a finite group, there are finitely many irreducible representations.  The set of irreducible representations is $\widehat G_0$, and the number of irreducible representations is $N_{\widehat G_0}$.  This set of representations $\widehat G_0$ can be thought of as a discrete set of points, or a lattice of points.  Give each point of $\widehat G_0$ a label, denoted by $l$.  Then the set $\widehat G_0$ looks like a list of points,
\begin{align}
\widehat G_0 = \{l_1, l_2, \ldots, l_{N_{\widehat G_0}} \}.
\end{align}
In the sum over possible representations, the variable $\tau_j$ will take on all $N_{\widehat G_0}$ possible values:
\begin{align}
\tau_j \in \{l_1, l_2, \ldots, l_{N_{\widehat G_0}} \}.
\end{align}
A path is then a choice of points $l$, one for each variable $\tau$.  We will sometimes use the notation $\{\tau\}$ to indicate a path:
\begin{align}
\{\tau\} = \{ \tau_1, \tau_2, \ldots, \tau_N \}.
\end{align}
Care must be taken so as not to confuse the subscripts.  The subscript of the $\tau$ variables runs from 1 to $N$, the length of the path, and it indicates how far along the path we are.  The subscript on the $l$'s runs from 1 to $N_{\widehat G_0}$, and indicates which point in the lattice $\widehat G_0$ we are discussing.  Figure \ref{fig:path_example} gives a picture of what we mean by this notation.  In that example, the ``path'' is an ordered list of three elements from $\widehat G_0$, $\{\tau\} = (\tau_1,\tau_2,\tau_3) = (l_2, l_5, l_3)$.

From this point, the derivation of the path integral for the particular case of the discrete Heisenberg-Weyl group will proceed in three steps.  First, we will use the structure of the group to find an explicit expression for the matrix $M_A$.  Second, this expression will be applied $k-1$ times in succession, in order to evaluate the $k^{th}$ star product formula above.  This will give an expression that can be interpreted either as a sum over ``paths'', or as a sum over measures.
The third step is to apply this sum over paths (or measures) to the limit formula in Equation (\ref{eq:star_exponent}) in order to find the symbol of the exponent.

\subsection{An expression for $M_A$ for the Heisenberg-Weyl group}

We are now considering the case of the star product for the Heisenberg-Weyl group $\mathfrak{H}_n$, which, as a set, is formed of $n^3$ points.  Using the definitions of the matrix $M_A$ [Equation (\ref{eq:def_M_A})] and the kernel of the star product [Equation (\ref{eq:matrix_kernel})], we have the following (up to normalization),
\begin{equation}
M_A(\tau_1,\tau_2) = \sum_{\tau'} \sum_{h_1} \sum_{h_2} \tau_1(h_1 \diamond h_2) \tau'(-h_1) \tau_2(-h_2) A(\tau').
\end{equation}
This is an $n^3 \times n^3$ matrix.  Since we are interested in calculating with symbols instead of functions on the group, we will rewrite this expression as sums over $\tau$'s instead of sums over group elements $h$.  Start by grouping the $\tau'$ terms:
\begin{align}
M_A(\tau_1,\tau_2) &= \sum_{h_1} \left[\sum_{h_2} \tau_1(h_1 \diamond h_2)  \tau_2(-h_2)\right] \sum_{\tau'}  \tau'(-h_1) A(\tau') \\
&= \sum_{h_1} \left[ \sum_{h_2} \tau_1(h_1 \diamond h_2)  \tau_2(-h_2) \right] \tilde A(h_1),
\end{align}
where $\tilde A$ is the (commutative) inverse Fourier transform of $A$.
This expression is now a sum over the group, of a product of two functions on the group.  We can use the Plancharel identity to write this as a sum of products of the Fourier transforms of the two functions.  First give a name to the sum in the square brackets:
\begin{equation}
\tilde{\mathfrak{M}}(h_1; \tau_1,\tau_2) = \sum_{h_2} \tau_1(h_1 \diamond h_2)  \tau_2(-h_2) .
\end{equation}
The Plancharel identity gives us this new expression:
\begin{align}
M_A(\tau_1,\tau_2) &= \sum_{\tau} \mathfrak{M}(\tau; \tau_1,\tau_2) A(\tau).
\end{align}
We now only have to find $\mathfrak{M}(\tau; \tau_1,\tau_2)$, which is the (commutative) Fourier transform of $\tilde{\mathfrak{M}}(h_1; \tau_1,\tau_2)$, with respect to the variable $h_1$.  In order to reduce subscripts, define $g=h_1$.  The Fourier transform is then
\begin{align}
\mathfrak{M}(\tau; \tau_1,\tau_2) &= \sum_g \tau(g) \sum_{h_2} \tau_1(g \diamond h_2)  \tau_2(-h_2) .
\end{align}
In order to continue, use the following notation for the representations:
\begin{equation}
\tau(g) = e^{\frac{2\pi i}{n}(\check{z}\cdot {z} + \check{\lambda} \lambda )}.
\end{equation}
Each $\tau$ is labeled by an element of the dual, $(\check{z},\check{\lambda}) \in \mathbb Z_n^{3}$.  Also, we will write the symplectic product as either a two-form $\omega({z_1 },{z_2 })$, or in matrix notation
\begin{equation}
z_1^{\text{T}} \cdot {\bf J} \cdot z_2 .
\end{equation}
These, together with the group product law, imply the following:
\begin{align}
\mathfrak{M}(\tau; \tau_1,\tau_2) &= 
\sum_{{z },\lambda} \sum_{{z}_2, \lambda_2 } \exp\left\{ \frac{2\pi i}{n} \left( 
\check{z}\cdot {z} + \check{\lambda} \lambda 
+ \check{z}_1\cdot ({z} +{z}_2)   \right)\right\}  \notag \\
& \quad \times \exp\left\{ \frac{2\pi i}{n} \left(  \check{\lambda}_1( \lambda_1 + \lambda_2 +z_1^{\text{T}} \cdot {\bf J} \cdot z_2 )
-\check{z}_2\cdot {z}_2 - \check{\lambda}_2 \lambda_2
 \right)\right\}  \\
 &= 
\sum_{{z },{z}_2}  \exp\left\{ \frac{2\pi i}{n} \left( 
{z}\cdot(\check{z}+ \check{z}_1) 
+ {z}_2\cdot(\check{z}_1-\check{z}_2)
+ \check{\lambda}_1 z_1^{\text{T}} \cdot {\bf J} \cdot z_2 
   \right)\right\} \notag \\
&   \quad \times \sum_{\lambda, \lambda_2 }
\exp\left\{ \frac{2\pi i}{n} \left(  \lambda (\check{\lambda} +\check{\lambda}_1)
+  \lambda_2 (\check{\lambda}_1 -\check{\lambda}_2)
 \right)\right\}.
\end{align}
The sums over $\lambda$ and $\lambda_2$ give us two Kronecker delta functions, and the sum over $z$ gives us another\footnote{In the limit of large $n$, the discrete group goes over to the continuous Heisenberg-Weyl group, and these Kronecker delta functions go over to Dirac delta functions}:
\begin{align}
\mathfrak{M}(\tau; \tau_1,\tau_2) &= \delta(\check{\lambda} +\check{\lambda}_1)
\delta(\check{\lambda}_1 -\check{\lambda}_2)
\sum_{{z}_2}
\delta\!\left( \check{z}+ \check{z}_1 + \check{\lambda}_1{\bf \cdot J} \cdot z_2 \right)
\exp\left\{ \frac{2\pi i}{n} \left( (\check{z}_1-\check{z}_2)\cdot {z}_2  \right)\right\} \\
&=\delta(\check{\lambda} +\check{\lambda}_1)
\delta(\check{\lambda}_1 -\check{\lambda}_2)
\exp\left\{ \frac{2\pi i}{n} \left(
- \frac{1}{ \check{\lambda}_1} (\check{z}_1-\check{z}_2) \cdot {\bf J }^{-1} 
\cdot (\check{z}+ \check{z}_1 )  \right)\right\} \\
&=\delta(\check{\lambda} +\check{\lambda}_1)
\delta(\check{\lambda}_1 -\check{\lambda}_2)
\exp\left\{ \frac{2\pi i}{n} \left(
 \frac{\omega(\check{z}_1-\check{z}_2 , \check{z}+ \check{z}_1 )}
 { \check{\lambda}_1}   \right)\right\}.
\end{align}
Now put this back into the equation for $M_A$:
\begin{align}\label{eq:unreduced_matrix_M_A}
M_A(\tau_1,\tau_2) = \sum_{\check{z},\check\lambda} A(\check{z},\check\lambda)
\delta(\check{\lambda} +\check{\lambda}_1)
\delta(\check{\lambda}_1 -\check{\lambda}_2)
\exp\left\{ \frac{2\pi i}{n} \left(
 \frac{\omega(\check{z}_1-\check{z}_2 , \check{z}+ \check{z}_1 )}
 { \check{\lambda}_1}   \right)\right\} .
\end{align}
This is the expression that we wanted to compute.  We have written $M_A$ without the sums over the group which we had previously used.

\subsubsection{Reductions of the matrix product}

In Section \ref{sec:weyl-zobin} we saw that a particular choice of section would give a symbol with a delta-like behavior in the $\lambda$ direction.  We can follow up on that observation by considering functions $A(\check{z},\check\lambda)$ which have certain properties.  For example, a delta-like behavior could be replicated here, by assuming that 
\begin{equation}\label{eq:symbol_phase_separation}
A(\check{z},\check\lambda) = A(\check{z})  \delta(\check\lambda-1).
\end{equation}
Inserting this into the equation for $M_A$ gives us a simplified formula:
\begin{align}\label{eq:simplified_M_A}
M_A(\tau_1,\tau_2) = \delta(1 +\check{\lambda}_1)\delta(1 +\check{\lambda}_2)
\sum_{\check{z}} A(\check{z})
\exp\left( -\frac{2\pi i}{n} \omega(\check{z}_1-\check{z}_2 , \check{z}+ \check{z}_1)\right).
\end{align}
While $M_A$ is an $n^3 \times n^3$ matrix, the form for $A$ given in Equation (\ref{eq:symbol_phase_separation}) has reduced $M_A$ essentially to an  $n^2 \times n^2$ matrix:
\begin{align}
M_A(\tau_1,\tau_2) = \delta(1 +\check{\lambda}_1)\delta(1 +\check{\lambda}_2) \tilde M_A(\check z_1, \check z_2), 
\end{align}
where the $n^2 \times n^2$ matrix $\tilde M_A$ is defined as
\begin{align}\label{eq:definition_of_tilde_M_A}
\tilde M_A(\check z_1, \check z_2)=\sum_{{z}} A({z})
\exp\left( \frac{2\pi i}{n} \omega({z}+ \check{z}_1,\check{z}_1-\check{z}_2 )\right).
\end{align}
This reduction is reminiscent of the reduction of the regular representation to the primaries, which also left us with functions only on phase space.

This suggests that similar such reductions are possible, by considering the reduction of the primary representations to the irreducible representations.  Since the regular representation and its reductions act on functions on the group, we need to perform an inverse Fourier transform of $A$ back onto the group.  This gives
\begin{align}
\tilde A(g) &= \sum_{\check z} \sum_{\check \lambda} e^{-\frac{2\pi i}{n} z \check z} e^{-\frac{2\pi i}{n} \lambda \check\lambda } A(\check z) \delta(\check\lambda -1)  \\
&= e^{-\frac{2\pi i}{n}\lambda} \sum_{\check z} e^{-\frac{2\pi i}{n} z \check z} A(\check z)  \\
&= \tilde A(z) e^{-\frac{2\pi i}{n}\lambda}.
\end{align}
This is a covariant function on the group, as described in Equation (\ref{eq:covariant_functions}).  This function lives in the $\rho_L$ invariant subspace associated with the primary representation.  We could instead consider functions which live in the subspace associated with an irreducible representation.  For example, we could consider functions in the subspace given in Equation (\ref{eq:cov_for_irreps}).  Such a function can be expanded on the basis functions given in Equation (\ref{eq:dhw_q_basis}):
\begin{align}
\tilde A(g) = \sum_q a_q \phi_q(g) .
\end{align}
We can now insert the expression for $\phi_q(g)$, and perform a Fourier transform to get back to a function of $\tau$.  This is done as follows:
\begin{align}
A(\tau) &= \sum_g \tau(g) \sum_q a_q \sum_p \sum_\lambda e^{-\frac{2\pi i}{n} \lambda} \delta_{(q,0,0)\diamond (0,p,\lambda)}(g) \\
&= \sum_q \sum_p \sum_\lambda a_q  e^{-\frac{2\pi i}{n} \lambda} \tau\big( (q,p,\lambda + qp) \big) \\
&= \sum_q \sum_p \sum_\lambda a_q \exp\!\left(\frac{2\pi i}{n} (-\lambda +\check q q +\check p p + \check \lambda (\lambda + qp) \right) .
\end{align}
The sum over $\lambda$ will give a delta function in $\check\lambda$ as we had before, while the sum over $p$ will give a delta function which can be used to evaluate the sum over $q$.  Thus we obtain
\begin{align}
A(\tau)&= \delta(\check\lambda -1) \sum_q a_q \exp\!\left(\frac{2\pi i}{n} \check q q \right) \sum_p \exp\!\left(\frac{2\pi i}{n} (\check p p  + qp) \right) \\
&= \delta(\check\lambda -1) \sum_q a_q \exp\!\left(\frac{2\pi i}{n} \check q q \right) \delta(\check p + q) \\
&= \delta(\check\lambda -1) \exp\!\left(\frac{2\pi i}{n} \check q \check p \right) a_{-\check p} \, .
\end{align}
If we think of the coefficients $a_{-\check p}$ as a function of $\check p$, then we can write $A(\tau)$ in a way analogous to Equation (\ref{eq:symbol_phase_separation}):
\begin{align}\label{eq:symbol_for_irrep_reduction}
A(\tau) = A(\check p) \delta(\check\lambda -1) \exp\!\left(\frac{2\pi i}{n} \check q \check p \right).
\end{align}
We can now see if the matrix $M_A$ will reduce any further, given this form of $A(\tau)$.  Inserting this expression for $A(\tau)$ into Equation (\ref{eq:unreduced_matrix_M_A}) gives
\begin{align}
M_A(\tau_1,\tau_2) &= \delta(1+\check\lambda_1)\delta(1+\check\lambda_2)
\sum_{\check z, \check \lambda} A(\check p) \exp\!\left(\frac{2\pi i}{n} \check q \check p \right) \exp\!\left(-\frac{2\pi i}{n} \omega(\check z_1-\check z_2, \check z -\check z_1)  \right) \\
&= \delta(1+\check\lambda_1)\delta(1+\check\lambda_2) e^{\frac{2\pi i}{n} \omega(\check z_1,\check z_2)}  \sum_{\check p,\check q} A(\check p)
\exp\!\left(\frac{2\pi i}{n}(\check q \check p- \omega(\check z_1-\check z_2, \check z )  )\right) .
\end{align}
Writing out the symplectic product lets us gather terms which involve $\check q$, and sum over them.  This gives us a delta function which can be used to evaluate the sum over $\check p$:
\begin{align}
M_A(\tau_1,\tau_2) &= \delta(1+\check\lambda_1)\delta(1+\check\lambda_2) e^{\frac{2\pi i}{n} \omega(\check z_1,\check z_2)} \notag \\
& \quad \times \sum_{\check p} \delta(\check p + (\check p_1 - \check p_2) ) A(\check p) \exp\!\left(\frac{2\pi i}{n}  (\check q_1 - \check q_2)(\check p_1 - \check p_2)  )\right) \\
&= \delta(1+\check\lambda_1)\delta(1+\check\lambda_2) \exp\!\left(\frac{2\pi i}{n} \big(\omega(\check z_1,\check z_2) +  (\check q_1 - \check q_2)(\check p_1 - \check p_2) \big)\right) A(\check p_2 - \check p_1) .
\end{align}
If not for the symplectic product in the phase, this expression would depend on $\check z_1$ and $\check z_2$ only through their difference.  It is important to note that the $\check q$'s appear only in the phases.  When the powers of this matrix are computed, this means that the sums over $\check q$'s which will appear can be evaluated, leaving only sums over $\check p$'s, which will lead to a path integral only in $\check p$.

\subsection{The $N^{th}$ power and the path integral on phase space}

We can now insert the expression for $M_A$ from Equation (\ref{eq:simplified_M_A}) into Equation (\ref{eq:matrix_product_as_path_sum}), the formula for the repeated star product of a symbol.  In the formula for the $N^{th}$ power of the symbol, we need the $(N-1)^{th}$ power of the matrix $M_A$.
\begin{align}
M_A^{N-1}(\tau_1,\tau_k) &= \sum_{\tau_2} \cdots \sum_{\tau_{N-1}} 
M_A(\tau_1,\tau_2) \cdots M_A(\tau_{N-1},\tau_{N}) \\
&= \delta(1 +\check{\lambda}_1)\delta(1 +\check{\lambda}_N) \sum_{\check z_2} \cdots \sum_{\check z_{N-1}} \tilde M_A(\check z_1,\check z_2) \cdots \tilde M_A(\check z_{N-1},\check z_{N}). \label{eq:M_A_matrix_product}
\end{align}
Because we are using the reduced form of the matrix $M_A$, these sums can be thought of as a sum over paths, where a path is a length-$N$ sequence of points in the discrete phase space $\mathbb Z_n^{2}$.

\comment{%
This repeated sum can be thought of as a sum over all possible ``paths'', where a path is given by the string which specifies the choice of the $\tau$'s.  For example, if each $\tau$ could be labeled by the values 0 or 1, then a ``path'' would be a binary string of length $N$.
\begin{align}
\{\tau \} = ( 1,1,...,0,0,1,0)
\end{align}
This string can be thought of as a path of length $k$ in the set of labels, $\{0, 1\}$.
For the discrete Heisenberg-Weyl group we are considering, the $\tau$'s can be labeled by the points in $\mathbb Z_n^{3}$, so there are $n^3$ choices for different ``letters'' in any given position in the string.  This means that for us, the string will correspond to a path in the lattice $\mathbb Z_n^{3}$, where steps along the path are jumps to different points on the lattice.  When we make the assumption that the symbol takes the form given in Equation (\ref{eq:symbol_phase_separation}), then the sum reduces to the phase space component only.  In this case, the path is a sequence of points of length $N$ in the discrete phase space $\mathbb Z_n^{2}$.
}%

Substituting Equation (\ref{eq:definition_of_tilde_M_A}) into Equation (\ref{eq:M_A_matrix_product}) gives
\begin{align}\label{eq:double_path_sum}
M_A^{N-1}(\tau_1,\tau_N) &= \delta(1 +\check{\lambda}_1)\delta(1 +\check{\lambda}_N)
\sum_{\check{z}_2} \cdots \sum_{\check{z}_{N-1}}  \notag \\
&\quad \left[
\sum_{{z}_1} \cdots \sum_{{z}_{N-1}} 
A({z}_1)
\exp\left( \frac{2\pi i}{n} \omega({z}_1+\check{z}_1,\check{z}_1-\check{z}_2)\right) \cdots  \right.  \notag \\
& \quad \left. \times  \,\,  A({z}_{N-1})
\exp\left(\frac{2\pi i}{n} \omega({z}_{N-1}+\check{z}_{N-1},\check{z}_{N-1}-\check{z}_N )\right) \right] \\
&= \delta(1 +\check{\lambda}_1)\delta(1 +\check{\lambda}_N)
\sum_{\{\check{z}\}} \sum_{\{{z}\}} 
\prod_{j=1}^{N-1} A({z}_j)
\exp\left(\frac{2\pi i}{n} \omega({z}_j+\check{z}_j,\check{z}_j-\check{z}_{j+1} )\right).
\end{align}
Here, we have grouped the sums into two sums over the paths defined as
\begin{align}
\{\check{z}\} &= (\check{z}_2,\check{z}_3, \ldots, \check{z}_{N-1} ) \\
\{{z}\} &= ( z_1, z_2, z_3, \ldots, z_{N-1} ).
\end{align}
This is now looking more like a ``path integral'', where we sum over a set of possible paths, except that there seem to be two different paths, $\{\check{z}\}$ and $\{{z}\}$.  

It turns out that this expression can be written in terms of only one path, $\{{z}\}$.   In the process of summing over $\{\check{z}\}$, however, the symplectic term in the phase becomes more complicated.  Careful algebra and a convenient change of variables simplify the resulting phase, but the details of the calculation are messy.  The variables $\{\check{z}\}$ provide a nice sort of bookkeeping, recording which of the ${z}$'s meet in the symplectic phase, and with which sign.  

We will first directly perform the summation over $\{\check{z}\}$, and then describe a new way to evaluate the resulting path integral by interpreting it as a sum over probability distributions of the points in phase space.  We then return to the expression above with the two sets of paths, and outline a way to evaluate the matrix $M_A^{N-1}$ in terms of a certain type of joint probability distribution of the points $\check z$ and $z$.  Because this approach uses two sets of paths, difficulties with the algebra in the direct calculation can be avoided.  This comes at the cost of having to introduce the joint probabilities (as will be discussed later), but these can be dealt with using standard tools from information theory.

\subsubsection{Direct summation over the paths $\{\check{z}\}$}

One way to proceed with the evaluation of the matrix $M_A^{N-1}$ is to sum over the $\{\check{z}\}$ paths.  The sums over each $\check{z}_j$ can be evaluated, starting with $j=N-1$.  This results in a delta function for $\check{z}_{N-2}$, which takes care of one more of the sums.  This procedure can be continued, unraveling, as it were, the symplectic phase starting at the end.   
\begin{align}
\sum_{\check{z}_{N-1}} \quad &\implies & \check{z}_{N-2} &= \check{z}_{N} + ({z}_{N-1} - {z}_{N-2}) \\
\sum_{\check{z}_{N-3}} \quad &\implies & \check{z}_{N-4} &= \check{z}_{N} + ({z}_{N-1} - {z}_{N-2}+{z}_{N-3} - {z}_{N-4}) \\
&\quad \vdots \notag \\
\sum_{\check{z}_{N-j}} \quad &\implies & \check{z}_{N-j-1} &= \check{z}_{N} + ({z}_{N-1} - {z}_{N-2}+\ldots +{z}_{N-j} - {z}_{N-j-1}) 
\end{align}
This pattern of grouping of the variables ${z}_{j}$ suggests a change of variables which may simplify the expressions obtained from this calculation.  Define new variables
\begin{align}
{z}'_{j} = \sum_{m=1}^{j} (-1)^{m-1} {z}_{N-m}, \quad j=1,\ldots, N-1.
\end{align}
The procedure used above, with summing over one variable giving a delta function for the next, is simplest to use if we assume that $N$ is odd.  This means that we will be left with one delta function at the end, which will relate ${z}'_{N-1}$ to the specified parameters $\check{z}_{1}$ and $\check{z}_{N}$.
Written in the new variables, the $(N-1)^{th}$ power of the matrix $M_A$ becomes
\begin{align}\label{eq:finite_path_integral}
&M_A^{N-1}(\check{z}_1,\check\lambda_1,\check{z}_N,\check\lambda_N) = \delta(1 +\check{\lambda}_1)\delta(1 +\check{\lambda}_N) 
 \exp\left(\frac{2\pi i}{n} \omega(\check{z}_N,\check{z}_{1})\right)  \notag \\
&\quad \times \sum_{\{{z}'\}}
\delta(\check{z}_{1}-\check{z}_{N}-{z}'_{N-1})
  \prod_{j=1}^{N-1} A\left((-1)^{j-1}({z}'_{j}-{z}'_{j-1})\right)  
 \exp\left(\frac{2\pi i}{n} \omega({z}'_j,{z}'_{j+1}) \right).
\end{align}
This form of the $(N-1)^{th}$ power of $M_A$ is valid when the symbol $A$ is delta-like in the $\lambda$ component [Equation (\ref{eq:symbol_phase_separation})].  If instead, we are considering the case in Equation (\ref{eq:symbol_for_irrep_reduction}), where the Fourier transform of the symbol is in the invariant subspace associated with an irreducible representation, then we get a different form for $M_A^{n-1}$.  Specifically, the $q'$ dependence of $A(z')$ will only be in the phase.  This means that the sums over $q'$ can be evaluated, leaving only the sums over $p'$.  These remaining sums can be grouped into paths in ``configuration'' space (actually ``momentum'' space, but we know these are just different choices of Lagrange subspace).  So the difference between the phase-space path integral and configuration-space path integral (a la Feynman) is that they arise from different reductions of the regular representation.  When considering the reduction to the primary representations, we are left with functions on phase space, and the repeated star product leads to the phase-space path integral.  On the other hand, when considering the reduction to irreducible representations, we must choose a Lagrange subspace in phase space.  The repeated star product then leads us to consider the path integral in the ``configuration'' space defined by the Lagrange plane.

\comment{
Note: this is slightly different than expected.  Continue with the expected form:
\begin{align}\label{eq:finite_path_integral}
M_A^{N-1}(\tau_1,\tau_N) 
&=\delta(1 +\check{\lambda})\delta(1 +\check{\lambda}_N) \sum_{\{{z }\}} 
\left(\prod_j A({z}_j) \right) \exp\left( \frac{2\pi i}{n} \sum_{j=2}^{N} 
\omega({z}_{j}, {z}_{j-1}  )
 \right) .
\end{align}
}

At this point, there are two ways to continue the calculation.  The first way (which is more appropriate for the continuous group) is to write this as a path integral in phase space.  This is the approach taken by previous authors, e.g., by Berezin and Shubin in \cite{Berezin:1991fk}, and we will use this approach in the following section when we discuss the path integral for the continuous Heisenberg-Weyl group.  A second, novel, way to proceed does not involve turning this sum into a path integral.  Instead, observe that for the discrete group we are working with, there are only finitely many values which the symplectic phase can take on.  Instead of summing over all possible paths $\{{z}\}$, we could group all the paths which give the same symplectic phase, and then sum over the possible values of the phase.  The tricky part of the calculation becomes trying to count all the ways to get the same phase.  Fortunately, maximum entropy arguments can be used to obtain expressions for the multiplicities of each phase.  Once these multiplicity functions are calculated --- which is work for the future --- the remaining sum becomes simpler.  This connection between path integrals and concepts from statistical mechanics is, we believe, new.

\subsection{From the path integral to the sum over measures}

We now proceed to calculating the repeated star product from a statistical point of view.  Start by considering the sum over two paths, as given in Equation (\ref{eq:double_path_sum}).  Instead of reducing this to a sum over only one path as was done in the previous section, we start directly with Equation  (\ref{eq:double_path_sum}).  The symplectic phase involves terms from both the path $\{\check z\}$ and from the path $\{ z \}$:
\begin{align}
\omega(z_j + \check z_j, \check z_j - \check z_{j+1} ) = \omega(z_j, \check z_j - \check z_{j+1} ) - \omega(\check z_j, \check z_{j+1} ).
\end{align}
Notice that, while the first terms involves points from both paths, the second involves terms from only one path.  We will consider this second term separately, before returning to the more complicated term which involves both paths.

Consider the term in the sum over paths in Equation (\ref{eq:double_path_sum}) which involves the symbol $A$.  It is simply the product of the values of $A$, evaluated along a path:
\begin{align}
\prod_j A(z_j) = A(z_1) A(z_2)\ldots A(z_j) \ldots A(z_{N-1}).
\end{align}
Each of the variables ${z}_j$ represents a point in the discrete phase space $\mathbb Z_n^{2}$, and therefore can take on only finitely many values.  We can think of the points in phase space as points in a lattice.  If we label the lattice points by $l$'s, then each of the variables ${z}_j$ will have some label as its value:
\begin{align}
{z}_j \in \{l_1, l_2,\ldots l_{n^2}  \}.
\end{align}
We can also consider the path $\{z\}$ formed by the variables $z_j$:
\begin{align}
\{z\} = (z_1, z_2, \ldots, z_j, \ldots z_N).
\end{align}
If we think of the index $j$ as a sort of discrete ``time'', then the path is a mapping from time to the points in phase space.
\begin{align}
z : \{2,3,\ldots, N \} \rightarrow \{l_1, l_2, \ldots, l_{n^2} \}
\end{align}
Similarly, a complex-valued function on phase space is simply a list of complex numbers, one for each point $l$ in phase space. 
\begin{align}
A: \{ l \} \rightarrow \mathbb{C} \\
A(l) = A_l
\end{align}
The phase space point label $l$ is used as a subscript for convenience.  When written this way, the discrete nature of phase space is emphasized, and $A$ can be thought of as a length $n^2$ vector, since there are $n^2$ points in phase space.  Now specialize to the case where $A$ is a term from the limit formula for the exponential [Equation (\ref{eq:operator_exponent})]:
\begin{align}
A(z) = \left(1+\frac{it}{N} H(z) \right)  \approx e^{it H(z)/N }. 
\end{align}
We can now rewrite the product over the path as a weighted sum over the points of phase space:
\begin{align}
\prod_{j=1}^{N-1} A({z}_j) &= A({z}_1) A({z}_2) \ldots A({z}_{N-1}) = e^{it N_1 H_{l_1}/N} e^{it N_2 H_{l_2}/N} \ldots e^{it N_{n^2} H_{l_{n^2} }/N}  \\
& =\exp\!\left( \frac{it}{N} \sum_{l \in \mathbb Z_n^{2}} N_l H_l \right) \label{eq:symbol_of_H_term}.
\end{align}
The incidence function $N_l$ simply counts how many times the variables ${z}_j$ take on the value $l$.  Another way to think of $N_l$ is in terms of the path $\{{z}\}$.  In this context, $N_l$ is the number of times that someone following the path $\{{z}\}$ would visit the location $l$, where $H(z)$ takes the value $H_l$.  Since the incidence function counts steps along a path, the sum over all values of $l$ must be equal to the length of the path:
\begin{align}
\sum_l N_l  = N .
\end{align}
Once the path is specified, the incidence $N_l$ is completely determined.  This change in the point of view allows us to switch the product of $H$'s accumulated along the path for a weighted sum over phase space.  Since the incidence $N_l$ is the weight in the sum, we can see that it is playing the role of a measure on phase space.  The product on the given path has been turned into a sum with a given measure.  With proper normalization, this measure can be treated as a probability distribution on phase space.

The phase term in the sum over paths can be manipulated in a similar way.  Start by considering only the term which involves the pair of points $(\check{z}_j,\check{z}_{j+1})$.  
\begin{align}
T(\{{\check z}\}) = \exp\!\left( - \frac{2\pi i}{n} \sum_{j=1}^{N-1} 
\omega(\check z_j, \check z_{j+1} ) 
 \right) .
\end{align}
Since the symplectic product takes two points as input, the phase has a matrix of possible values.
\begin{align}
[{\bm \omega}]_{l,m} = \omega(l,m), \quad 0\leq l,m < n^2
\end{align}
We can now rewrite the phase, by converting the sum along the path into a sum over possible values of $\omega$.  Now instead of just counting how many times each point appears in the path, we need to count how many times each possible pair of points appears as neighbors in the path.  This gives us a coincidence table, which we can write as a matrix:
\begin{align}
\check{\bf N}_{l,m} = \text{\# times the pair } (l,m) \text{ appears in the path}.
\end{align}
For a path of length $N$, there are $N-1$ pairs of points.  This implies that the sum of all the elements in the coincidence table is
\begin{align}
\sum_{l,m} \check{\bf N}_{l,m} = N-1.
\end{align}
Also, this coincidence table is derived from a path $\{\check z\}$, which has a particular incidence function $\check N_l$.  The coincidence table must be consistent with $\check N_l$, and this consistency constraint can be written in terms of the column and row sums of the table $\check{\bf N}$:
\begin{align}
\check N_l = \sum_m  \check{\bf N}_{l,m} = \sum_m  \check{\bf N}_{m,l} .
\end{align}
With the matrices $\check{\bf N}$ and ${\bm \omega}$ defined, the phase can now be written as a contraction of the matrices.
\begin{align}
\sum_{j=1}^{N-1} \omega(\check{z}_j, \check{z}_{j+1}) = \sum_{l,m} \check{\bf N}_{l,m}{\bm \omega}_{l,m}
\equiv \check{\bf N}:{\bm \omega}
\end{align}
In order to simplify notation, the symbol ``$:$'' is used to denote the contraction of the matrices.  Note that the matrix ${\bm \omega}$ is antisymmetric, ${\bm \omega}^{\rm T} = -{\bm \omega}$.  That means that this contraction looks like an inner product:
\begin{align}
\check{\bf N}:{\bm \omega} = - \text{Tr}(\check{\bf N} {\bm \omega}^{\rm T}) = ({\bm \omega}, \check{\bf N}).
\end{align}

We can now rewrite $T(\{\check{z}\})$ in terms of the coincidence table $ \check{\bf N}$, which is calculated for the particular path $\{\check{z}\}$.  
\begin{align}\label{eq:simple_symp_terms}
T(\{\check{z}\}) =  \exp\!\left(  \frac{2\pi i}{n}  \check{\bf N}:{\bm \omega}    \right)
\end{align}
This simplification of $T$ will help give us a way to organize the paths in the sum over all paths in Equation (\ref{eq:double_path_sum}), by grouping paths with the same coincidence table $ \check{\bf N}$.

\subsubsection{The double sum over paths and joint coincidence tables}

We now turn to the calculation of the phase which involves both paths, $\{\check z\}$ and $\{ z\}$.  The phase which involved only one path was written this in terms of the two-point coincidence table $\check{\bf N}_{l,m}$, which counted how many times a particular pair of points appeared in one path.  Now we have to count how many times the triple $z_j, \check z_l, \check z_m$ appears in  the two paths.  Introduce a triple-coincidence table, ${\bf N}_{l,m_1,m_2}$, which counts how many times $z_l$ appears along with the pair $\check z_{m_1}, \check z_{m_2}$.  This triple-coincidence table must agree with the coincidence table $\check{\bf N}_{m_1,m_2}$ for the path $\{\check z\}$ and the incidence function $N_l$ for the path $\{ z \}$:
\begin{align}
\sum_l {\bf N}_{l,m_1,m_2} = \check{\bf N}_{l,m}, 
\quad \text{and } \quad  \sum_{m_1,m_2} {\bf N}_{l,m_1,m_2} = N_l .
\end{align}
In order to compute the phase from this triple-coincidence table, we need to be able to combine it with the matrix $\bm \omega$ properly.  The two terms in the symplectic product can be calculated from two different contractions of $\bm \omega$ with ${\bf N}_{l,m_1,m_2}$.  These contractions are
\begin{align}
\omega(z_j, \check z_j - \check z_{j+1} ) \rightarrow  \sum_{l,m} {\bm \omega}_{l,m} \sum_{m_1} {\bf N}_{l,m_1,m - m_1} = {\bm \omega}:\tilde{\bf N}
\end{align}
and
\begin{align}
\omega(\check z_j, \check z_{j+1} ) \rightarrow \sum_{l,m} {\bm \omega}_{l,m} \sum_l {\bf N}_{l,m_1,m_2} = {\bm \omega}:\check{\bf N}.
\end{align}
Here we have defined a new coincidence table $\tilde{\bf N}$ as the sum
\begin{align}
\tilde{\bf N}_{l,m} = \sum_{m_1} {\bf N}_{l,m_1,m - m_1}.
\end{align}
Note that since this is not the coincidence table for only one path, it does not need to satisfy the same marginal sums as the ordinary coincidence table for one path.  Specifically, the row and column sums give incidence functions for the two different paths:
\begin{align}
N_l = \sum_m \tilde{\bf N}_{l,m},
\end{align}
and \comment{Tracy conjecture: max entropy -> \tilde N is the outer product of N_l and \check N_m. }
\begin{align}
\check N_m = \sum_l \tilde{\bf N}_{l,m}.
\end{align}

With these definitions, and the expressions in Equations (\ref{eq:symbol_of_H_term}) and (\ref{eq:simple_symp_terms}), we can rewrite Equation (\ref{eq:double_path_sum}) as
\begin{align}\label{eq:discrete_entropy_sum}
M_H^{N-1}(\tau_1,\tau_N) &= \delta(1 +\check{\lambda}_1)\delta(1 +\check{\lambda}_N)
\sum_{N_l} \exp\!\left( \frac{it}{N}\sum N_l H_l  \right)  \notag \\
&\quad \times \sum_{\check N_m} \left( \widetilde{\sum_{\bf N}} 
\mathcal{M}({\bf N})
\exp\left(\frac{2\pi i}{n} \left(  {\bm \omega}:\tilde{\bf N} - {\bm \omega}:\check{\bf N}  \right)   \right)  \right).
\end{align}
The tilde on the inner sum indicates that it is a sum over all possible triple-coincidence tables $\bf N$ which agree with the incidence functions $N_l$ and $\check N_m$ for the two paths.  We have introduced the multiplicity function $\mathcal{M}({\bf N})$ which counts the number of triple-coincidence tables which give the same phase.  This multiplicity function is necessary since there are generically many pairs of paths $\{ z\}$ and $\{\check z\}$ which give the same $\bf N$.  Since each triple-coincidence table can only be consistent with one pair of incidence functions $N_l$ and $\check N_m$, the sum over $N_l$ and $\check N_m$ together with the weighted sum over ${\bf N}$ gives the same result as the sum over all paths $\{{z}\}$ and $\{\check{z}\}$.

It is at this stage that we are ready to use some tools from statistical mechanics.  We would like to use a maximum entropy calculation in order to evaluate the sum over coincidence tables.  This can be done, and will give an expression for the sum in the limit of longer and longer paths.  As the length of the path goes to infinity, we can use, for example, the generating function for digram frequencies given in Chapter 22 of Jaynes \cite{Jaynes:2003lr}.  This generating function can be used to calculate the probability of a pair of paths given that the path have a particular triple-coincidence table, $p(\{{z}\},\{\check{z}\} | {\bf N}$).  From this probability distribution in ``path space'' we can calculate the multiplicity function $\mathcal{M}({\bf N})$.  We are most interested in finding those pairs of paths associated with the maximum of $\mathcal{M}({\bf N})$.  Such calculation is the subject of ongoing research, and is just one of the many new research directions based on the group theory point of view developed in this dissertation.

\section{The Path Integral for the Continuous Heisenberg-Weyl Group}

The phase space path integral arises when the symbol of the exponential of an operator is computed in the context of the Heisenberg-Weyl group \cite{Berezin:1991fk}.  As with the calculation for the discrete Heisenberg-Weyl group in the previous section, we start by computing the star product of a symbol with itself, repeated $N$ times, and then apply this to Equation (\ref{eq:operator_exponent}) to find the exponent.

In this section, the following notation will be used.  An operator $\hat A$ will be embedded into a section $S_A$.  Then the group Fourier transforms will be applied, giving the symbol of $S_A$ and the commutative Fourier transform of the symbol.
\begin{equation}
S_A(\rho) \stackrel{\mathcal{F}_{\mathfrak H}^{-1}}{\longrightarrow}
\tilde a (g) \stackrel{\mathcal{F}_{\mathbb{R}^{2n+1} }}{\longrightarrow} a(\tau)
\end{equation}
Here $a(\tau)$ is the symbol of the section $S_A(\rho)$, and the function on the group, $\tilde a(g)$, is the (commutative) inverse Fourier transform of $a(\tau)$ obtained at the intermediate step in the calculation.

We now proceed to the $N$-fold star product calculation.  This star product is obtained by finding the symbol of an $N$-fold product of operators.
\begin{align}
(a_1 \star a_2 \star \ldots \star a_N)(\tau)&=[Q^{-1}( S_{A_1} S_{A_2} \ldots S_{A_N})](\tau) \\
&= [\mathcal{F}_{\mathbb{R}^{2n+1}} \circ \mathcal{F}_{\mathfrak H}^{-1}( S_{A_1} S_{A_2} \ldots S_{A_N})](\tau)
\end{align}
We start by using the convolution theorem to calculate the inverse transform $\mathcal{F}_{\mathfrak H}^{-1}$.
\begin{equation}
\big(\mathcal{F}_{\mathfrak H}^{-1}( S_{A_1} S_{A_2} \ldots S_{A_N}) \big)(g) = 
\idotsint_{\mathfrak{H}^N} a_1(g_1) a_2(g_2) \ldots a_N(g_N) \delta_g(g_1\diamond g_2\diamond \ldots \diamond g_N) \prod_{j=1}^N dg_j
\end{equation}
This form of the convolution is nice, since it isolates $g$ and provides a delta function which is useful for evaluating the next Fourier transform integral.
\begin{align}
[Q^{-1}(S_{A_1} \ldots S_{A_N})](\tau) &= \int_{\mathbb{R}^{2n+1}} dg\, \tau(g)
\idotsint_{\mathfrak{H}^N} a_1(g_1)  \ldots a_N(g_N) \delta_g(g_1\diamond \ldots g_N) \prod_{j=1}^N dg_j \\
&= \idotsint_{\mathfrak{H}^N} a_1(g_1)  \ldots a_N(g_N) \tau(g_1\diamond \ldots \diamond g_N) \prod_{j=1}^N dg_j \label{eq:repeated_star_integral}
\end{align}
We can now use our knowledge of the groups $\mathfrak{H}$ and $\mathbb{R}^{2n+1}$ to rewrite the term involving $\tau$.  As a set, these two groups are the same, and can be separated into a $2n$-dimensional phase space component, $\mathbf z$, and an additional 1-dimensional component, $\lambda$.  The dual $\mathbb{R}^{*(2n+1)}$ can also be separated into components.  Use the $\check{~}$ notation to signify an element of the dual.  I.e., write an irreducible representation of $\mathbb{R}^{2n+1}$ as
\begin{equation}
\tau(g) = e^{i(\check{\bf z}\cdot {\bf z} + \check{\lambda} \lambda )}.
\end{equation}
With this notation, and the form of the group product law for the Heisenberg-Weyl group, we can now write the $\tau$ term from the above integral as
\begin{align}
\tau(g_1\diamond \ldots \diamond g_N) &= \exp\left\{ i\left( 
\check{\bf z}\cdot ({\bf z}_1 + \ldots  {\bf z}_N) 
+ \check{\lambda}( \lambda_1 + \ldots  \lambda_N + \frac{1}{2}\Omega({\bf z}_1, \ldots, {\bf z}_N)) \right)\right\} \\
&= e^{i \check{\lambda}( \lambda_1 + \ldots + \lambda_N)}
\exp\left\{ i\left( 
\check{\bf z}\cdot ({\bf z}_1 + \ldots  {\bf z}_N) 
+  \frac{\check{\lambda}}{2} \Omega({\bf z}_1, \ldots, {\bf z}_N)
\right)\right\} \label{eq:tau_rewrite}
\end{align}
The symplectic products from $g_1\diamond \ldots \diamond g_N$ have been grouped together into one term, which can be directly calculated, and is
\begin{equation}
\Omega({\bf z}_1, {\bf z}_2, \ldots, {\bf z}_N) = \omega({\bf z}_1, {\bf z}_2) + 
\omega({\bf z}_1+ {\bf z}_2,  {\bf z}_3) + \ldots + 
\omega({\bf z}_1+ {\bf z}_2 + \ldots+ {\bf z}_{N-1},  {\bf z}_N).
\end{equation}
From Equation (\ref{eq:tau_rewrite}), it looks like the variables $\lambda_j$ are separating out nicely.  We might therefore try to integrate over these variables in Equation (\ref{eq:repeated_star_integral}).  We can do this if we assume that the functions $a_j(g_j)$ are separable.  In a slight abuse of notation, write these functions as products.
\begin{equation}
a_j(g_j) = b_j(\lambda_j) a_j({\bf z}_j)
\end{equation}
This form is motivated by calculation in Section \ref{sec:weyl-zobin}, where it is shown that the Weyl symbol is a special case of the Zobin symbol, which arises when the function $a(g)$ satisfies exactly this sort of separability condition.  (Notice that this condition is similar to the covariance condition given in Equation (\ref{eq:covariant_functions}), and used to decompose the regular representation.)

Putting all of this together, and plugging it into Equation (\ref{eq:repeated_star_integral}), gives us a new expression.
\begin{align}
[Q^{-1}(S_{A_1} &\ldots S_{A_N})](\check{\bf z},\check \lambda) \notag \\
&= \idotsint \left( \prod_j a_j({\bf z}_j) \right) 
\exp\left\{ i\left( 
\check{\bf z}\cdot \sum_j {\bf z}_j  
+  \frac{\check{\lambda}}{2} \Omega({\bf z}_1, \ldots, {\bf z}_N)
\right)\right\}   \prod_{j=1}^N d{\bf z}_j  \notag \\
&\quad \times  \idotsint e^{i \check{\lambda}( \lambda_1 + \ldots + \lambda_N)} b_1(\lambda_1) \ldots  b_N(\lambda_N)    \prod_{j=1}^N d\lambda_j
\end{align}
The integrals over the $\lambda_j$'s are ordinary Fourier transforms.  The difficult part of the calculation arises when trying to simplify the ${\bf z}_j$ integrals.  Berezin and Shubin \cite{Berezin:1991fk} evaluate an integral such as this and obtain the phase space path integral, while Ozorio de Almeida \cite{dealmeida-1992} interprets the phase $\Omega$ in terms of the area of polygons in phase space.  These approaches are similar to the calculation for the discrete group, which was done in the previous section.  The details of one approach (due to N. Zobin) are given in Appendix \ref{app:path_int}, and result in the expression (NB the notation has been changed so it matches the notation used in this section):
\begin{align}
&\left[Q^{-1}\left(1+ \frac{itS_H}{N}\right)^N\right](\check{\bf z}_j,\check\lambda_j)  \notag \\
&\quad = \int\limits_{\widehat G_0^N}\! 
 \exp\!\left(- \frac{i\check \lambda}{2} \sum_{k=2}^{N+1} \omega(\check {\bf z}_{k+1}-\check {\bf z}_{k}, \check {\bf z}_{k} - \check {\bf z}_{k-1}) 
+ \frac{it}{N} \sum_{j=1}^N H(\check{\bf z}_j,\check\lambda_j)  \right) \prod_{j=1}^N d\check{\bf z}_j
\prod_{j=1}^N d\check\lambda_j .
\end{align}
In the limit $N\rightarrow\infty$, the sums in the phase go over to integrals, and the integrals over $\widehat G_0$ become a path integral.

Alternatively, we could rewrite this path integral in terms of measures on phase space, just as the discrete ``path integral'' was rewritten in the previous section.  This idea will be explored further in the next section.

\section{The path integral as a Fourier transform on the space of measures}

The group theoretical perspective that has been developed in this section leads naturally to an interpretation of the path integral as a Fourier transform on the space of measures.  In order to motivate this interpretation, we will start by examining the formula for the discrete ``path integral'' given in Equation (\ref{eq:discrete_entropy_sum}).  Consider the phase space portion of that formula:
\begin{align}
\tilde M_H^{N-1}(\check{z}_1, \check{z}_N) = \sum_{N_l} \exp\!\left( \frac{it}{N} \sum_l N_l H_l \right) \left( \widetilde{\sum_{\bf N}} \mathcal{M}({\bf N}) \exp\!\left( \frac{2\pi i}{n} {\bf N}:{\bm \omega} \right)  \right).
\end{align}
Recall that the incidence function $N_l$ counted how many times a path of length $N$ (with endpoints $\check z_1$ and $\check z_N$) visited each point $l$ in phase space.  We then converted the sum over all such paths into a sum over all possible incidence functions $N_l$.  Now consider the limit $N\rightarrow \infty$.  In this limit, the endpoints of the path should not contribute to the sum, so let them be anything.  Also, the term $N_l/N$ becomes a probability density $p_l$, and the sum over $p_l$ goes over to an integral:
\begin{align}\label{eq:sum_p_l}
M_H = \int  \exp\!\left( it \sum_l p_l H_l \right) \, d\mu(p_l).
\end{align}
Here we have defined the measure $d\mu(p_l)$ as
\begin{align}
d\mu(p_l) = \lim_{N\rightarrow \infty} \frac{1}{(n^2)^N} \widetilde{\sum_{\bf N}} \mathcal{M}({\bf N}) \exp\!\left( \frac{2\pi i}{n} {\bf N}:{\bm \omega} \right)  .
\end{align}
This measure depends on $p_l$ through the constraint on the sum over coincidence tables $\bf N$.   The normalization factor $\frac{1}{(n^2)^N}$ has been introduced in order to keep $d\mu(p_l)$ finite.  The multiplicity $\mathcal{M}({\bf N})$ is the number of paths with coincidence table ${\bf N}$, and $(n^2)^N$ is the total number of paths of length $N$.  The ratio then looks like a probability on the space of probabilities $p_l$:
\begin{align}
p(p_l)  = \lim_{N\rightarrow \infty} \widetilde{\sum_{\bf N}} \frac{\mathcal{M}({\bf N}) }{(n^2)^N}.
\end{align}
This probability is nearly the measure $d\mu(p_l)$, except that it does not include the phase.  We are now set up to interpret $M_H$ as a Fourier transform.

First, notice that the phase in $M_H$ involves two functions of the discrete phase space.  There are $n^2$ points in phase space, so $H_l$ and $p_l$ can be thought of as vectors of length $n^2$, i.e., they are points in an $n^2$-dimensional space.  The function $\Phi(p_l)$ is also a function on $\mathbb R^{n^2}$:
\begin{align}
\Phi : \mathbb R^{n^2} \rightarrow \mathbb C .
\end{align}
The (ordinary) Fourier transform of such a function is usually written
\begin{align}
\tilde \Phi (k_l) = \int e^{i \sum_l k_l p_l} \Phi(p_l) \,d^{n^2}p .
\end{align}
Comparing this to Equation (\ref{eq:sum_p_l}), shows that $M_H$ is actually the Fourier transform of $\Phi(p_l)$, evaluated at the point $t H_l$.  Notice that this is a Fourier transform with respect to the variable $p_l$, which can also be thought of as probability density, or measure, on the discrete phase space.  So the path integral can be viewed as a Fourier transform over the space of measures.

This interpretation of the discrete path integral as a Fourier transform over the space of measures extends to the continuous path integral.  This can be seen by letting the number of points $n^2$ in the discrete phase space become large.  The hamiltonian function $H_l$ will go over to the hamiltonian on the continuous phase space, $H(z)$, and the probability $p_l$ will go over to a probability density, or measure, on phase space.  The dot product will then become an integral:
\begin{align}
\sum_l p_l H_l \rightarrow \int H(z) p(z)dz \equiv \langle H(z) , p(z) \rangle.
\end{align}
This lets us write the path integral as an infinite-dimensional Fourier transform.
\begin{align}
M[H(z)] = \int e^{ i  \langle t H(z) , p(z) \rangle } \Phi[p(z)] \, d\mu[p(z)].
\end{align}
This can either be viewed as an alternative interpretation of the path integral, or as the definition of the infinite-dimensional Fourier transform.  This insight is due to Zobin, and is a new way to understand the path integral.

\section{Summary}

In this chapter, the group theoretical foundations of path integrals were described.  Starting with the theory of symbols presented in Chapter \ref{chp:Symbols}, we considered the symbol of a function of an operator.  In particular, the symbol of the exponential of an operator was calculated using the power series definition for the exponential function.
This results in an expression which requires the repeated application of the star product for the calculation of the $N^{th}$ order term in the series.  In the limit of large $N$, the star products in this series can be recast as a path integral.  While this star-product approach to the path integral has been considered previously \cite{Berezin:1991fk,citeulike:712823,dealmeida-1992}, we have shown that this approach is very general, as it is based on the non-commutative Fourier transform for an (almost) arbitrary group.  

As an example of the flexibility of this approach, we show how to calculate the ``path-integral'' for the finite Heisenberg-Weyl group.  The resulting sum over paths looks quite different than the ordinary path integral, since there are no continuous ``paths'' on finite groups.  By examining this finite example, we were able to show how the path integral can be connected to ideas from statistical mechanics.  The sum over all paths was converted into a sum over probability distributions in phase space.  This opens up several avenues of future research, including the possibility of using maximum entropy arguments to evaluate the sum over distributions.  Bringing these ideas back to the continuous path integral leads to the formulation of an ``infinite-dimensional'' Fourier transform; the Fourier transform over the space of measures.

This group theoretical framework for understanding the path integral has also shed new light on the connection between the phase-space path integral and the configuration-space path integral.  The generic path integral which arises from this theory relies heavily on the Fourier transform for groups, in both its commutative and non-commutative versions.  The Fourier transform itself is related to the reduction of the regular representation into irreducible representations, so it is natural to consider the effect of such reductions on the path integral.  We show how the phase-space path integral arises when considering the reduction to the primary representation, since (for the Heisenberg-Weyl group) the primary representation acts on functions on phase space.  The further reduction to the irreducible representations requires choosing a subspace of phase space as the ``configuration space''.  This choice leads to a reduction of the path integral to an integral over paths in configuration space.  Thus, by recasting the symbol of an operator as a double Fourier transform for the Heisenberg-Weyl group, we have provided a new group theoretical framework for phase-space and configuration-space path integrals.

%% file: Chapter-ModeConversion.tex
%%%%%%%%%%%%%%%%%%%%%%%%%%%%%%%%%%%%%%%%%%%%%%%%%%%%%%%%%%%%%%%%%%%%%%%%%%
%
% Ph.D. dissertation manuscript
% Chapter 8: Applications
%
% Andrew Stephen Richardson (Fall 2007)
% College of William and Mary
% Department of Physics
% Prof. Eugene Tracy, advisor
%
% Based on Paul King and Andrew Norman's template (modified by Wirawan Purwanto)
%
%%%%%%%%%%%%%%%%%%%%%%%%%%%%%%%%%%%%%%%%%%%%%%%%%%%%%%%%%%%%%%%%%%%%%%%%%%

\chapter{A Brief Survey of Connections to Mode Conversion Theory}
\label{chp:ModeConversion}

This chapter presents a cursory discussion of several areas of investigations suggested by the results of the previous chapters.  This chapter is not meant to be a final report on finished research, but rather it is meant to convey the wealth of new questions and ideas suggested by this work.  It is our hope that the ideas presented in this chapter convey the richness of this field, and the variety of the avenues of research which are now open for further study.  

There are three main ideas presented in this chapter.  First, we will see that it might be useful to use the diagonals of the dispersion matrix as ray hamiltonians when using WKB methods for vector problems, rather than using the eigenvalues as is typically done.  This requires, however, that the matrix operator be put into ``normal form''.  The second idea is that it may be possible to combine the discrete and continuous Heisenberg-Weyl groups to form a ``double symbol'' for the operator-valued dispersion matrix which arises in vector wave problems.  The third idea presented in this chapter relates to using the Wigner function to include the effects of turbulent fluctuations on mode conversions, by marginalizing the Wigner function over a system parameter whose value is uncertain.

\section{Review of Vector WKB}

All of these ideas presented in this chapter relate to vector wave problems.  Recall that such a problem can be written in matrix form:
\begin{align}\label{eq:vector_wave_equation}
\hat{\bf D} \cdot {\bm \psi}(x)=
\left(
\begin{array}{cc}
 \hat D_{11}  &  \hat D_{12}  \\
\hat D_{21}   &   \hat D_{22}
\end{array}
\right)
\left(
\begin{array}{c}
  \psi_1(x)   \\
  \psi_2(x) 
\end{array}
\right) =0.
\end{align}
This equation is written as a matrix of operators acting on the field ${\bm \psi}(x)$.  If we want to use the mathematics of symbols to write this operator as a function in phase space, we would usually calculate the symbol of each element of the matrix separately.  This defines the dispersion matrix, which is a matrix-valued function on phase space:
\begin{align}
{\bf D}(x,k) = \left(
\begin{array}{cc}
 D_{11}(x,k)  &  D_{12}(x,k)  \\
 D_{21}(x,k)   &  D_{22}(x,k)
\end{array}
\right) .
\end{align}
This matrix is the starting point for the construction of the ray-tracing approximation.  First we recall how ray-tracing methods can be derived from the phase space path integral, in the case of scalar wave equations.

For a scalar wave equation, the path integral is:
\begin{align}
\int e^{i\int_0^T \left[\frac{1}{2} (q(t)\dot p(t) - p(t)\dot q(t))   - H(q(t),p(t)) \right]\, dt} \, Dq(t) \,Dp(t).
\end{align}
Hamilton's equations for the rays can be derived from this path integral by evaluating it using the stationary phase approximation.  Write the phase in this integral as
\begin{align}
S[{\bf z}(t)] = \int_0^T \left( \frac{1}{2} {\bf z}\cdot {\bf J}\cdot \dot{\bf z} - H({\bf z}) \right) dt.
\end{align}
Variation of this with respect to the path, ${\bf z}(t)\rightarrow {\bf z}(t) +\epsilon {\bm \zeta}(t)$ gives, 
\begin{align}
\frac{\delta}{\delta {\bf z}} S[{\bf z}(t)] =
 \int _0^T \left( \frac{1}{2} {\bf z}\cdot {\bf J}\cdot \dot{\bm \zeta} + \frac{1}{2} {\bm \zeta}\cdot {\bf J}\cdot \dot{\bf z} - {\bm \zeta}\cdot \nabla H({\bf z}) \right) dt.
\end{align}
Use integration by parts to move the time derivative from ${\bm \zeta}$ to ${\bf z}$:
\begin{align}
\frac{\delta}{\delta {\bf z}} S[{\bf z}(t)] &=
 \int _0^T \left( -\frac{1}{2} \dot{\bf z}\cdot {\bf J}\cdot {\bm \zeta} + \frac{1}{2} {\bm \zeta}\cdot {\bf J}\cdot \dot{\bf z} - {\bm \zeta}\cdot \nabla H({\bf z}) \right) dt \\
 &=
 \int _0^T \left( {\bm \zeta}\cdot {\bf J}\cdot \dot{\bf z} - {\bm \zeta}\cdot \nabla H({\bf z}) \right) dt .
\end{align}
For this to equal zero for any variation, we must have
\begin{align}\label{eq:review_hamiltons_equations}
{\bf J}\cdot \dot{\bf z} - \nabla H({\bf z}) = 0,
\end{align}
which can be rearranged to obtain Hamilton's equations:
\begin{align}
\dot q &= \frac{\partial H}{\partial p} \\
\dot p &= -\frac{\partial H}{\partial q}.
\end{align}
These equations can be solved to find the rays, which are then used to construct approximate solutions.  The function $H(q,p)$ here is the symbol of the hamiltonian operator.  When the wave equation to be solved is a vector wave equation, instead of a scalar equation, the symbol of the dispersion operator $\hat{\bf D}$ is no longer a scalar function, and cannot be directly used to generate rays.

The usual way to get around this is to use one of the eigenvalues of the dispersion matrix as the ray hamiltonian.  Write the eigenvalues as functions of phase space: $D_\alpha(x,k), \alpha = 1,2$.  From Equation (\ref{eq:vector_wave_equation}), we can see that in order for a solution to be non-zero, it must be a zero eigenvector of the dispersion matrix.  Using an eigenvalue $D_\alpha$ as the ray hamiltonian ensures that the numerical value of $D_\alpha$ will remain constant along a ray.  So if the ray starts with initial conditions $(x_0,k_0)$ such that $D_\alpha(x_0,k_0)=0$, then $D_\alpha$ will remain zero along a ray.  The WKB method can then be used to construct an approximate solution by tracing out a family of such rays.

\section{Diagonals as ray Hamiltonians for vector problems}

As described above, the derivation of the ray-tracing equations works fine for problems with scalar Hamiltonian operators, such as we have in the scalar Schr\"odinger equation.  For vector problems, we need to exponentiate the matrix of operators, $\hat{\bf D}$.  In the derivation of the path integral from the theory of symbols, we considered the limit expression for the exponent,
\begin{align}
e^{it \hat H} = \lim_{N\rightarrow \infty} \left( 1+ \frac{it\hat H}{N}\right)^N .
\end{align}
For the matrix of operators $\hat{\bf D}$, we might consider a similar expression
\begin{align}
e^{it \hat{\bf D}} = \lim_{N\rightarrow \infty} \left( {\bf 1}+ \frac{it\hat{\bf D}}{N}\right)^N,
\end{align}
where ${\bf 1}$ is the $2\times 2$ identity matrix.  It is this formula which suggests a possible new approach for a ray-tracing algorithm.  

\comment{%
Consider the first term on the diagonal of the matrix above.  It will contain terms which have products of the elements of the matrix $\hat{\bf D}$.  For example, with $N=3$, the upper left diagonal element will contain the terms 
\begin{align}
\hat D_{11}^3 + \hat D_{11} \hat D_{12} \hat D_{21} + \hat D_{12}\hat D_{21}\hat D_{11} + \hat D_{12}\hat D_{22}\hat D_{21} .
\end{align}
This expression could be simplified if the diagonal terms commute with the off diagonal terms:
\begin{align}
\hat D_{11}^3 + \hat D_{11} (\hat D_{12} \hat D_{21} + \hat D_{12}\hat D_{21}) + \hat D_{22}\hat D_{12}\hat D_{21} .
\end{align}
}%

Consider the symbol of this expression for the case where $N=2$.  This can be written using the star product:
\begin{align}
\left({\bf 1} + \frac{it {\bf D}(x,k)}{2} \right) \star \left({\bf 1} + \frac{it {\bf D}(x,k)}{2} \right).
\end{align}
This star product will look like the ordinary matrix product, plus terms involving Moyal corrections whose leading terms involve the Poisson bracket of the elements of the matrix.  If the diagonal elements commute with the off-diagonals (e.g., $\{ D_{11}, D_{12} \}=0$), then the Moyal correction terms get pushed to one higher order.  For example, the upper right off-diagonal term of this product, with Moyal corrections, can be written (using $\epsilon$ as an ordering parameter) as:
\begin{align}
&\left[\left({\bf 1} + \frac{it {\bf D}}{2} \right) \star \left({\bf 1} + \frac{it {\bf D}}{2} \right)\right]_{12}  \notag \\
&= \frac{it D_{12}}{2} + \frac{it D_{21}}{2} + \left(\frac{it}{2}\right)^2  D_{11}\star D_{12} + \left(\frac{it}{2}\right)^2  D_{12}\star D_{22} \\
& = \frac{it}{2}(D_{12} + D_{21}) + \left(\frac{it}{2}\right)^2 \left[  D_{11}D_{12} + D_{12}D_{22}
+\epsilon\left( \{ D_{11}, D_{12} \} + \{ D_{12}, D_{22} \}     \right) + \mathcal{O}(\epsilon^2) \right] \\
& = \frac{it}{2}(D_{12} + D_{21}) + \left(\frac{it}{2}\right)^2 \left[  D_{11}D_{12} + D_{12}D_{22}
+ \mathcal{O}(\epsilon^2) \right]
\end{align}

Additionally, as we know from the work on mode conversion problems, a nice polarization to choose is one where the diagonal terms represent the ``uncoupled'' wave modes, and the off diagonal terms represent a locally constant coupling.  

These observations suggest that it might be useful to choose the local polarization basis along the ray to be one where the the diagonal elements commute with the off diagonal elements.  In terms of the symbols of the matrix elements, this means that the symbols of the diagonals would Poisson commute with the symbols of the off-diagonals:
\begin{align}
\{ D_{11}(x,k), D_{12}(x,k) \} =\{ D_{11}(x,k), D_{21}(x,k) \} =0
\end{align}
and 
\begin{align}
\{ D_{22}(x,k), D_{12}(x,k) \} =\{ D_{22}(x,k), D_{21}(x,k) \} =0.
\end{align}
This defines a type of ``normal form'' for the dispersion matrix \cite{Tracy:2007yq}.  In this normal form, we want to interpret the diagonals as uncoupled modes, so we will propose that the diagonals be used as the ray hamiltonians for the rays of the two different modes.  We would then update the polarization along the ray so as to keep the matrix in normal form.  With the diagonals as the ray hamiltonians, the off-diagonals would remain constant along the ray:
\begin{align}
\frac{d}{d\sigma} D_{12} = \{ D_{11}, D_{12} \} = 0.
\end{align}

This proposed normal form differs from the approach of Emmrich and Weinstein \cite{Emmrich:1996lr}, where the full Moyal corrections are treated order by order, but the expansion is taken about a {\em fixed} (but arbitrary) point in phase space.

This normal form approach may be especially useful for solving mode conversion problems.  Instead of the rays forming an avoided crossing as they do when using the eigenvalues as ray hamiltonians, we would instead find the mode conversion point where the two dispersion surfaces intersect.  A mode conversion point (or mode conversion surface in higher dimensions) is then a solution to 
\begin{align}
D^{NF}_{11}(x,k) = D^{NF}_{22}(x,k) = 0,
\end{align}
where the superscript is a reminder that the matrix must be put into normal form.

\begin{figure}
\begin{center}
\includegraphics[width=2.2in,height=2.2in]{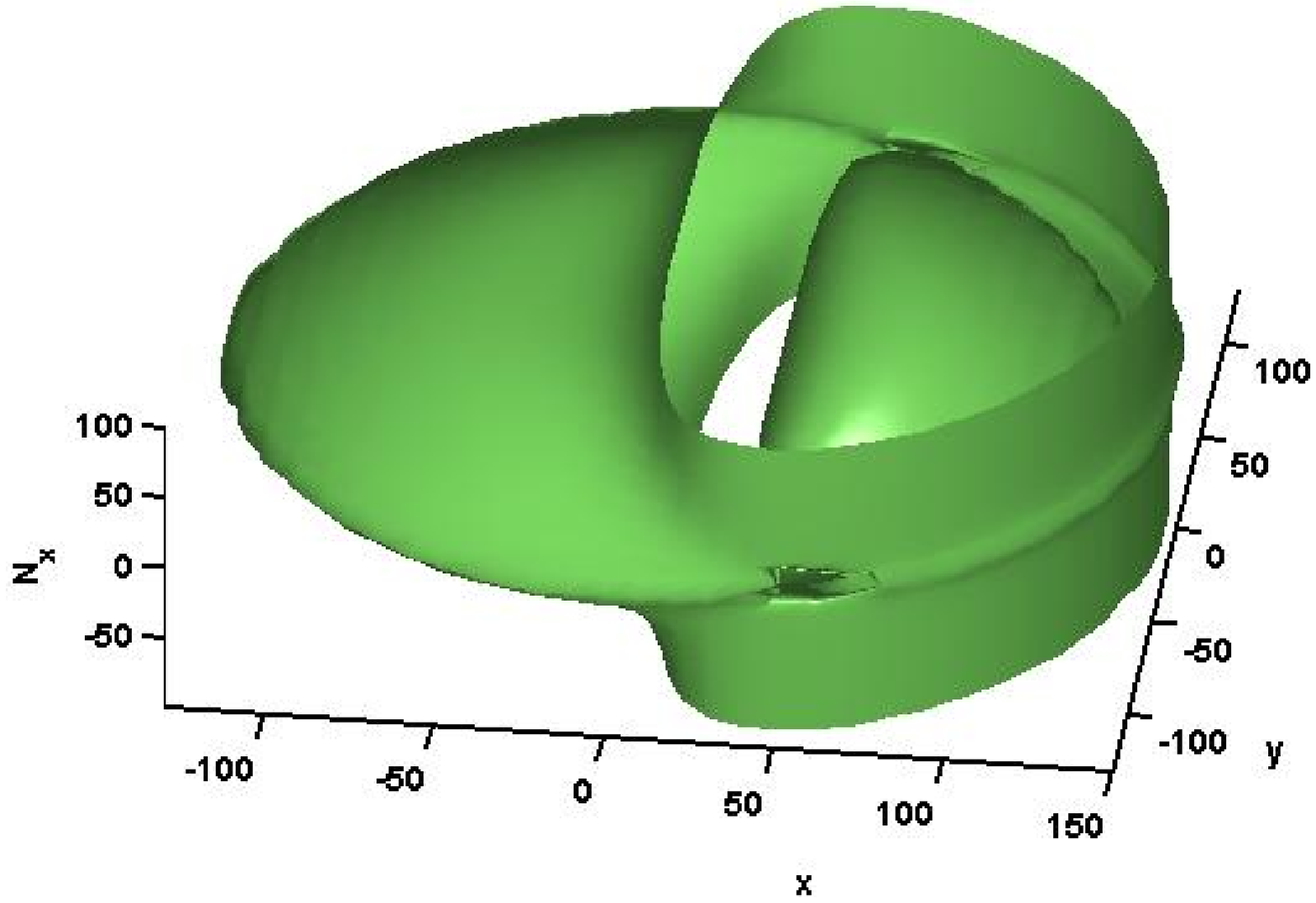}\qquad
\includegraphics[scale=0.35]{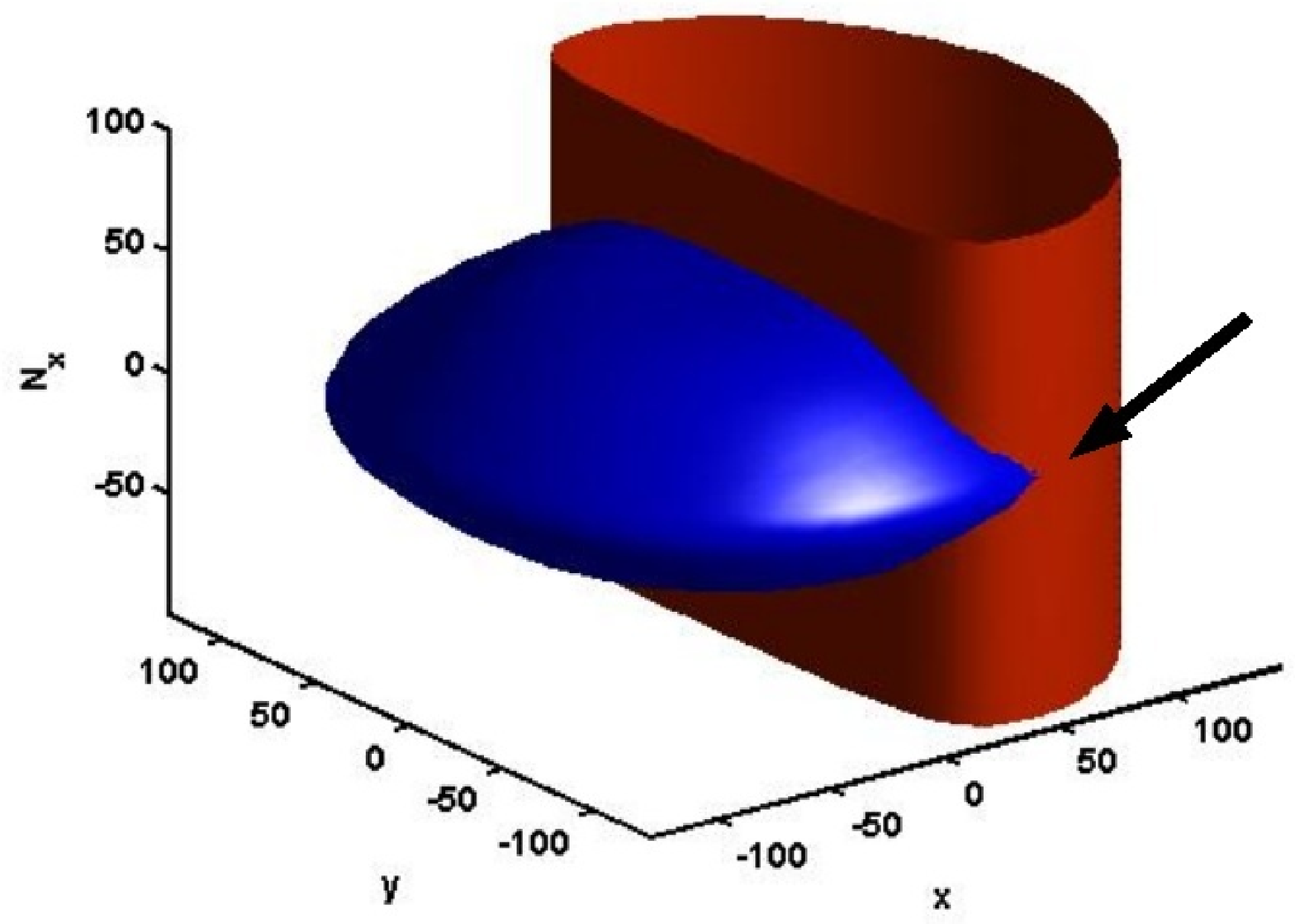}
\end{center}
\caption{\label{fig:normal_form}
A simple calculation which motivates the normal form approach.  The dispersion surface on the left is calculated using the determinant of the dispersion matrix, while the surfaces on the right are calculated using the diagonals of the matrix, in ``normal form''.  The coupling causes the surface on the left to reconnect in complicated ways, while mode conversion regions in the figure on the right can be quickly identified by the intersections of the two surfaces.  The arrow indicates a region where the intersection is non-transverse, and therefore not solvable using standard mode conversion theory.
}
\end{figure}

Figure \ref{fig:normal_form} shows an example which motivates this normal form approach.  In that figure is shown the dispersion surfaces for waves in a simple cold-plasma model of a tokamak \cite{citeulike:472573}.  This simple model can be put into something approximating normal form everywhere in phase space by a particular change of polarization basis.  The dispersion surface calculated from the determinant of ${\bf D}(x,k)$ has a complicated structure (left), which is a result of the ``avoided crossings'' near mode conversions.  However, the ``crossings'' are not avoided when the zeros of the diagonals are plotted, and the mode conversion regions can be identified by the intersection of the two surfaces.  Note that there are regions where the intersections are {\em not} transverse (indicated by the arrow in the figure on the right), and hence the usual mode conversion theory will not work.

\section{``Double'' symbol for vector wave problems}

There is a second approach to vector wave problems which is suggested by the theory of symbols as described in this work.  This goes back to the basic idea underlying the Zobin theory of symbols, which says that the symbol of an operator is a double Fourier transform defined on an appropriate non-commutative group.  The dispersion operator that we are working with has both a continuous part --- for example the usual symbols of each matrix entry are smooth functions on phase space --- and a discrete part ---  the matrix structure itself.  So according to the construction given in Chapter \ref{chp:GroupTheory}, we can compute the symbol of this operator-valued matrix if we have a group with an irreducible representation which is also an operator-valued matrix.  This suggests that we should form a sort of ``double'' symbol.  We can combine the continuous Heisenberg-Weyl group with the discrete Heisenberg-Weyl group (or any other group with finite matrix representations), and we will get the type of irreducible representations we need.  
\begin{align} 
\Gamma(\Lambda(\widehat{\mathfrak H \otimes \mathfrak{H}_n}),dP)
\xrightarrow{\mathcal{F}_{\mathfrak H \otimes \mathfrak{H}_n}^{-1}}
L_2(\mathfrak H \otimes \mathfrak{H}_n &, dg\,dN) \notag \\
&\stackrel{\text{as a set}}{=} \notag \\
L_2(\mathbb R^{2n+1} &\otimes \mathbb Z_n , dg\,dN) 
\xrightarrow{\mathcal{F}_{\mathbb R^{2n+1} \otimes \mathbb Z_n}}
L_2(\widehat{\mathbb R^{2n+1} \otimes \mathbb Z_n}, dP)
\end{align}
The double symbol of $\hat{\bf D}$ calculated using this approach will be a function on phase space (corresponding to the symbol of the differential part of $\hat{\bf D}$), with an additional discrete parameter (corresponding to the discrete symbol of the matrix part of $\hat{\bf D}$):
\begin{align}
\hat{\bf D} \rightarrow D(x,k; j).
\end{align}
This can be thought of as a function on several copies of phase space, each of which is labeled by $j$.  This new, double, symbol of the matrix operator $\hat{\bf D}$ has an interesting mathematical structure, and its use in vector wave problems is an area of future research.

\section{Applications of the phase space picture: Entropy, Mixed States, and Averaged Wigner Functions}

The problem of turbulent fluctuations in plasmas presents significant theoretical and experimental challenges.  We describe here how the idea of the mixed state from quantum mechanics could possibly be used to model the uncertainty introduced by turbulence in a plasma.  The Wigner function for the mixed state can look much more classical because of decoherence.  Similarly, we might expect that the Wigner function for a pair of interacting wave modes would become more localized to their respective dispersion surfaces when the uncertainty introduced by turbulence is taken into account.

\subsection{Mode Conversion Model}
As in Chapter \ref{chp:coupled_osc}, we use a simple model of two coupled oscillators to describe the mode conversion processes;  
\begin{eqnarray}
\ddot x_1 + \omega_1^2(t) \, x_1 +\eta \, (x_2-x_1) &=&0 \\
\ddot x_2 + \omega_2^2(t) \, x_2 +\eta \, (x_1-x_2) &=&0 .
\end{eqnarray}
When the time dependent natural frequencies of the two oscillators are nearly equal, they can exchange energy.  The dispersion functions for the uncoupled modes are given by
\begin{align}
D_\alpha(t,\omega) \equiv \omega^2 -\omega_\alpha^2(t) , \quad \alpha=\{1,2\}
\end{align}
and the dispersion curves for the uncoupled modes are defined as the solutions to the equations $D_\alpha = 0$.  Here we assume these curves cross in the time-frequency plane at the point $(t_0, \omega_0)$.  This is the mode conversion point, and the amount of conversion will depend on how long the oscillators will stay in resonance.  This will depend on how fast the dispersion function for one mode changes when following rays of the other mode; expressed using the Possion bracket this becomes $\mathcal{B} \equiv \vert\{D_1,D_2\}\vert_{(t_0,\omega_0)}$.  Geometrically, this quantity is related to the angle between the two dispersion curves where they cross at the mode conversion point.

\begin{figure}
\begin{center}
\includegraphics[scale=0.9]{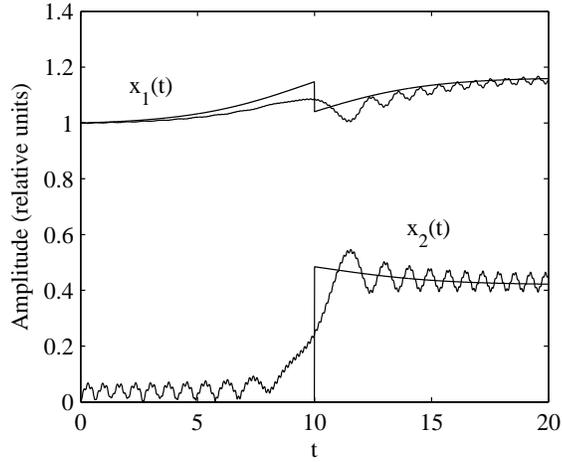}
\end{center}
\caption{\label{cap:coeffs}
Absolute value of $x_1$ and $x_2$, from numerical solution, superimposed on the WKB amplitude.  Note that the jump in the WKB amplitude at the mode conversion point predicts the amount of mode conversion well.
}
\end{figure}

As has been demonstrated using a variety of techniques \cite{metaplectic_formulation,Kaufman:PhysLettA1993,littlejohn:149,PhysRevLett.70.1799,doi:10.1063/1.2098213,0741-3335-49-1-004,tracy:082102} and discussed in Chapter \ref{chp:coupled_osc} of this dissertation, the mode conversion can be described as a scattering process, with incoming and outgoing wave amplitudes and phases.  The transmission and conversion coefficients are 
\begin{align}
\tau = \exp\left(-\pi \vert \eta\vert^2/\mathcal{B} \right) 
\end{align}
\begin{align}
\beta = \frac{\sqrt{ 2\pi\tau \mathcal{B}} } {\eta \Gamma(-i\vert \eta\vert^2/\mathcal{B} ) } 
\end{align}
These predictions agree well with numerical solutions, as seen in Figure \ref{cap:coeffs}.

\subsection{Mixed State Wigner Function}
In the density matrix formulation of quantum mechanics, a mixed state is one which is written as
\begin{align}
\rho = \sum_j \mathrm{p}_j \vert \psi_j \rangle \langle \psi_j \vert
\end{align}
where the probabilities $0 \leq \mathrm{p}_j \leq 1$ sum to one, but are otherwise unconstrained.  The Wigner function for this state is simply the weighted sum of the Wigner functions for the constituent states.
\begin{align}
W_j(q,p)=\int dq' \; e^{\text{i}pq' } \psi_j^*(q+\frac{q'}{2}) \psi_j(q-\frac{q'}{2})
\end{align}
\begin{align}
W(q,p) = \sum_j \mathrm{p}_j  W_j(q,p)
\end{align}
This construction of a ``mixed state'' Wigner function can be thought of as a marginalization over some parameters in the system which are not well known.

\subsection{Coupled Oscillator Wigner Function}

\begin{figure}
\begin{center}
\includegraphics[scale=0.7]{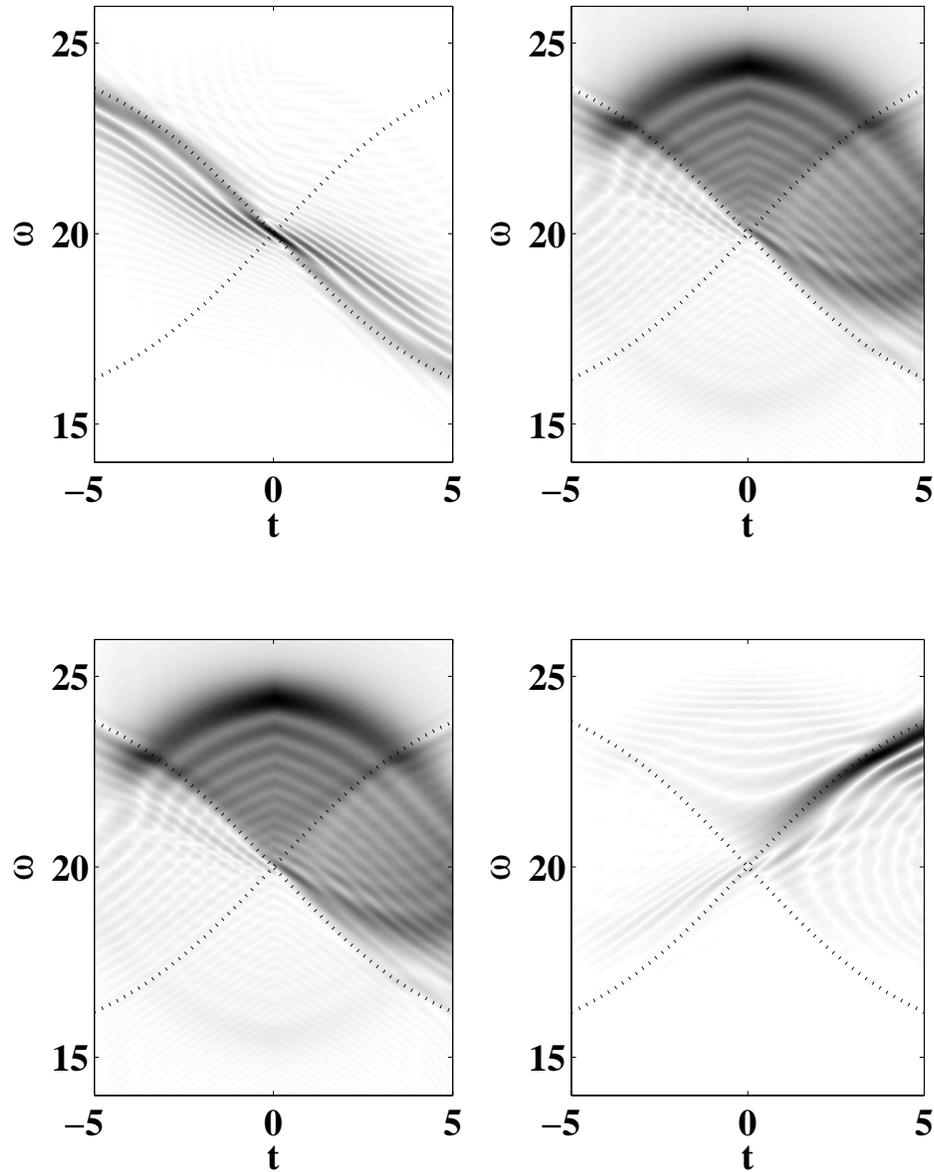}
\end{center}
\caption{\label{cap:pre}
Absolute value of the Wigner matrix and the dispersion curves for the coupled oscillator model of mode conversion in a plasma.  In order to isolate the positive frequencies, the numerical solution plotted here was calculated using a complex solution.  The Wigner function for a real-valued solution would also have support for negative frequencies, as well as an interference pattern near zero frequency.
}
\end{figure}

We can calculate the Wigner function for the coupled oscillator model by using the time-frequency version of the Wigner function.  Also, since the states in the model are vectors, then the Wigner function becomes a matrix, with greek indices labeling the modes.
\begin{align}
W_{\alpha \beta}(t,\omega)=\int d\tau \; e^{\text{i}\omega \tau} x_\alpha^*(t+\frac{\tau}{2}) x_\beta(t-\frac{\tau}{2})
\end{align}
The four components of this Wigner matrix are shown in Figure \ref{cap:pre}, along with the dispersion curves.  Note the complicated interference patterns which extend well away from the dispersion curves.  This is due to the fact that the Wigner function is non-local in time, and there is phase coherence for long times.  Physically we expect turbulence to destroy this phase coherence.

\subsection{Turbulent Uncertainties}

\begin{figure}
\begin{center}
\includegraphics[scale=0.7]{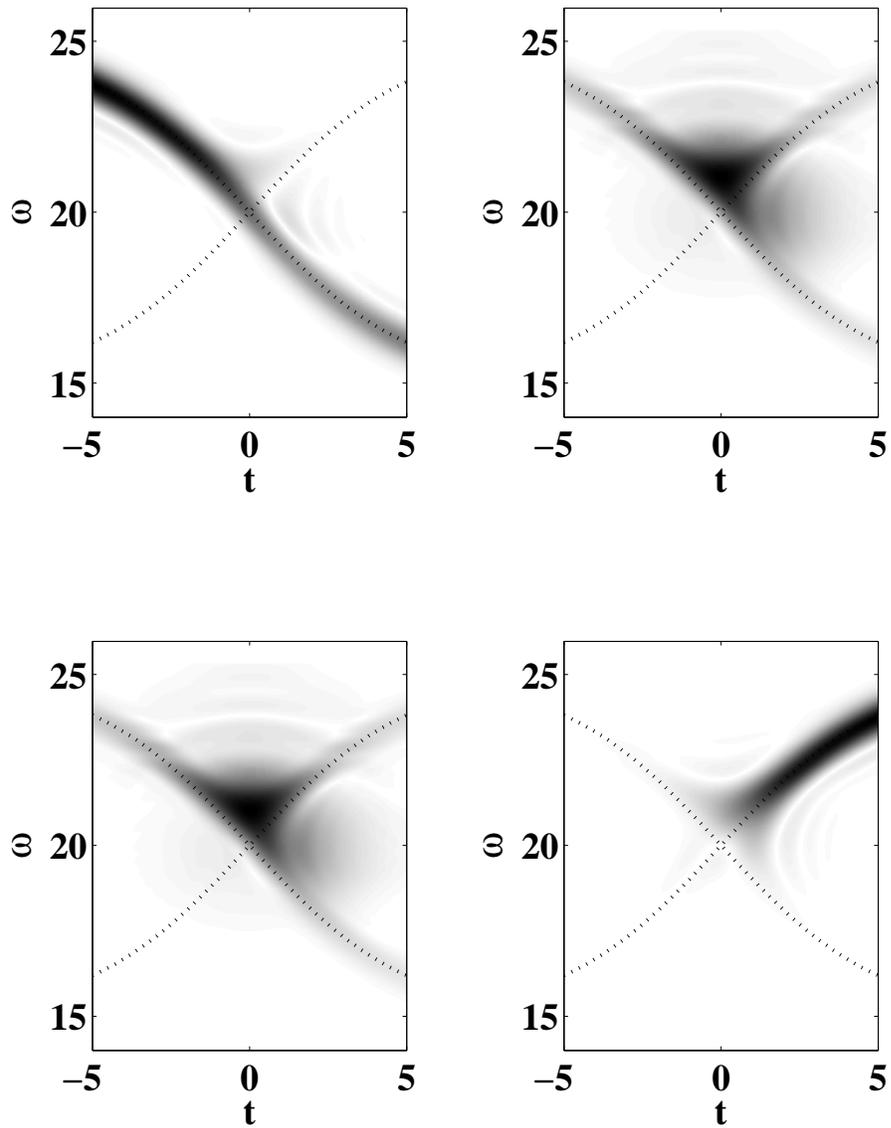}
\end{center}
\caption{\label{cap:mix}
Absolute value of the Wigner matrix for the mixed state.  This shows some broadening, but the function is now much better confined to the dispersion curves.  The interference patterns seen in Figure (\ref{cap:pre}) have mostly been averaged out.
}
\end{figure}

Here we model the effect of turbulence as an uncertainty in the parameters of the model.  Specifically, we take the mode conversion point from a Gaussian distribution centered at the original $(t_0, \omega_0)$.  This uncertainty is introduced into the model by modifying the parameters of the dispersion functions so that the dispersion curves rigidly shift from their original location.  Hence, the elements of the Wigner matrix also rigidly shift, and the resulting ensemble of functions can simply be summed with the appropriate Gaussian weights.  The resulting Wigner matrix will be a ``mixed state'' function, and shows much better confinement to the dispersion curves.  The averaging procedure effectively washes out the fine scale structure in the Wigner function, resulting in a more ``classical'' distribution on phase space (see Figure \ref{cap:mix}).  There is a natural physical connection between such turbulence effects and the path integral approach described in this dissertation.  This is an area for future investigation.

This calculation illustrates the potential that the mixed state Wigner function has for analyzing mode conversion in plasmas with turbulent fluctuations.  There are several important aspects which remain to be considered, however.  First, the physical interpretation of the mixed state calculation should be clarified in this context.  Here the analogy with semi-classical quantum systems will probably be fruitful.  Second, there is the issue of finding an equation of motion for the mixed state Wigner function.  In many quantum problems this is not too hard, since the Hamiltonian is the same for each state in the ensemble.  In our case, however, each of the ``states'' is a solution to a slightly different equation.  Do these somehow average into a ``master equation'' for the mixed state Wigner function?

%% file: Chapter-Conclusion.tex
%%%%%%%%%%%%%%%%%%%%%%%%%%%%%%%%%%%%%%%%%%%%%%%%%%%%%%%%%%%%%%%%%%%%%%%%%%
%
% Ph.D. dissertation manuscript
% Chapter 9: Conclusion
%
% Andrew Stephen Richardson (Fall 2007)
% College of William and Mary
% Department of Physics
% Prof. Eugene Tracy, advisor
%
% Based on Paul King and Andrew Norman's template (modified by Wirawan Purwanto)
%
%%%%%%%%%%%%%%%%%%%%%%%%%%%%%%%%%%%%%%%%%%%%%%%%%%%%%%%%%%%%%%%%%%%%%%%%%%

\chapter{Summary and Conclusion\label{chp:Conclusion}}

\section{Summary of Part I}

In the first part of this dissertation, we described the phase space theory of mode conversion.  Chapter \ref{chp:PhaseSpaceIntro} gives an introduction to the phase space point of view for solving generic wave equations.  The WKB method for construction of approximate solutions is reviewed, and connected to phase-space ray-tracing algorithms.  Then, in Chapter \ref{chp:coupled_osc}, the example of two coupled oscillators is given.  If the natural frequencies of the oscillators are time dependant, and cross at some time, then this problem can be recast into a form which is mathematically very similar to a mode conversion problem.  The complete description of the solution of this coupled oscillator problem provides a pedagogical introduction to the phase space techniques used in the theory of mode conversion.

In Chapter \ref{chp:higher_order}, we apply these tools to a standard avoided crossing mode conversion.  Usually, the solution of such a problem involves the linearization of the dispersion function about the mode conversion point.  A local solution is then constructed so that incoming and outgoing WKB solutions can be asymptotically matched, allowing the problem to be treated as a sort of ``ray splitting''.  We analyze the effects of the next order terms, and show how to construct a local solution which takes these quadratic terms into account.  By including the effects of the quadratic order terms, the region in which the matching can be performed has been enlarged substantially.

\section{Summary of Part II}

The phase space theory described in Part I of this dissertation makes extensive use of the theory of symbols of operators.  In Part II, the mathematical foundations of the theory of symbols is described.  Because these mathematical foundations rely heavily on the theory of representations of groups, we first give a review of group theory in Chapter \ref{chp:GroupTheory}.  The example of the Heisenberg-Weyl group is examined in detail, showing how the relationship between phase space and configuration space arises from the reduction of the regular representation of this group.  

Then, in Chapter \ref{chp:Symbols}, we describe how the symbol of an operator can be calculated by a double Fourier transform.  The operator is first embedded into a section of the dual bundle, which is like an operator-valued ``function'' on the set of irreducible representations of a non-commutative group.  The non-commutative Fourier transform is then applied to convert this section into a function on the group, where the group is considered as a set.  Then, using a commutative group structure on this same set, we perform another Fourier transform.  The result is an ordinary complex-valued function, which is the Zobin symbol of the operator we started with.

Using this definition of the symbol, as developed by Zobin, we proceeded to calculate the symbol of a function of an operator in Chapter \ref{chp:PathIntegral}.  In particular, we considered the exponential of an operator, defined using a power series.  The symbol of the $N^{th}$ power of an operator was computed by using the $N^{th}$ star product of the symbol of the operator.  In the limit of large $N$, this repeated application of the star product can be written as a path integral, where the paths live in the dual to the commutative group.  This general theory for path integrals was then illustrated by explicitly calculating the repeated star product for the discrete Heisenberg-Weyl group.  Since there are irreducible matrix representations of this group, the calculation presented can be used to calculate functions of a matrix.  In particular, the exponential of a matrix leads to a discrete ``path integral'', which by grouping similar paths can be written in terms of a multiplicity function.  This led to the consideration of the connections between path integrals and statistical mechanics.  In addition, the multiplicity function gives rise to a probability distribution, or measure, on the space of all possible paths.  This leads to an interpretation of the path integral as a Fourier transform on the space of measures, which for the continuous Heisenberg-Weyl group becomes an infinite-dimensional Fourier transform.  Considering the path integral for the Heisenberg-Weyl group also shows which aspect of group theory underlies the connection between the phase-space path integral and the configuration-space path integral.  Specifically, the reduction of the regular representation to the primary representations leads to consideration of functions on phase space.  This leads to the phase-space path integral.  Further reduction of the primary representations to irreducible representations involves functions on configuration space.  This reduces the phase-space path integral to the configuration-space path integral.

The new group-theoretical approach to path integrals developed in Chapter \ref{chp:PathIntegral} has many potential applications.  In Chapter \ref{chp:ModeConversion} we outline several ways which this new point of view could be applied to mode conversion theory.  This chapter points out several avenues of current and future research.  

We first see how consideration of the star product formulation of the path integral may lead to using the diagonals of the dispersion matrix as ray hamiltonians for constructing WKB solutions to vector wave problems.  This requires the definition of a new ``normal form'' for the dispersion matrix, where the symbols of the diagonal elements Poisson commute with the symbols of the off-diagonal elements.  In addition to simplifying ray-tracing algorithms, this could also help provide physical insight for vector wave problems with non-standard mode conversion geometries.  

A second avenue of research based on our new group theory perspective involves calculating a ``double'' symbol for vector wave problems.  The wave operator for vector wave problems can be written as a matrix of (pseudo)differential operators.  The ordinary symbol of each element of the matrix can be calculated, giving the dispersion matrix as a function of phase space.  However, as we saw in Chapter \ref{chp:Symbols}, it is possible to calculate the ``symbol'' of a matrix.  So we can calculate the discrete ``symbol'' of the dispersion matrix at each point in phase space.  This gives a new ``double symbol'' of the wave operator, which is a function of several discrete variables in addition to being a function of the phase space variables.

As a final potential topic of further research, we discuss in Chapter \ref{chp:ModeConversion} the possibility of using an averaged Wigner function to model the effects of turbulent plasma fluctuations on mode conversion.  This approach is based on the connection between the Wigner function and the density matrix in quantum mechanics.  In quantum mechanics, mixed state density matrices are used to model the decoherence of a quantum state due to interaction with the environment.  This decoherence makes the Wigner function for a mixed state look like a more classical probability distribution on phase space.  We suggest that a ``mixed state'' Wigner function could be used to describe mode-converting waves in a turbulent plasma.  Preliminary calculations suggest that this would make the Wigner function appear more ``classical'', with amplitude confined to regions near the dispersion curves for the various wave modes.  

In conclusion, this dissertation explores the depth and richness of the phase space perspective.  Traditional asymptotics can refine solutions to mode conversion problems, as was done with the higher order corrections in Chapter \ref{chp:higher_order}.
Additionally, because of the group theoretical foundations provided by the Zobin theory of symbols, many new areas of research have been opened, with a wide range of potential applications.

%% file: Appendix-Path-Integral.tex
%%%%%%%%%%%%%%%%%%%%%%%%%%%%%%%%%%%%%%%%%%%%%%%%%%%%%%%%%%%%%%%%%%%%%%%%%%
%
% Ph.D. dissertation manuscript
% Appendix - Path Integral Derivation
%
% Andrew Stephen Richardson (Fall 2007)
% College of William and Mary
% Department of Physics
% Prof. Eugene Tracy, advisor
%
% Based on Paul King and Andrew Norman's template (modified by Wirawan Purwanto)
%
%%%%%%%%%%%%%%%%%%%%%%%%%%%%%%%%%%%%%%%%%%%%%%%%%%%%%%%%%%%%%%%%%%%%%%%%%%

\chapter{Derivation of the Path Integral Formula\label{app:path_int}}

The symbol of the exponential of an operator can be calculated from the symbol of an operator by making use of the properties of the Fourier transform for groups.  The resulting formula is in the form of a path integral.  In this appendix we give the details of the calculation leading up to the path integral.  The resulting formula is defined for large $N$, and in the limit $N \rightarrow \infty$, this formula becomes a path integral (if a particular ordering is chosen).  This calculation is due to Zobin, and we thank him for allowing us to reproduce it here.

\section{Definition of a Symbol}

We start with the definition of a symbol in terms of the Fourier transform for groups. 
\begin{align}
\tau \in \widehat G_0 \qquad\qquad
g \in G_0 = G \qquad\qquad
\pi \in \widehat G
\end{align}
\begin{align} \begin{CD}
L_2(\widehat G_0,d\hat\mu)@
<\mathcal{F}_{G_0} << L_2(G_0,d\mu)=L_2(G,d\mu) @
> \mathcal{F}_G >> L_2(\widehat G,dP)
\end{CD} \end{align}
\begin{align}
\text{A symbol: } & S(\tau) \in L_2(\widehat G_0,d\hat\mu) \\
\text{A section: } & s(\pi)=(\mathcal{F}_G \mathcal{F}_{G_0}^{-1} S(\tau))(\pi) \equiv Q S  \in L_2(\widehat G,dP)
\end{align}

\section{Symbol of the exponential of a section}

We are interested in functions of operators.  In particular, we want the exponential of an operator.  $Q^{-1} e^{its} (\tau)= \text{ ?}$  We will make use of the limit formula for the exponential, $e^{its} \approx (1+ \frac{its}{N})^N$.

\begin{align}
Q^{-1}\left(1+ \frac{its}{N}\right)^N \left(\tau\right) 
&= \mathcal{F}_{G_0} \left( \mathcal{F}_G^{-1} \left(1+ \frac{its}{N}\right)^N\right) \left(\tau\right) \\
&= \mathcal{F}_{G_0} \left( \mathcal{F}_G^{-1} \left(1+ \frac{its}{N}\right)\right)^{\star_G N} \left(\tau\right) \\
&= \mathcal{F}_{G_0} \left(\delta_e+ \frac{it\mathcal{F}_G^{-1}s}{N}\right)^{\star_G N} \left(\tau\right)\\
&=\left(\mathcal{F}_{G_0} \int_{G^N} \delta_g\left(\mathop{\diamond}_{j=N}^1 g_j \right) \prod_{j=1}^N \left(\delta_e\left(g_j\right)+\frac{it \left(\mathcal{F}_G^{-1} s\right)\left(g_j\right)}{N} \right) \prod_{j=1}^N dg_j\right)(\tau)\\
&=\int \left(\mathcal{F}_{G_0}  \delta_g\left(\mathop{\diamond}_{N}^1 g_j\right) \right)(\tau) \prod_{j=1}^N \left(\delta_e\left(g_j\right)+\frac{it \left(\mathcal{F}_G^{-1} s\right)\left(g_j\right)}{N} \right) \prod_{j=1}^N dg_j\\
&=\int \left(\int \tau(g) \delta_g\left(\mathop{\diamond}_{N}^1 g_j\right) dg \right) \prod_{j=1}^N \left(\delta_e\left(g_j\right)+\frac{it \left(\mathcal{F}_G^{-1} s\right)\left(g_j\right)}{N} \right) \prod_{j=1}^N dg_j\\
&=\int\limits_{G^N=G_0^N} \!\!\!\!\!\tau(\mathop{\diamond}_{N}^1 g_j) \prod_{j=1}^N \left(\delta_e\left(g_j\right)+\frac{it \left(\mathcal{F}_G^{-1} s\right)\left(g_j\right)}{N} \right) \prod_{j=1}^N dg_j
\end{align}
At this point, we invoke the Plancharel theorem in $G_0^N$ in order to change the integral from one over the group to one over its dual $\widehat G_0^N$.
\begin{align}
&=\int\limits_{\widehat G_0^N}\! \left(\mathcal{F}_{G^N}\tau(g_N \ldots g_1) \right) (\tau_1,\ldots,\tau_N) \left(\mathcal{F}_{G}\prod_{j=1}^N \left(\delta_e\left(g_j\right)+\frac{it \left(\mathcal{F}_G^{-1} s\right)\left(g_j\right)}{N} \right) \right)(\tau_1,\ldots,\tau_N) \prod_{j=1}^N d\tau_j\\
&=\int\limits_{\widehat G_0^N}\! \Theta_N (\tau ;\tau_1,\ldots,\tau_N) \prod_{j=1}^N \left(1+\frac{itS(\tau_j)}{N} \right) \prod_{j=1}^N d\tau_j
\end{align}
At this point, we apply the formula for the exponential again, to get that 
\begin{align}
\prod_{j=1}^N \left(1+\frac{itS(\tau_j)}{N} \right) \approx 
\exp \left( \frac{it}{N} \sum_{j=1}^N S(\tau_j)  \right).
\end{align}
In the limit $N \rightarrow \infty$, the sum in the exponential can be interpreted as an integral.

\section{Calculation of $\Theta_N (\tau ;\tau_1,\ldots,\tau_N)$}
By definition, we have
\begin{align}
\Theta_N (\tau ;\tau_1,\ldots,\tau_N)&=
\left(\mathcal{F}_{G^N}\tau(g_N \ldots g_1) \right) (\tau_1,\ldots,\tau_N) \\
&= \int\limits_{G_0^N} \tau_1(g_1^{-1})\ldots\tau_N(g_N^{-1}) \tau(g_N \ldots g_1) \prod_{j=1}^N dg_j .
\end{align}
Now make a change of variables:
\begin{align}
h_k \equiv g_k \diamond g_{k-1} \cdots \diamond g_1 \implies\quad g_k = h_k\diamond h_{k-1}^{-1} \text{\quad and \quad} g_k^{-1} = h_{k-1}\diamond h_k^{-1}
\end{align}
Putting this back into the equation above gives
\begin{align}
\Theta_N (\tau ;\tau_1,\ldots,\tau_N)&=
\int \tau(h_N) \prod_{k=1}^N \tau_k(h_{k-1} \diamond h_k^{-1}) \prod_{k=1}^N dh_k \\
&= \int \tau(h_N) \prod_{k=1}^N \tau_k\left(h_{k-1} - h_k - \frac{1}{2}\omega(h_{k-1},h_k) \right) \prod_{k=1}^N dh_k.
\end{align}
(Note: the $\frac{1}{2}\omega(h_{k-1},h_k)$ term comes from the group product law, and depends on the choice of group.  For example, if we are computing with the discrete HW group, then we do not have the factor of $\frac{1}{2}$.)
We now introduce some new notation so that this integration becomes easier to perform.  We will write the $\tau$'s in exponential form, and we will also separate the $2n$-dim component from the 1-dim component in both the $\tau$'s and the $h$'s.
\begin{align}
\tau_k &= e^{i \langle z_k, \cdot \rangle} \\
z_k &= (\hat z_k, \tilde z_k)\\
h_k &= (\hat h_k, \tilde h_k)
\end{align}
Here, the $\hat\ $ refers to the $2n$-dim component and the $\tilde\ $ refers to the 1-dim component.
\begin{align}
\Theta_N &(\tau ;\tau_1,\ldots,\tau_N)  \notag \\
&=
\int \tau(h_N) 
\exp\left[ i\sum_{k=1}^N \langle  z_k, h_{k-1} - h_k \rangle - \frac{i}{2}\sum_{k=1}^N \tilde z_k \omega(\hat h_{k-1}, \hat h_k) \right]
\prod_{k=1}^N dh_k\\
&=
\int \tau(h_N) 
\exp\left[ i\sum_{k=1}^N \langle  z_{k+1}- z_k, h_k \rangle - \frac{i}{2}\sum_{k=1}^N \tilde z_k \omega(\hat h_{k-1}, \hat h_k) \right]
\prod_{k=1}^N dh_k\\
&=
\int \hat \tau(\hat h_N) \tilde \tau(\tilde h_N)  \notag \\
&\times\exp\left[ i\sum_{k=1}^N \langle \hat z_{k+1}-\hat z_k, \hat h_k \rangle  
+(\tilde  z_{k+1}-\tilde  z_k)\tilde h_k 
- \frac{i}{2}\sum_{k=1}^N \tilde z_k \omega(\hat h_{k-1}, \hat h_k) \right]
\prod_{k=1}^N d\hat h_k \, d\tilde h_k
\end{align}
Now evaluate the integrals over $\tilde h_k$.
\begin{align}
\int  &\tilde \tau(\tilde h_N) 
\exp\left[ i\sum_{k=1}^N (\tilde  z_{k+1}-\tilde  z_k)\tilde h_k  \right]
\prod_{k=1}^N d\tilde h_k \\
&=\int  \exp\left[ i\sum_{k=1}^{N-1} (\tilde  z_{k+1}-\tilde  z_k)\tilde h_k  \right]
\prod_{k=1}^{N-1} d\tilde h_k \int \exp\left[ -i \tilde z_N \tilde h_N +i \tilde z \tilde h_N \right] d\tilde h_n \\
&= \delta( \tilde z -\tilde z_N ) \int \exp\left[ i\sum_{k=1}^{N-1} (\tilde  z_{k+1}-\tilde  z_k)\tilde h_k  \right] \prod_{k=1}^{N-1}  d\tilde h_k  \\
&= \prod_{k=1}^{N}\delta( \tilde z -\tilde z_k ) 
\end{align}
Now insert this back into the equation for $\Theta_N$.
\begin{align}
\Theta_N& (\tau ;\tau_1,\ldots,\tau_N)  \notag \\
&=\prod_{k=1}^{N}\delta( \tilde z -\tilde z_k ) 
\int \hat \tau(\hat h_N)  
\exp\left[ i\sum_{k=1}^N \langle \hat z_{k+1}-\hat z_k, \hat h_k \rangle  
- \frac{i \tilde z}{2}\sum_{k=1}^N \omega(\hat h_{k-1}, \hat h_k) \right]
\prod_{k=1}^N d\hat h_k  \\
&=\prod_{k=1}^{N}\delta( \tilde z -\tilde z_k ) \notag \\
&\quad\times \int 
\exp\left[- i\left(\sum_{k=1}^N \langle \hat z_{k}-\hat z_{k+1}, \hat h_k \rangle  -\langle\hat z,\hat h_N\rangle \right) 
+ \frac{i \tilde z}{2}\sum_{k=1}^N \omega(\hat h_k, \hat h_{k-1}) \right]
\prod_{k=1}^N d\hat h_k  \\
&=\left[ \mathcal{F}_{G_0^N} \exp \left(  \frac{i \tilde z}{2}\sum_{k=2}^{N+1} \omega(\hat h_k, \hat h_{k-1}) \right)\right]
\left( \hat z_1-\hat z_2,\hat z_2-\hat z_3,\ldots,\hat z_{N-1}-\hat z_N,\hat z_N-\hat z \right)
\end{align}
This integral is in the form of a Fourier transform of a multidimensional Gaussian.  This can be seen by organizing the variables into vectors:
\begin{align}
\vec  z = \left (
\begin{matrix}
\hat  z_1 \\
\vdots \\
\hat  z_N
\end{matrix}
\right)
\qquad
\vec  z^s = \left (
\begin{matrix}
\hat  z_2 \\
\vdots \\
\hat  z_N \\
\hat  z
\end{matrix}
\right)
\qquad
\vec h = \left (
\begin{matrix}
\hat h_1 \\
\vdots \\
\hat h_N
\end{matrix}
\right)
\end{align}
(The $\hat  z$ in $\vec  z^s$ takes care of the $\tau(\hat h_N)$ term.)  We will also need the matrix given by the two-form $\omega$:
\begin{align}
W\left(\vec h, \vec l \;\right) = -\frac{i\tilde  z}{4} \sum \omega(\hat h_{k-1}, \hat l_k) - \frac{i\tilde  z}{4} \sum \omega(\hat l_{k-1}, \hat h_k).
\end{align}
In matrix form, this looks like
\begin{align}
W = \left (
\begin{matrix}
0 & i\omega &  &  &  \\
- i\omega & 0 & \ddots & 0 &  \\
 & \ddots & 0 & \ddots &  \\
& 0 & \ddots & 0 & i\omega \\
& & & -i\omega & 0  \\
\end{matrix}
\right).
\end{align}
The equation for $\Theta_N$ becomes
\begin{align}
\Theta_N (\tau ;\tau_1,\ldots,\tau_N)&=
\int  
e^{ i \langle  \vec  z^s -\vec  z, \vec h \rangle }
e^{ \langle W \vec h, \vec h \rangle }
 d\vec h  \\
&= \frac{ \exp{\left( \langle W^{-1} (\vec  z^s -\vec  z), \vec  z^s -\vec  z\rangle \right)} }{\sqrt{\det (W)}} 
=\pm \, \exp{\left(- \langle W (\vec  z^s -\vec  z), \vec  z^s -\vec  z\rangle\right)}.
\end{align}

\section{Putting it back together}
Putting everything back together we get a formula for the symbol of the exponential of a section:
\begin{align}
Q^{-1}\left(1+ \frac{its}{N}\right)^N \left(\tau\right) 
&=\pm\int\limits_{\widehat G_0^N}\! 
 e^{- \langle W (\vec  z^s -\vec  z), \vec  z^s -\vec  z\rangle}
\exp \left( \frac{it}{N} \sum_{j=1}^N S(\tau_j)  \right) \prod_{j=1}^N d\tau_j \\
&=\pm\int\limits_{\widehat G_0^N}\! 
 \exp \left(- \frac{i\tilde z}{2} \sum_{k=2}^{N+1} \omega(\hat z_{k+1}-\hat z_{k}, \hat z_{k} - \hat z_{k-1}) 
+ \frac{it}{N} \sum_{j=1}^N S(\tau_j)  \right) \prod_{j=1}^N d\tau_j .
\end{align}
Notice that the expression in the exponential is a function of the $\tau_j$'s and $\tau$, which is written in terms of the matrix $W$ as defined in the last section.

%% file: Vita.tex
%%%%%%%%%%%%%%%%%%%%%%%%%%%%%%%%%%%%%%%%%%%%%%%%%%%%%%%%%%%%%%%%%%%%%%%%%%
%
% Ph.D. dissertation manuscript
% Vita
%
% Andrew Stephen Richardson (Fall 2007)
% College of William and Mary
% Department of Physics
% Prof. Eugene Tracy, advisor
%
% Based on Paul King and Andrew Norman's template (modified by Wirawan Purwanto)
%
%%%%%%%%%%%%%%%%%%%%%%%%%%%%%%%%%%%%%%%%%%%%%%%%%%%%%%%%%%%%%%%%%%%%%%%%%%

\vitaauthor{Andrew Stephen Richardson}

\begin{thesisauthorvita}

Andrew Stephen Richardson was born on August 20, 1980, in Alexandria, Virginia.  He was educated at home in Virginia from kindergarten through high school.  In 1998 he entered George Mason University in Fairfax, Virginia, to study physics.  He graduated with honors in May 2002, and in Fall 2002, he entered the physics graduate program at the College of William and Mary in Williamsburg, Virginia.
In 2003 he joined Professor Tracy's research group, studying nonlinear dynamics.  He then moved to studying mathematical physics and theoretical plasma physics with Professor Tracy, and on November 30, 2007, he successfully defended this dissertation.  

\end{thesisauthorvita}